\documentclass{article}

\usepackage{longtable}

\usepackage{multirow} 

\usepackage[normalem]{ulem}
\usepackage{amsmath}
\usepackage{amssymb}
\usepackage{bm}
\usepackage[usenames]{color}
\usepackage{epubtk}
\usepackage[pdftex]{graphicx}
\usepackage{booktabs}

\DeclareMathOperator{\arccot}{arccot} 
\newcommand{\comment}[1]{}

\newcommand{\bra}[1]{\langle #1|}
\newcommand{\ket}[1]{|#1\rangle}
\newcommand{\braket}[2]{\langle #1|#2 \rangle}
\newcommand{\mean}[1]{\langle #1\rangle}
\newcommand{\tmean}[1]{\overline{#1}}

\renewcommand{\in}{\mathrm{in}} 
\newcommand{\out}{\mathrm{out}} 
\newcommand{\vac}{\mathrm{vac}} 
\newcommand{\sqz}{\mathrm{sqz}} 

\newcommand{\intinfty}{\int_{-\infty}^{\infty}\!}
\newcommand{\intOinfty}{\displaystyle\int_{0}^{\infty}\!}

\newcommand{\vb}[1]{\pmb{#1}}

\newcommand{\smatrix}[4]{\begin{bmatrix}#1 & #2\\#3 & #4\end{bmatrix}}
\newcommand{\partdiff}[2]{\dfrac{\partial#1}{\partial#2}}
\newcommand{\hc}{\mathrm{h.c.}}
\newcommand{\svector}[2]{\begin{bmatrix}#1\\#2\end{bmatrix}}
\newcommand{\svup}{\begin{bmatrix}1\\0\end{bmatrix}}
\newcommand{\svdown}{\begin{bmatrix}0\\1\end{bmatrix}}
\renewcommand{\Re}{\mathop{\mathrm{Re}}\nolimits}
\renewcommand{\Im}{\mathop{\mathrm{Im}}\nolimits}

\begin{document}

\title{Quantum Measurement Theory in Gravitational-Wave Detectors}

\author{%
\epubtkAuthorData{Stefan L.\ Danilishin}{%
School of Physics, University of Western Australia,\\
35 Stirling Hwy, Crawley 6009, Australia\\
and\\
Faculty of Physics, Moscow State University\\
Moscow 119991, Russia}{%
shtefan.danilishin@uwa.edu.au}{%
}%
\and
\epubtkAuthorData{Farid Ya.\ Khalili}{%
Faculty of Physics, Moscow State University\\
Moscow 119991, Russia}{%
khalili@phys.msu.ru}{%
}
}

\date{}
\maketitle

\begin{abstract}
The fast progress in improving the sensitivity of the
gravitational-wave detectors, we all have witnessed in the recent
years, has propelled the scientific community to the point at which
quantum behavior of such immense measurement devices as
kilometer-long interferometers starts to matter. The time when their
sensitivity will be mainly limited by the quantum noise of light is
around the corner, and finding ways to reduce it will
become a necessity. Therefore, the primary goal we pursued in this
review was to familiarize a broad spectrum of readers with the theory
of quantum measurements in the very form it finds application in the
area of gravitational-wave detection.
We focus on how quantum noise arises in gravitational-wave
interferometers and what limitations it imposes on the achievable
sensitivity. We start from the very basic concepts and gradually
advance to the general linear quantum measurement theory and its
application to the calculation of quantum noise in the contemporary
and planned interferometric detectors of gravitational radiation of
the first and second generation. Special attention is paid to the
concept of the Standard Quantum Limit and the methods of its surmounting.
\end{abstract}

\epubtkKeywords{gravitational-wave detectors, quantum measurement
  theory, quantum noise, standard quantum limit, quantum
  non-demolition measurement, optical rigidity, quantum speedmeter,
  squeezed light, filter cavities, back-action evasion}

\newpage

\section{Introduction}
\label{section:introduction}

The more-than-ten-years-long history of the large-scale laser gravitation-wave (GW) detectors (the first one, TAMA~\cite{TAMAsite} started to operate in 1999, and the most powerful pair, the two detectors of the LIGO project~\cite{LIGOsite}, in 2001, not to forget about the two European members of the international interferometric GW detectors network, also having a pretty long history, namely, the German-British interferometer GEO\,600 \cite{GEOsite} located near Hannover, Germany, and the joint European large-scale detector Virgo \cite{VIRGOsite}, operating near Pisa, Italy) can be considered both as a great success and a complete failure, depending on the point of view. On the one hand, virtually all technical requirements for these detectors have been met, and the planned sensitivity levels have been achieved. On the other hand, no GWs have been detected thus far.

The possibility of this result had been envisaged by the community, and during the same last ten years, plans for the second-generation detectors were developed~\cite{Thorne2000, Fritschel2002, Acernese2006-2, Willke2006, AdvLIGOsite, LCGTsite}. Currently (2012), both LIGO detectors are shut down, and their upgrade to the Advanced LIGO, which should take about three years, is underway. The goal of this upgrade is to increase the detectors' sensitivity by about one order of magnitude~\cite{Smith2009}, and therefore the rate of the detectable events by three orders of magnitude, from some `half per year' (by the optimistic astrophysical predictions) of the second generation detectors to, probably, hundreds per year.

This goal will be achieved, mostly, by means of quantitative improvements (higher optical power, heavier mirrors, better seismic isolation, lower loss, both optical and mechanical) and evolutionary changes of the interferometer configurations, most notably, by introduction of the signal recycling mirror. As a result, the second-generation detectors will be \textit{quantum noise limited}. At higher GW frequencies, the main sensitivity limitation will be due to phase fluctuations of light inside the interferometer (shot noise). At lower frequencies, the random force created by the amplitude fluctuations (radiation-pressure noise) will be the main or among the major contributors to the sum noise.

It is important that these noise sources both have the same quantum origin, stemming from the fundamental quantum uncertainties of the electromagnetic field, and thus that they obey the Heisenberg uncertainty principle and can not be reduced simultaneously~\cite{81a1Ca}. In particular, the shot noise can (and will, in the second generation detectors) be reduced by means of the optical power increase. However, as a result, the radiation-pressure noise will increase. In the `naively' designed measurement schemes, built on the basis of a Michelson interferometer, kin to the first and the second generation GW detectors, but with sensitivity chiefly limited by quantum noise, the best strategy for reaching a maximal sensitivity at a given spectral frequency would be to make these noise source contributions (at this frequency) in the total noise budget equal. The corresponding sensitivity point is known as the Standard Quantum Limit (SQL)~\cite{Sov.Phys.JETP_26.831_1968, 92BookBrKh}.

This limitation is by no means an absolute one, and can be evaded using more sophisticated measurement schemes. Starting from the first pioneering works oriented on solid-state GW detectors~\cite{77a1eBrKhVo, 78a1eBrKhVo, Thorne_PRL_40_667_1978}, many methods of overcoming the SQL were proposed, including the ones suitable for practical implementation in laser-interferometer GW detectors. The primary goal of this review is to give a comprehensive introduction of these methods, as well as into the underlying theory of \emph{linear quantum measurements}, such that it remains comprehensible to a broad audience.

The paper is organized as follows. In
Section~\ref{sec:Interferometry}, we give a classical (that is,
non-quantum) treatment of the problem, with the goal to familiarize
the reader with the main components of laser GW
detectors. In Section~\ref{sec:quantum_light} we provide the necessary
basics of quantum optics. In
Section~\ref{sec:linear_quantum_measurement} we demonstrate the main
principles of linear quantum measurement theory, using simplified
toy examples of the quantum optical position meters. In
Section~\ref{sec:QN_in_GW_interferometers}, we provide the full-scale
quantum treatment of the standard Fabry--P{\'e}rot--Michelson topology
of the modern optical GW detectors. At last, in
Section~\ref{sec:sub-SQL_schemes}, we consider three methods of
overcoming the SQL, which are viewed now as the most probable
candidates for implementation in future laser GW
detectors. Concluding remarks are presented in
Section~\ref{sec:Conclusion}. Throughout the review we use the
notations and conventions presented in Table~\ref{table:notations}
below.

\ifpdf
  \renewcommand{\arraystretch}{1.3}
  \begin{longtable}{l p{9cm}}
\else
  \begin{table}[htbp]
\fi
\caption[Notations and conventions]{Notations and conventions, used in
  this review, given in alphabetical order for both, greek (first) and
  latin (after greek) symbols.}
\label{table:notations}
\ifpdf\\\else
  \begin{tabular}{l p{9cm}}
\fi
\toprule
\textbf{Notation and value} & \textbf{Comments} \\
 \midrule
\ifpdf
  \endfirsthead
  \multicolumn{2}{c}{\small\textbf{\tablename~\thetable{}} -- \emph{Continued}}
  \\[4mm]
  \toprule
  \textbf{Notation and value} & \textbf{Comments} \\ 
  \midrule
  \endhead
\fi
$\ket{\alpha}$ & coherent state of light with dimensionless complex amplitude $\alpha$\\
$\beta=\arctan\gamma/\delta$ & normalized detuning \\
  $\gamma$  &  interferometer half-bandwidth \\
  $\varGamma=\sqrt{\gamma^2+\delta^2}$ & effective bandwidth \\
  $\delta=\omega_p-\omega_0$ & optical pump detuning from the cavity resonance frequency $\omega_0$ \\
  $\epsilon_d=\sqrt{\dfrac{1}{\eta_d}-1}$ & excess quantum noise due to optical losses in the detector readout system with quantum efficiency $\eta_d$ \\
  $\zeta = t - x/c$ & space-time-dependent argument of the field strength of a light wave, propagating in the positive direction of the $x$-axis \\
  $\eta_d$    & quantum efficiency of the readout system (e.g., of a photodetector)\\
  $\theta$  & squeeze angle \\
  $\vartheta,\,\varepsilon$  &  some short time interval \\
  $\lambda$ & optical wave length \\
  $\mu$     & reduced mass \\
  $\nu=\Omega-\Omega_0$ & mechanical detuning from the resonance frequency \\
  $\xi=\sqrt{\dfrac{S}{S_{\mathrm{SQL}}}}$ & SQL beating factor \\
  $\rho$     & signal-to-noise ratio \\
  $\tau=L/c$     & miscellaneous time intervals; in particular, $L/c$ \\
  $\phi_{\mathrm{LO}}$  & homodyne angle \\
  $\varphi=\phi_{\mathrm{LO}}-\beta$ & \\
  $\chi_{AB}(t,t') = \frac{i}{\hbar}[\hat{A}(t),\,\hat{B}(t')]$ & general linear time-domain susceptibility \\
  $\chi_{xx}$     & probe body mechanical succeptibility \\
  $\omega$  &  optical band frequencies \\
  $\omega_0$ &  interferometer resonance frequency \\
  $\omega_p$ &  optical pumping frequency \\
  $\Omega$   &  mechanical band frequencies; typically, $\Omega=\omega-\omega_p$ \\
  $\Omega_0$ &  mechanical resonance frequency \\
  $\Omega_q=\sqrt{\dfrac{2S_\mathcal{FF}}{\hbar M}}$ & quantum noise
  ``corner frequency'' \\
  \midrule
  $A$ &  power absorption factor in Fabry--P{\'e}rot cavity per bounce \\
  $\hat{a}(\omega),\,\hat{a}^\dag(\omega)$ & annihilation and creation operators of photons with frequency $\omega$\\
  $\hat{a}_c(\Omega) = \dfrac{\hat{a}(\omega_0+\Omega)+\hat{a}^\dag(\omega_0-\Omega)}{\sqrt{2}}$ & two-photon amplitude quadrature operator\\
  $\hat{a}_s(\Omega) = \dfrac{\hat{a}(\omega_0+\Omega)-\hat{a}^\dag(\omega_0-\Omega)}{i\sqrt{2}}$ & two-photon phase quadrature operator\\
  $\mean{\hat{a}_i(\Omega)\circ\hat{a}_j(\Omega')}\equiv$ &
  \multirow{2}{9cm}{Symmetrised (cross) correlation of the field quadrature operators ($i,j=c,s$)} \\ 
  $\frac12\mean{\hat{a}_i(\Omega)\hat{a}_j(\Omega')+\hat{a}_j(\Omega')\hat{a}_i(\Omega)}$ & \\
  $\mathcal{A}$ &  light beam cross section area \\
  $c$ &  speed of light \\
  $\mathcal{C}_0=\sqrt{\dfrac{4\pi\hbar\omega_p}{\mathcal{A}c}}$ & light quantization normalization constant\\
  $\mathcal{D}=(\gamma-i\Omega)^2 + \delta^2$ & Resonance denominator of the optical cavity transfer function, defining its characteristic conjugate frequencies (``cavity poles'') \\
  $E$ &   electric field strength \\
  $\mathcal{E}$ & classical complex amplitude of the light\\
  $\mathcal{E}_{c} = \sqrt{2}\mathrm{Re}[\mathcal{E}],\,\mathcal{E}_{s} = \sqrt{2}\mathrm{Im}[\mathcal{E}]$ & classical quadrature amplitudes of the light \\
  $\vb{\mathcal{E}} = \begin{bmatrix}
                        \mathcal{E}_c\\
                        \mathcal{E}_s
                      \end{bmatrix}$ & vector of classical quadrature amplitudes\\
  $\hat{F}_{\mathrm{b.a.}}$   & back-action force of the meter \\
  $G$ &  signal force \\
  $h$ &  dimensionless GW signal (a.k.a.\ metrics variation) \\
  $H=\begin{bmatrix} \cos\phi_{\mathrm{LO}} \\ \sin\phi_{\mathrm{LO}} \end{bmatrix}$ & homodyne vector \\
  $\hat{\mathcal{H}}$ & Hamiltonian of a quantum system \\
  $\hbar$ &  Planck's constant \\
  $\mathbb{I}$ & identity matrix\\
  $\mathcal{I}$ &  optical power \\
  $\mathcal{I}_c$ &  circulating optical power in a cavity \\
  $\mathcal{I}_{\mathrm{arm}}$ &  circulating optical power per interferometer arm cavity\\
  $J=\dfrac{4\omega_0\mathcal{I}_c}{McL}$ & normalized circulating power \\
  $k_p=\omega_p/c$ & optical pumping wave number \\
  $K$ &  rigidity, including optical rigidity \\
  $\mathcal{K}=\dfrac{2J\gamma}{\Omega^2(\gamma^2+\Omega^2)}$ & Kimble's optomechanical coupling factor \\
  $\mathcal{K}_{\mathrm{SM}}=\dfrac{4J\gamma}{(\gamma^2+\Omega^2)^2}$ & optomechanical coupling factor of the Sagnac speedmeter \\
  $L$ &  cavity length \\
  $M$ &  probe-body mass \\
  $O$ &  general linear meter readout observable \\
  $\mathbb{P}[\alpha]=\begin{bmatrix}
  \cos\alpha & -\sin\alpha\\
  \sin\alpha & \cos\alpha
\end{bmatrix}$ &  matrix of counterclockwise rotation (pivoting) by angle $\alpha$ \\
  $r$ &  amplitude squeezing factor ($e^r$)\\
  $r_{\mathrm{dB}} = 20r\log_{10}e$ & power squeezing factor in decibels\\
  $R$ &  power reflectivity of a mirror \\
  $\mathbb{R}(\Omega)$ &  reflection matrix of the Fabry--P{\'e}rot cavity \\
  $S(\Omega)$ &  noise power spectral density (double-sided) \\
  $S_\mathcal{XX}(\Omega)$ &  measurement noise power spectral density (double-sided) \\
  $S_\mathcal{FF}(\Omega)$ &  back-action noise power spectral density (double-sided) \\
  $S_\mathcal{XF}(\Omega)$ &  cross-correlation power spectral density (double-sided) \\
  $\mathbb{S}_{\vac}(\Omega) = \frac12\mathbb{I}$ & vacuum quantum state power spectral density matrix\\
  $\mathbb{S}_{\sqz}(\Omega)$ & squeezed quantum state power spectral density matrix\\
  $\mathbb{S}_{\sqz}[r,\theta]=\mathbb{P}[\theta]
  \begin{bmatrix}
    e^r & 0\\
    0 & e^{-r}
  \end{bmatrix}\mathbb{P}[-\theta]$  & squeezing matrix \\
  $T$ &  power transmissivity of a mirror \\
  $\mathbb{T}$ &  transmissivity matrix of the Fabry--P{\'e}rot cavity \\
  $v$ & test-mass velocity \\
  $\mathcal{W}$ &  optical energy \\
  $W_{\ket{\psi}}(X,\,Y)$ &  Wigner function of the quantum state $\ket{\psi}$ \\
  $x$ &  test-mass position \\
  $\hat{X} = \frac{\hat{a}+\hat{a}^\dag}{\sqrt{2}}$ & dimensionless oscillator (mode) displacement operator\\
  $\hat{Y} = \frac{\hat{a}-\hat{a}^\dag}{i\sqrt{2}}$ & dimensionless oscillator (mode) momentum operator\\ 
  \bottomrule
\ifpdf
\end{longtable}
\else\end{tabular}
\end{table}
\fi

\newpage
\section{Interferometry for GW Detectors: Classical Theory}
\label{sec:Interferometry}

\subsection{Interferometer as a weak force probe}

In order to have a firm basis for understanding how quantum noise
influences the sensitivity of a GW detector it would be illuminating
to give a brief description of the interferometers as weak force/tiny
displacement meters. It is by no means our intention to give a
comprehensive survey of this ample field that is certainly worthy of a
good book, which there are in abundance, but rather to provide the
reader with the wherewithal for grasping the very principles of the GW
interferometers operation as well as of other similar ultrasensitive
optomechanical gauges. The reader interested in a more detailed
description of the interferometric techniques being used in the field
of GW detectors might enjoy reading this book~\cite{91Book_Blair} or
the comprehensive Living Reviews on the subject by Freise and
Strain~\cite{lrr-2010-1} and by Pitkin et al.~\cite{lrr-2011-5}.

\subsubsection{Light phase as indicator of a weak force}
\label{sec:phasemeter}

\epubtkImage{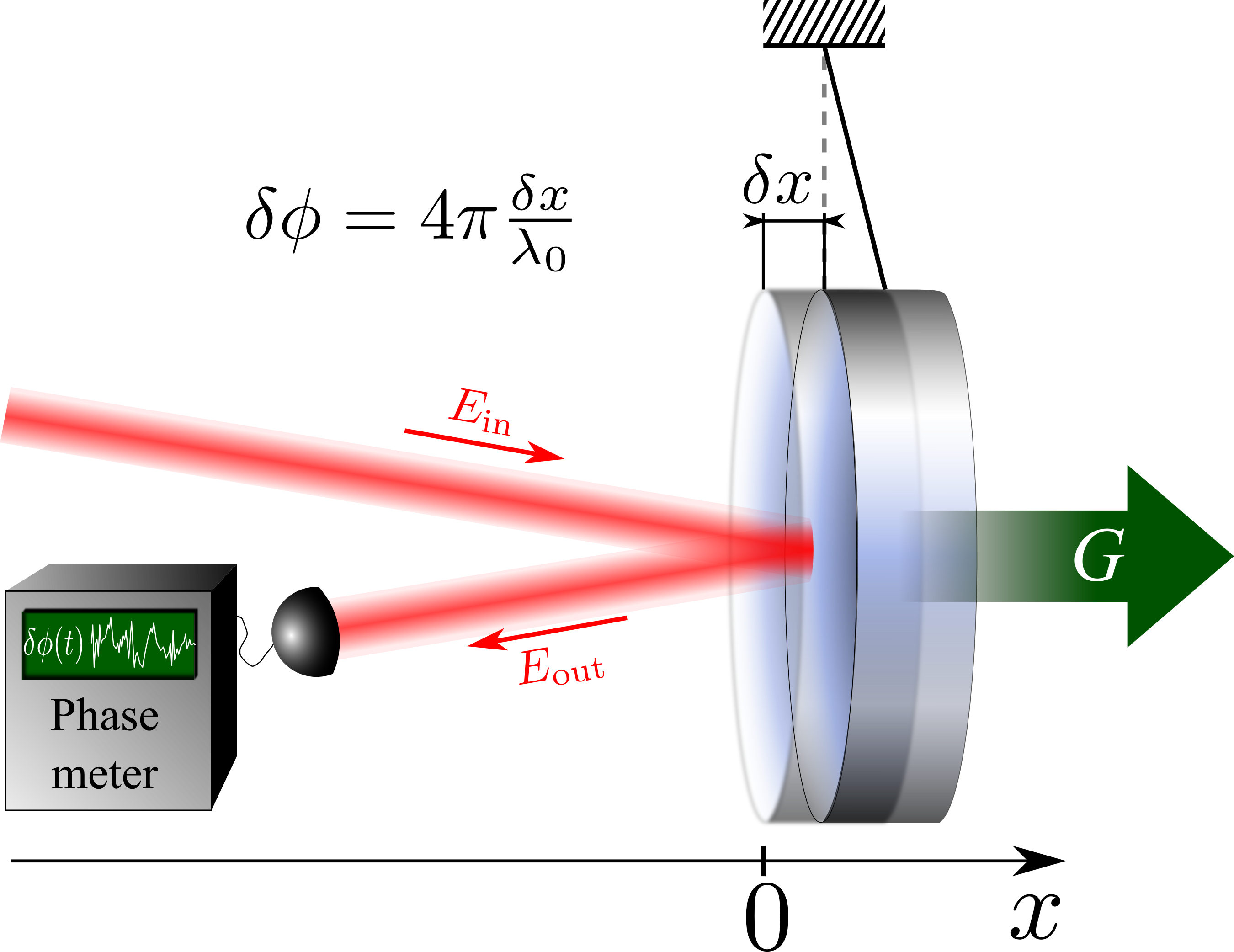}{%
\begin{figure}[htb]
\centerline{\includegraphics[width=.5\textwidth]{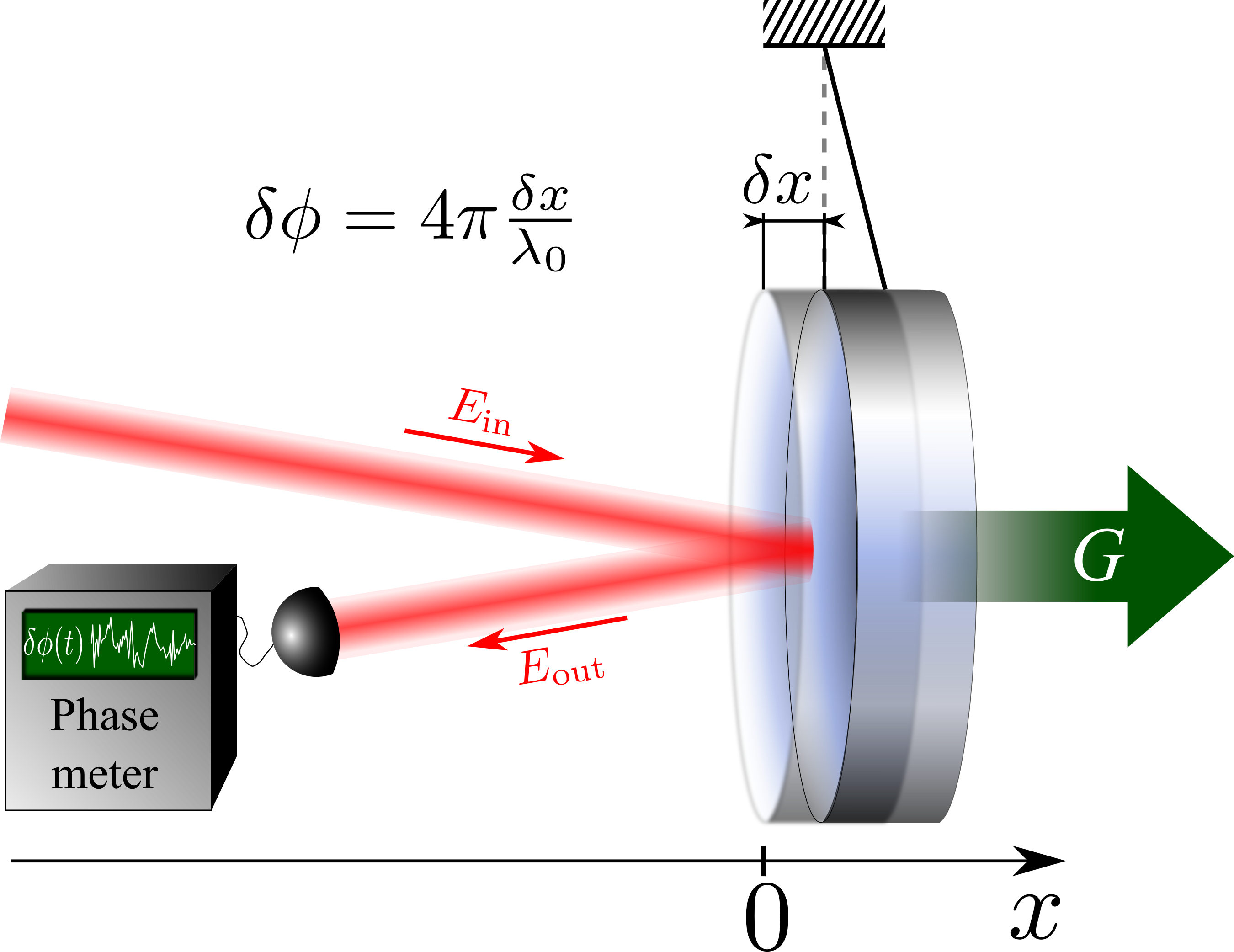}}
  \caption{Scheme of a simple weak force measurement: an external signal force $\boldsymbol{G}$ pulls the mirror from its equilibrium position $x=0$, causing displacement $\delta x$. The signal displacement is measured by monitoring the phase shift of the light beam, reflected from the mirror.}
\label{fig:phase_meas_force}
\end{figure}}

Let us, for the time being, imagine that we are capable of measuring an electromagnetic (e.g., light) wave-phase shift $\delta\phi$ with respect to some coherent reference of the same frequency. Having such a hypothetical tool, what would be the right way to use it, if one had a task to measure some tiny classical force? The simplest device one immediately conjures up is the one drawn in Figure~\ref{fig:phase_meas_force}. It consists of a movable totally-reflective mirror with mass $M$ and a coherent paraxial light beam, that impinges on the mirror and then gets reflected towards our hypothetical phase-sensitive device. The mirror acts as a \textit{probe} for an external force $G$ that one seeks to measure. The response of the mirror on the external force $G$ depends upon the details of its dynamics. For definiteness, let the mirror be a harmonic oscillator with mechanical eigenfrequency $\Omega_m = 2\pi f_m$. Then the mechanical equation of motion gives a connection between the mirror displacement $x$ and the external force $G$ in the very familiar form of the harmonic oscillator equation of motion:
\begin{equation}
\label{eq:oscillator_EOM}
    M\ddot x+M\Omega_m^2 x = G(t)\,,\ \Longrightarrow\ x(t) = x_0(t)+\int_0^t dt'\, \chi_{xx}(t-t')G(t'),
\end{equation}
where $x_0(t) = x(0)\cos\Omega_m t+p(0)/(M\Omega_m)\sin\Omega_mt$ is the free motion of the mirror defined by its initial displacement $x(0)$ and momentum $p(0)$ at $t=0$ and
\begin{equation}
\label{eq:oscillator_Green's_fcn}
    \chi_{xx}(t-t')=\frac{\sin\Omega_m(t-t')}{M\Omega_m}\,,\quad t\geqslant t'\,,
\end{equation}
is the oscillator Green's function. It is easy to see that the reflected light beam carries in its phase the information about the displacement $\delta x(t) = x(t)-x(0)$ induced by the external force $G$. Indeed, there is a phase shift between the incident and reflected beams, that matches the additional distance the light must propagate to the new position of the mirror and back, i.e.,
\begin{equation}
\label{eq:dphi_to_dx_relation}
    \delta\phi = \frac{2\omega_0\delta x}{c} = 4\pi\frac{\delta x}{\lambda_0}\,,
\end{equation}
with $\omega_0 = 2\pi c/\lambda_0$ the incident light frequency, $c$ the speed of light and $\lambda_0$ the light wavelength. Here we implicitly assume mirror displacement to be much smaller than the light wavelength.

Apparently, the information about the signal force $G(t)$ can be obtained from the measured phase shift by \textit{post-processing} of the measurement data record $\delta\tilde{\phi}(t)\propto \delta x(t)$ by substituting it into Eq.~\eqref{eq:oscillator_EOM} instead of $x$. Thus, the estimate of the signal force $\tilde{G}$ reads:
\begin{equation}
\label{eq:GW_to_dphi_relation}
    \tilde{G} = \frac{Mc}{2\omega_0}\left[\delta\ddot{\tilde{\phi}}+\Omega_m^2 \delta\tilde{\phi}\right]\,.
\end{equation}
This kind of post-processing pursues an evident goal of getting rid of any information about the eigenmotion of the test object while keeping only the signal-induced part of the total motion. The above time-domain expression can be further simplified by transforming it into a Fourier domain, since it does not depend anymore on the initial values of the mirror displacement $x(0)$ and momentum $p(0)$:
\begin{equation}
\label{eq:GW_to_dphi_spectral}
    \tilde{G}(\Omega) = \frac{Mc}{2\omega_0}\left[\Omega_m^2-\Omega^2\right]\delta\tilde{\phi}(\Omega)\,,
\end{equation}
where
\begin{equation}
\label{eq:Fourier_transform}
A(\Omega) = \int_{-\infty}^\infty dt\,A(t)e^{i\Omega t}
\end{equation}
denotes a Fourier transform of an arbitrary time-domain function $A(t)$. If the expected signal spectrum occupies a frequency range that is much higher than the mirror-oscillation frequency $\Omega_m$ as is the case for ground based interferometric GW detectors, the oscillator behaves as a free mass and the term proportional to $\Omega_m^2$ in the equation of motion can be omitted yielding:
\begin{equation}
\label{eq:GW_to_dphi_free_mass_spectral}
    \tilde{G}^{\mathrm{f.m.}}(\Omega) = -\frac{Mc\Omega^2}{2\omega_0}\delta\tilde{\phi}_\Omega\,.
\end{equation}

\subsubsection{Michelson interferometer}
\label{sec:MI}

Above, we assumed a direct light phase measurement with a hypothetical device in order to detect a weak external force, possibly created by a GW. However, in reality, direct phase measurement are not so easy to realize at optical frequencies. At the same time, physicists know well how to measure light intensity (amplitude) with very high precision using different kinds of photodetectors ranging from ancient--yet--die-hard reliable photographic plates to superconductive photodetectors capable of registering individual photons~\cite{goltsman:705}. How can one transform the signal, residing in the outgoing light phase, into amplitude or intensity variation? This question is rhetorical for physicists, for interference of light as well as the multitude of interferometers of various design and purpose have become common knowledge since a couple of centuries ago. Indeed, the amplitude of the superposition of two coherent waves depends on the relative phase of these two waves, thus transforming phase variation into the variation of the light amplitude.

\epubtkImage{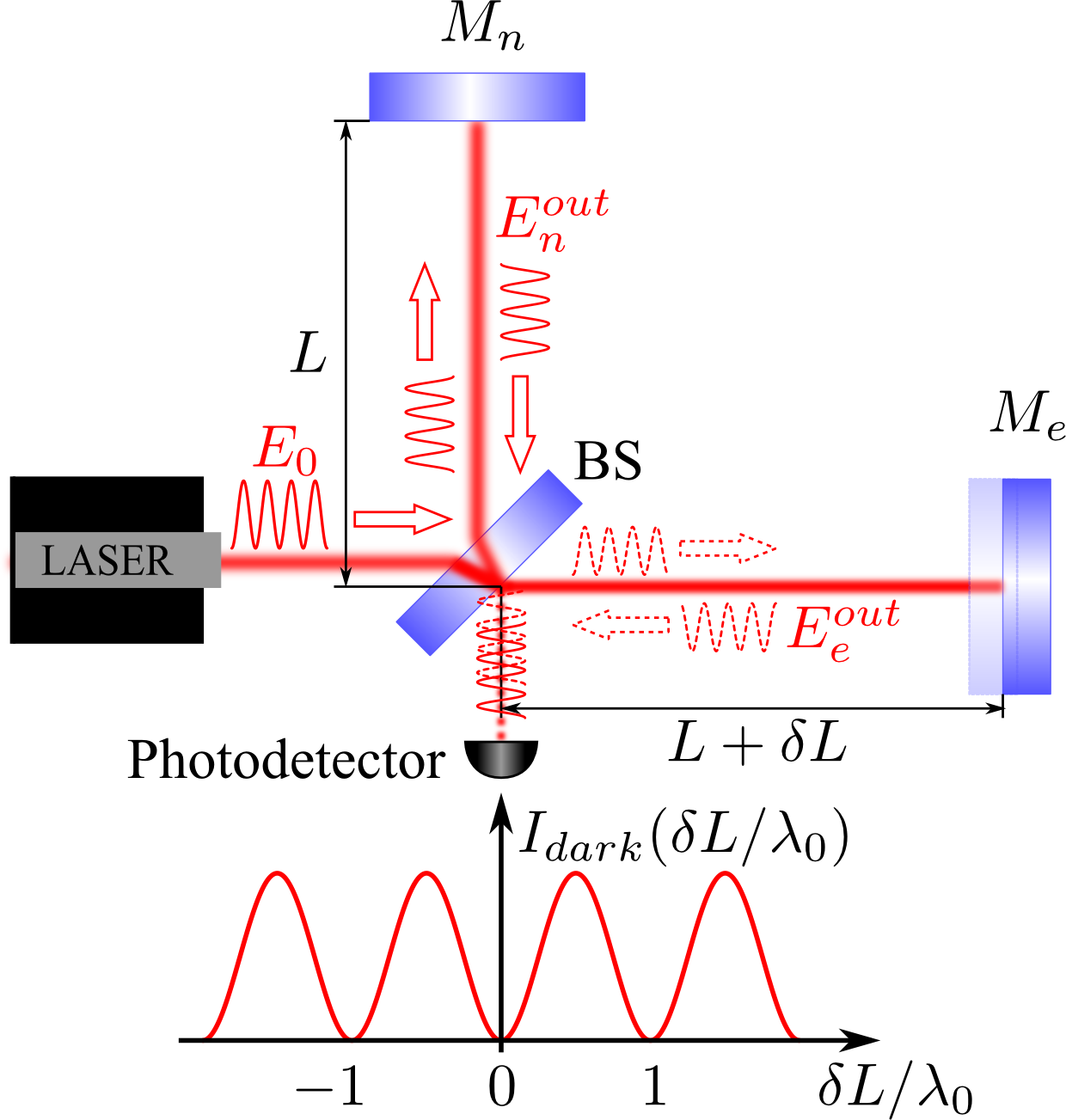}{%
\begin{figure}[htb]
\centerline{\includegraphics[width=.5\textwidth]{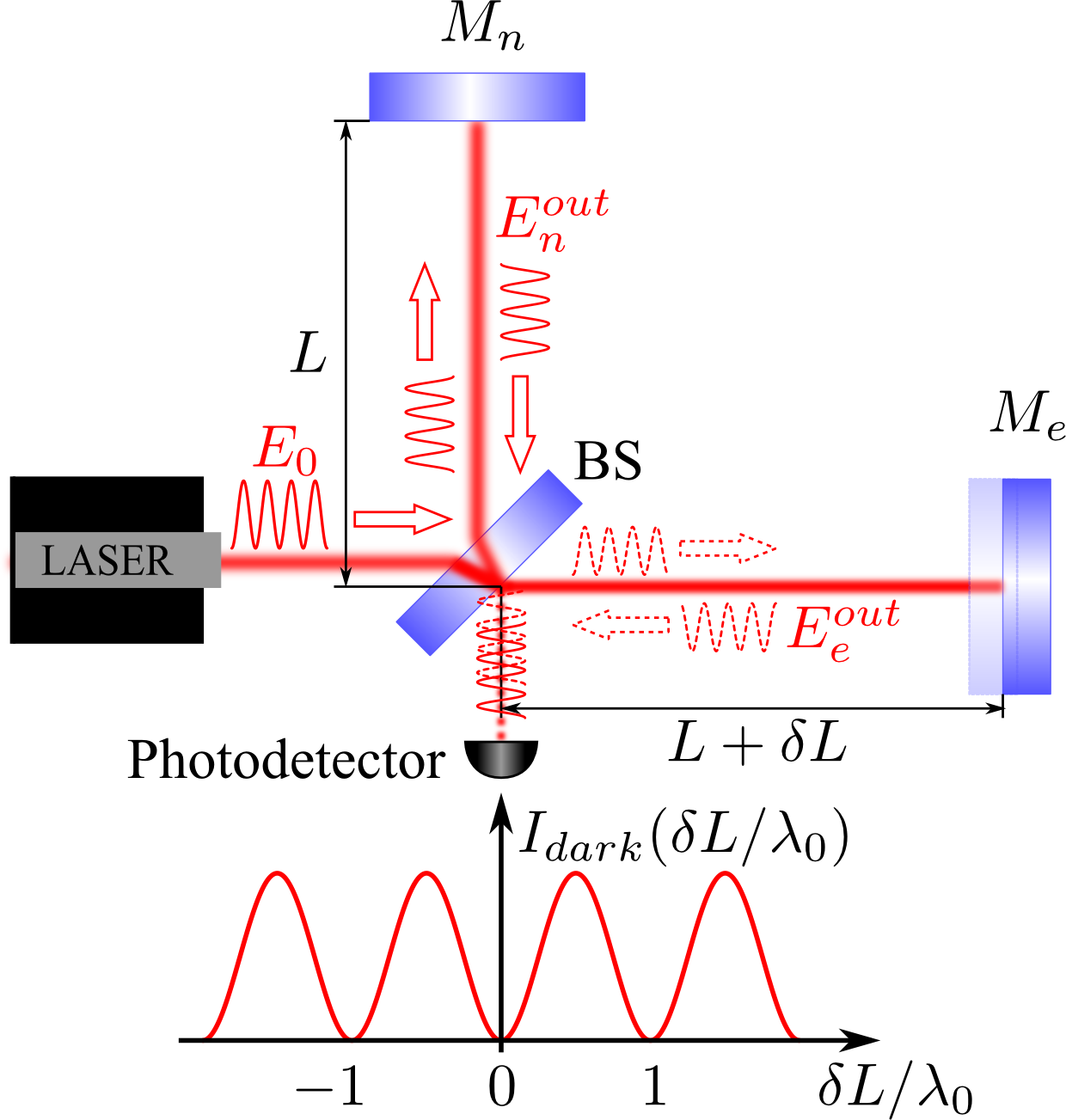}}
    \caption{Scheme of a Michelson interferometer. When the end mirrors of the interferometer arms $M_{n,e}$ are at rest the length of the arms $L$ is such that the light from the laser gets reflected back entirely (bright port), while at the dark port the reflected waves suffer destructive interference keeping it really dark. If, due to some reason, e.g., because of GWs, the lengths of the arms changed in such a way that their difference was $\delta L$, the photodetector at the dark port should measure light intensity $I_{\mathrm{dark}}(\delta L) = \frac{I_0}{2}(1-\cos4\pi\frac{\delta L}{\lambda})$.}
\label{fig:Michelson_interferometer}
\end{figure}}

For the detection of GWs, the most popular design is the Michelson interferometer~\cite{80BookeBoWo, 91Book_Blair,lrr-2010-1}, which schematic view is presented in Figure~\ref{fig:Michelson_interferometer}. Let us briefly discuss how it works. Here, the light wave from a laser source gets split by a semi-transparent mirror, called a \emph{beamsplitter}, into two waves with equal amplitudes, travelling towards two highly-reflective mirrors $M_{n,e}$\epubtkFootnote{Here, we adopt the system of labeling parts of the interferometer by the cardinal directions, they are located with respect to the interferometer central station, e.g., $M_n$ and $M_e$ in Figure~\ref{fig:Michelson_interferometer} stand for `northern' and `eastern' end mirrors, respectively.} to get reflected off them, and then recombine at the beamsplitter. The readout is performed by a photodetector, placed in the signal port. The interferometer is usually tuned in such a way as to operate at a \emph{dark fringe}, which means that by default the lengths of the arms are taken so that the optical paths for light, propagating back and forth in both arms, are equal to each other, and when they recombine at the signal port, they interfere destructively, leaving the photodetector unilluminated. On the opposite, the two waves coming back towards the laser, interfere constructively. The situation changes if the end mirrors get displaced by some external force in a differential manner, i.e., such that the difference of the arms lengths is non-zero: $\delta L = L_e-L_n \neq 0$. Let a laser send to the interferometer a monochromatic wave that, at the beamsplitter, can be written as
$$E_{\mathrm{laser}}(t) = E_0\cos(\omega_0 t)\,.$$
Hence, the waves reflected off the interferometer arms at the beamsplitter (before interacting with it for the second time) are\epubtkFootnote{Here and below we keep to a definition of the reflectivity coefficient of the mirrors that implies that the reflected wave acquires a phase shift equal to $\pi$ with respect to the incident wave if the latter impinged the reflective surface from the less optically dense medium (air or vacuum). In the opposite case, when the incident wave encounters reflective surface from inside the mirror, i.e., goes from the optically more dense medium (glass), it is assumed to acquire no phase shift upon reflection.}:
$$E^{\out}_{n,e}(t) = -\frac{E_0}{\sqrt2}\cos(\omega_0t-2\omega_0L_{n,e}/c)\,,$$
and after the beamsplitter:
\begin{eqnarray*}
E_{\mathrm{dark}}(t) &=& \dfrac{E_{n}^{\out}(t)-E_{e}^{\out}(t)}{\sqrt2} = E_0\sin\frac{\omega_0\delta L}{c}\sin\left(\omega_0t-\omega_0[L_n+L_e]/c\right)\,,\\
E_{\mathrm{bright}}(t) &=& \dfrac{E_{n}^{\out}(t)+E_{e}^{\out}(t)}{\sqrt2} = -E_0\cos\frac{\omega_0\delta L}{c}\cos\left(\omega_0t-\omega_0[L_n+L_e]/c\right)\,.
\end{eqnarray*}
And the intensity of the outgoing light in both ports can be found using a relation $\mathcal{I}\propto\overline{E^2}$ with overline meaning time-average over many oscillation periods:
\begin{equation}
\mathcal{I}_{\mathrm{dark}}(\delta L/\lambda_0) = \frac{\mathcal{I}_0}{2}\left(1-\cos4\pi\frac{\delta L}{\lambda_0}\right)\,,\quad\mbox{and}\quad \mathcal{I}_{\mathrm{bright}}(\delta L/\lambda_0) = \frac{\mathcal{I}_0}{2}\left(1+\cos4\pi\frac{\delta L}{\lambda_0}\right)\,.
\end{equation}
Apparently, for small differential displacements $\delta L\ll\lambda_0$, the Michelson interferometer tuned to operate at the dark fringe has a sensitivity to $\sim(\delta L/\lambda_0)^2$ that yields extremely weak light power on the photodetector and therefore very high levels of dark current noise. In practice, the interferometer, in the majority of cases, is slightly detuned from the dark fringe condition that can be viewed as an introduction of some constant small bias $\delta L_0$ between the arms lengths. By this simple trick experimentalists get linear response to the signal nonstationary displacement $\delta x(t)$:
\begin{eqnarray}
\label{eq:MI_out_intensity}
\mathcal{I}_{\mathrm{dark}}(\delta
x/\lambda_0) &=& \frac{\mathcal{I}_0}{2}\left(1-\cos4\pi\frac{\delta
  L_0+\delta x}{\lambda_0}\right)\simeq \nonumber\\ 8\pi^2
\mathcal{I}_0\frac{\delta L_0\delta x}{\lambda_0^2} +
\mathcal{O}\left(\frac{\delta x^2}{\lambda_0^2},\,\frac{\delta
  L_0^2}{\lambda_0^2}\right) &=& \mbox{const.} \times 4\pi\frac{\delta x}{\lambda_0}+ \mathcal{O}\left(\frac{\delta x^2}{\lambda_0^2},\,\frac{\delta L_0^2}{\lambda_0^2}\right)\,.
\end{eqnarray}
Comparison of this formula with Eq.~\eqref{eq:dphi_to_dx_relation}
should immediately conjure up the striking similarity between the
response of the Michelson interferometer and the single moving
mirror. The nonstationary phase difference of light beams in two
interferometer arms $\delta\phi(t) = 4\pi\delta x(t)/\lambda_0$ is
absolutely the same as in the case of a single moving mirror
(cf.\ Eq.~\eqref{eq:dphi_to_dx_relation}). It is no coincidence,
though, but a manifestation of the internal symmetry that all
Michelson-type interferometers possess with respect to coupling
between mechanical displacements of their arm mirrors and the optical
modes of the outgoing fields. In the next section \ref{sec:GW_interaction_with_IF}, we show how this symmetry displays itself in GW interferometers.

\subsubsection{Gravitational waves' interaction with interferometer}
\label{sec:GW_interaction_with_IF}

Let us see how a Michelson interferometer interacts with the GW. For this purpose we need to understand, on a very basic level, what a GW is. Following the poetic, yet precise, definition by Kip Thorne, `gravitational waves are \emph{ripples in the curvature of spacetime} that are emitted by violent astrophysical events, and that propagate out from their source with the speed of light'~\cite{04BookBlTh, 73BookeMiThWh}. A weak GW far away from its birthplace can be most easily understood from analyzing its action on the probe bodies motion in some region of spacetime. Usually, the deformation of a circular ring of free test particles is considered (see Chapter~26: Section~26.3.2 of~\cite{04BookBlTh} for more rigorous treatment) when a GW impinges it along the $z$-direction, perpendicular to the plane where the test particles are located. Each particle, having plane coordinates $(x,\,y)$ with respect to the center of the ring, undergoes displacement $\delta\boldsymbol{r}\equiv(\delta x,\,\delta y)$ from its position at rest, induced by GWs:
\begin{eqnarray}
\label{eq:GW_action_xy}
\delta x = \dfrac12 h_+x\,, & \delta y = -\dfrac12 h_+y\,,\\
\delta x = \dfrac12 h_\times y\,, & \delta y = \dfrac12 h_\times x\,.
\end{eqnarray}
Here, $h_+\equiv h_+(t-z/c)$ and $h_\times\equiv h_\times(t-z/c)$ stand for two independent polarizations of a GW that creates an acceleration field resulting in the above deformations. The above expressions comprise a solution to the equation of motion for free particles in the \emph{tidal acceleration field} created by a GW:
$$\delta\ddot{\boldsymbol{r}} = \dfrac12\left[(\ddot h_+ x+\ddot h_\times y)\boldsymbol{e}_x+(-\ddot h_+ y+\ddot h_\times x)\boldsymbol{e}_y\right]\,,$$
with $\boldsymbol{e}_{x}=\{1,0\}^{\mathsf{T}}$ and $\boldsymbol{e}_{y}=\{0,1\}^{\mathsf{T}}$ the unit vectors pointing in the $x$ and $y$ direction, respectively.

\epubtkImage{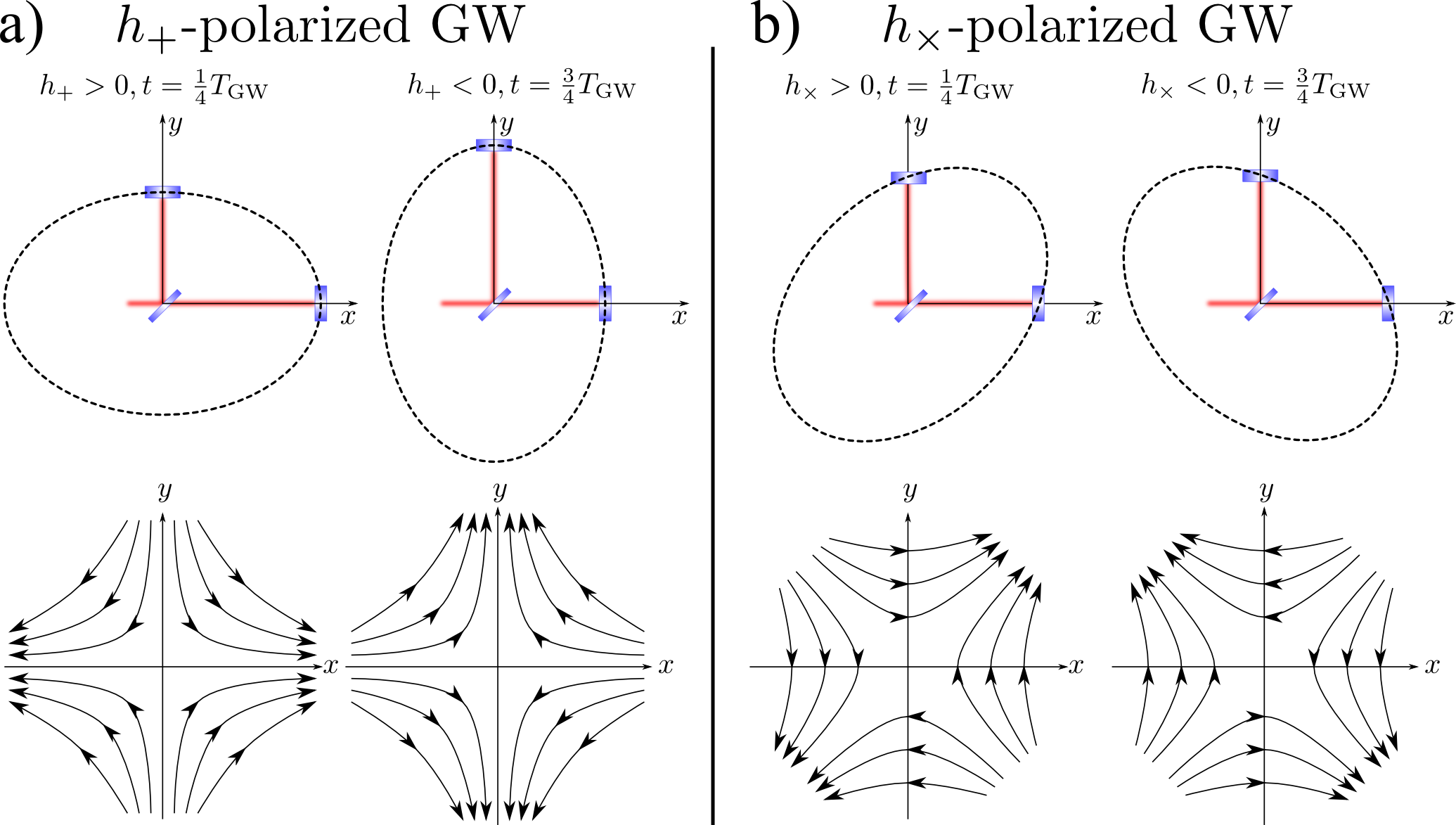}{%
\begin{figure}[htb]
  \centerline{\includegraphics[width=\textwidth]{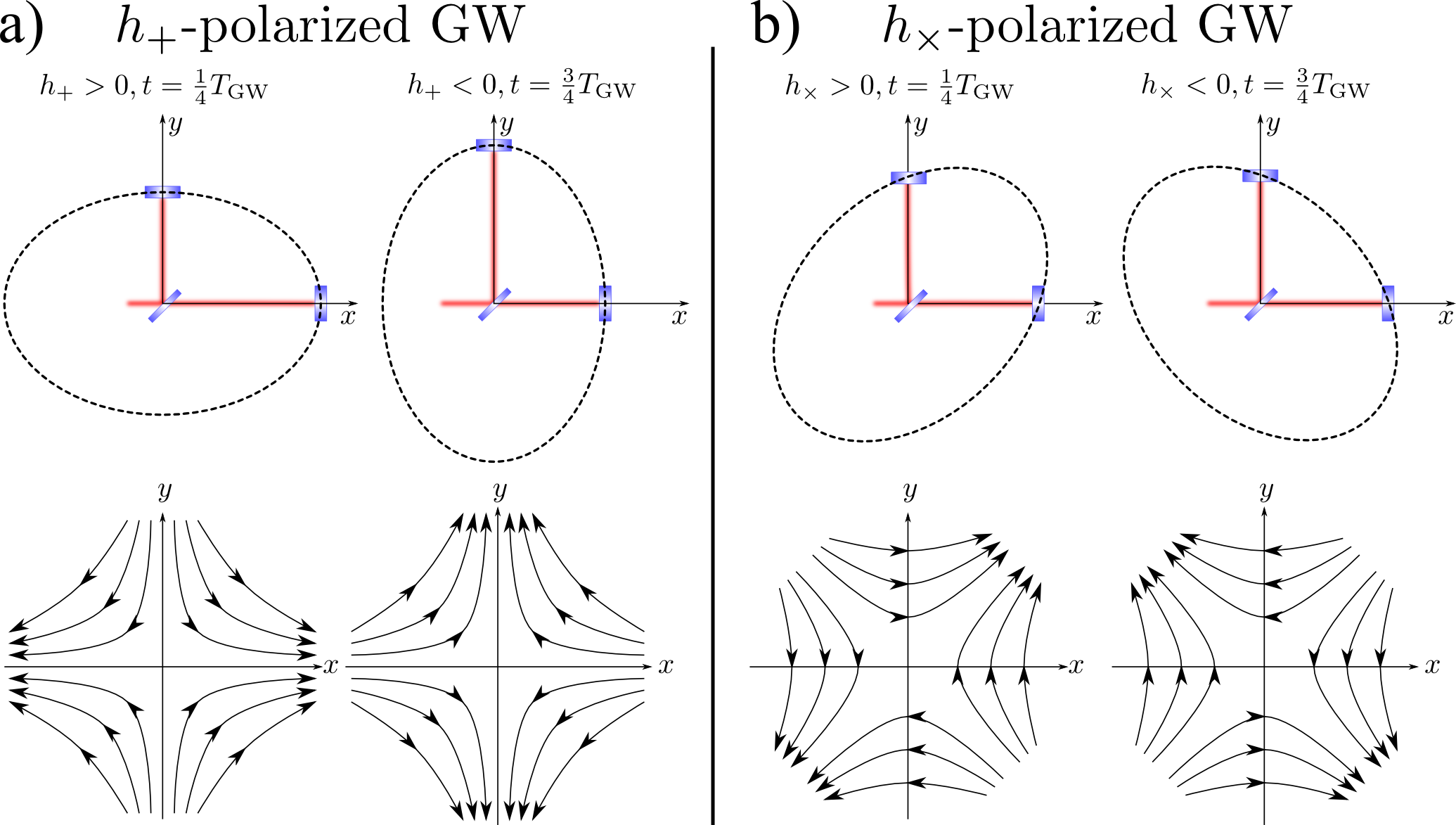}}
  \caption{Action of the GW on a Michelson interferometer: (a) $h_+$-polarized GW periodically stretch and squeeze the interferometer arms in the $x$- and $y$-directions, (b) $h_\times$-polarized GW though have no impact on the interferometer, yet produce stretching and squeezing of the imaginary test particle ring, but along the directions, rotated by $45^\circ$ with respect to the $x$ and $y$ directions of the frame. The lower pictures feature field lines of the corresponding tidal acceleration fields $\propto\ddot h_{+,\times}$.}
\label{fig:GW_action}
\end{figure}}

For our Michelson interferometer, one can consider the end mirrors to be those test particles that lie on a circular ring with beamsplitter located in its center. One can choose arms directions to coincide with the frame $x$ and $y$ axes, then the mirrors will have coordinates $(0,\,L_n)$ and $(L_e,\,0)$, correspondingly. For this case, the action of the GW field on the mirrors is featured in Figure~\ref{fig:GW_action}. It is evident from this picture and from the above formulas that an $h_\times$-polarized component of the GW does not change the relative lengths of the Michelson interferometer arms and thus does not contribute to its output signal; at the same time, $h_+$-polarized GWs act on the end masses of the interferometer as a pair of tidal forces of the same value but opposite in direction:
$$G_n =-\dfrac12M_n\ddot h_+L_n\,,\quad G_e =\dfrac12M_e\ddot h_+L_e\,.$$

Assuming $G_e=-G_n=G$, $M_n=M_e=M$, and $L_e=L_n=L$, one can write down the equations of motion for the interferometer end mirrors that are now considered free ($\Omega_m\ll\Omega_{\mathrm{GW}}$) as:
$$M \ddot x = G\,,\quad M \ddot y = -G\,,$$
and for the differential displacement of the mirrors $\delta L = L_e-L_n = x-y$, which, we have shown above, the Michelson interferometer is sensitive to, one gets the following equation of motion:
\begin{equation}
\label{eq:GW_force_to_h_rel}
M\delta\ddot L = 2 G(t) = M\ddot h_+(t) L
\end{equation}
that is absolutely analogous to Eq.~\eqref{eq:oscillator_EOM} for a single free mirror with mass $M$. Therefore, we have proven that a Michelson interferometer has the same dynamical behavior with respect to the tidal force $G(t)=M\ddot h_+(t) L/2$ created by GWs, as the single movable mirror with mass $M$ to some external generic force $G(t)$.

The foregoing conclusion can be understood in the following way: for GWs are inherently quadruple and, when the detector's plane is orthogonal to the wave propagation direction, can only excite a differential mechanical motion of its mirrors, one can reduce a complicated dynamics of the interferometer probe masses to the dynamics of a single effective particle that is the differential motion of the mirrors in the arms. This useful observation appears to be invaluably helpful for calculation of the real complicated interferometer responses to GWs and also for estimation of its optical quantum noise, that comprises the rest of this review.

\subsection{From incident wave to outgoing light: light transformation in the GW interferometers}

To proceed with the analysis of quantum noise in GW interferometers we first need to familiarize ourselves with how a light field is transformed by an interferometer and how the ability of its mirrors to move modifies the outgoing field. In the following paragraphs, we endeavor to give a step-by-step introduction to the mathematical description of light in the interferometer and the interaction with its movable mirrors.

\subsubsection{Light propagation}
\label{sec:propagation}

We first consider how the light wave is described and how its characteristics transform, when it propagates from one point of free space to another. Yet the real light beams in the large scale interferometers have a rather complicated inhomogeneous transverse spatial structure, the approximation of a plane monochromatic wave should suffice for our purposes, since it comprises all the necessary physics and leads to right results. Inquisitive readers could find abundant material on the field structure of light in real optical resonators in particular, in the introductory book~\cite{yariv1990optical} and in the Living Review by Vinet~\cite{lrr-2009-5}.

So, consider a plane monochromatic linearly polarized light wave propagating in vacuo in the positive direction of the $x$-axis. This field can be fully characterized by the strength of its electric component $E(t-x/c)$ that should be a sinusoidal function of its argument $\zeta=t-x/c$ and can be written in three equivalent ways:
\begin{equation}
\label{eq:EMW_classic_E}
    E(\zeta) = {\cal E}_0\cos\left[\omega_0\zeta-\phi_0\right]
  \equiv \mathcal{E}_c\cos\omega_0\zeta + \mathcal{E}_s\sin\omega_0\zeta
  \equiv \frac{\mathcal{E}e^{-i\omega_0\zeta}+\mathcal{E}^*e^{i\omega_0\zeta}}{\sqrt{2}}\,,
\end{equation}
where ${\cal E}_0$ and $\phi_0$ are called \emph{amplitude} and \emph{phase}, ${\cal E}_c$ and ${\cal E}_s$ take names of cosine and sine \emph{quadrature amplitudes}, and complex number $\mathcal{E} = |\mathcal{E}|e^{i\arg \mathcal{E}}$ is known as the \emph{complex amplitude} of the electromagnetic wave. Here, we see that our wave needs two real or one complex parameter to be fully characterized in the given location $x$ at a given time $t$. The `amplitude-phase' description is traditional for oscillations but is not very convenient since all the transformations are nonlinear in phase. Therefore, in optics, either quadrature amplitudes or complex amplitude description is applied to the analysis of wave propagation. All three descriptions are related by means of straightforward transformations:
\begin{equation}
\begin{array}{ll}
\label{eq:EMW_quadrature_def}
     {\cal E}_0 = \sqrt{{\cal E}_c^2+{\cal E}_s^2} = \sqrt{2}|\mathcal{E}| & \qquad \tan\phi_0 = {\cal E}_s/{\cal E}_c = \arg \mathcal{E},\quad \phi_0\in[0,\,2\pi]\,,\\
     {\cal E}_c = \frac{\mathcal{E}+\mathcal{E}^*}{\sqrt 2} = \sqrt{2}\mathrm{Re}\left[\mathcal{E}\right] = {\cal E}_0\cos\phi_0 & \qquad {\cal E}_s = \frac{\mathcal{E}-\mathcal{E}^*}{i\sqrt 2} = \sqrt{2}\mathrm{Im}\left[\mathcal{E}\right] = {\cal E}_0\sin\phi_0\,,\\
     \mathcal{E} = \frac{{\cal E}_c+i{\cal E}_s}{\sqrt2} = \frac{{\cal E}_0}{\sqrt{2}}e^{i\phi_0} & \qquad \mathcal{E}^* = \frac{{\cal E}_c-i{\cal E}_s}{\sqrt2} = \frac{{\cal E}_0}{\sqrt{2}}e^{-i\phi_0}\,.
\end{array}
\end{equation}
The aforesaid means that for complete understanding of how the light field transforms in the optical device, knowing the rules of transformation for only two characteristic real numbers -- real and imaginary parts of the complex amplitude suffice. Note also that the electric field of a plane wave is, in essence, a function of a single argument $\zeta=t-x/c$ (for a forward propagating wave) and thus can be, without loss of generality, substituted by a time dependence of electric field in some fixed point, say with $x=0$, thus yielding $E(\zeta)\equiv E(t)$. We will keep to this convention throughout our review.

Now let us elaborate the way to establish a link between the wave electric field strength values taken in two spatially separated points, $x_1 = 0$ and $x_2 = L$. Obviously, if nothing obscures light propagation between these two points, the value of the electric field in the second point at time $t$ is just the same as the one in the first point, but at earlier time, i.e., at $t'=t-L/c$: $E^{(L)}(t) = E^{(0)}(t-L/c)$. This allows us to introduce a transformation that propagates EM-wave from one spatial point to another. For complex amplitude $\mathcal{E}$, the transformation is very simple:
\begin{equation}
\label{eq:EMW_free_prop_CAmp_transform}
  \mathcal{E}^{(L)} = e^{i\omega_0L/c} \mathcal{E}^{(0)}\,.
\end{equation}
Basically, this transformation is just a counterclockwise rotation of a wave complex amplitude vector on a complex plane by an angle $\phi_L=\left[\frac{\omega_0 L}{c}\right]_{\mathrm{mod}\ 2\pi}$. This fact becomes even more evident if we look at the transformation for a 2-dimensional vector of quadrature amplitudes $\vb{\mathcal{E}} = \left\{{\cal E}_c,{\cal E}_s\right\}^{\mathsf{T}}$, that are:
\begin{equation}
\label{eq:EMW_free_prop_QAmp_transform}
 \vb{\cal E}^{(L)} = \begin{bmatrix}
                    \cos\phi_L & -\sin\phi_L\\
                    \sin\phi_L & \cos\phi_L
                  \end{bmatrix}\cdot
                  \begin{bmatrix}
                    {\cal E}_c^{(0)}\\
                    {\cal E}_s^{(0)}
                  \end{bmatrix} =
 \mathbb{P}\left[\phi_L\right]\vb{\cal E}^{(0)}\,,
\end{equation}
where
\begin{equation}
\label{eq:CCW_rotation_matrix}
\mathbb{P}[\theta] =
\begin{bmatrix}
  \cos\theta & -\sin\theta\\
  \sin\theta & \cos\theta
\end{bmatrix}
\end{equation}
stands for a standard counterclockwise rotation (pivoting) matrix on a 2D plane.
In the special case when the propagation distance is much smaller than the light wavelength $L\ll\lambda$, the above two expressions can be expanded into Taylor's series in $\phi_L=2\pi L/\lambda\ll1$ up to the first order:
\begin{equation}
\label{eq:EMW_free_prop_small_x}
    E^{(L\ll\lambda)} = (1+i\phi_L)E^{(0)}\,
\end{equation}
 and
\begin{equation}
\label{eq:EMW_free_prop_small_x_matrix}
  \vb{\cal E}^{(L\ll\lambda)} =
                  \begin{bmatrix}
                    1 & -\phi_L\\
                    \phi_L & 1
                  \end{bmatrix}
                  \cdot
                  \begin{bmatrix}
                    {\cal E}_c^{(0)}\\
                    {\cal E}_s^{(0)}
                  \end{bmatrix} =
                  \left(\begin{bmatrix}
                    1 & 0\\
                    0 & 1
                  \end{bmatrix}
                  +
                  \begin{bmatrix}
                    0 & -\phi_L\\
                    \phi_L & 0
                  \end{bmatrix}\right)
                  \cdot
                  \begin{bmatrix}
                    {\cal E}_c^{(0)}\\
                    {\cal E}_s^{(0)}
                  \end{bmatrix} =
 (\mathbb{I}+\delta\mathbb{P}\left[\phi_L\right])\vb{\cal E}^{(0)}\,,
\end{equation}
where $\mathbb{I}$ stands for an identity matrix and $\delta\mathbb{P}[\phi_L]$ is an \emph{infinitesimal} increment matrix that generate the difference between the field quadrature amplitudes vector $\vb{\mathcal{E}}$ after and before the propagation, respectively.

It is worthwhile to note that the quadrature amplitudes representation is used more frequently in literature devoted to quantum noise calculation in GW interferometers than the complex amplitudes formalism and there is a historical reason for this. Notwithstanding the fact that these two descriptions are absolutely equivalent, the quadrature amplitudes representation was chosen by Caves and Schumaker as a basis for their two-photon formalism for the description of quantum fluctuations of light~\cite{85a1CaSch, 85a2CaSch} that became from then on the workhorse of quantum noise calculation. More details about this extremely useful technique are given in the sections \ref{sec:2photon_formalism} and \ref{sec:light_quantum_states} of this review. Unless otherwise specified, we predominantly keep ourselves to this formalism and give all results in terms of it.

\subsubsection{Modulation of light}
\label{sec:modulation}

Above, we have seen that a GW signal displays itself in the modulation of the phase of light, passing through the interferometer. Therefore, it is illuminating to see how the modulation of the light phase and/or amplitude manifests itself in a transformation of the field complex amplitude and quadrature amplitudes.
Throughout this section we assume our carrier field is a monochromatic light wave with frequency $\omega_0$, amplitude $\mathcal{E}_0$ and initial phase $\phi_0=0$:
$$E_{\mathrm{car}}(t) = \mathcal{E}_0\cos\omega_0 t=\mathrm{Re}\left[\mathcal{E}_0e^{-i\omega_0 t}\right]\,.$$

\paragraph*{Amplitude modulation.} The modulation of light amplitude is straightforward to analyze. Let us do it for pedagogical sake: imagine one managed to modulate the carrier field amplitude slow enough compared to the carrier oscillation period, i.e., $\Omega\ll\omega_0$, then:
$$E_{\mathrm{AM}}(t) = \mathcal{E}_0(1+\epsilon_m\cos(\Omega t+\phi_m))\cos\omega_0 t\,,$$
where $\epsilon_m\ll1$ and $\phi_m$ are some constants called modulation depth and relative phase, respectively.
The complex amplitude of the modulated wave equals to
$$\mathcal{E}_{\mathrm{AM}}(t)=\frac{\mathcal{E}_0}{\sqrt{2}}\left(1+\epsilon_m\cos(\Omega t+\phi_m)\right)\,,$$
and the carrier quadrature amplitudes are, apparently, transformed as follows:
$$\mathcal{E}_{c,\mathrm{AM}}(t) = \mathcal{E}_0\left(1+\epsilon_m\cos(\Omega t+\phi_m)\right)\quad\mbox{and}\quad\mathcal{E}_{s,\mathrm{AM}}(t) = 0\,.$$
The fact that the amplitude modulation shows up only in the quadrature that is in phase with the carrier field sets forth why this quadrature is usually named \emph{amplitude quadrature} in the literature. In our review, we shall also keep to this terminology and refer to cosine quadrature as amplitude one.

Illuminating also is the calculation of the modulated light spectrum, that in our simple case of single frequency modulation is straightforward:
$$E_{\mathrm{AM}}(t) = \mathrm{Re}\left[\mathcal{E}_0e^{-i\omega_0 t}+\frac{\mathcal{E}_0\epsilon_m}{2}e^{-i\phi_m}e^{-i(\omega_0+\Omega) t}+\frac{\mathcal{E}_0\epsilon_m}{2}e^{i\phi_m}e^{-i(\omega_0-\Omega) t}\right]\,.$$
Apparently, the spectrum is discrete and comprises three components, i.e., the harmonic at carrier frequency $\omega_0$ with amplitude $A_{\omega_0}=\mathcal{E}_0$ and two satellites at frequencies $\omega_0\pm\Omega$, also referred to as \emph{modulation sidebands}, with (complex) amplitudes $A_{\omega_0\pm\Omega}=\epsilon_m\mathcal{E}_0e^{\mp i\phi_m}/2$. The graphical interpretation of the above considerations is given in the left panel of Figure~\ref{fig:modulation}. Here, carrier fields as well as sidebands are represented by rotating vectors on a complex plane. The carrier field vector has length $\mathcal{E}_0$ and rotates clockwise with the rate $\omega_0$, while sideband components participate in two rotations at a time. The sum of these three vectors yields a complex vector, whose length oscillates with time, and its projection on the real axis represents the amplitude-modulated light field.

\begin{sloppypar}
The above can be generalized to an arbitrary periodic modulation function $A(t)=\sum_{k=1}^\infty A_k\cos(k\Omega+\phi_k)$, with $E_{\mathrm{AM}}(t) = \mathcal{E}_0(1+\epsilon_mA(t))\cos\omega_0 t$. Then the spectrum of the modulated light consists again of a carrier harmonic at $\omega_0$ and an infinite discrete set of sideband harmonics at frequencies $\omega_0\pm k\Omega$ ($k=\overline{1,\infty}$):
\end{sloppypar}
\begin{equation}
\label{eq:AM_discrete}
 E_{\mathrm{AM}}(t) = \mathcal{E}_0\cos\omega_0 t+\frac{\epsilon_m \mathcal{E}_0}{2}\sum_{k=1}^\infty A_k\left\{\cos[(\omega_0-k\Omega) t-\phi_k]+\cos[(\omega_0+k\Omega) t+\phi_k]\right\}
 \,.
\end{equation}
\begin{sloppypar}
Further generalization to an arbitrary (real) non-periodic modulation function $A(t) = \intinfty\frac{d\omega}{2\pi}A(\Omega)e^{-i\Omega t}$ is apparent:
\end{sloppypar}
\begin{eqnarray}
\label{eq:AM_continuous}
E_{\mathrm{AM}}(t) &=& \mathrm{Re}\left[\mathcal{E}_0e^{-i\omega_0
    t}+\epsilon_m \mathcal{E}_0e^{-i\omega_0
    t}\intinfty\frac{d\Omega}{2\pi}A(\Omega)e^{-i\Omega t}\right] = \nonumber\\
 && \mathcal{E}_0\cos\omega_0 t+\frac{\epsilon_m \mathcal{E}_0}{2}\intinfty\frac{d\omega}{2\pi}\left\{A(\omega-\omega_0)+A(\omega+\omega_0)\right\}e^{-i\omega t}\,.
\end{eqnarray}
From the above expression, one readily sees the general structure of the modulated light spectrum, i.e., the central carrier peaks at frequencies $\pm\omega_0$ and the modulation sidebands around it, whose shape retraces the modulation function spectrum $A(\omega)$ shifted by the carrier frequency $\pm\omega_0$.

\epubtkImage{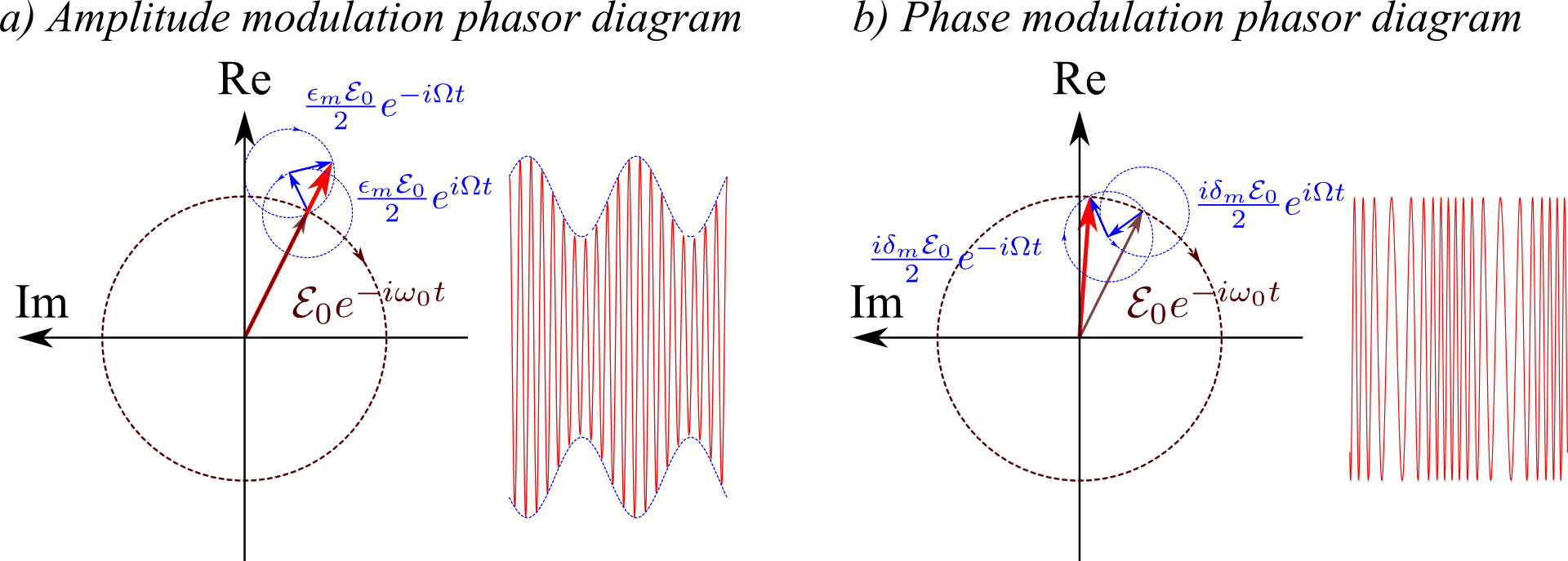}{%
\begin{figure}[htb]
  \centerline{\includegraphics[width=\textwidth]{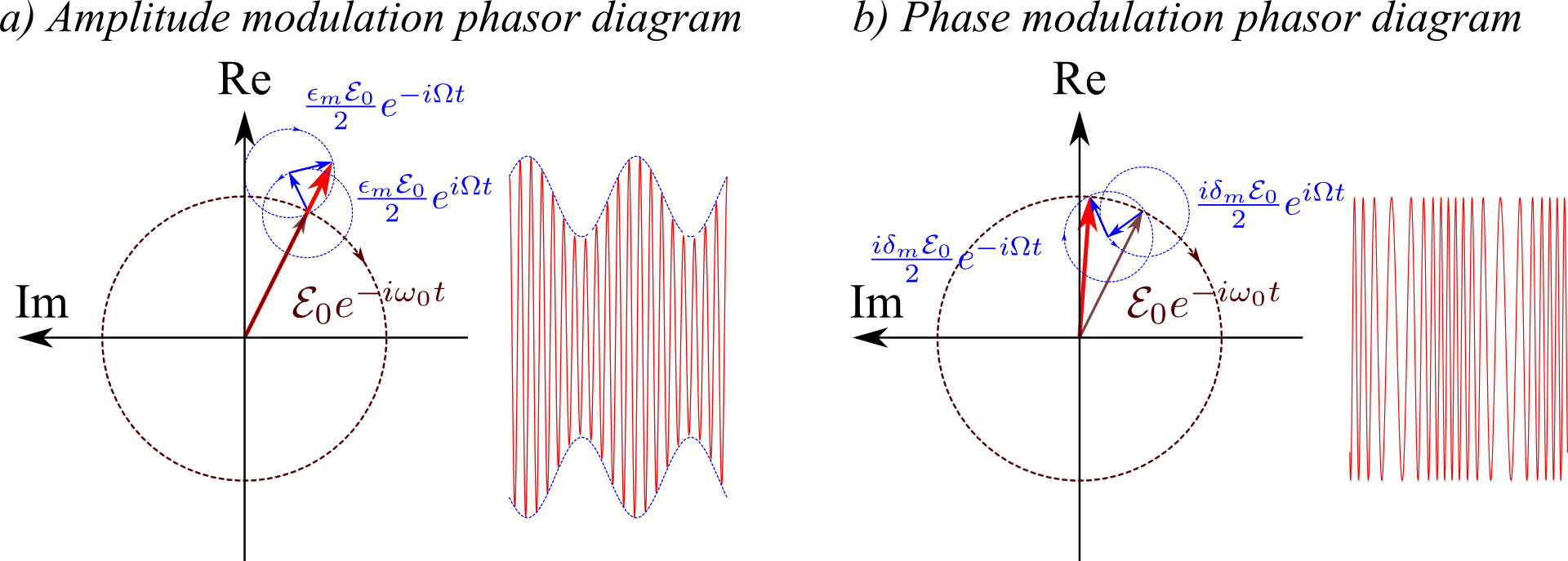}}
  \caption{Phasor diagrams for amplitude (\emph{Left panel}) and phase (\emph{Right panel}) modulated light. Carrier field is given by a brown vector rotating clockwise with the rate $\omega_0$ around the origin of the complex plane frame. Sideband fields are depicted as blue vectors. The lower ($\omega_0-\Omega$) sideband vector origin rotates with the tip of the carrier vector, while its own tip also rotates with respect to its origin counterclockwise with the rate $\Omega$. The upper ($\omega_0+\Omega$) sideband vector origin rotates with the tip of the upper sideband vector, while its own tip also rotates with respect to its origin counterclockwise with the rate $\Omega$. Modulated oscillation is a sum of these three vectors an is given by the red vector. In the case of amplitude modulation (AM), the modulated oscillation vector is always in phase with the carrier field while its length oscillates with the modulation frequency $\Omega$. The time dependence of its projection onto the real axis that gives the AM-light electric field strength is drawn to the right of the corresponding phasor diagram. In the case of phase modulation (PM), sideband fields have a $\pi/2$ constant phase shift with respect to the carrier field (note factor $i$ in front of the corresponding terms in Eq.~\eqref{eq:PM_discrete}; therefore its sum is always orthogonal to the carrier field vector, and the resulting modulated oscillation vector (red arrow) has approximately the same length as the carrier field vector but outruns or lags behind the latter periodically with the modulation frequency $\Omega$. The resulting oscillation of the PM light electric field strength is drawn to the right of the PM phasor diagram and is the projection of the PM oscillation vector on the real axis of the complex plane.}
\label{fig:modulation}
\end{figure}}

\paragraph*{Phase modulation.} The general feature of the modulated signal that we pursued to demonstrate by this simple example is the creation of the modulation sidebands in the spectrum of the modulated light. Let us now see how it goes with a phase modulation that is more related to the topic of the current review. The simplest single-frequency phase modulation is given by the expression:
$$E_{\mathrm{PM}}(t) = \mathcal{E}_0\cos(\omega_0 t+\delta_m\cos(\Omega t+\phi_m))\,,$$
where $\Omega\ll\omega_0$, and the phase deviation $\delta_m$ is assumed to be much smaller than 1. Using Eqs.~\eqref{eq:EMW_quadrature_def}, one can write the complex amplitude of the phase-modulated light as:
$$\mathcal{E}_{\mathrm{PM}}(t)=\frac{\mathcal{E}_0}{\sqrt{2}}e^{i\delta_m\cos(\Omega t+\phi_m)}\,,$$
and quadrature amplitudes as:
$$\mathcal{E}_{c,\mathrm{PM}}(t) = \mathcal{E}_0\cos\left[\delta_m\cos(\Omega t+\phi_m)\right]\quad\mbox{and}\quad\mathcal{E}_{s,\mathrm{PM}}(t) =\mathcal{E}_0\sin\left[\delta_m\cos(\Omega t+\phi_m)\right]\,.$$
Note that in the weak modulation limit ($\delta_m\ll1$), the above equations can be approximated as:
$$\mathcal{E}_{c,\mathrm{PM}}(t) \simeq \mathcal{E}_0\quad\mbox{and}\quad\mathcal{E}_{s,\mathrm{PM}}(t) \simeq \delta_m\mathcal{E}_0\cos(\Omega t+\phi_m)\,.$$
This testifies that for a weak modulation only the sine quadrature, which is $\pi/2$ out-of-phase with respect to the carrier field, contains the modulation signal. That is why this sine quadrature is usually referred to as \emph{phase quadrature}. It is also what we will call this quadrature throughout the rest of this review.

In order to get the spectrum of the phase-modulated light it is necessary to refer to the theory of Bessel functions that provides us with the following useful relation (known as the Jacobi--Anger expansion):
\begin{equation*}
  e^{i\delta_m\cos(\Omega t+\phi_m)} = \sum_{k=-\infty}^\infty i^k J_k(\delta_m)e^{ik(\Omega t+\phi_m)},
\end{equation*}
where $J_k(\delta_m)$ stands for the $k$-th Bessel function of the first kind. This looks a bit intimidating, yet for $\delta_m\ll1$ these expressions simplify dramatically, since near zero Bessel functions can be approximated as:
$$J_0(\delta_m)\simeq 1-\frac{\delta_m^2}{4}+\mathcal{O}(\delta_m^4)\,,\quad J_1(\delta_m) = \frac{\delta_m}{2}+\mathcal{O}(\delta_m^3)\,,\quad J_k(\delta_m) = \frac{1}{k!}\left(\frac{\delta_m}{2}\right)^k+\mathcal{O}(\delta_m^{k+2})\ (k\geqslant2)\,.$$

Thus, for sufficiently small $\delta_m$, we can limit ourselves only to the terms of order $\delta_m^0$ and $\delta_m^1$, which yields:
\begin{equation}
\label{eq:PM_discrete}
E_{\mathrm{PM}}(t) \simeq \mathrm{Re}\left[\mathcal{E}_0e^{i\omega_0 t}+i\frac{\delta_m \mathcal{E}_0}{2}\left(e^{i[(\omega_0+\Omega) t+\phi_m]}+e^{i[(\omega_0-\Omega) t-\phi_m]}\right)\right]\,,
\end{equation}
and we again face the situation in which modulation creates a pair of sidebands around the carrier frequency. The difference from the amplitude modulation case is in the way these sidebands behave on the complex plane. The corresponding phasor diagram for phase modulated light is drawn in Figure~\ref{fig:modulation}. In the case of PM, sideband fields have $\pi/2$ constant phase shift with respect to the carrier field (note factor $i$ in front of the corresponding terms in Eq.~\eqref{eq:PM_discrete}); therefore its sum is always orthogonal to the carrier field vector, and the resulting modulated oscillation vector has approximately the same length as the carrier field vector but outruns or lags behind the latter periodically with the modulation frequency $\Omega$. The resulting oscillation of the PM light electric field strength is drawn to the right of the PM phasor diagram and is the projection of the PM oscillation vector on the real axis of the complex plane.

Let us now generalize the obtained results to an arbitrary modulation function
$\Phi(t)$:
$$E_{\mathrm{PM}}(t)=\mathcal{E}_0\cos(\omega_0 t+\delta_m\Phi(t))\,.$$
In the most general case of arbitrary modulation index $\delta_m$, the corresponding formulas are very cumbersome and do not give much insight. Therefore, we again consider a simplified situation of sufficiently small $\delta_m\ll1$. Then one can approximate the phase-modulated oscillation as follows:
$$
E_{\mathrm{PM}}(t)=\mathrm{Re}\left[\mathcal{E}_0e^{-i\omega_0 t}e^{i\delta_m\Phi(t)}\right]\simeq\mathrm{Re}\left[\mathcal{E}_0e^{-i\omega_0 t}\left\{1+i\delta_m\Phi(t)\right\}\right]\,.
$$
When $\Phi(t)$ is a periodic function: $\Phi(t) = \sum_{k=1}^\infty\Phi_k\cos{k\Omega+\phi_k}$, and in weak modulation limit $\delta_m\ll1$, the spectrum of the PM light is apparent from the following expression:
\begin{eqnarray}
\label{eq:PM_discrete2}
  E_{\mathrm{PM}}(t) &\simeq&
    \mathcal{E}_0\cos\omega_0 t -
    \frac{\delta_m\mathcal{E}_0}{2}\sum_{k=1}^\infty \Phi_k\left\{\sin\left[(\omega_0-k\Omega) t-\phi_k\right]+\sin\left[(\omega_0+k\Omega) t+\phi_k\right]\right\} \nonumber\\
    &=& \mathrm{Re}\left[\mathcal{E}_0e^{-i\omega_0 t}+
      \frac{i\delta_m\mathcal{E}_0}{2}\sum_{k=1}^\infty \Phi_k\left\{e^{-i\left[(\omega_0-k\Omega) t-\phi_k\right]}+
      e^{-i\left[(\omega_0+k\Omega) t+\phi_k\right]}\right\}\right]\,,
 \end{eqnarray}
while for the real non-periodic modulation function $\Phi(t) = \intinfty\frac{d\omega}{2\pi}\Phi(\Omega)e^{-i\Omega t}$ the spectrum can be obtained from the following relation:
\begin{eqnarray}
\label{eq:PM_continuous}
E_{\mathrm{PM}}(t) &\simeq& \mathrm{Re}\left[\mathcal{E}_0e^{-i\omega_0
    t}+i\delta_m \mathcal{E}_0e^{-i\omega_0
    t}\intinfty\frac{d\Omega}{2\pi}\Phi(\Omega)e^{-i\Omega t}\right] \nonumber\\ 
 &=& \mathcal{E}_0\cos\omega_0 t+\frac{\delta_m \mathcal{E}_0}{2}\intinfty\frac{d\omega}{2\pi}\left\{i\Phi(\omega-\omega_0)-i\Phi(\omega+\omega_0)\right\}e^{-i\omega t}\,.
\end{eqnarray}
And again we get the same general structure of the spectrum with carrier peaks at $\pm\omega_0$ and shifted modulation spectra $i\Phi(\omega\pm\omega_0)$ as sidebands around the carrier peaks. The difference with the amplitude modulation is an additional $\pm\pi/2$ phase shifts added to the sidebands.

\subsubsection{Laser noise}

Thus far we have assumed the carrier field to be perfectly monochromatic having a single spectral component at carrier frequency $\omega_0$ fully characterized by a pair of classical quadrature amplitudes represented by a 2-vector $\vb{\mathcal{E}}$. In reality, this picture is no good at all; indeed, a real laser emits not a monochromatic light but rather some spectral line of finite width with its central frequency and intensity fluctuating. These fluctuations are usually divided into two categories: (i) \emph{quantum noise} that is associated with the spontaneous emission of photons in the gain medium, and (ii) \emph{technical noise} arising, e.g., from excess noise of the pump source, from vibrations of the laser resonator, or from temperature fluctuations and so on. It is beyond the goals of this review to discuss the details of the laser noise origin and methods of its suppression, since there is an abundance of literature on the subject that a curious reader might find interesting, e.g., the following works~\cite{Paschotta_APB_2004-79-2-153, Paschotta_APB_2004-79-2-163, Paschotta_APB_2006-82-2-265, Willke_LPR_2010-4-6-780, Harb_JOSAB_1997-14-11-2936, Heurs_APB_2006-85-1-79}.

For our purposes, the very existence of the laser noise is important as it makes us to reconsider the way we represent the carrier field. Apparently, the proper account for laser noise prescribes us to add a random time-dependent modulation of the amplitude (for intensity fluctuations) and phase (for phase and frequency fluctuations) of the carrier field~\eqref{eq:EMW_classic_E}:
$$E(t) = (\mathcal{E}_0+\hat{e}_{n}(t))\cos\left[\omega_0 t+\phi_0+\hat{\phi}_{n}(t)\right]\,,$$
where we placed hats above the noise terms on purpose, to emphasize that quantum noise is a part of laser noise and its quantum nature has to be taken into account, and that the major part of this review will be devoted to the consequences these hats lead to. However, for now, let us consider hats as some nice decoration.

Apparently, the corrections to the amplitude and phase of the carrier light due to the laser noise are small enough to enable us to use the weak modulation approximation as prescribed above. In this case one can introduce a more handy amplitude and phase quadrature description for the laser noise contribution in the following manner:
\begin{equation}
\label{eq:EMW_laser_noise}
E(t) = \left(\mathcal{E}_c + \hat{e}_c(t)\right)\cos\omega_0t + \left(\mathcal{E}_s + \hat{e}_s(t)\right)\sin\omega_0t\,,
\end{equation}
where $\hat{e}_{c,s}$ are related to $\hat{e}_{n}$ and $\hat{\phi}_{n}$ in the same manner as prescribed by Eqs.~\eqref{eq:EMW_quadrature_def}. It is convenient to represent a noisy laser field in the Fourier domain:
$$
\hat{e}_{c,s}(t) = \intinfty \frac{d\Omega}{2\pi}\hat{e}_{c,s}(\Omega)e^{-i\Omega t}\,.
$$

Worth noting is the fact that $\hat{e}_{c,s}(\Omega)$ is a spectral representation of a real quantity and thus satisfies an evident equality $\hat{e}_{c,s}^\dag(-\Omega)=\hat{e}_{c,s}(\Omega)$ (by $\dag$ we denote the  Hermitian conjugate that for classical functions corresponds to taking the complex conjugate of this function). What happens if we want to know the light field of our laser with noise at some distance $L$ from our initial reference point $x=0$? For the carrier field component at $\omega_0$, nothing changes and the corresponding transform is given by Eq.~\eqref{eq:EMW_free_prop_QAmp_transform}, yet for the noise component $$\delta\hat{E}_{\mathrm{noise}}(t) = \hat{e}_c(t)\cos\omega_0t+\hat{e}_s(t)\sin\omega_0t$$
there is a slight modification. Since the field continuity relation holds for the noise field to the same extent as for the carrier field:
 $$\delta\hat{E}^{(L)}_{\mathrm{noise}}(t) = \delta\hat{E}^{(0)}_{\mathrm{noise}}(t-L/c)\,,$$
 the following modification applies:
 \begin{equation}
\label{eq:EMW_sidebands_free_prop_matrix}
\hat{\boldsymbol{e}}^{(L)}(\Omega)=\begin{bmatrix}
  \hat{e}_c^{(L)}(\Omega)\\
  \hat{e}_s^{(L)}(\Omega)
\end{bmatrix} = e^{i\Omega L/c}
\begin{bmatrix}
  \cos \dfrac{\omega_0L}{c} & -\sin\dfrac{\omega_0L}{c}\\
 \sin\dfrac{\omega_0L}{c} & \cos\dfrac{\omega_0L}{c}
\end{bmatrix}\cdot\begin{bmatrix}
  \hat{e}_c^{(0)}(\Omega)\\
  \hat{e}_s^{(0)}(\Omega)
\end{bmatrix} = e^{i\Omega L/c}\mathbb{P}\left[\dfrac{\omega_0L}{c}\right]\hat{\boldsymbol{e}}^{(0)}(\Omega)\,.
 \end{equation}
Therefore, for sideband field components the propagation rule shall be modified by adding a frequency-dependent phase factor $e^{i\Omega L/c}$ that describes an extra phase shift acquired by a sideband field relative to the carrier field because of the frequency difference $\Omega=\omega-\omega_0$.

\subsubsection{Light reflection from optical elements}
\label{sec:light_reflection}

So, we are one step closer to understanding how to calculate the quantum
noise of the light coming out of the GW interferometer. It is necessary
to understand what happens with light when it is reflected from such
optical elements as mirrors and beamsplitters. Let us first consider
these elements of the interferometer fixed at their positions. The
impact of mirror motion will be considered in the next subsection \ref{sec:Mirror_motion}. One
can also refer to Section~2 of the Living Review by Freise
and Strain~\cite{lrr-2010-1} for a more detailed treatment of this
topic.

\epubtkImage{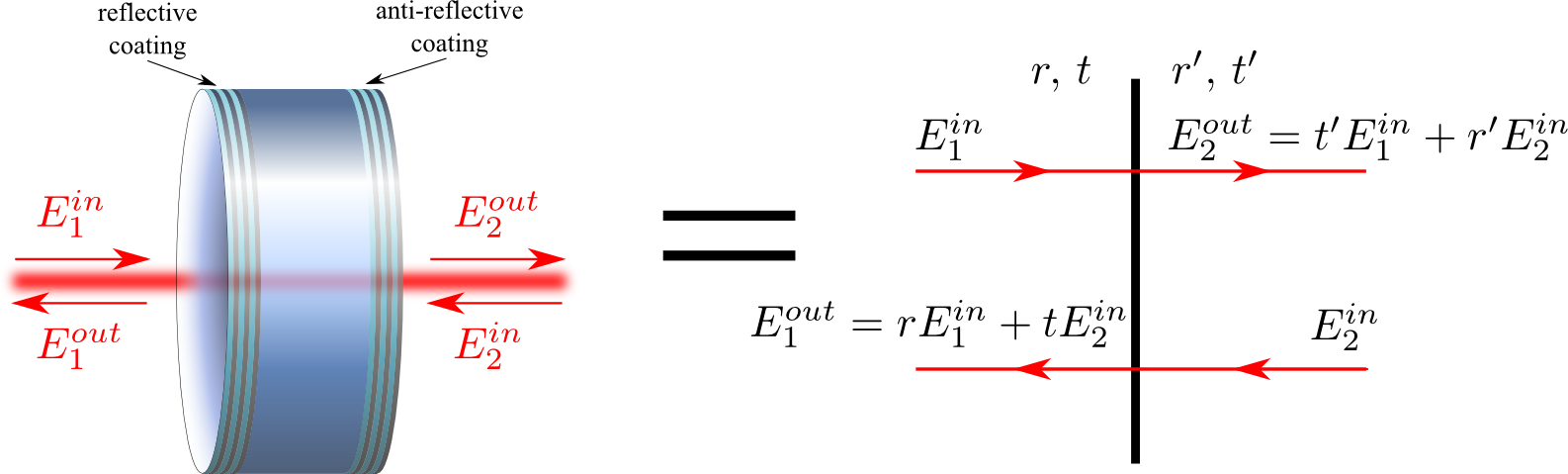}{%
\begin{figure}[htbp]
  \centerline{\includegraphics[width=\textwidth]{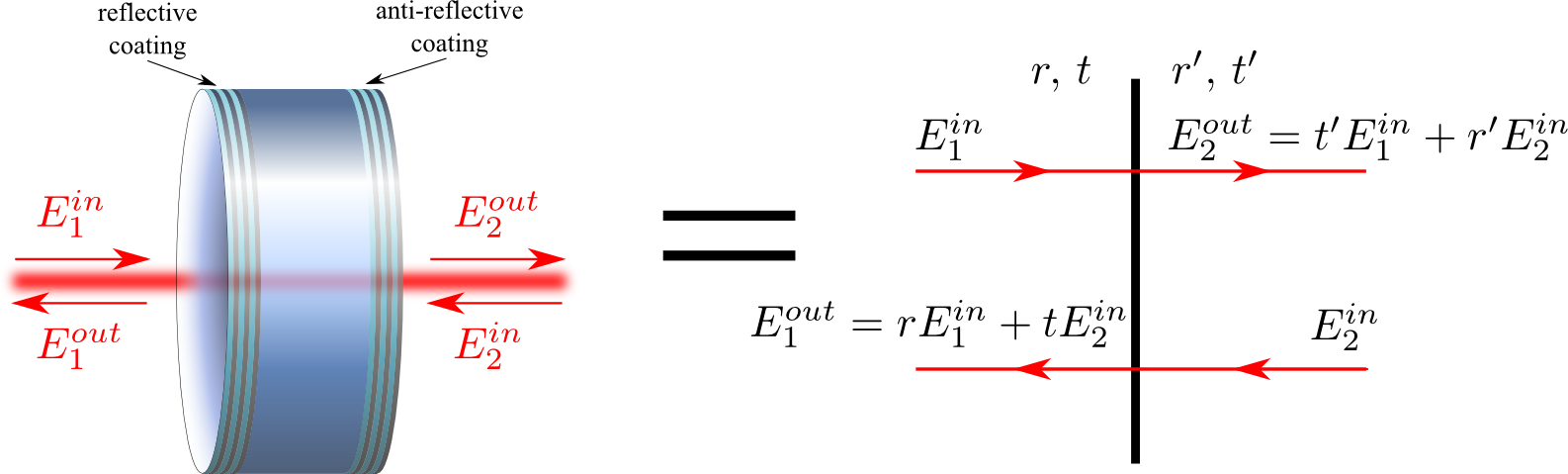}}
  \caption{Scheme of light reflection off the coated mirror.}
\label{fig:mirror}
\end{figure}}

Mirrors of the modern interferometers are rather complicated optical systems usually consisting of a dielectric slab with both surfaces covered with multilayer dielectric coatings. These coatings are thoroughly constructed in such a way as to make one surface of the mirror highly reflective, while the other one is anti-reflective. We will not touch on the aspects of coating technology in this review and would like to refer the interested reader to an abundant literature on this topic, e.g., to the following book~\cite{2012_unpublished_Harry} and reviews and articles~\cite{lrr-2009-5, CQG.19.897_Harry, CQG.26.15.155012_2009_Martin, CQG.24.2.405_2006_Harry, CQG.20.13.2917_2003_Penn, CQG.19.5.883_2002_Crooks, PhysRevD.57.659_1998_Levin, PhysRevLett.93.250602_2004_Numata, PhysRevD.78.102003_2008_Evans}. For our purposes, assuming the reflective surface of the mirror is flat and lossless should suffice. Thus, we represent a mirror by a reflective plane with (generally speaking, complex) coefficients of reflection $r$ and $r'$ and transmission $t$ and $t'$ as drawn in Figure~\ref{fig:mirror}. Let us now see how the ingoing and outgoing light beams couple on the mirrors in the interferometer.

\paragraph*{Mirrors:}
From the general point of view, the mirror is a linear system with 2 input and 2 output ports. The way how it transforms input signals into output ones is featured by a $2\times2$ matrix that is known as the \emph{transfer matrix} of the mirror $\mathbb{M}$:
\begin{equation}
\label{eq:mirror_IOrel}
\begin{bmatrix}
  E^{\out}_1(t)\\
  E^{\out}_2(t)
\end{bmatrix}= \mathbb{M}\cdot
\begin{bmatrix}
  E^{\in}_1(t)\\
  E^{\in}_2(t)
\end{bmatrix}=
\begin{bmatrix}
  r & t\\
  t' & r'
\end{bmatrix}\cdot
\begin{bmatrix}
  E^{\in}_1(t)\\
  E^{\in}_2(t)
\end{bmatrix}\,.
\end{equation}
Since we assume no absorption in the mirror, reflection and transmission coefficients should satisfy Stokes' relations~\cite{1849a1St, 80BookeBoWo} (see also Section~12.12 of~\cite{95BookMaWo}):
\begin{equation}
\label{eq:Stokes_rel}
 |r|=|r'|\,, \qquad |t|=|t'|\,, \qquad |r|^2+|t|^2 = 1\,, \qquad r^*t'+r't^* = 0\,, \qquad r^*t+r't'^{*} = 0\,,
\end{equation}
that is simply a consequence of the conservation of energy. This conservation of energy yields that the optical transfer matrix $\mathbb{M}$ must be unitary: $\mathbb{M}^\dag = \mathbb{M}^{-1}$. Stokes' relations leave some freedom in defining complex reflectivity and transmissivity coefficients. Two of the most popular variants are given by the following matrices:
\begin{equation}
\label{eq:IO_mirror_matrix}
\mathbb{M}_{\mathrm{sym}} = \begin{bmatrix}
  \sqrt{R} & i\sqrt{T}\\
  i\sqrt{T} & \sqrt{R}
\end{bmatrix}\,,\quad\mbox{and}\quad \mathbb{M}_{\mathrm{real}} = \begin{bmatrix}
  -\sqrt{R} & \sqrt{T}\\
  \sqrt{T} & \sqrt{R}
\end{bmatrix}\,,
\end{equation}
where we rewrote transfer matrices in terms of real power reflectivity and transmissivity coefficients $R=|r|^2$ and $T=|t|^2$ that will find extensive use throughout the rest of this review.

The transformation rule, or putting it another way, \emph{coupling relations} for the quadrature amplitudes can easily be obtained from Eq.~\eqref{eq:mirror_IOrel}. Now, we have two input and two output fields. Therefore, one has to deal with 4-dimensional vectors comprising of quadrature amplitudes of both input and output fields, and the transformation matrix become $4\times4$-dimensional, which can be expressed in terms of the outer product of a $2\times2$ matrix $\mathbb{M}_{\mathrm{real}}$ by a $2\times2$ identity matrix $\mathbb{I}$:
\begin{equation}
\label{eq:IO_mirror_matrix_4x4_zero}
\begin{bmatrix}
  \mathcal{E}^{\out}_{1c}\\
  \mathcal{E}^{\out}_{1s}\\
  \mathcal{E}^{\out}_{2c}\\
  \mathcal{E}^{\out}_{2s}
\end{bmatrix}=
\begin{bmatrix}
  \vb{\mathcal{E}}^{\out}_{1}\\
  \vb{\mathcal{E}}^{\out}_{2}
\end{bmatrix}=
\mathbb{M}_{\mathrm{real}}\otimes\mathbb{I}
\cdot
\begin{bmatrix}
  \vb{\mathcal{E}}^{\in}_{1}\\
  \vb{\mathcal{E}}^{\in}_{2}
\end{bmatrix}=
\begin{bmatrix}
  -\sqrt{R} &  0 &\sqrt{T} & 0\\
  0 & -\sqrt{R} & 0 & \sqrt{T} \\
  \sqrt{T} & 0 & \sqrt{R} & 0\\
  0 & \sqrt{T} & 0 & \sqrt{R}
\end{bmatrix}\cdot
\begin{bmatrix}
  \mathcal{E}^{\in}_{1c}\\
  \mathcal{E}^{\in}_{1s}\\
  \mathcal{E}^{\in}_{2c}\\
  \mathcal{E}^{\in}_{2s}
\end{bmatrix}\,.
\end{equation}
The same rules apply to the sidebands of each carrier field:
\begin{equation}
\label{eq:IO_mirror_matrix_4x4_first}
\begin{bmatrix}
  \hat{\boldsymbol{e}}^{\out}_{1}(\Omega)\\
  \hat{\boldsymbol{e}}^{\out}_{2}(\Omega)
\end{bmatrix}=
\mathbb{M}_{\mathrm{real}}\otimes\mathbb{I}
\cdot
\begin{bmatrix}
  \hat{\boldsymbol{e}}^{\in}_{1}(\Omega)\\
  \hat{\boldsymbol{e}}^{\in}_{2}(\Omega)
\end{bmatrix}=
\begin{bmatrix}
  -\sqrt{R} &  0 &\sqrt{T} & 0\\
  0 & -\sqrt{R} & 0 & \sqrt{T} \\
  \sqrt{T} & 0 & \sqrt{R} & 0\\
  0 & \sqrt{T} & 0 & \sqrt{R}
\end{bmatrix}\cdot
\begin{bmatrix}
  \hat{e}^{\in}_{1c}(\Omega)\\
  \hat{e}^{\in}_{1s}(\Omega)\\
  \hat{e}^{\in}_{2c}(\Omega)\\
  \hat{e}^{\in}_{2s}(\Omega)
\end{bmatrix}\,.
\end{equation}

\epubtkImage{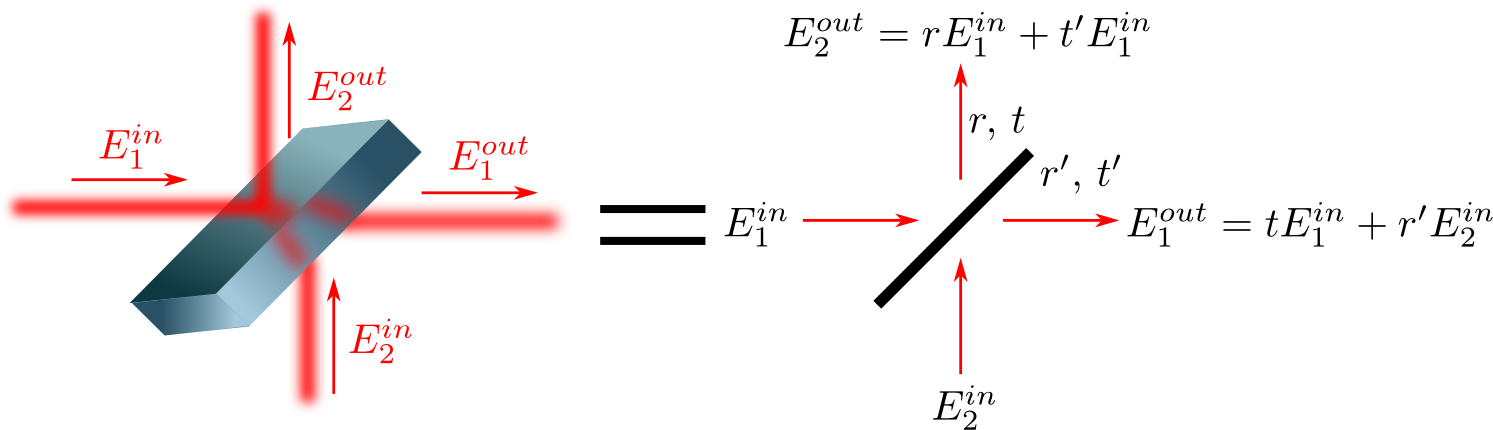}{%
\begin{figure}[htbp]
  \centerline{\includegraphics[width=\textwidth]{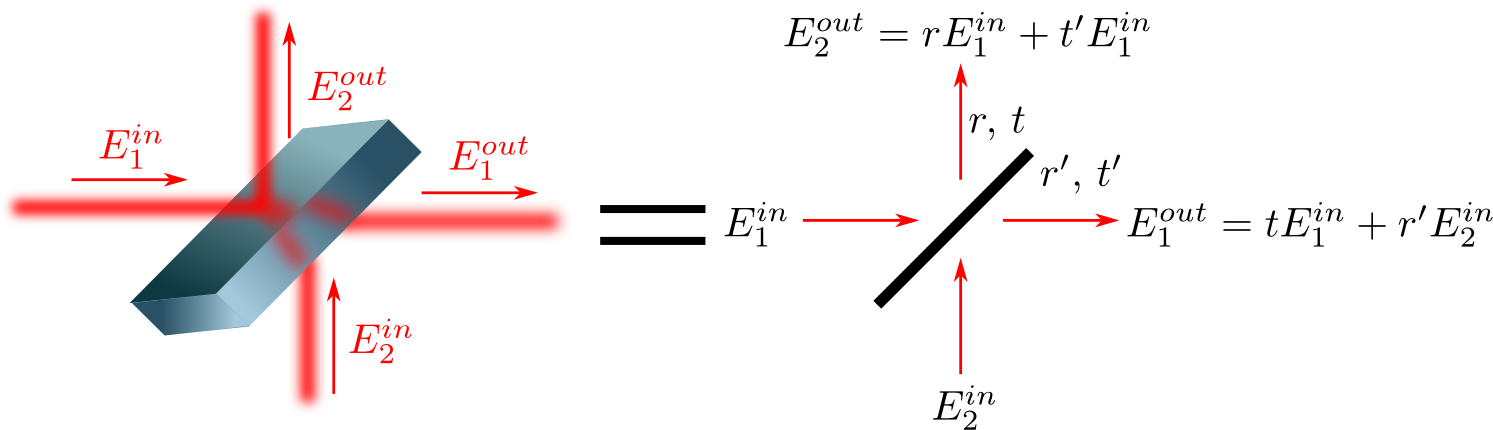}}
  \caption{Scheme of a beamsplitter.}
\label{fig:beamsplitter}
\end{figure}}

In future, for the sake of brevity, we reduce the notation for matrices like $\mathbb{M}_{\mathrm{real}}\otimes\mathbb{I}$ to simply $\mathbb{M}_{\mathrm{real}}$.
\paragraph*{Beam splitters:}
Another optical element ubiquitous in the interferometers is a beamsplitter (see Figure~\ref{fig:beamsplitter}). In fact, it is the very same mirror considered above, but the angle of input light beams incidence is different from 0 (if measured from the normal to the mirror surface). The corresponding scheme is given in Figure~\ref{fig:beamsplitter}. In most cases, symmetric 50\%/50\% beamsplitter are used, which imply $R=T=1/2$ and the coupling matrix $\mathbb{M}_{50/50}$ then reads:
\begin{equation}
\label{eq:IO_beamsplitter_matrix}
\mathbb{M}_{50/50} =
\begin{bmatrix}
  -1/\sqrt{2} &  0 &1/\sqrt{2} & 0\\
  0 & -1/\sqrt{2} & 0 & 1/\sqrt{2} \\
  1/\sqrt{2} & 0 & 1/\sqrt{2} & 0\\
  0 & 1/\sqrt{2} & 0 & 1/\sqrt{2}
\end{bmatrix}\,.
\end{equation}

\paragraph*{Losses in optical elements:}
\label{sec:losses_in_OE}
Above, we have made one assumption that is a bit idealistic. Namely,
we assumed our mirrors and beamsplitters to be lossless, but it could
never come true in real experiments; therefore, we need some way to
describe losses within the framework of our formalism. Optical loss is a
term that comprises a very wide spectrum of physical processes, including
scattering on defects of the coating, absorption of light photons in
the mirror bulk and coating that yields heating and so on. A full
description of loss processes is rather complicated. However, the most
important features that influence the light fields, coming off the
lossy optical element, can be summarized in the following two simple
statements:

\begin{enumerate}
  \item Optical loss of an optical element can be characterized by a single number (possibly, frequency dependent) $\epsilon$ (usually, $|\epsilon|\ll 1$) that is called the \emph{absorption coefficient}.  $\epsilon$ is the fraction of light power being lost in the optical element:
  $$E^{\out}(t)\to \sqrt{1-\epsilon}E^{\out}(t).$$
  \item Due to the fundamental law of nature summarized by the Fluctuation Dissipation Theorem (FDT)~\cite{PhysRev.83.34, Landau_Lifshitz_v5}, optical loss is always accompanied by additional noise injected into the system. It means that additional noise field $\hat{n}$ uncorrelated with the original light is mixed into the outgoing light field in the proportion of $\sqrt{\epsilon}$ governed by the absorption coefficient.
\end{enumerate}

These two rules conjure up a picture of an effective system comprising of
a lossless mirror and two imaginary non-symmetric beamsplitters with
reflectivity $\sqrt{1-\epsilon}$ and transmissivity $\sqrt{\epsilon}$
that models optical loss for both input fields, as drawn in
Figure~\ref{fig:lossy_mirror}.

\epubtkImage{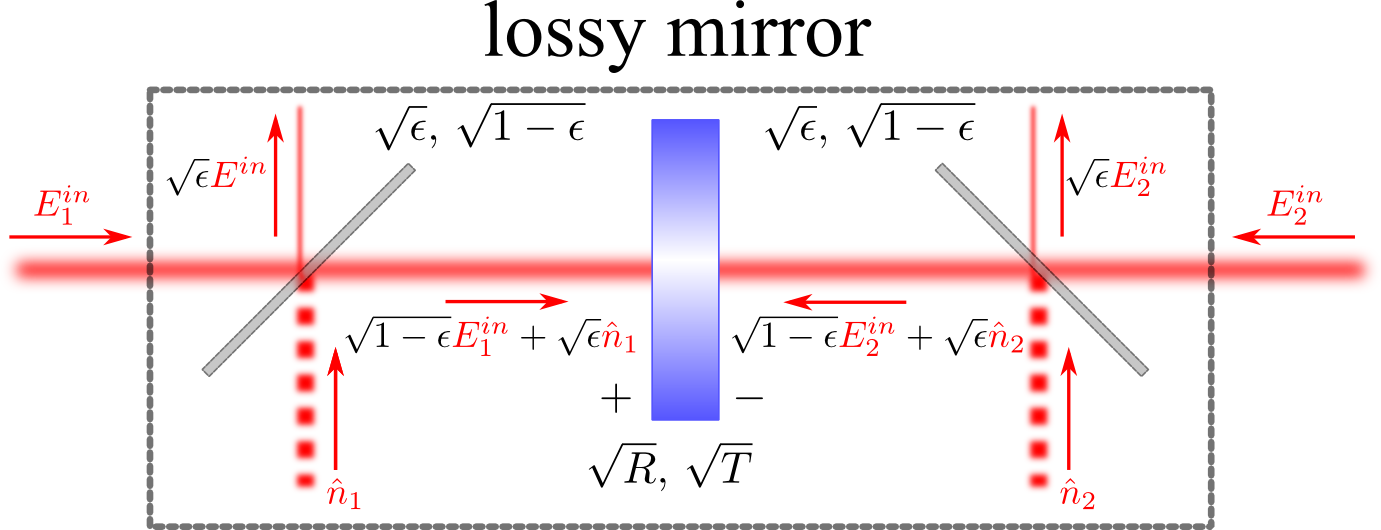}{%
\begin{figure}[htbp]
  \centerline{\includegraphics[width=\textwidth]{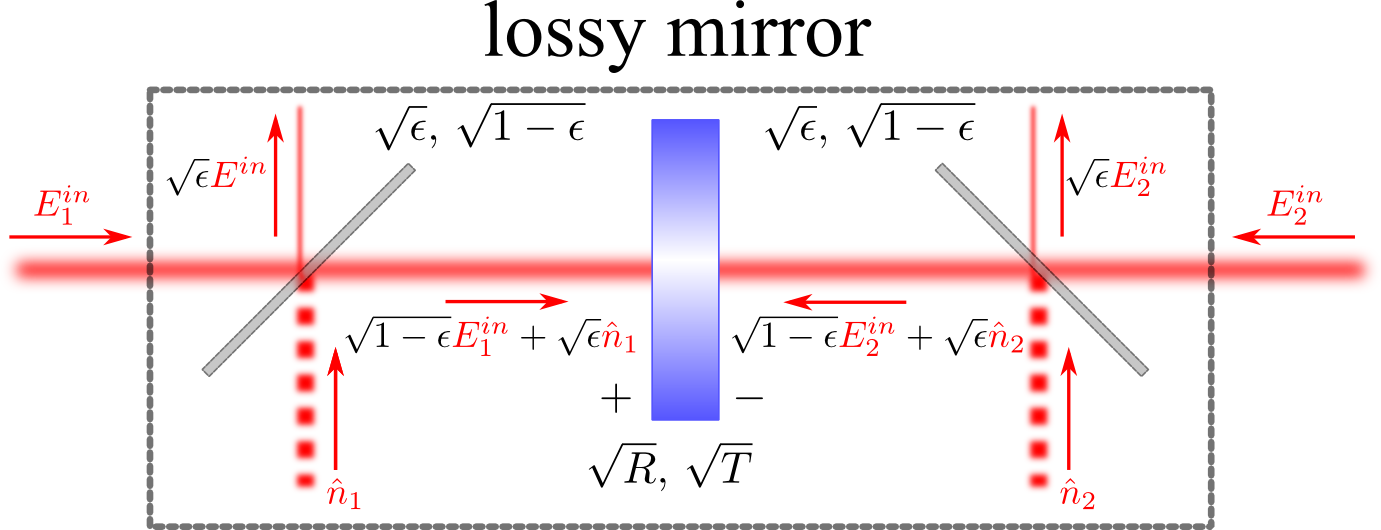}}
  \caption{Model of lossy mirror.}
\label{fig:lossy_mirror}
\end{figure}}

Using the above model, it is possible to show that for a lossy mirror the transformation of carrier fields given by Eq.~\eqref{eq:IO_mirror_matrix_4x4_zero} should be modified by simply multiplying the output fields vector by a factor $1-\epsilon$:
\begin{equation}
\label{eq:IO_lossy_mirror_relations_zero}
\begin{bmatrix}
  \vb{\mathcal{E}}^{\out}_{1}\\
  \vb{\mathcal{E}}^{\out}_{2}
\end{bmatrix}=
(1-\epsilon)\mathbb{M}_{\mathrm{real}}
\cdot
\begin{bmatrix}
  \vb{\mathcal{E}}^{\in}_{1}\\
  \vb{\mathcal{E}}^{\in}_{2}
\end{bmatrix} \simeq \mathbb{M}_{\mathrm{real}}
\cdot
\begin{bmatrix}
  \vb{\mathcal{E}}^{\in}_{1}\\
  \vb{\mathcal{E}}^{\in}_{2}
\end{bmatrix}
\,,
\end{equation}
where we used the fact that for low loss optics in use in GW interferometers, the absorption coefficient might be as small as $\epsilon\sim 10^{-5}$--$10^{-4}$. Therefore, the impact of optical loss on classical carrier amplitudes is negligible. Where the noise sidebands are concerned, the transformation rule given by Eq.~\eqref{eq:IO_mirror_matrix_4x4_first} changes a bit more:
\begin{eqnarray}
\label{eq:IO_lossy_mirror_relations_first}
  \begin{bmatrix}
  \hat{\boldsymbol{e}}^{\out}_{1}(\Omega)\nonumber\\
  \hat{\boldsymbol{e}}^{\out}_{2}(\Omega)
\end{bmatrix} &=&
(1-\epsilon)\mathbb{M}_{\mathrm{real}}
\cdot
\begin{bmatrix}
  \hat{\boldsymbol{e}}^{\in}_{1}(\Omega)\nonumber\\
  \hat{\boldsymbol{e}}^{\in}_{2}(\Omega)
\end{bmatrix} + \sqrt{\epsilon(1-\epsilon)}\mathbb{M}_{\mathrm{real}}\cdot
\begin{bmatrix}
  \hat{\boldsymbol{n}}_{1}(\Omega)\nonumber\\
  \hat{\boldsymbol{n}}_{2}(\Omega)
\end{bmatrix} \nonumber\\ &\simeq& 
(1-\epsilon)\mathbb{M}_{\mathrm{real}}
\cdot
\begin{bmatrix}
  \hat{\boldsymbol{e}}^{\in}_{1}(\Omega)\nonumber\\
  \hat{\boldsymbol{e}}^{\in}_{2}(\Omega)
\end{bmatrix} + \sqrt{\epsilon}
\begin{bmatrix}
  \hat{\boldsymbol{n}}_{1}'(\Omega)\\
  \hat{\boldsymbol{n}}_{2}'(\Omega)
\end{bmatrix}\,.
\end{eqnarray}
Here, we again used the smallness of $\epsilon\ll1$ and also the fact that matrix $\mathbb{M}_{\mathrm{real}}$ is unitary, i.e., we redefined the noise that enters outgoing fields due to loss as $\left\{\hat{\boldsymbol{n}}_{1}',\,\hat{\boldsymbol{n}}_{2}'\right\}^{\mathsf{T}} = \mathbb{M}\cdot\left\{\hat{\boldsymbol{n}}_{1},\,\hat{\boldsymbol{n}}_{2}\right\}^{\mathsf{T}}$, which keeps the new noise sources $\hat{n}_1'(t)$ and $\hat{n}_2'(t)$ uncorrelated: $\mean{\hat{n}_1(t)\hat{n}_2(t')}=\mean{\hat{n}_1'(t)\hat{n}_2'(t')}=0$.

\subsubsection{Light modulation by mirror motion}
\label{sec:Mirror_motion}

For full characterization of the light transformation in the GW
interferometers, one significant aspect remains untouched, i.e., the
field transformation upon reflection off the movable mirror. Above
(see Section~\ref{sec:phasemeter}), we have seen that motion
of the mirror yields phase modulation of the reflected wave. Let us
now consider this process in more detail.

\epubtkImage{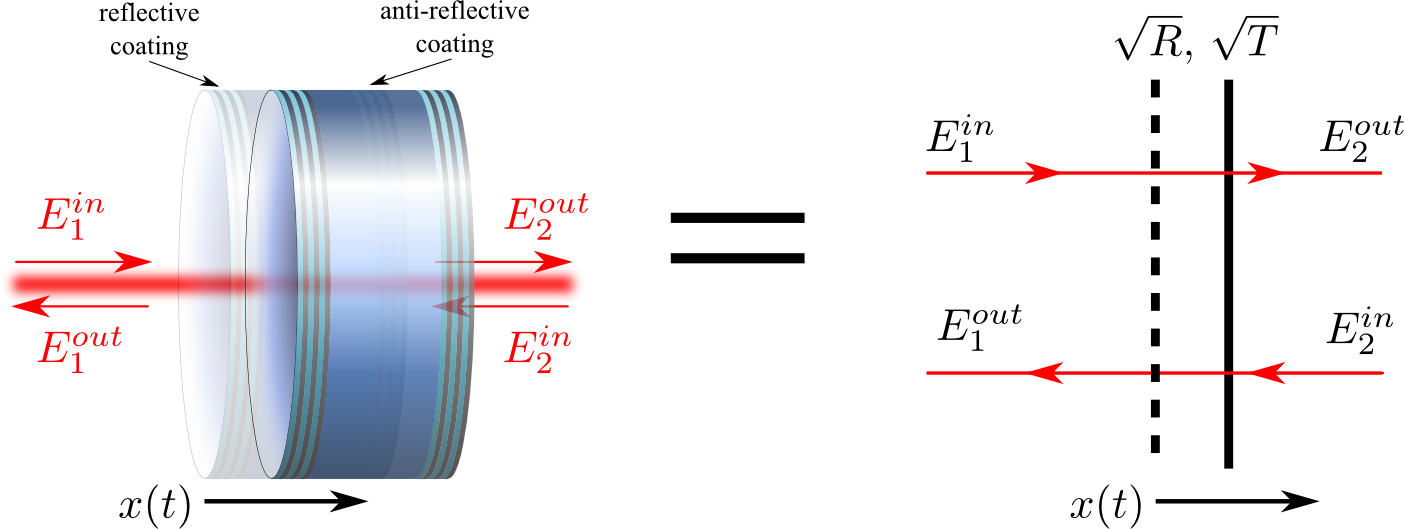}{%
\begin{figure}[htbp]
\centerline{\includegraphics[width=\textwidth]{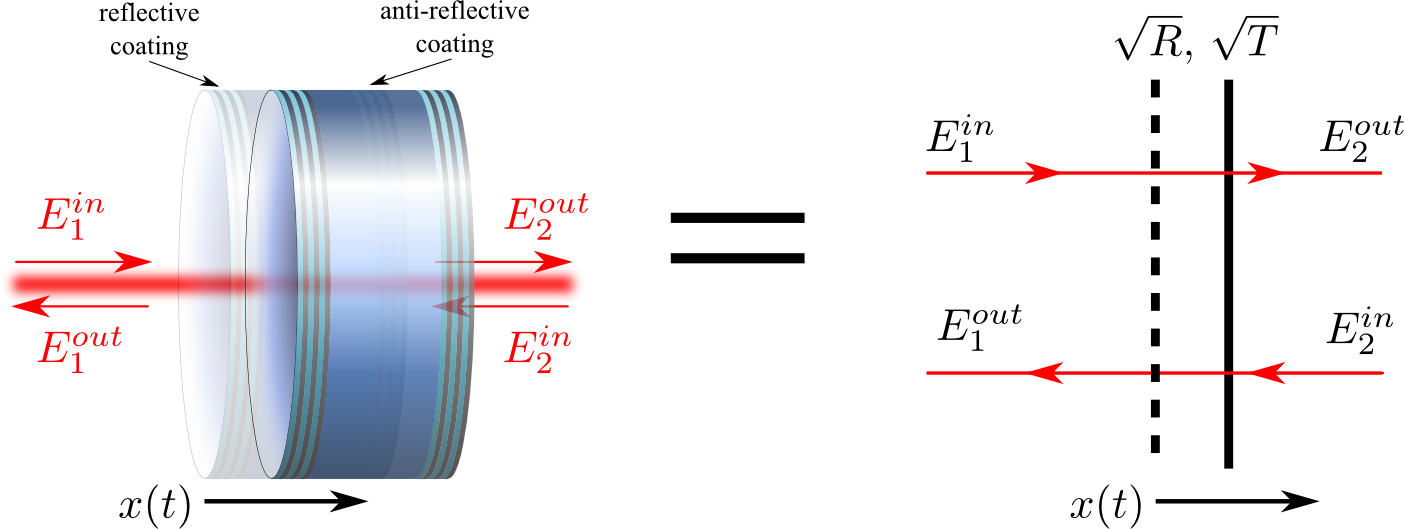}}
\caption{Reflection of light from the movable mirror.}
\label{fig:mov_mirr_refl}
\end{figure}}

Consider the mirror described by the matrix $\mathbb{M}_{\mathrm{real}}$, introduced above. Let us set the convention that the relations of input and output fields is written for the initial position of the movable mirror reflective surface, namely for the position where its displacement is $x=0$ as drawn in Figure~\ref{fig:mov_mirr_refl}. We assume the sway of the mirror motion to be much smaller than the optical wavelength: $x/\lambda_0\ll1$. The effect of the mirror displacement $x(t)$ on the outgoing field $E^{\out}_{1,2}(t)$ can be straightforwardly obtained from the propagation formalism. Indeed, considering the light field at a fixed spatial point, the reflected light field at any instance of time $t$ is just the result of propagation of the incident light by twice the mirror displacement taken at time of reflection and multiplied by reflectivity $\pm\sqrt{R}$\epubtkFootnote{In fact, the argument of $x$ should be written as $t_*$, that is the moment when the actual reflection takes place and is the solution to the equation: $c(t-t_*)=x(t_*)$, but since the mechanical motion is much slower than that of light one has $\delta x/c\ll 1$. This fact implies $t\simeq t_*$.}:
\begin{eqnarray}
\label{eq:IO_mirror_relations_x}
E^{\out}_1(t) = -\sqrt{R} E^{\in}_1(t-2x(t)/c)+\sqrt{T}E^{\in}_2(t)\,,\nonumber\\
E^{\out}_2(t) = \sqrt{T}E^{\in}_1(t)+\sqrt{R} E^{\in}_2(t+2x(t)/c)\,.
\end{eqnarray}
Remember now our assumption that $x\ll\lambda_0$; according to Eq.~\eqref{eq:EMW_free_prop_small_x_matrix} the mirror motion modifies the quadrature amplitudes in a way that allows one to separate this effect from the reflection. It means that the result of the light reflection from the moving mirror can be represented as a sum of two independently calculable effects, i.e., the reflection off the fixed mirror, as described above in Section~\ref{sec:light_reflection}, and the response to the mirror displacement (see Section~\ref{sec:propagation}), i.e., the signal presentable as a sideband vector $\left\{\boldsymbol{s}_1(\Omega),\boldsymbol{s}_2(\Omega)\right\}^{\mathsf{T}}$. The latter is convenient to describe in terms of the response vector $\{\boldsymbol{R}_1\,,\boldsymbol{R}_2\}^{\mathsf{T}}$ that is defined as:
\begin{eqnarray}
\boldsymbol{s}_1(\Omega)=\boldsymbol{R}_1 x(t) = -\sqrt{R}\delta\mathbb{P}\left[2\omega_0 x(t)/c\right]\cdot\vb{\mathcal{E}}^{\in}_{1} = -\dfrac{2\omega_0\sqrt{R}}{c}
\begin{bmatrix}
\mathcal{E}^{\in}_{1s}\\
-\mathcal{E}^{\in}_{1c}
\end{bmatrix} x(t) \nonumber\\
\boldsymbol{s}_2(\Omega) = \boldsymbol{R}_2 x(t) = -\sqrt{R}\delta\mathbb{P}\left[2\omega_0x(t)/c\right]\cdot\vb{\mathcal{E}}^{\in}_{2} = -\dfrac{2\omega_0\sqrt{R}}{c}
\begin{bmatrix}
\mathcal{E}^{\in}_{2s}\\
-\mathcal{E}^{\in}_{2c}
\end{bmatrix} x(t)\,.\label{eq:IO_motion_induced_sidebands}
\end{eqnarray}

Note that we did not include sideband fields $\hat{\boldsymbol{e}}^{\in}_{1,2}(\Omega)$ in the definition of the response vector. In principle, sideband fields also feel the motion induced phase shift; however, as far as it depends on the product of one very small value of $2\omega_0x(t)/c=4\pi x(t)/\lambda_0\ll1$ by a small sideband amplitude $|\hat{\boldsymbol{e}}^{\in}_{1,2}(\Omega)|\ll |\mathcal{E}^{\in}_{1,2}|$, the resulting contribution to the final response will be dwarfed by that of the classical fields. Moreover, the mirror motion induced contribution~\eqref{eq:IO_motion_induced_sidebands} is itself a quantity of the same order of magnitude as the noise sidebands, and therefore we can claim that the classical amplitudes of the carrier fields are not affected by the mirror motion and that the relations~\eqref{eq:IO_mirror_matrix_4x4_zero} hold for a moving mirror too. However, the relations for sideband amplitudes must be modified. In the case of a lossless mirror, relations~\eqref{eq:IO_mirror_matrix_4x4_first} turn:
\begin{equation}
\label{eq:IO_mirror_relations_first+x}
\begin{bmatrix}
  \hat{\boldsymbol{e}}^{\out}_{1}(\Omega)\\
  \hat{\boldsymbol{e}}^{\out}_{2}(\Omega)
\end{bmatrix}=
\mathbb{M}_{\mathrm{real}}
\cdot
\begin{bmatrix}
  \hat{\boldsymbol{e}}^{\in}_{1}(\Omega)\\
  \hat{\boldsymbol{e}}^{\in}_{2}(\Omega)
\end{bmatrix} +
\begin{bmatrix}
\boldsymbol{R}_1\\
\boldsymbol{R}_2
\end{bmatrix}x(\Omega)\,,
\end{equation}
where $x(\Omega)$ is the Fourier transform of the mirror displacement $x(t)$
$$x(\Omega) = \intinfty dt\, x(t)e^{i\Omega t}\,.$$

It is important to understand that signal sidebands characterized by a vector $\left\{\boldsymbol{s}_1(\Omega),\boldsymbol{s}_2(\Omega)\right\}^{\mathsf{T}}$, on the one hand, and the noise sidebands $\left\{\vb{\hat{e}}_1(\Omega),\,\vb{\hat{e}}_2(\Omega)\right\}^{\mathsf{T}}$, on the other hand, have the same order of magnitude in the real GW interferometers, and the main role of the advanced quantum measurement techniques we are talking about here is to either increase the former, or decrease the latter as much as possible in order to make the ratio of them, known as the \emph{signal-to-noise ratio} (SNR), as high as possible in as wide as possible a frequency range.

\subsubsection{Simple example: the reflection of light from a perfect moving mirror}

All the formulas we have derived here, though being very simple in essence, look cumbersome and not very transparent in general. In most situations, these expressions can be simplified significantly in real schemes. Let us consider a simple example for demonstration purposes, i.e., consider the reflection of a single light beam from a perfectly reflecting ($R=1$) moving mirror as drawn in Figure~\ref{fig8}. The initial phase $\phi_0$ of the incident wave does not matter and can be taken as zero. Then $\mathcal{E}^{\in}_c = \mathcal{E}_0$ and $\mathcal{E}^{\in}_s = 0$. Putting these values into Eq.~\eqref{eq:IO_mirror_matrix_4x4_zero} and accounting for $\vb{\mathcal{E}}^{\in}_2=0$, quite reasonably results in the amplitude of the carrier wave not changing upon reflection off the mirror, while the phase changes by $\pi$:
$$\mathcal{E}^{\out}_c = -\mathcal{E}^{\out}_c = -\mathcal{E}_0,\quad \mathcal{E}^{\out}_s = 0\,.$$
Since we do not have control over the laser noise, the input light has laser fluctuations in both quadratures $\hat{\boldsymbol{e}}^{\in}_1 = \{\hat{e}^{\in}_{1c},\hat{e}^{\in}_{1s}\}$ that are transformed in full accordance with Eq.~\eqref{eq:IO_mirror_matrix_4x4_first}):
$$\hat{\boldsymbol{e}}^{\out}_1(\Omega) = -\hat{\boldsymbol{e}}^{\in}_1(\Omega)\,.$$
Again, nothing surprising. Let us see what happens with a mechanical motion induced component of the reflected wave: according to Eq.~\eqref{eq:IO_mirror_relations_first+x}, the reflected light will contain a motion-induced signal in the s-quadrature:
$$\boldsymbol{s}(\Omega) = \dfrac{2 \omega_0}{c}
\begin{bmatrix}
0\\
\mathcal{E}_0
\end{bmatrix}x(\Omega)\,.
$$
This fact, i.e., that the mirror displacement that just causes phase modulation of the reflected field, enters only the s-quadrature, once again justifies why this quadrature is usually referred to as \emph{phase quadrature} (cf. section \ref{sec:modulation}).

\epubtkImage{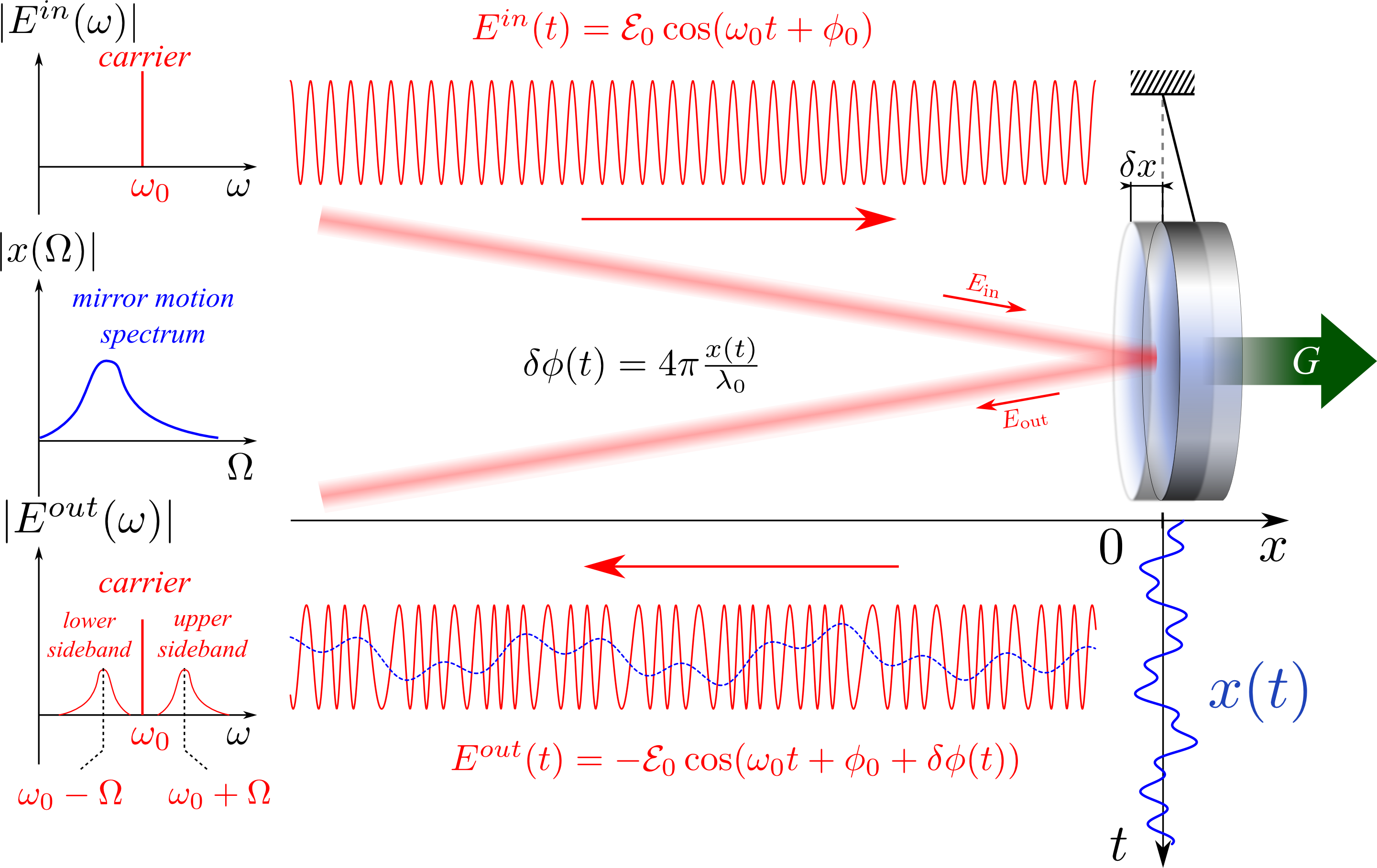}{%
\begin{figure}[htbp]
\centerline{\includegraphics[width=\textwidth]{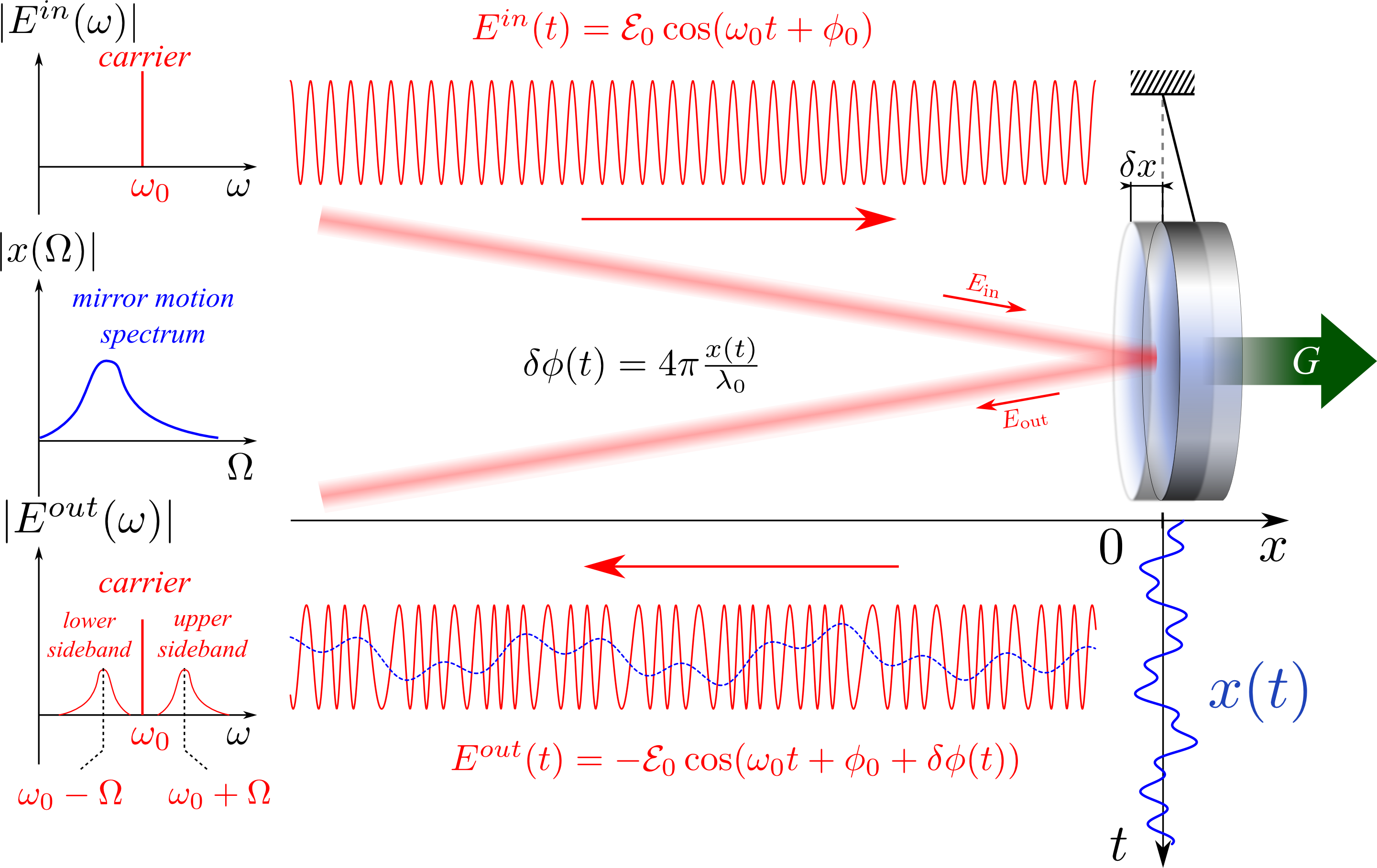}}
\caption{Schematic view of light modulation by perfectly reflecting mirror motion.
An initially monochromatic laser field $E^{\in}(t)$ with frequency $\omega_0 = 2\pi c/\lambda_0$ gets reflected from the mirror that commits slow (compared to optical oscillations) motion $x(t)$ (blue line) under the action of external force $\boldsymbol{G}$. Reflected the light wave phase is modulated by the mechanical motion so that the spectrum of the outgoing field $E^{\out}(\omega)$ contains two sidebands carrying all the information about the mirror motion. The left panel shows the spectral representation of the initial monochromatic incident light wave (upper plot), the mirror mechanical motion amplitude spectrum (middle plot) and the spectrum of the phase-modulated by the mirror motion, reflected light wave (lower plot).}
\label{fig8}
\end{figure}}

It is instructive to see the spectrum of the outgoing light
in the above considered situation. It is, expectedly, the spectrum of
a phase modulated monochromatic wave that has a central peak at the
carrier wave frequency and the two sideband peaks on either sides of
the central one, whose shape follows the spectrum of the modulation
signal, in our case, the spectrum of the mechanical displacement of
the mirror $x(t)$. The left part of Figure~\ref{fig8} illustrates
the aforesaid. As for laser noise, it enters the outgoing light in an
additive manner and the typical (though simplified) amplitude spectrum
of a noisy light reflected from a moving mirror is given in
Figure~\ref{fig9}.

\epubtkImage{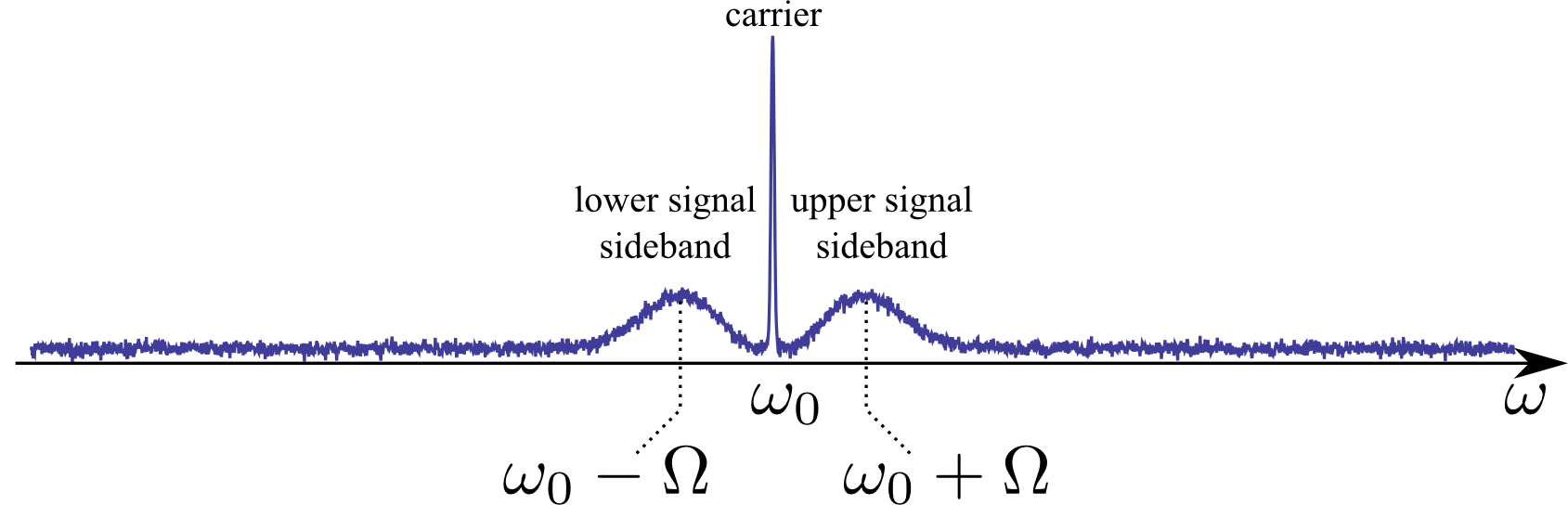}{%
\begin{figure}[htbp]
  \centerline{\includegraphics[width=\textwidth]{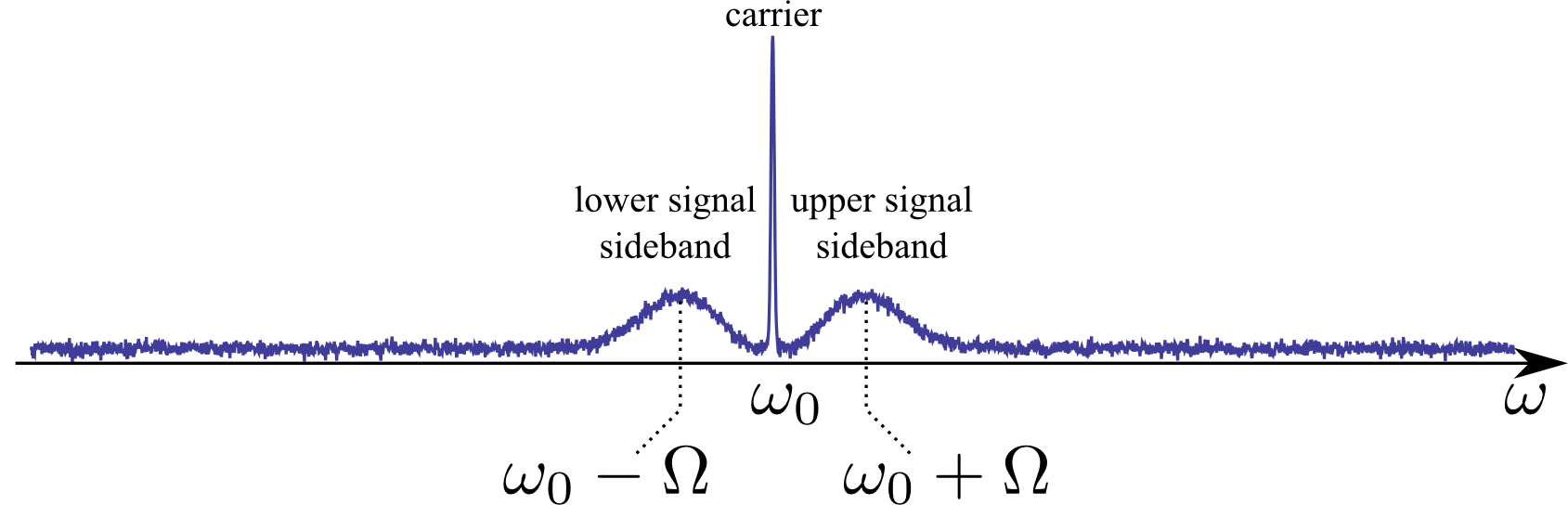}}
  \caption{The typical spectrum (amplitude spectral density) of the light leaving the interferometer with movable mirrors. The central peak corresponds to the carrier light with frequency $\omega_0$, two smaller peaks on either side of the carrier represent the signal sidebands with the shape defined by the mechanical motion spectrum $x(\Omega)$; the noisy background represents laser noise.}
\label{fig9}
\end{figure}}

\subsection{Basics of Detection: Heterodyne and homodyne readout techniques}
\label{subsec:Detection}

Let us now address the question of how one can detect a GW signal imprinted onto the parameters of the light wave passing through the interferometer. The simple case of a Michelson interferometer considered in Section~\ref{sec:MI} where the GW signal was encoded in the phase quadrature of the light leaking out of the signal(dark) port, does not exhaust all the possibilities. In more sophisticated interferometer setups that are covered in sections \ref{sec:QN_in_GW_interferometers} and \ref{sec:sub-SQL_schemes}, a signal component might be present in both quadratures of the outgoing light and, actually, to different extent at different frequencies; therefore, a detection method that allows measurement of an arbitrary combination of amplitude and phase quadrature is required. The two main methods are in use in contemporary GW detectors: these are \emph{homodyne} and \emph{heterodyne} detection~\cite{03a1BuChMa, 06a1SoChKaMi, 08a1Wa_etal, 09a1Hi_etal}. Both are common in radio-frequency technology as methods of detection of phase- and frequency-modulated signals. The basic idea is to mix a faint signal wave with a strong \emph{local oscillator} wave, e.g., by means of a beamsplitter, and then send it to a detector with a quadratic non-linearity that shifts the spectrum of the signal to lower frequencies together with amplification by an amplitude of the local oscillator. This topic is also discussed in Section~4 of the Living Review by Freise and Strain~\cite{lrr-2010-1} with more details relevant to experimental implementation.

\subsubsection{Homodyne and DC readout}
\label{sec:homodyne}

\paragraph*{Homodyne readout.}

Homodyne detection uses local oscillator light with the same carrier frequency as the signal. Write down the signal wave as:
\begin{equation}
\label{eq:homodyne_signal_wave}
S(t) = S_c(t)\cos\omega_0t+S_s(t)\sin\omega_0t
\end{equation}
and the local oscillator wave as:
\begin{equation}
\label{eq:homodyne_LO}
L(t) = L_c(t)\cos\omega_0t+L_s(t)\sin\omega_0t\,.
\end{equation}
Signal light quadrature amplitudes $S_{c,s}(t)$ might contain GW signal $G_{c,s}(t)$ as well as quantum noise $n_{c,s}(t)$ in both quadratures:
$$S_{c,s}(t) = G_{c,s}(t)+n_{c,s}(t)\,,$$
while the local oscillator is a laser light with classical amplitudes: 
$$(L^{(0)}_c,\,L^{(0)}_s)=(L_0\cos\phi_{LO},\,L_0\sin\phi_{LO})\,,$$ 
where we introduced a \emph{homodyne angle} $\phi_{LO}$, and laser noise $l_{c,s}(t)$:
$$L_{c,s}(t) = L^{(0)}_{c,s}+l_{c,s}(t)\,.$$
Note that the local oscillator classical amplitude $L_0$ is much larger than all other signals:
$$L_0\gg\max\left[l_{c,s},\,G_{c,s},\,n_{c,s}\right]\,.$$
Let mix these two lights at the beamsplitter as drawn in the left panel of Figure~\ref{fig:homodyne_readout_scheme} and detect the two resulting outgoing waves with two photodetectors. The two photocurrents $i_{1,2}$ will be proportional to the intensities $I_{1,2}$ of these two lights:
\begin{eqnarray*}
  i_1\propto I_1\propto \frac{\tmean{(L+S)^2}}{2} =\frac{L_0^2}{2}+L_0(G_c+l_c+n_c)\cos\phi_{LO}+L_0(G_s+l_s+n_s)\sin\phi_{LO}+\mathcal{O}\left[G_{c,s}^2,l_{c,s}^2,n_{c,s}^2\right]\,,\\
 i_2\propto I_2\propto \frac{\tmean{(L-S)^2}}{2} =\frac{L_0^2}{2}-L_0(G_c-l_c+n_c)\cos\phi_{LO}-L_0(G_s-l_s+n_s)\sin\phi_{LO}+\mathcal{O}\left[G_{c,s}^2,l_{c,s}^2,n_{c,s}^2\right]\,,
\end{eqnarray*}
where $\tmean{A}$ stands for time averaging of $A(t)$ over many optical oscillation periods, which reflects the inability of photodetectors to respond at optical frequencies and thus providing natural low-pass filtering for our signal. The last terms in both expressions gather all the terms quadratic in GW signal and both noise sources that are of the second order of smallness compared to the local oscillator amplitude $L_0$ and thus are omitted in further consideration.

\epubtkImage{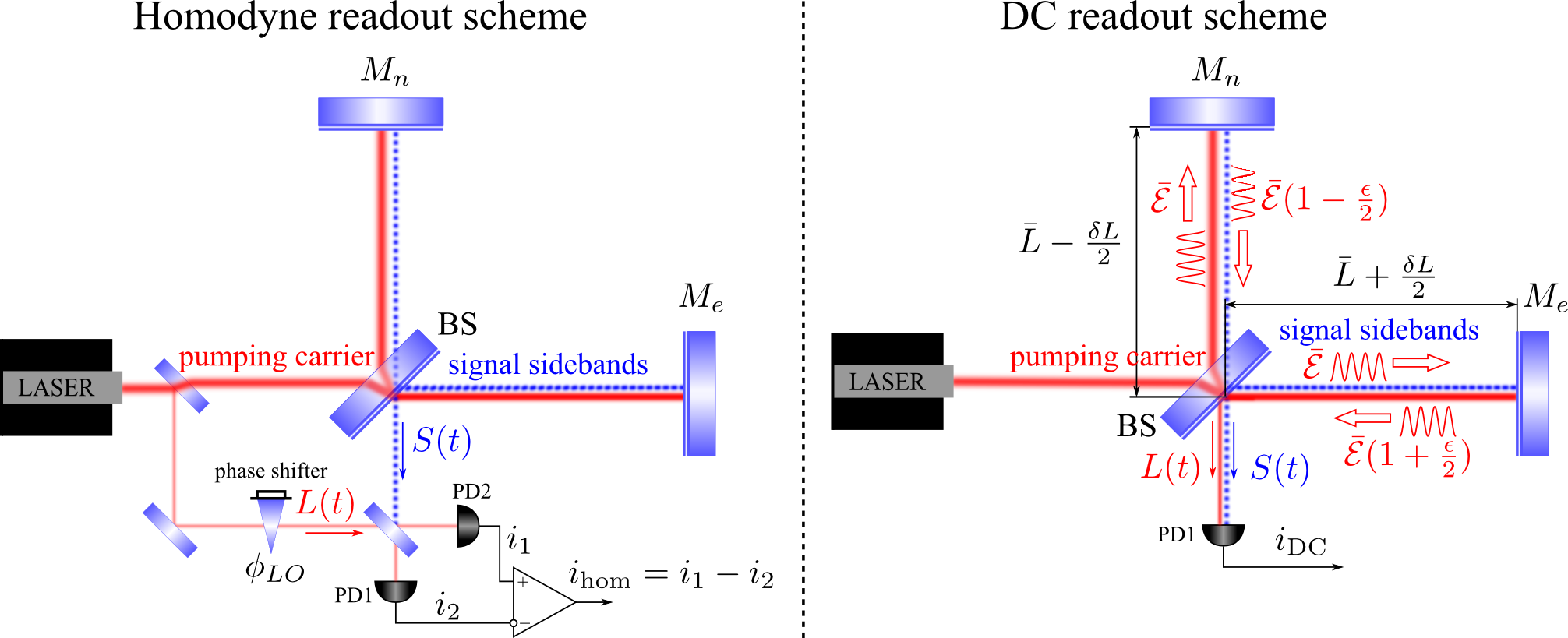}{%
\begin{figure}[htb]
  \centerline{\includegraphics[width=\textwidth]{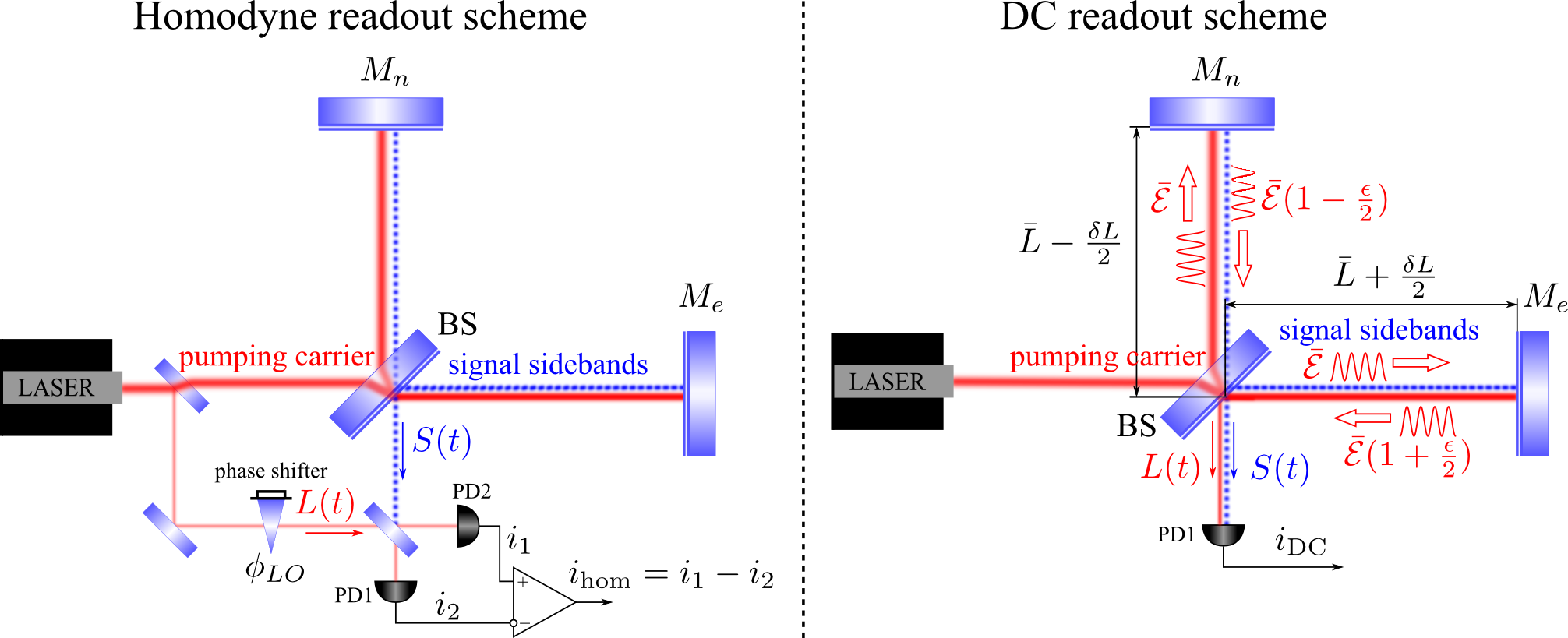}}
  \caption{Schematic view of homodyne readout (\emph{left panel}) and DC readout (\emph{right panel}) principle implemented by a simple Michelson interferometer.}
\label{fig:homodyne_readout_scheme}
\end{figure}}

In a classic homodyne balanced scheme, the difference current is read out that contains only a GW signal and quantum noise of the dark port:
\begin{equation}
\label{eq:homodyne_photocurrent}
i_{\mathrm{hom}}=i_1-i_2\propto 2L_0\left[(G_c+n_c)\cos\phi_{LO}+(G_s+n_s)\sin\phi_{LO}\right]\,.
\end{equation}
Whatever quadrature the GW signal is in, by proper choice of the homodyne angle $\phi_{LO}$ one can recover it with minimum additional noise. That is how homodyne detection works in theory.

However, in real interferometers, the implementation of a homodyne readout appears to be fraught with serious technical difficulties. In particular, the local oscillator frequency has to be kept extremely stable, which means its optical path length and alignment need to be actively stabilized by a low-noise control system~\cite{09a1Hi_etal}. This inflicts a significant increase in the cost of the detector, not to mention the difficulties in taming the noise of stabilising control loops, as the experience of the implementation of such stabilization in a Garching prototype interferometer has shown \cite{PhysLettA.277.3.135_Freise,PhysRevLett.81.5493_Heinzel,99pth1Heinzel}.

\paragraph*{DC readout.}

These factors provide a strong motivation to look for another way to implement homodyning. Fortunately, the search was not too long, since the suitable technique has already been used by Michelson and Morley in their seminal experiment~\cite{1887a1MiMo}. The technique is known as DC-readout and implies an introduction of a constant arms length difference, thus pulling the interferometer out of the dark fringe condition as was mentioned in Section~\ref{sec:MI}. The advantage of this method is that the local oscillator is furnished by a part of the pumping carrier light that leaks into the signal port due to arms imbalance and thus shares the optical path with the signal sidebands. It automatically solves the problem of phase-locking the local oscillator and signal lights, yet is not completely free of drawbacks. The first suggestion to use DC readout in GW interferometers belongs to Fritschel~\cite{2003LSC_Fritschel} and then got comprehensive study by the GW community~\cite{06a1SoChKaMi, 08a1Wa_etal, 09a1Hi_etal}.

Let us discuss how it works in a bit more detail. The schematic view of a Michelson interferometer with DC readout is drawn in the right panel of Figure~\ref{fig:homodyne_readout_scheme}. As already mentioned, the local oscillator light is produced by a deliberately-introduced constant difference $\delta L$ of the lengths of the interferometer arms. It is also worth noting that the component of this local oscillator created by the asymmetry in the reflectivity of the arms that is always the case in a real interferometer and attributable mostly to the difference in the absorption of the `northern' and `eastern' end mirrors as well as asymmetry of the beamsplitter. All these factors can be taken into account if one writes the carrier fields at the beamsplitter after reflection off the arms in the following symmetric form:
\begin{eqnarray*}
  E^{\out}_n(t) = -\frac{E_0}{\sqrt{2}}(1-\epsilon_n)\cos\omega_0(t-2L_n/c)= -\frac{\bar{\mathcal{E}}}{\sqrt{2}}(1-\Delta\epsilon)\cos\omega_0(\tmean{t}+\Delta L/c)\,,\\
  E^{\out}_e(t) = -\frac{E_0}{\sqrt{2}}(1-\epsilon_e)\cos\omega_0(t-2L_e/c)= -\frac{\bar{\mathcal{E}}}{\sqrt{2}}(1+\Delta\epsilon)\cos\omega_0(\tmean{t}-\Delta L/c)\,,
\end{eqnarray*}
where $\epsilon_{n,e}$ and $L_{n,e}$ stand for optical loss and arm lengths of the corresponding interferometer arms, $\Delta\epsilon = \frac{\epsilon_n-\epsilon_e}{2(1-\bar{\epsilon})}$ is the optical loss relative asymmetry with $\bar{\epsilon}=(\epsilon_n+\epsilon_e)/2$, $\tmean{\mathcal{E}}= E_0(1-\bar{\epsilon})$ is the mean pumping carrier amplitude at the beamsplitter, $\Delta L = L_n-L_e$ and $\bar{t} = t+\frac{L_n+L_e}{2c}$. Then the classical part of the local oscillator light in the signal (dark) port will be given by the following expression:
 \begin{equation}
\label{eq:DCreadout_LO}
L^{(0)}_{\mathrm{DC}}(t) =\frac{ E^{\out}_n(t)-E^{\out}_e(t)}{\sqrt{2}} = \underbrace{\bar{\mathcal{E}}\Delta\epsilon\cos\frac{\omega_0\Delta L}{c}}_{\sqrt{2}L_c^{(0)}}\cos\omega_0\bar{t}+\underbrace{\bar{\mathcal{E}}\sin\frac{\omega_0\Delta L}{c}}_{\sqrt{2}L_s^{(0)}}\sin\omega_0\bar{t}\,,
\end{equation}
where one can define the local oscillator phase and amplitude through the apparent relations:
\begin{equation}
\label{eq:DCreadout_phiLO}
\tan\phi_{\mathrm{DC}} = \frac{1}{\Delta\epsilon}\tan\frac{\omega_0\Delta L}{c}\,\quad L_{\mathrm{DC}}^{(0)} \simeq \bar{\mathcal{E}}\sqrt{(\Delta\epsilon)^2+\left(\frac{\omega_0\Delta L}{c}\right)^2}\simeq \frac{\omega_0E_0\Delta L}{c}\,,
\end{equation}
where we have taken into account that $\omega_0\Delta L/c\ll1$ and the rather small absolute value of the optical loss coefficient $\max\left[\epsilon_n,\epsilon_e\right]\sim 10^{-4}\ll1$ available in contemporary interferometers.
One sees that were there no asymmetry in the arms optical loss, there would be no opportunity to change the local oscillator phase. At the same time, the GW signal in the considered scheme is confined to the phase quadrature since it comprises the time-dependent part of $\Delta L$ and thus the resulting photocurrent will be proportional to:
\begin{equation}
\label{eq:DCreadout_photocurrent}
i_{\mathrm{DC}}\propto\tmean{(L+S)^2}\simeq \left(L_{\mathrm{DC}}^{(0)}\right)^2+2L_{\mathrm{DC}}^{(0)}(l^{\out}_c+n_c)\cos\phi_{\mathrm{DC}}+2L_{\mathrm{DC}}^{(0)}(G_s+l^{\out}_s+n_s)\sin\phi_{\mathrm{DC}},
\end{equation}
where $l_{c,s}^{\out}$ denote the part of the input laser noise that leaked into the output port:
\begin{eqnarray}
l_c^{\out}(t) \simeq l_c^{\in}\Delta\epsilon\cos\frac{\omega_0\Delta L}{c}-l_s^{\in}\sin\frac{\omega_0\Delta L}{c}\simeq l_c^{\in}\Delta\epsilon-l_s^{\in}\frac{\omega_0\Delta L}{c}\,,\label{eq:DCreadout_laser_noise_cos}\\
l_s^{\out}(t) \simeq l_c^{\in}\sin\frac{\omega_0\Delta L}{c}+l_s^{\in}\Delta\epsilon\cos\frac{\omega_0\Delta L}{c}\simeq l_c^{\in}\frac{\omega_0\Delta L}{c}+l_s^{\in}\Delta\epsilon\label{eq:DCreadout_laser_noise_sin}\,,
\end{eqnarray}
and $n_{c,s}$ stand for the quantum noise associated with the signal sidebands and entering the interferometer from the signal port.

In the case of a small offset of the interferometer from the dark fringe condition, i.e., for $\omega_0\Delta L/c=2\pi\Delta L/\lambda_0\ll1$, the readout signal scales as local oscillator classical amplitude, which is directly proportional to the offset itself: $L_{\mathrm{DC}}^{(0)}\simeq 2\pi E_0\frac{\Delta L}{\lambda_0}$. The laser noise associated with the pumping carrier also leaks to the signal port in the same proportion, which might be considered as the main disadvantage of the DC readout as it sets rather tough requirements on the stability of the laser source, which is not necessary for the homodyne readout. However, this problem, is partly solved in more sophisticated detectors by implementing power recycling and/or Fabry--P\'erot cavities in the arms. These additional elements turn the Michelson interferometer into a resonant narrow-band cavity for a pumping carrier with effective bandwidth determined by transmissivities of the power recycling mirror (PRM) and/or input test masses (ITMs) of the arm cavities divided by the corresponding cavity length, which yields the values of bandwidths as low as $\sim$~10~Hz. Since the target GW signal occupies higher frequencies, the laser noise of the local oscillator around signal frequencies turns out to be passively filtered out by the interferometer itself.

DC readout has already been successfully tested at the LIGO 40-meter interferometer in Caltech~\cite{08a1Wa_etal} and implemented in GEO\,600~\cite{07pth1Hi, 09a1Hi_etal, 2010a1De_etal} and in Enhanced LIGO~\cite{2011LSC_Fricke, 2009LSC_TechPaper_T0900023}. It has proven a rather promising substitution for the previously ubiquitous heterodyne readout (to be considered below) and has become a baseline readout technique for future GW detectors~\cite{09a1Hi_etal}.

\subsubsection{Heterodyne readout}
\label{sec:heterodyne}

Up until recently, the only readout method used in terrestrial GW detectors has been the heterodyne readout. Yet with more and more stable lasers being available for the GW community, this technique gradually gives ground to a more promising DC readout method considered above. However, it is instructive to consider briefly how heterodyne readout works and learn some of the reasons, that it has finally given way to its homodyne adversary.

\epubtkImage{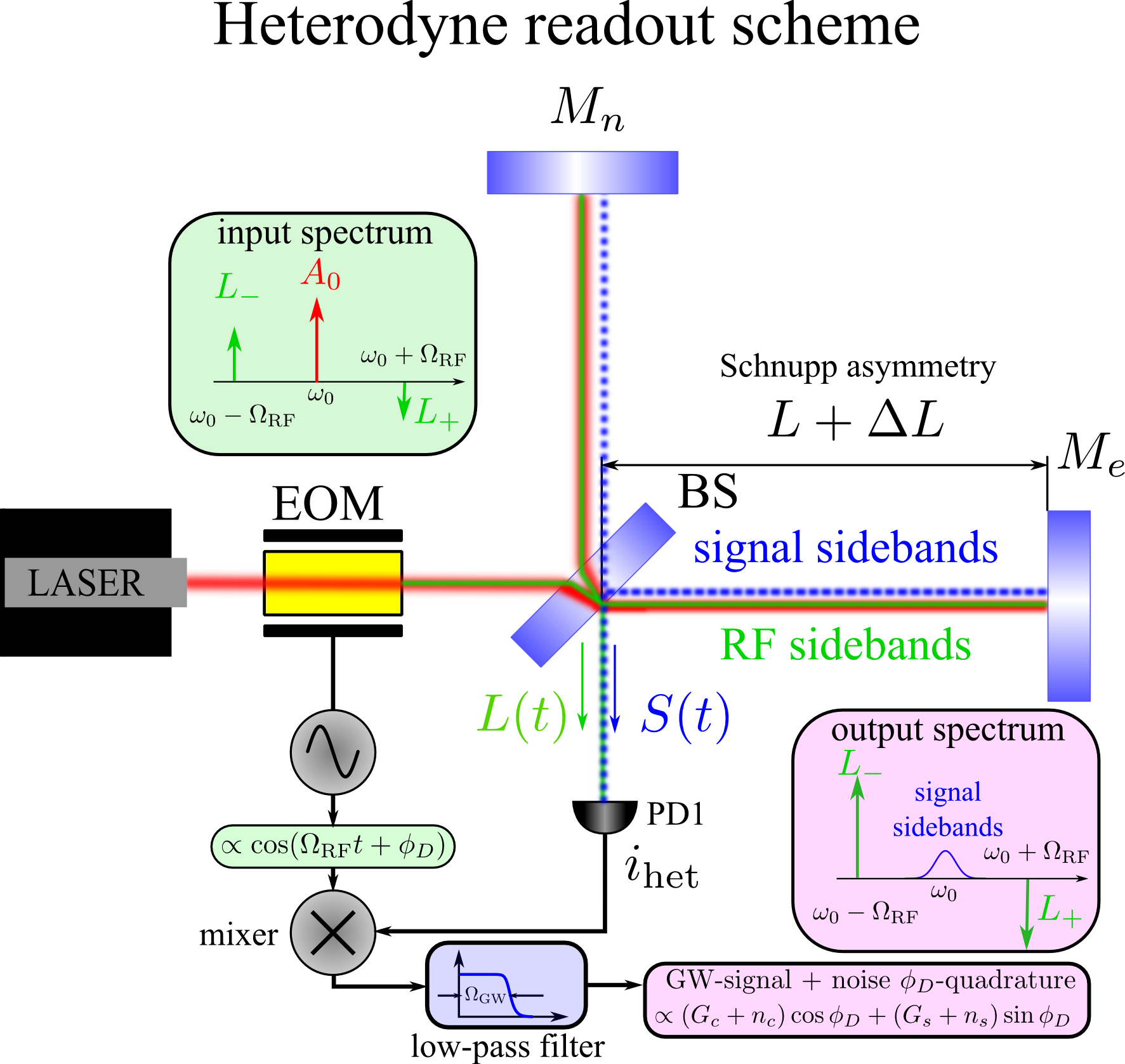}{%
\begin{figure}[htb]
  \centerline{\includegraphics[width=.75\textwidth]{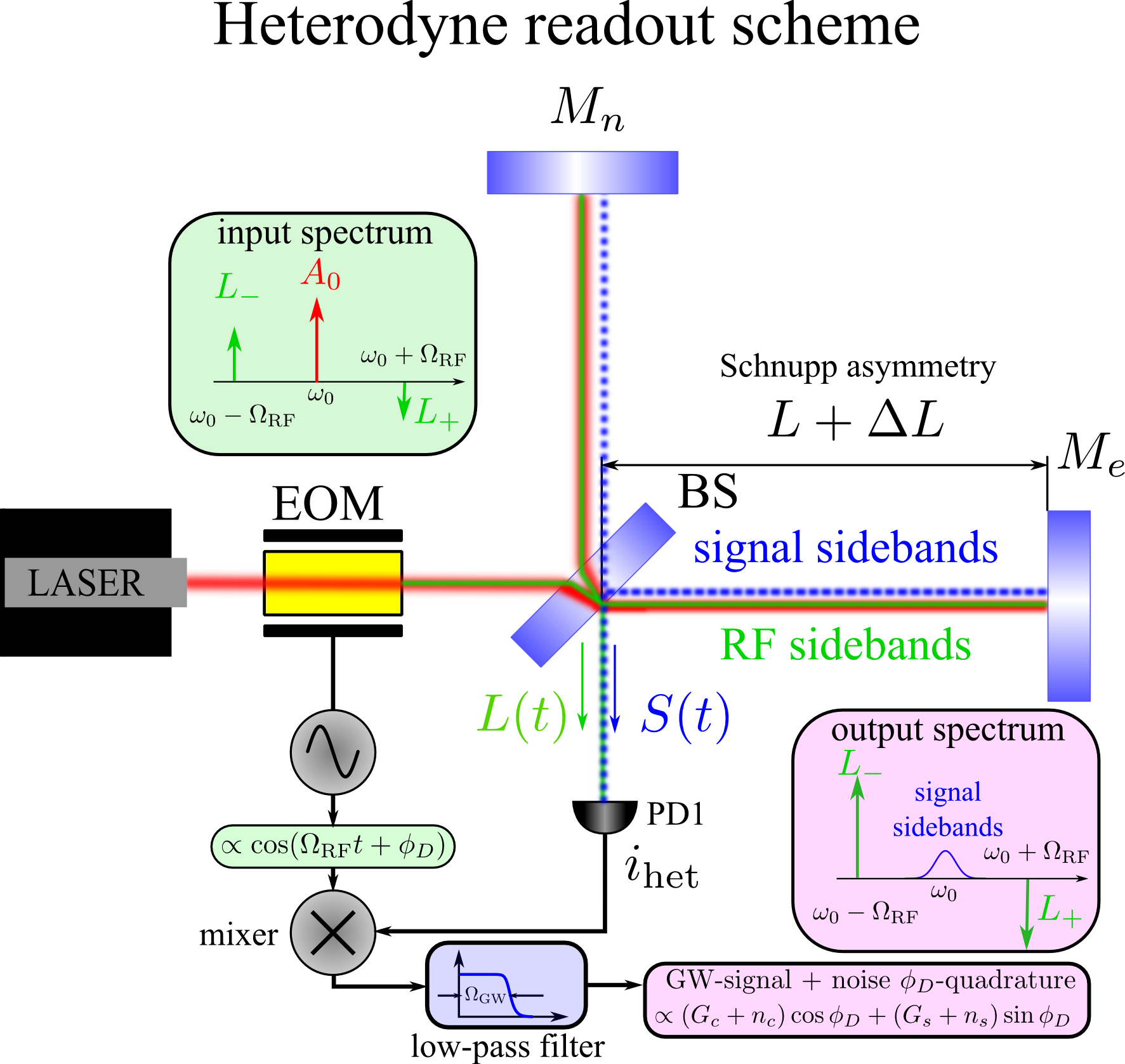}}
  \caption{Schematic view of heterodyne readout principle implemented by a simple Michelson interferometer. Green lines represent modulation sidebands at radio frequency $\Omega_{\mathrm{RF}}$ and blue dotted lines feature signal sidebands}
\label{fig:heterodyne_readout_scheme}
\end{figure}}

The idea behind the heterodyne readout principle is the generalization of the homodyne readout, i.e., again, the use of strong local oscillator light to be mixed up with the faint signal light leaking out the dark port of the GW interferometer save the fact that local oscillator light frequency is shifted from the signal light carrier frequency by $\Omega_{\mathrm{RF}}\sim$ several megahertz. In GW interferometers with heterodyne readout, local oscillator light of different than $\omega_0$ frequency is produced via phase-modulation of the pumping carrier light by means of electro-optical modulator (EOM) before it enters the interferometer as drawn in Figure~\ref{fig:heterodyne_readout_scheme}. The interferometer is tuned so that the readout port is dark for the pumping carrier. At the same time, by introducing a macroscopic (several centimeters) offset $\Delta L$ of the two arms, which is known as Schnupp asymmetry~\cite{1988Conf_Schnupp}, the modulation sidebands at radio frequency $\Omega_{\mathrm{RF}}$ appear to be optimally transferred from the pumping port to the readout one. Therefore, the local oscillator at the readout port comprises two modulation sidebands, $L_{\mathrm{het}}(t) = L_+(t)+L_-(t)$, at frequencies $\omega_0+\Omega_{\mathrm{RF}}$ and $\omega_0-\Omega_{\mathrm{RF}}$, respectively. These two are detected together with the signal sidebands at the photodetector, and then the resulting photocurrent is demodulated with the RF-frequency reference signal yielding an output current proportional to GW-signal.

This method was proposed and studied in great detail in the following works~\cite{JMO.34.6.793_1987, 1988Conf_Schnupp, PhysLettA.277.3.135_Freise,PhysRevLett.81.5493_Heinzel,99pth1Heinzel,PhysRevA.44.4693, PhysRevA.43.5022} where the heterodyne technique for GW interferometers tuned in resonance with pumping carrier field was considered and, therefore, the focus was made on the detection of only phase quadrature of the outgoing GW signal light. This analysis was further generalized to detuned interferometer configurations in~\cite{03a1BuChMa, 06a1SoChKaMi} where the full analysis of quantum noise in GW dual-recycled interferometers with heterodyne readout was done.

Let us see in a bit more detail how the heterodyne readout works as exemplified by a simple Michelson interferometer drawn in Figure~\ref{fig:heterodyne_readout_scheme}. The equation of motion at the input port of the interferometer creates two phase-modulation sideband fields ($L_+(t)$ and $L_-(t)$) at frequencies $\omega_0\pm\Omega_{\mathrm{RF}}$:
\begin{eqnarray*}
  L_+(t) = \left[L^{(0)}_{c+}+l_{c+}\right]\cos(\omega_0+\Omega_{\mathrm{RF}})t+\left[L^{(0)}_{s+}+l_{s+}\right]\sin(\omega_0+\Omega_{\mathrm{RF}})t\,,\\
 L_-(t) = \left[L^{(0)}_{c-}+l_{c-}\right]\cos(\omega_0-\Omega_{\mathrm{RF}})t+\left[L^{(0)}_{s-}+l_{s-}\right]\sin(\omega_0-\Omega_{\mathrm{RF}})t\,,
\end{eqnarray*}
where $L^{(0)}_{(c,s)\pm}$ stand for classical quadrature amplitudes of the modulation (upper and lower) sidebands\epubtkFootnote{In the resonance-tuned case, the phase modulation of the input carrier field creates equal magnitude sideband fields as discussed in Section~\ref{sec:modulation}, and these sideband fields are transmitted to the output port thanks to Schnupp asymmetry in the same state, i.e., they remain equal in magnitude and reside in the phase quadrature. In detuned configurations of GW interferometers, the upper and lower RF-sideband fields are transformed differently, which influences both their amplitudes and phases at the readout port.} and $l_{(c,s)\pm}(t)$ represent laser noise around the corresponding modulation frequency.

Unlike the homodyne readout schemes, in the heterodyne ones, not only the quantum noise components $n_{c,s}^{\omega_0}$ falling into the GW frequency band around the carrier frequency $\omega_0$ has to be accounted for but also those rallying around twice the RF modulation frequencies $\omega_0\pm2\Omega_{\mathrm{RF}}$:
\begin{eqnarray}
  S_{\mathrm{het}}(t) &=&
  (G_c+n_c^{\omega_0})\cos\omega_0t+(G_s+n_s^{\omega_0})\sin\omega_0t \nonumber\\
  && + n_c^{\omega_0+2\Omega_{\mathrm{RF}}}\cos(\omega_0+2\Omega_{\mathrm{RF}})t+n_s^{\omega_0+2\Omega_{\mathrm{RF}}}\sin(\omega_0+2\Omega_{\mathrm{RF}})t \\
  && + n_c^{\omega_0-2\Omega_{\mathrm{RF}}}\cos(\omega_0-2\Omega_{\mathrm{RF}})t+n_s^{\omega_0-2\Omega_{\mathrm{RF}}}\sin(\omega_0-2\Omega_{\mathrm{RF}})t\,.
\end{eqnarray}
The analysis of the expression for the heterodyne photocurrent
\begin{equation}
\label{eq:heterodyne_photocurrent}
i_{\mathrm{het}}\propto\tmean{(L_{\mathrm{het}}+S)^2} = \tmean{S^2}+\tmean{L_+^2}+\tmean{L_-^2}+2\tmean{(L_++L_-)S}+2\tmean{L_+L_-}\,
\end{equation}
gives a clue to why these additional noise components emerge in the outgoing signal. It is easier to perform this kind of analysis if we represent the above trigonometric expressions in terms of scalar products of the vectors of the corresponding quadrature amplitudes and a special unit-length vector $\boldsymbol{H}[\phi] = \left\{\cos\phi,\,\sin\phi\right\}^{\mathsf{T}}$, e.g.:
$$S_{\mathrm{het}}\equiv(\boldsymbol{G}+\boldsymbol{n}_{\omega_0})^{\mathsf{T}}\cdot\boldsymbol{H}[\omega_0t]+\boldsymbol{n}_{\omega_0+2\Omega_{\mathrm{RF}}}^{\mathsf{T}}\cdot\boldsymbol{H}[(\omega_0+\Omega_{\mathrm{RF}})t]+\boldsymbol{n}_{\omega_0-2\Omega_{\mathrm{RF}}}^{\mathsf{T}}\cdot\boldsymbol{H}[(\omega_0-2\Omega_{\mathrm{RF}})t]$$
where $\boldsymbol{G}=\left\{G_c,\,G_s\right\}^{\mathsf{T}}$ and $\boldsymbol{n}_{\omega\alpha}=\left\{n^\omega_c,\,n_s^\omega\right\}^{\mathsf{T}}$. Another useful observation, provided that $\omega_0\gg\max[\Omega_1,\,\Omega_2]$, gives us the following relation:
$$\tmean{\boldsymbol{H}[(\omega_0+\Omega_1)t]\boldsymbol{H}^{\mathsf{T}}[(\omega_0+\Omega_2)t]}=\frac12
\begin{bmatrix}
  \cos(\Omega_1-\Omega_2)t & -\sin(\Omega_1-\Omega_2)t\\
  \sin(\Omega_1-\Omega_2)t & \cos(\Omega_1-\Omega_2)t
\end{bmatrix} = \frac12\mathbb{P}\left[(\Omega_1-\Omega_2)t\right] \,.
 $$
Using this relation it is straightforward to see that the first three terms in Eq.~\eqref{eq:heterodyne_photocurrent} give DC components of the photocurrent, while the fifth term oscillates at double modulation frequency $2\Omega_{\mathrm{RF}}$. It is only the term $2\tmean{(L_++L_-)S}$ that is linear in GW signal and thus contains useful information:
\begin{equation*}
  2\tmean{(L_++L_-)S} \simeq I_c(t)\cos\Omega_{\mathrm{RF}}t+I_s(t)\sin\Omega_{\mathrm{RF}}t+\left\{\mbox{oscillations at frequency }3\Omega_{\mathrm{RF}}\right\}
\end{equation*}
where
\begin{eqnarray*}
  I_c(t) &=& (\boldsymbol{G}+\boldsymbol{n}_{\omega_0}+\boldsymbol{n}_{\omega_0+2\Omega_{\mathrm{RF}}})^{\mathsf{T}}\cdot\boldsymbol{L}^{(0)}_++(\boldsymbol{G}+\boldsymbol{n}_{\omega_0}+\boldsymbol{n}_{\omega_0-2\Omega_{\mathrm{RF}}})^{\mathsf{T}}\cdot\boldsymbol{L}^{(0)}_-\,,\\
  I_s(t) &=& -i(\boldsymbol{G}+\boldsymbol{n}_{\omega_0}+\boldsymbol{n}_{\omega_0+2\Omega_{\mathrm{RF}}})^{\mathsf{T}}\cdot\sigma_y\cdot\boldsymbol{L}^{(0)}_++i(\boldsymbol{G}+\boldsymbol{n}_{\omega_0}+\boldsymbol{n}_{\omega_0-2\Omega_{\mathrm{RF}}})^{\mathsf{T}}\cdot\sigma_y\cdot\boldsymbol{L}^{(0)}_-\,,
\end{eqnarray*}
and $\sigma_y$ is the 2nd Pauli matrix:
$$\sigma_y =
\begin{bmatrix}
  0 & -i\\
  i & 0
\end{bmatrix}\,.
$$
In order to extract the desired signal quadrature the photodetector readout current $i_{\mathrm{het}}$ is mixed with (multiplied by) a demodulation function $D(t) = D_0\cos(\Omega_{\mathrm{RF}}t+\phi_D)$ with the resulting signal filtered by a low-pass filter with upper cut-off frequency $\Lambda\ll\Omega_{\mathrm{RF}}$ so that only components oscillating at GW frequencies $\Omega_{\mathrm{GW}}\ll\Omega_{\mathrm{RF}}$ appear in the output signal (see Figure~\ref{fig:heterodyne_readout_scheme}).

It is instructive to see what the above procedure yields in the simple case of the Michelson interferometer tuned in resonance with RF-sidebands produced by pure phase modulation: $L_{c+}^{(0)}=L_{c-}^{(0)}=0$ and $L_{s+}^{(0)}=L_{s-}^{(0)}=L_0$. The foregoing expressions simplify significantly to the following:
$$I_c(t) = 2L_0\left(G_s+n_s^{\omega_0}+\frac{n_s^{\omega_0-2\Omega_{\mathrm{RF}}}}{2}+\frac{n_s^{\omega_0+2\Omega_{\mathrm{RF}}}}{2}\right)\quad\mbox{and}\quad I_s(t) = -L_0\left(n_s^{\omega_0-2\Omega_{\mathrm{RF}}}-n_s^{\omega_0+2\Omega_{\mathrm{RF}}}\right)\,.$$
Apparently, in this simple case of equal sideband amplitudes (\emph{balanced heterodyne detection}), only single phase quadrature of the GW signal can be extracted from the output photocurrent, which is fine, because the Michelson interferometer, being equivalent to a simple movable mirror with respect to a GW tidal force as shown in Section~\ref{sec:phasemeter} and \ref{sec:MI}, is sensitive to a GW signal only in phase quadrature. Another important feature of heterodyne detection conspicuous in the above equations is the presence of additional noise from the frequency bands that are twice the RF-modulation frequency away from the carrier. As shown in~\cite{03a1BuChMa} this noise contributes to the total quantum shot noise of the interferometer and makes the high frequency sensitivity of the GW detectors with heterodyne readout 1.5 times worse compared to the ones with homodyne or DC readout.

For more realistic and thus more sophisticated optical configurations, including Fabry--P\'{e}rot cavities in the arms and additional recycling mirrors in the pumping and readout ports, the analysis of the interferometer sensitivity becomes rather complicated. Nevertheless, it is worthwhile to note that with proper optimization of the modulation sidebands and demodulation function shapes the same sensitivity as for frequency-independent homodyne readout schemes can be obtained~\cite{03a1BuChMa}. However, inherent additional frequency-independent quantum shot noise brought by the heterodyning process into the detection band hampers the simultaneous use of advanced quantum non-demolition (QND) techniques and heterodyne readout schemes significantly.

\newpage
\section{Quantum Nature of Light and Quantum Noise}
\label{sec:quantum_light}

Now is the time to remind ourselves of the word `quantum' in the title of our review. Thus far, the quantum nature of laser light being used in the GW interferometers has not been accounted for in any way. Nevertheless, quantum mechanics predicts striking differences for the variances of laser light amplitude and phase fluctuations, depending on which quantum state it is in. Squeezed vacuum~\cite{1995BookWaMi, 95BookMaWo, 97BookScZu, 81a1Ca, 02a1KiLeMaThVy} injection that has been recently implemented in the GEO\,600 detector and has pushed the high-frequency part of the total noise down by 3.5~dB~\cite{Vahlbruch_CQG_27_084027_2010,Nat.Phys.7.12.962_2011_LSC} serves as a perfect example of this. In this section, we provide a brief introduction into the quantization of light and the typical quantum states thereof that are common for the GW interferometers.

\subsection{Quantization of light: Two-photon formalism}
\label{sec:2photon_formalism}

From the point of view of quantum field theory, a freely propagating electromagnetic wave can be characterized in each spatial point with location vector $\boldsymbol{r}=(x,y,z)$ at time $t$ by a Heisenberg operator of an electric field strength $\hat{E}(\boldsymbol{r},\,t)$.\epubtkFootnote{Insofar as the light beams in the interferometer can be well approximated as paraxial beams, and the polarization of the light wave does not matter in most of the considered interferometers, we will omit the vector nature of the electric field and treat it as a scalar field with strength defined by a scalar operator-valued function $\hat{E}(x,y,z,t)$.} The electric field Heisenberg operator of a light wave traveling along the positive direction of the $z$-axis can be written as a sum of a positive- and negative-frequency parts:
\begin{equation}
\label{eq:EMW_quantum}
\hat{E}(x,y,z;t) = u(x,y,z)\left\{\hat{E}^{(+)}(t)+\hat{E}^{(-)}(t)\right\}\,,
\end{equation}
where $u(x,y,z)$ is the spatial mode shape, slowly changing on a wavelength $\lambda$ scale, and
\begin{equation}
\label{eq:EMW_expansion_in_modes}
\hat{E}^{(+)}(t) = \int_0^\infty\dfrac{d\omega}{2\pi}\sqrt{\dfrac{2\pi \hbar\omega}{\mathcal{A}c}}\hat{a}_\omega e^{-i\omega t}\,,\quad\mbox{and}\quad \hat{E}^{(-)}(t)=\left[\hat{E}^{(+)}(t)\right]^\dag\,.
\end{equation}
Here, $\mathcal{A}$ is the effective cross-section area of the light beam, and $\hat{a}_\omega$ ($\hat{a}_\omega^\dag$) is the single photon \emph{annihilation} (\emph{creation}) operator in the mode of the field with frequency $\omega$. The meaning of Eq.~\eqref{eq:EMW_expansion_in_modes} is that the travelling light wave can be represented by an expansion over the continuum of harmonic oscillators -- modes of the electromagnetic field, -- that are, essentially, independent degrees of freedom. The latter implies the commutation relations for the operators $\hat{a}_\omega$ and $\hat{a}_\omega^\dag$:
\begin{equation}
\label{eq:EMW_commutator_modes}
\left[\hat{a}_\omega,\,\hat{a}_{\omega'}^\dag\right] = 2\pi\delta(\omega-\omega')\,,\quad\mbox{and}\quad \left[\hat{a}_\omega,\,\hat{a}_{\omega'}\right] = \left[\hat{a}^\dag_\omega,\,\hat{a}^\dag_{\omega'}\right] = 0\,.
\end{equation}

In GW detectors, one deals normally with a close to monochromatic laser light with carrier frequency $\omega_0$, and a pair of modulation sidebands created by a GW signal around its frequency in the course of parametric modulation of the interferometer arm lengths. The light field coming out of the interferometer cannot be considered as the continuum of independent modes anymore. The very fact that sidebands appear in pairs implies the two-photon nature of the processes taking place in the GW interferometers, which means the modes of light at frequencies $\omega_{1,2}=\omega_0\pm\Omega$ have correlated complex amplitudes and thus the new two-photon operators and related formalism is necessary to describe quantum light field transformations in GW interferometers. This was realized in the early 1980s by Caves and Schumaker who developed the  two-photon formalism~\cite{85a1CaSch, 85a2CaSch}, which is widely used in GW detectors as well as in quantum optics and optomechanics.

One starts by defining modulation sideband amplitudes as
$$\hat{a}_+ = \hat{a}_{\omega_0+\Omega}\,,\quad\hat{a}_- = \hat{a}_{\omega_0-\Omega}\,,$$
and factoring out the oscillation at carrier frequency $\omega_0$ in Eqs.~\eqref{eq:EMW_quantum}, which yields:
 \begin{eqnarray} \label{eq:2photon_Eplus}
\hat{E}^{(+)}(t) = \frac{\mathcal{C}_0 e^{-i\omega_0t}}{\sqrt{2}}\int_{-\omega_0}^\infty\frac{d\Omega}{2\pi}\lambda_+(\Omega)\hat{a}_+e^{-i\Omega t}\simeq\frac{\mathcal{C}_0}{\sqrt{2}}e^{-i\omega_0t}\intinfty\frac{d\Omega}{2\pi}\lambda_+(\Omega)\hat{a}_+e^{-i\Omega t}\,,\nonumber\\
\hat{E}^{(-)}(t) = \frac{\mathcal{C}_0e^{i\omega_0t}}{\sqrt{2}}\int_{-\infty}^{\omega_0}\frac{d\Omega}{2\pi}\lambda_-(\Omega)\hat{a}^\dag_-e^{-i\Omega t}\simeq\frac{\mathcal{C}_0}{\sqrt{2}}e^{i\omega_0t}\intinfty\frac{d\Omega}{2\pi}\lambda_-(\Omega)\hat{a}^\dag_-e^{-i\Omega t}\,,
\end{eqnarray}
where we denote $\mathcal{C}_0\equiv\sqrt{\frac{4\pi\hbar\omega_0}{\mathcal{A}c}}$ and define functions $\lambda_{\pm}(\Omega)$ following~\cite{85a1CaSch} as
\begin{equation*}
\lambda_{\pm}(\Omega) = \sqrt{\frac{\omega_0\pm\Omega}{\omega_0}}\,,
\end{equation*}
and use the fact that $\omega_0\gg\Omega_{\mathrm{GW}}$ enables us to expand the limits of integrals to $\omega_0\to\infty$.
The operator expressions in front of $e^{\pm i\omega_0t}$ in the foregoing Eqs.~\eqref{eq:2photon_Eplus} are quantum analogues to the complex amplitude $\mathcal{E}$ and its complex conjugate $\mathcal{E}^*$ defined in Eqs.~\eqref{eq:EMW_quadrature_def}:
\begin{equation*}
\hat{\mathcal{E}}(t)=\frac{\mathcal{C}_0}{\sqrt{2}}\hat{a}(t)\equiv\frac{\mathcal{C}_0}{\sqrt{2}}\intinfty\frac{d\Omega}{2\pi}\lambda_+(\Omega)\hat{a}_+e^{-i\Omega
  t}\,,\ \mbox{and}\ \hat{\mathcal{E}}^\dag(t)=\frac{\mathcal{C}_0}{\sqrt{2}}\hat{a}^\dag(t)\equiv\frac{\mathcal{C}_0}{\sqrt{2}}\intinfty\frac{d\Omega}{2\pi}\lambda_-(\Omega)\hat{a}^\dag_-e^{-i\Omega t}\,.
\end{equation*}
Again using Eqs.~\eqref{eq:EMW_quadrature_def}, we can define two-photon quadrature amplitudes as:
  \begin{eqnarray}
\label{eq:2photon_quadratures}
  \hat{\mathcal{E}}_c(t) = \frac{\hat{\mathcal{E}}(t)+\hat{\mathcal{E}}^\dag(t)}{\sqrt{2}} = \frac{\mathcal{C}_0}{\sqrt{2}}\intinfty\frac{d\Omega}{2\pi}\frac{\lambda_+\hat{a}_++\lambda_-\hat{a}^\dag_-}{\sqrt{2}}e^{-i\Omega t} \equiv \frac{\mathcal{C}_0}{\sqrt{2}}\intinfty\frac{d\Omega}{2\pi}\hat{a}_c(\Omega)e^{-i\Omega t} \nonumber\\
   \hat{\mathcal{E}}_s(t) = \frac{\hat{\mathcal{E}}(t)-\hat{\mathcal{E}}^\dag(t)}{i\sqrt{2}} = \frac{\mathcal{C}_0}{\sqrt{2}}\intinfty\frac{d\Omega}{2\pi}\frac{\lambda_+\hat{a}_+-\lambda_-\hat{a}^\dag_-}{i\sqrt{2}}e^{-i\Omega t} \equiv \frac{\mathcal{C}_0}{\sqrt{2}}\intinfty\frac{d\Omega}{2\pi}\hat{a}_s(\Omega)e^{-i\Omega t}.
\end{eqnarray}
Note that so-introduced operators of two-photon quadrature amplitudes $\hat{a}_{c,s}(t)$ are Hermitian and thus their frequency domain counterparts satisfy the relations for the spectra of Hermitian operator:
\begin{equation*}
 \hat{a}^\dag_{c,s}(t) = \hat{a}_{c,s}(t)\ \Longrightarrow\ \hat{a}^\dag_{c,s}(\Omega) = \hat{a}_{c,s}(-\Omega)\,.
\end{equation*}
Now we are able to write down commutation relations for the quadrature operators, which can be derived from Eq.~\eqref{eq:EMW_commutator_modes}:
\begin{eqnarray}
  \left[\hat{a}_c(\Omega),\,\hat{a}^\dag_c(\Omega')\right] &=& \left[\hat{a}_s(\Omega),\,\hat{a}^\dag_s(\Omega')\right] = 2\pi\frac{\Omega}{\omega_0}\delta(\Omega-\Omega')\,,\label{eq:2photon_commutator1}\\
  \left[\hat{a}_c(\Omega),\,\hat{a}^\dag_s(\Omega')\right] &=& \left[\hat{a}^\dag_c(\Omega),\,\hat{a}_s(\Omega')\right] = 2\pi i\delta(\Omega-\Omega')\,,\label{eq:2photon_commutator2}
\end{eqnarray}

The commutation relations represented by
Eqs.~\eqref{eq:2photon_commutator1} indicate that quadrature
amplitudes do not commute at different times,
i.e., $[\hat{a}_c(t),\,\hat{a}_c(t')]=[\hat{a}_s(t),\,\hat{a}_s(t')]\neq0$,
which imply they could not be considered for proper output observables
of the detector, for a nonzero commutator, as we would see later,
means an additional quantum noise inevitably contributes to the final
measurement result. The detailed explanation of why it is so can be
found in many works devoted to continuous linear quantum measurement
theory, in particular, in Chapter~6 of~\cite{92BookBrKh}, Appendix~2.7
of~\cite{03pth1Ch} or in~\cite{03a1BrGoKhMaThVy}. Where GW
detection is concerned, all the authors are agreed on the point that
the values of GW frequencies $\Omega$
$(1\mathrm{\ Hz} \leqslant\Omega/2\pi\leqslant 10^{3}\mathrm{\ Hz})$, being much smaller than
optical frequencies $\omega_0/2\pi\sim10^{15}\mathrm{\ Hz}$, allow one to neglect
such weak commutators as those of Eqs.~\eqref{eq:2photon_commutator1}
in all calculations related to GW detectors output quantum noise. This
statement has gotten an additional ground in the calculation conducted in
Appendix~2.7 of~\cite{03pth1Ch} where the value of the additional
quantum noise arising due to the nonzero value of commutators
\eqref{eq:2photon_commutator1} has been derived and its extreme
minuteness compared to other quantum noise sources has been
proven. Braginsky et~al. argued in~\cite{03a1BrGoKhMaThVy} that the
two-photon quadrature amplitudes defined by
Eqs.~\eqref{eq:2photon_quadratures} are not the real measured
observables at the output of the interferometer, since the
photodetectors actually measure not the energy flux
\begin{equation}
\hat{\mathcal{I}}(t)=\int\int_0^\infty\frac{d\omega d\omega'}{(2\pi)^2}\,\hbar\sqrt{\omega\omega'}\hat{a}^\dag_\omega\hat{a}_{\omega'}e^{i(\omega-\omega')t}
\end{equation}
but rather the photon number flux:
\begin{equation}
\hat{\mathcal{N}}(t)=\int\int_0^\infty\frac{d\omega d\omega'}{(2\pi)^2}\,\hat{a}^\dag_\omega\hat{a}_{\omega'}e^{i(\omega-\omega')t}\,.
\end{equation}
The former does not commute with itself: $[\hat{\mathcal{I}}(t),\,\hat{\mathcal{I}}(t')]\neq0$, while the latter apparently does $[\hat{\mathcal{N}}(t),\,\hat{\mathcal{N}}(t')]=0$ and therefore
is the right observable for a self-consistent quantum description of the GW interferometer output signal.

In the course of our review, we shall adhere to the approximate quadrature amplitude operators that can be obtained from the exact ones given by Eqs.~\eqref{eq:2photon_quadratures} by setting $\lambda_\pm(\Omega)\to1$, i.e.,
  \begin{eqnarray}
\label{eq:2photon_quads_def}
\hat{a}_c(t) &=& \intinfty\frac{d\Omega}{2\pi}\frac{\hat{a}_++\hat{a}_-^\dag}{\sqrt{2}}e^{-i\Omega t}\ \Longleftrightarrow\ \hat{a}_c(\Omega) = \frac{\hat{a}_++\hat{a}_-^\dag}{\sqrt{2}}\,,\nonumber\\
\hat{a}_s(t) &=& \intinfty\frac{d\Omega}{2\pi}\frac{\hat{a}_+-\hat{a}_-^\dag}{i\sqrt{2}}e^{-i\Omega t}\ \Longleftrightarrow\ \hat{a}_s(\Omega) = \frac{\hat{a}_+-\hat{a}_-^\dag}{i\sqrt{2}}\,.
\end{eqnarray}

The new approximate two-photon quadrature operators satisfy the following commutation relations in the frequency domain:
\begin{equation}
\label{eq:2photon_quads_commutator_spectral}
\left[\hat{a}_c(\Omega),\,\hat{a}_s(\Omega')\right] = 2\pi i\delta(\Omega+\Omega')\,,\quad\mbox{and}\quad \left[\hat{a}_c(\Omega),\,\hat{a}_c(\Omega')\right] = \left[\hat{a}_s(\Omega),\,\hat{a}_s(\Omega')\right] = 0\,,
\end{equation}
and in the time domain:
\begin{equation}
\label{eq:2photon_quads_commutator_time}
\left[\hat{a}_c(t),\,\hat{a}_s(t')\right] = i\delta(t-t')\,,\quad\mbox{and}\quad \left[\hat{a}_c(t),\,\hat{a}_c(t')\right] = \left[\hat{a}_s(t),\,\hat{a}_s(t')\right] = 0\,
\end{equation}

Then the electric field strength operator~\eqref{eq:EMW_quantum} can be rewritten in terms of the two-photon quadrature operators as:
\begin{equation}
\label{eq:2photon_EMW_quadratures}
\hat{E}(x,y,z;t) = u(x,y,z)\mathcal{C}_0\left[\hat{a}_c(t)\cos\omega_0t+\hat{a}_s(t)\sin\omega_0t\right].
\end{equation}
Hereafter, we will omit the spatial mode factor $u(x,y,z)$ since it does not influence the final result for quantum noise spectral densities. Moreover, in order to comply with the already introduced division of the optical field into classical carrier field and to the 1st order corrections to it comprising of laser noise and signal induced sidebands, we adopt the same division for the quantum fields, i.e., we detach the mean values of the corresponding quadrature operators via the following redefinition
$\hat{a}^{\mathrm{old}}_{c_s}\to A_{c,s}+\hat{a}^{\mathrm{new}}_{c,s}$ with $A_{c,s}\equiv\mean{\hat{a}^{\mathrm{old}}_{c,s}}$. Here, by $\mean{\hat{a}^{\mathrm{old}}_{c,s}}$ we denote an ensemble average over the quantum state $\ket{\psi}$ of the light wave: $\mean{\hat{A}}\equiv\bra{\psi}\hat{A}\ket{\psi}$. Thus, the electric field strength operator for a plain electromagnetic wave will have the following form:
\begin{equation}
\label{eq:2photon_E_strain}
\hat{E}(x,y;t) = \mathcal{C}_0\left[(A_c+\hat{a}_c(t))\cos\omega_0t+(A_s+\hat{a}_s(t))\sin\omega_0t\right]\,.
\end{equation}
Further, we combine the two-photon quadratures into vectors in the same manner as we used to do for classical fields:
$$\boldsymbol{A}\equiv
\begin{bmatrix}
A_c\\
A_s
\end{bmatrix}\,,\quad\mbox{and}\quad
\hat{\boldsymbol{a}}\equiv\begin{bmatrix}
\hat{a}_c\\
\hat{a}_s
\end{bmatrix}\,.
$$

Now, when we have defined a quantum Heisenberg operator of the electric field of a light wave, and introduced quantum operators of two-photon quadratures, the last obstacle on our way towards the description of quantum noise in GW interferometers is that we do not know the quantum state the light field finds itself in. Since it is the quantum state that defines the magnitude and mutual correlations of the amplitude and phase fluctuations of the outgoing light, and through it the total level of quantum noise setting the limit on the future GW detectors' sensitivity. In what follows, we shall consider vacuum and coherent states of the light, and also squeezed states, for they comprise the vast majority of possible states one could encounter in GW interferometers.

\subsection{Quantum states of light}
\label{sec:light_quantum_states}

\subsubsection{Vacuum state}
\label{sec:vacuum_state}

The quantum state of the travelling wave is a subtle structure, for the system it describes comprises a continuum of modes. However, each of these modes can be viewed at as a quantum oscillator with its own generalized coordinate $\hat{X}_\omega = (\hat{a}_\omega+\hat{a}^\dag_\omega)/\sqrt{2}$ and momentum $\hat{Y}_\omega=(\hat{a}_\omega-\hat{a}^\dag_\omega)/i\sqrt{2}$. The ground state of this system, known as a \emph{vacuum} state $\ket{\vac}$, is straightforward and is simply the direct product of the ground states $\ket{0}_\omega$ of all modes over all frequencies $\omega$:
\begin{equation}\label{eq:field_vac_state}
  \ket{\vac}\equiv\bigotimes\limits_\omega\ket{0}_\omega\,.
\end{equation}

By definition, the ground state of a mode with frequency $\omega$ is the state with minimum energy $E_{\vac}=\hbar\omega/2$ and no excitation:
\begin{equation}
\label{eq:annihilation_operator}
\hat{a}_\omega\ket{0}_\omega = 0\quad \mbox{and}\quad \bra{0}_\omega\hat{a}^\dag_\omega = 0\,.
\end{equation}

Consider now statistical properties of the vacuum state. The mean values of annihilation and creation operators as well as any linear combination thereof that includes quadrature amplitudes, are zero:
$$\bra{\vac}\hat{a}_\omega\ket{\vac}\equiv\mean{\hat{a}_\omega}=\mean{\hat{a}_\omega^\dag}=0\ \Rightarrow\ \mean{\hat{a}_c(\Omega)}=\mean{\hat{a}_s(\Omega)}=0\,.$$
Apparently, this also holds for time domain operators:
$$\bra{\vac}\hat{a}(t)\ket{\vac}\equiv\mean{\hat{a}(t)}=\mean{\hat{a}^\dag(t)}=0\ \Rightarrow\ \mean{\hat{a}_c(t)}=\mean{\hat{a}_s(t)}=0\,.$$
That the ground state of the oscillator is Gaussian is evident from its q-representation~\cite{97BookScZu}, namely
$$\psi_{\vac}(X_\omega) \equiv \braket{X_\omega}{0}_\omega = \dfrac{1}{\sqrt[4]{\pi}}\exp\left\{-\dfrac{X_\omega^2}{2}\right\}.$$
It means that knowing the second moments of quadrature amplitudes suffices for full characterization of the state $\ket{\vac}$. For this purpose, let us introduce a quadrature amplitudes matrix of spectral densities\epubtkFootnote{Herein, we make use of a double-sided power spectral density defined on a whole range of frequencies, both negative and positive, that yields the following connection to the variance of an arbitrary observable $\hat{o}(t)$:
$$\mathrm{Var}\left[\hat{o}(t)\right]\equiv\mean{\hat{o}^2(t)}-\mean{\hat{o}(t)}^2 = \intinfty\dfrac{d\Omega}{2\pi}S_o(\Omega)\,,$$
It is worth noting that in the GW community, the sensitivity of GW detectors as well as the individual noise sousces are usually characterized by a single-sided power spectral density $S^+_o(\Omega)$, that is simply defined on positive frequencies $\Omega\geqslant0$. The connection between these two is straightforward: $S^+_o(\Omega)=2S_o(\Omega)$ for $\Omega\geqslant0$ and $0$ otherwise.
} $\mathbb{S}(\Omega)$ according to the rule:
\begin{equation}
\label{eq:SpDens_matrix_def}
\smatrix{\mean{\hat{a}_c(\Omega)\circ\hat{a}_c(\Omega')}}
    {\mean{\hat{a}_c(\Omega)\circ\hat{a}_s(\Omega')}}
    {\mean{\hat{a}_s(\Omega)\circ\hat{a}_c(\Omega')}}
    {\mean{\hat{a}_s(\Omega)\circ\hat{a}_s(\Omega')}}
  = 2\pi\delta(\Omega+\Omega')\smatrix{S_{cc}(\Omega)}{S_{cs}(\Omega)}{S_{sc}(\Omega)}{S_{ss}(\Omega)}
  = 2\pi\mathbb{S}\delta(\Omega+\Omega') \,.
\end{equation}
where $S_{ij}(\Omega)$ ($i,j=c,s$) denote (cross) power spectral densities of the corresponding quadrature amplitudes ${\mean{\hat{a}_i(\Omega)\circ\hat{a}_j(\Omega')}}$ standing for the symmetrized product of the corresponding quadrature operators, i.e.:
$$\mean{\hat{a}_i(\Omega)\circ\hat{a}_j(\Omega')}\equiv\frac12\mean{\hat{a}_i(\Omega)\hat{a}_j(\Omega')+\hat{a}_j(\Omega')\hat{a}_i(\Omega)} \equiv 2\pi S_{ij}(\Omega)\delta(\Omega-\Omega')\,.$$
For a vacuum state, this matrix of spectral densities can easily be obtained from the commutation relations~\eqref{eq:2photon_quads_commutator_spectral} and equals to:
\begin{equation}
\label{eq:SpDens_matrix_vac}
\mathbb{S}_{\vac}(\Omega) = \smatrix{1/2}{0}{0}{1/2}\,,
\end{equation}
which implies that the (double-sided) power spectral densities of the quadrature amplitudes as well as their cross-spectral density are equal to:
$$S_{cc}(\Omega) = S_{ss}(\Omega) = \frac{1}{2}\qquad\mbox{and}\qquad S_{cs}(\Omega) = 0\,.$$
In time domain, the corresponding matrix of second moments, known as a covariance matrix with elements defined as $\mathbb{V}_{ij}\delta(t-t') = \mean{\hat{a}_i(t)\circ\hat{a}_j(t')}$, is absolutely the same as $\mathbb{S}_{\vac}(\Omega)$ :
\begin{equation}
\label{eq:Cov_matrix_vac}
\mathbb{V}_{\vac} = \smatrix{1/2}{0}{0}{1/2}\,.
\end{equation}

It is instructive to discuss the meaning of these matrices, $\mathbb{S}$ and $\mathbb{V}$, and of the values they comprise. To do so, let us think of the light wave as a sequence of very short square-wave light pulses with infinitesimally small duration $\varepsilon\to0$. The delta function of time in Eq.~\eqref{eq:Cov_matrix_vac} tells us that the noise levels at different times, i.e., the amplitudes of the different square waves, are statistically independent. To put it another way, this noise is Markovian. It is also evident from Eq.~\eqref{eq:SpDens_matrix_vac} that quadrature amplitudes' fluctuations are stationary, and it is this stationarity, as noted in~\cite{85a1CaSch} that makes quadrature amplitudes such a convenient language for describing the quantum noise of light in parametric systems exemplified by GW interferometers.

It is instructive to pay some attention to a pictorial representation of the quantum noise described by the covariance and spectral density matrices $\mathbb{V}$ and $\mathbb{S}$. With this end in view let us introduce quadrature operators for each short light pulse as follows:
\begin{equation}
\label{eq:X_and_Y_quadrature_def}
\hat{X}_\varepsilon(t)\equiv\frac{1}{\sqrt{\varepsilon}}\int_{t-\varepsilon/2}^{t+\varepsilon/2}d\tau\,\hat{a}_c(\tau)\,,
\qquad\mbox{and}\qquad
\hat{Y}_\varepsilon(t)\equiv\frac{1}{\sqrt{\varepsilon}}\int_{t-\varepsilon/2}^{t+\varepsilon/2}d\tau\,\hat{a}_s(\tau)\,.
\end{equation}

These operators $\hat{X}_\varepsilon(t)$ and $\hat{Y}_\varepsilon(t)$ are nothing else than dimensionless displacement and momentum of the corresponding mode (called quadratures in quantum optics), normalized by zero point fluctuation amplitudes $X_0$ and $P_0$: $\hat{X}_\varepsilon(t)\equiv\hat{x}_\varepsilon/X_0$ and $\hat{X}_\varepsilon(t)\equiv\hat{p}_\varepsilon/P_0$.
This fact is also justified by the value of their commutator:
$$\left[\hat{X}_\varepsilon(t),\,\hat{Y}_\varepsilon(t)\right]=i\,.$$
There is no difficulty in showing that diagonal elements of the covariance matrix $\mathbb{V}_{ii}$ are equal to the variances of the corresponding mode displacement $\hat{X}_\varepsilon$ and momentum $\hat{Y}_\varepsilon$:
$$\mathbb{V}_{cc} = \mean{\hat{X}^2_\varepsilon(t)} = 1/2\,,\quad \mbox{and} \quad \mathbb{V}_{ss} = \mean{\hat{Y}^2_\varepsilon(t)} = 1/2\,, $$
while non-diagonal terms represent correlations between these operators (zero in our case):
$\mathbb{V}_{cs} = \mathbb{V}_{sc} = \mean{\hat{X}_\varepsilon(t)\circ\hat{Y}_\varepsilon(t)} = 0$.
At the same time, we see that there is no correlation between the pulses, justifying the Markovianity of the quantum noise of light in vacuum state: $$\mean{\hat{X}_\varepsilon(t)\hat{X}_\varepsilon(t')} = \mean{\hat{Y}_\varepsilon(t)\hat{Y}_\varepsilon(t')} = \mean{\hat{X}_\varepsilon(t)\circ\hat{Y}_\varepsilon(t')}=0\,,\ t\neq t'\,.$$
An attempt to measure the light field amplitude as a function of time will
give the result depicted in Figure~\ref{fig:vac_meas_result}.

\epubtkImage{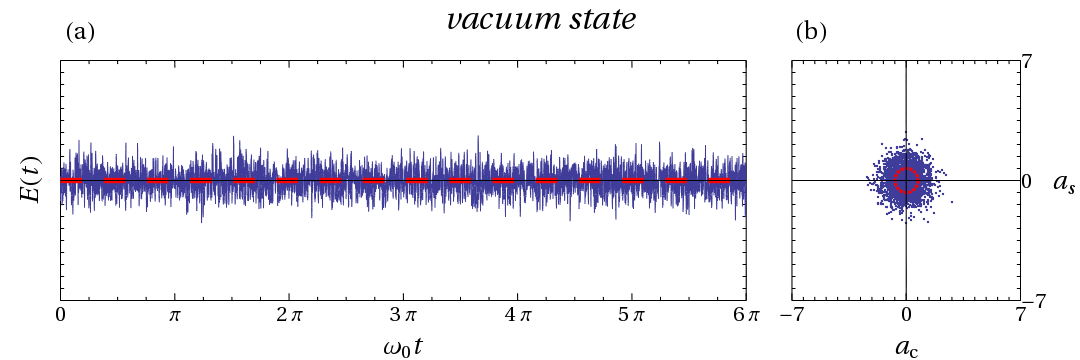}{%
\begin{figure}[htbp]
\centerline{\includegraphics[width=\textwidth,]{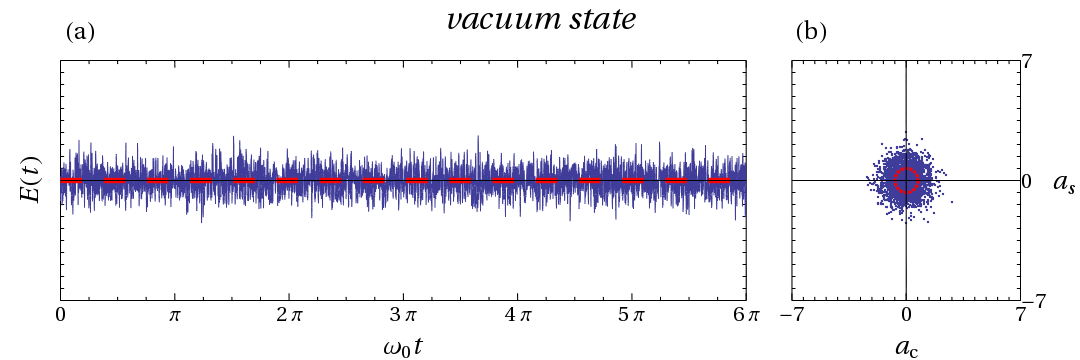}}
\caption{Light field in a vacuum quantum state $\ket{\vac}$. \emph{Left panel (a)} features a typical result one could get measuring the (normalized) electric field strength of the light wave in a vacuum state as a function of time. \emph{Right panel (b)} represents a phase space picture of the results of measurement. A red dashed circle displays the error ellipse for the state $\ket{\vac}$ that encircles the area of single standard deviation for a two-dimensional random vector $\hat{\boldsymbol{a}}$ of measured light quadrature amplitudes. The principal radii of the error ellipse (equal in vacuum state case) are equal to square roots of the covariance matrix $\mathbb{V}_{\vac}$ eigenvalues, i.e., to $1/\sqrt{2}$.}
\label{fig:vac_meas_result}
\end{figure}}

\epubtkImage{fig14.png}{%
\begin{figure}[htbp]
  \centerline{\includegraphics[width=.5\textwidth]{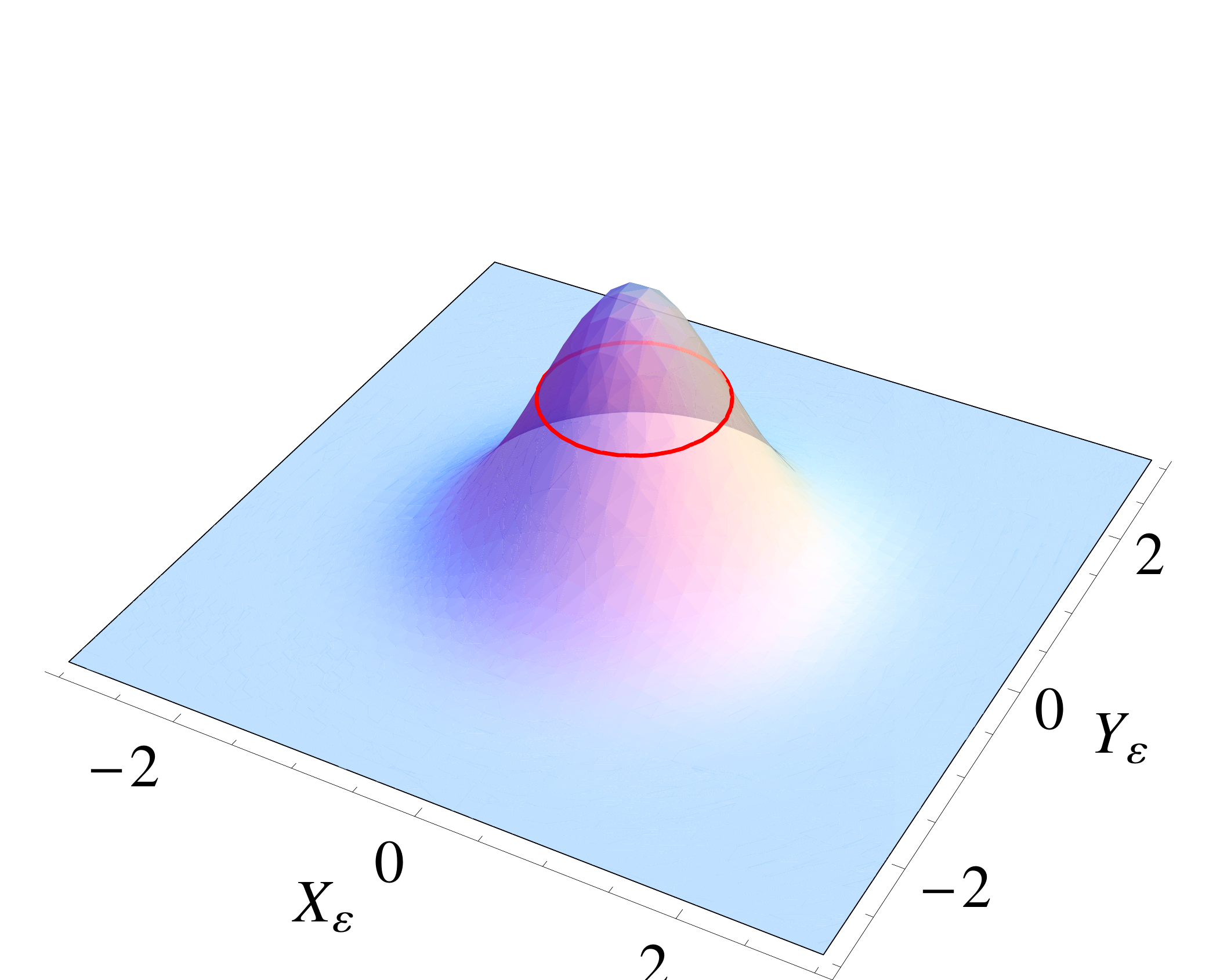}\hfill\includegraphics[width=.33\textwidth]{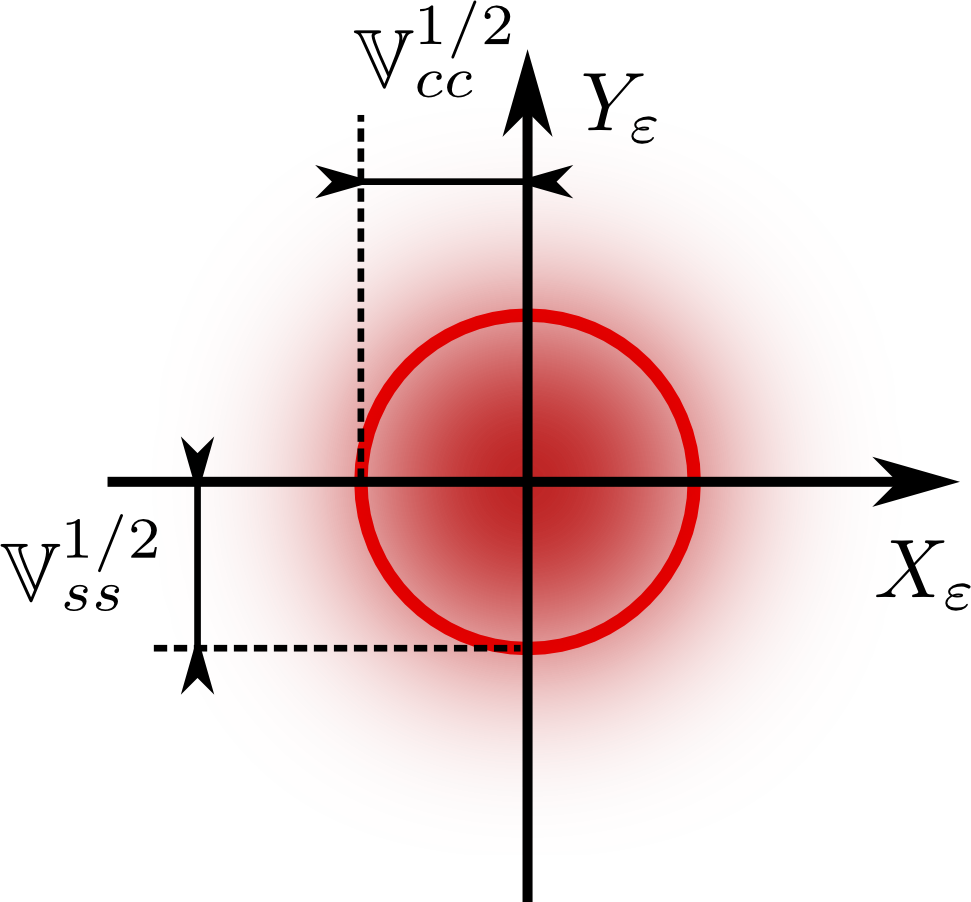}}
  \caption{Wigner function $W_{\ket{\vac}}(X_\varepsilon,\,Y_\varepsilon)$ of a ground state of harmonic oscillator (\emph{left panel}) and its representation in terms of the noise ellipse (\emph{right panel}).}
\label{fig:vac_Wigner_fcn}
\end{figure}}

The measurement outcome at each instance of time will be a random variable with zero mean and variance defined by a covariance matrix $\mathbb{V}_{\vac}$ of Eq.~\eqref{eq:Cov_matrix_vac}:
$$\mathrm{Var}[\hat{E}(t)] = \{\cos\omega_0t,\,\sin\omega_0t\}\mathbb{V}_{\vac}\{\cos\omega_0t,\,\sin\omega_0t\}^{\mathsf{T}} = \frac12\,.$$

In quantum mechanics, it is convenient to describe a quantum state in terms of a \emph{Wigner function}, a quantum version of joint (quasi) probability distribution for particle displacement and momentum ($X_\varepsilon$ and $Y_\varepsilon$ in our case):
\begin{eqnarray}
\label{eq:WignerFcn_vac}
W_{\ket{\vac}}(X_\varepsilon,\,Y_\varepsilon) &=& \intinfty\frac{d\xi}{2\pi}\exp\left\{-i\xi Y_\varepsilon\right\}\braket{X_\varepsilon+\xi/2}{\vac}\braket{\vac}{X_\varepsilon-\xi/2}\nonumber\\
&=& \dfrac{1}{2\pi\sqrt{\det\mathbb{V}_{\vac}}}\exp\left\{-\frac12\{X_\varepsilon,\,Y_\varepsilon\}^{\mathsf{T}}\mathbb{V}_{\vac}^{-1}\{X_\varepsilon,\,Y_\varepsilon\}\right\} = \frac{1}{\pi}\exp\left\{-(X_\varepsilon^2+Y_\varepsilon^2)\right\},
\end{eqnarray}
where $\xi$ is simply the variable of integration.
The above Wigner function describes a Gaussian state, which is simply the ground state of a harmonic oscillator represented by a mode with displacement $\hat{X}_\varepsilon$ and momentum $\hat{Y}_\varepsilon$. The corresponding plot is given in the left panel of Figure~\ref{fig:vac_Wigner_fcn}. Gaussian states are traditionally pictured by error ellipses on a phase plane, as drawn in the right panel of Figure~\ref{fig:vac_Wigner_fcn} (cf.\ right panel of Figure~\ref{fig:vac_meas_result}). Here as well as in Figure~\ref{fig:vac_meas_result}, a red line in both plots circumscribes all the values of $X_\varepsilon$ and $Y_\varepsilon$ that fall inside the standard deviation region of the Wigner function, i.e., the region where all pertinent points are within 1 standard deviation from the center of the distribution. For a vacuum state, such a region is a circle with radius $\sqrt{\mathbb{V}_{cc}}=\sqrt{\mathbb{V}_{ss}}=1/\sqrt{2}$. The area of this circle, equal to $1/2$ in dimensionless units and to $\hbar/2$ in case of dimensional displacement and momentum, is the smallest area a physical quantum state can occupy in a phase space. This fact yields from a very general physical principle, the Heisenberg uncertainty relation, that limits the minimal uncertainty product for canonically conjugate observables (displacement $X_\varepsilon$ and momentum $Y_\varepsilon$, in our case) to be less than $1/2$ in $\hbar$-units:
$$\left(\mathrm{Var}[\hat{X}_\varepsilon]\right)^{1/2}\left(\mathrm{Var}[\hat{Y}_\varepsilon]\right)^{1/2}\geqslant\frac12\,.$$
The fact that for a ground state this area is exactly equal to $1/2$ is due to the fact that it is a pure quantum state, i.e., the state of the particle that can be described by a wave function $\ket{\psi}$, rather than by a density operator $\hat{\rho}$. For more sophisticated Gaussian states with a non-diagonal covariance matrix $\mathbb{V}$, the Heisenberg uncertainty relation reads:
\begin{equation}
\label{eq:CovMat_HUP}
\det\mathbb{V}\geqslant\frac14\,,
\end{equation}
and noise ellipse major semi-axes are given by the square root of the matrix $\mathbb{V}$ eigenvalues.

Note the difference between Figures~\ref{fig:vac_meas_result} and \ref{fig:vac_Wigner_fcn}; the former features the result of measurement of an ensemble of oscillators (subsequent light pulses with infinitesimally short duration $\varepsilon$), while the latter
gives the probability density function for a single oscillator displacement and momentum.

\subsubsection{Coherent state}
\label{sec:coherent_state}

Another important state of light is a \emph{coherent state} (see, e.g.,~\cite{1995BookWaMi, 97BookScZu, 95BookMaWo, 01BookSchleich}). It is straightforward to introduce a coherent state $\ket{\alpha}$ of a single mode or a harmonic oscillator as a result of its ground state $\ket{0}$ shift on a complex plane by the distance and in the direction governed by a complex number $\alpha=|\alpha|e^{i\arg(\alpha)}$. This can be caused, e.g., by the action of a classical effective force on the oscillator. Such a shift can be described by a unitary operator called a displacement operator, since its action on a ground state $\ket{0}$ inflicts its shift in a phase plane yielding a state that is called a coherent state:
$$\ket{\alpha} = \hat{D}[\alpha]\ket{0} \equiv e^{\alpha\hat{a}^\dag-\alpha^*\hat{a}}\ket{0}\,,$$
or, more vividly, in q-representation of a corresponding mode of the field~\cite{01BookSchleich}
$$\psi_{coh}(X_\omega) \equiv \braket{X_\omega}{\alpha} = \frac{1}{\sqrt[4]{\pi}}\exp\left\{-\frac{(X_\omega-\sqrt{2}\alpha)^2}{2}\right\}\,.$$
The shift described by $\hat{D}[\alpha]$ is even more apparent if one writes down its action on an annihilation (creation) operator:
$$\hat{D}^\dag[\alpha]\hat{a}\hat{D}[\alpha] = \hat{a}+\alpha\,, \quad\left(\hat{D}^\dag[\alpha]\hat{a}^\dag\hat{D}[\alpha]= \hat{a}^\dag+\alpha^*\right)\,.$$
Moreover, a coherent state is an eigenstate of the annihilation operator:
\begin{equation*}
  \hat{a}\ket{\alpha} = \alpha\ket{\alpha}\,.
\end{equation*}

Using the definitions of the mode quadrature operators $\hat{X}_\omega\equiv\hat{X}$ and $\hat{Y}_\omega\equiv\hat{Y}$ (dimensionless oscillator displacement and momentum normalized by zero-point oscillations amplitude) given above, one immediately obtains for their mean values in a coherent state:
\begin{equation*}
  \bra{\alpha}\hat{X}\ket{\alpha} = \bra{0}\hat{D}^\dag[\alpha]\hat{X}\hat{D}[\alpha]\ket{0} = \sqrt{2}\mathrm{Re}[\alpha]\,,\qquad
  \bra{\alpha}\hat{Y}\ket{\alpha} = \bra{0}\hat{D}^\dag[\alpha]\hat{Y}\hat{D}[\alpha]\ket{0} = \sqrt{2}\mathrm{Im}[\alpha]\,.
\end{equation*}
Further calculation shows that quadratures variances:
\begin{equation*}
  \mathrm{Var}[\hat{X}] = \bra{\alpha}\hat{X}^2\ket{\alpha}-\left(\bra{\alpha}\hat{X}\ket{\alpha}\right)^2 =\frac12\,,\qquad \mathrm{Var}[\hat{Y}] = \bra{\alpha}\hat{Y}^2\ket{\alpha}-\left(\bra{\alpha}\hat{Y}\ket{\alpha}\right)^2 =\frac12
\end{equation*}
have the same values as those for a ground state. These two facts unequivocally testify in favour of the statement that a coherent state is just the ground state shifted from the origin of the phase plane to the point with coordinates $(\mean{\hat{X}}_\alpha,\,\mean{\hat{Y}}_\alpha) = \sqrt{2}(\mathrm{Re}[\alpha]\,,\mathrm{Re}[\alpha])$. It is instructive to calculate a Wigner function for the coherent state using a definition of Eq.~\eqref{eq:WignerFcn_vac}:
\begin{eqnarray*}
  W_{\ket{\alpha}}(X,\,Y) = \intinfty\frac{d\xi}{2\pi}\exp\left\{-i\xi Y\right\}\braket{X+\xi/2}{\alpha}\braket{\alpha}{X-\xi/2}=\\
\dfrac{1}{2\pi\sqrt{\det\mathbb{V}_{\vac}}}\exp\left\{-\frac12\{X-\sqrt{2}\mathrm{Re}[\alpha],\,Y-\sqrt{2}\mathrm{Im}[\alpha]\}^{\mathsf{T}}\mathbb{V}_{\vac}^{-1}\{X-\sqrt{2}\mathrm{Re}[\alpha],\,Y-\sqrt{2}\mathrm{Im}[\alpha]\}\right\} =\\ \frac{1}{\pi}\exp\left\{-\left[(X-\sqrt{2}\mathrm{Re}[\alpha])^2+(Y-\sqrt{2}\mathrm{Im}[\alpha])^2\right]\right\}\,,
\end{eqnarray*}
which once again demonstrates the correctness of the former statement.

Generalization to the case of continuum of modes comprising a light
wave is straightforward~\cite{PhysRevA.42.4102_1990} and goes along
the same lines as the definition of the field vacuum state, namely
(see Eq.~\eqref{eq:field_vac_state}):
\begin{equation}
  \ket{\alpha(\omega)} \equiv \bigotimes\limits_\omega\ket{\alpha}_\omega = \bigotimes\limits_\omega\hat{D}[\alpha(\omega)]\ket{0}_\omega =
   \exp\left\{\intinfty\frac{d\omega}{2\pi}(\alpha(\omega)\hat{a}_\omega^\dag-\alpha^*(\omega)\hat{a}_\omega)\right\}\ket{\vac}\,,
\end{equation}
where $\ket{\alpha}_\omega$ is the coherent state that the mode of the
field with frequency $\omega$ is in, and $\alpha(\omega)$ is the
distribution of complex amplitudes $\alpha$ over frequencies
$\omega$. Basically, $\alpha(\omega)$ is the spectrum of normalized
complex amplitudes of the field, i.e., $\alpha(\omega)\propto
\mathcal{E}(\omega)$. For example, the state of a free light wave
emitted by a perfectly monochromatic laser with emission frequency
$\omega_p$ and mean optical power $\mathcal{I}_0$ will be defined by
$\alpha(\omega) =
\pi\sqrt{\frac{2\mathcal{I}_0}{\hbar\omega_p}}\delta(\omega-\omega_p)$,
which implies that only the mode at frequency $\omega_p$ will be in a
coherent state, while all other modes of the field will be in their
ground states.

\epubtkImage{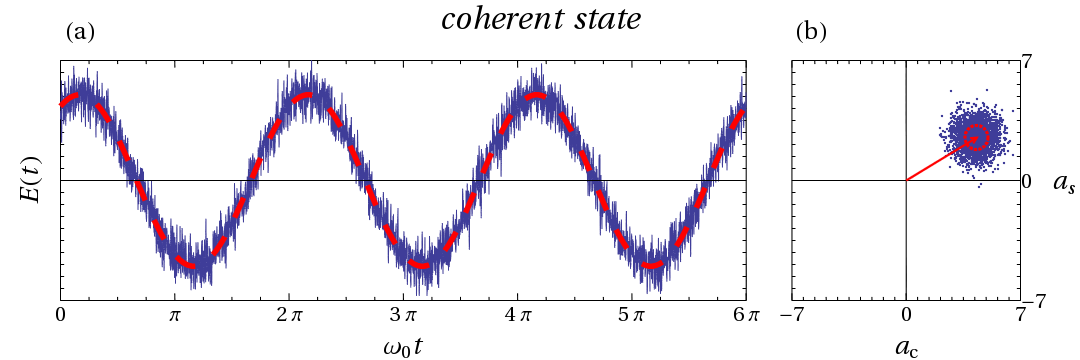}{%
\begin{figure}[htbp]
  \centerline{\includegraphics[width=\textwidth]{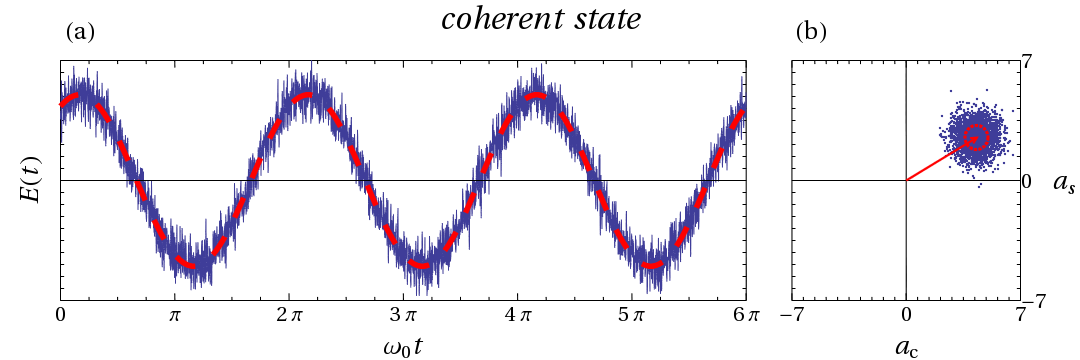}}
  \caption{Light field in a coherent quantum state $\ket{\alpha(\omega)}$. \emph{Left panel a)} features a typical result one could get measuring the (normalized) electric field strength of the light wave in a coherent state as a function of time. \emph{Right panel b)} represents a phase space picture of the results of measurement. The red dashed line in the left panel marks the mean value $\mean{\hat{E}(t)}$. The red arrow in the right panel features the vector of the mean values of quadrature amplitudes, i.e., $\boldsymbol{A}$, while the red dashed circle displays the error ellipse for the state $\ket{\alpha(\omega)}$ that encircles the area of single standard deviation for a two-dimensional random vector $\hat{\boldsymbol{a}}$ of quadrature amplitudes. The principle radii of the error ellipse (equal in the coherent state case) are equal to square roots of the covariance matrix $\mathbb{V}_{\mathrm{coh}}$, i.e., to $1/\sqrt{2}$.}
\label{fig:coh_meas_result}
\end{figure}}

Operator $\hat{D}[\alpha]$ is unitary, i.e., $\hat{D}^\dag[\alpha]\hat{D}[\alpha]=\hat{D}[\alpha]\hat{D}^\dag[\alpha]=\hat{I}$ with $\hat{I}$ the identity operator, while the physical meaning is in the translation and rotation of the Hilbert space that keeps all the physical processes unchanged. Therefore, one can simply use vacuum states instead of coherent states and subtract the mean values from the corresponding operators in the same way we have done previously for the light wave classical amplitudes, just below Eq.~\eqref{eq:2photon_EMW_quadratures}.
The covariance matrix and the matrix of power spectral densities for the quantum noise of light in a coherent state is thus the same as that of a vacuum state case.

The typical result one can get measuring the electric field strength of light emitted by the aforementioned ideal laser is drawn in the left panel of Figure~\ref{fig:coh_meas_result}.

\subsubsection{Squeezed state}
\label{sec:squeezed_state}

One more quantum state of light that is worth consideration is a squeezed state. To put it in simple words, it is a state where one of the oscillator quadratures variance appears decreased by some factor compared to that in a vacuum or coherent state, while the conjugate quadrature variance finds itself swollen by the same factor, so that their product still remains Heisenberg-limited. Squeezed states of light are usually obtained as a result of a parametric down conversion (PDC) process~\cite{67a1Kl, 69a1Kl} in optically nonlinear crystals. This is the most robust and experimentally elaborated way of generating squeezed states of light for various applications, e.g., for GW detectors~\cite{PhysRevLett.95.211102_2005_Vahlbruch, PhysRevLett.100.033602_2008, Takeno:07}, or for quantum communications and computation purposes~\cite{RevModPhys.77.513_2005}. However, there is another way to generate squeezed light by means of a ponderomotive nonlinearity inherent in such optomechanical devices as GW detectors. This method, first proposed by Corbitt et~al.~\cite{PhysRevA.73.023801_2006}, utilizes the parametric coupling between the resonance frequencies of the optical modes in the Fabry--P\'{e}rot cavity and the mechanical motion of its mirrors arising from the quantum radiation pressure fluctuations inflicting random mechanical motion on the cavity mirrors. Further, we will see that the light leaving the signal port of a GW interferometer finds itself in a ponderomotively squeezed state (see, e.g.,~\cite{02a1KiLeMaThVy} for details). A dedicated reader might find it illuminating to read the following review articles on this topic~\cite{Schnabel2010, LPOR:LPOR201000034}.

Worth noting is the fact that generation of squeezed states of light is the process that inherently invokes two modes of the field and thus naturally calls for usage of the two-photon formalism contrived by Caves and Schumaker~\cite{85a1CaSch, 85a2CaSch}. To demonstrate this let us consider the physics of a squeezed state generation in a nonlinear crystal. Here photons of a pump light with frequency $\omega_p=2\omega_0$ give birth to pairs of correlated photons with frequencies $\omega_1$ and $\omega_2$ (traditionally called \emph{signal} and \emph{idler}) by means of the nonlinear dependence of polarization in a birefringent crystal on electric field. Such a process can be described by the following Hamiltonian, provided that the pump field is in a coherent state $\ket{\alpha}_{\omega_p}$ with strong classical amplitude $|\alpha_p|\gg1$ (see, e.g., Section~5.2 of~\cite{1995BookWaMi} for details):
\begin{equation}
\label{eq:PDC_hamiltonian}
  \frac{\hat{H}_{\mathrm{PDC}}}{\hbar} = \omega_1\hat{a}^\dag_1\hat{a}_1 + \omega_2\hat{a}^\dag_2\hat{a}_2 + i\left(\chi\hat{a}^\dag_1\hat{a}^\dag_2e^{-2i\omega_0t}-\chi^*\hat{a}_1\hat{a}_2e^{2i\omega_0t}\right)\,,
\end{equation}
where $\hat{a}_{1,2}$ describe annihilation operators for the photons of the signal and idler modes and $\chi=\rho e^{2i\phi}$ is the complex coupling constant that is proportional to the second-order susceptibility of the crystal and to the pump complex amplitude. Worth noting is the meaning of $t$ in this Hamiltonian: it is a parameter that describes the duration of a pump light interaction with the nonlinear crystal, which, in the simplest situation, is either the length of the crystal divided by the speed of light $c$, or, if the crystal is placed between the mirrors of the optical cavity, the same as the above but multiplied by an average number of bounces of the photon inside this cavity, which is, in turn, proportional to the cavity finesse $\mathcal{F}$. It is straightforward to obtain the evolution of the two modes in the interaction picture (leaving apart the obvious free evolution time dependence $e^{-i\omega_{s,i}t}$) solving the Heisenberg equations:
\begin{equation}
\label{eq:PDC_HEqs_solution}
  \hat{a}_1(t) = \hat{a}_1\cosh \rho t+\hat{a}^\dag_2e^{2i\phi}\sinh \rho t\,,\qquad
  \hat{a}_2(t) = \hat{a}_2\cosh \rho t+\hat{a}^\dag_1e^{2i\phi}\sinh \rho t\,.
\end{equation}
Let us then assume the signal and idler mode frequencies symmetric with respect to the half of pump frequency $\omega_0=\omega_p/2$: $\omega_1\to\omega_+=\omega_0+\Omega$ and $\omega_2\to\omega_-=\omega_0-\Omega$ ($\hat{a}_1\to\hat{a}_+$ and $\hat{a}_2\to\hat{a}_-$). Then the electric field of a two-mode state going out of the nonlinear crystal will be written as (we did not include the pump field here assuming it can be ruled out by an appropriate filter):
\begin{equation*}
  \hat{E}(t) = \mathcal{C}_0\left[\hat{X}(t)\cos\omega_0t+\hat{Y}(t)\sin\omega_0t\right],
\end{equation*}
where two-mode quadrature amplitudes $\hat{X}(t)$ and $\hat{Y}(t)$ are defined along the lines of Eqs.~\eqref{eq:2photon_quads_def}, keeping in mind that only idler and signal components at the frequencies $\omega_\pm-\omega_0 = \pm\Omega$ should be kept in the integral, which yields:
\begin{eqnarray*}
  \hat{X}(t) &=& \frac{1}{\sqrt{2}}\left[\hat{a}_+(t)e^{i\Omega t}+\hat{a}^\dag_+(t)e^{-i\Omega t}+\hat{a}_-(t)e^{-i\Omega t}+\hat{a}^\dag_-(t)e^{i\Omega t}\right] = \hat{a}_c^{\sqz} e^{-i\Omega t}+\hat{a}_c^{\sqz\dag} e^{i\Omega t}\,,\\
  \hat{Y}(t) &=& \frac{1}{i\sqrt{2}}\left[\hat{a}_+(t)e^{i\Omega t}-\hat{a}^\dag_+(t)e^{-i\Omega t}+\hat{a}_-(t)e^{-i\Omega t}-\hat{a}^\dag_-(t)e^{i\Omega t}\right] = \hat{a}_s^{\sqz} e^{-i\Omega t}+\hat{a}_s^{\sqz\dag} e^{i\Omega t}\,,
\end{eqnarray*}
where $\hat{a}_c^{\sqz} = (\hat{a}_+(t)+\hat{a}^\dag_-(t))/\sqrt{2}$ and $\hat{a}_s^{\sqz} = (\hat{a}_+(t)-\hat{a}^\dag_-(t))/(i\sqrt{2})$ are the spectral quadrature amplitudes of the two-mode field at sideband frequency $\Omega$ (cf.\ Eqs.~\eqref{eq:2photon_quadratures}) after it leaves the nonlinear crystal.
Substituting Eqs.~\eqref{eq:PDC_HEqs_solution} into the above expressions yields transformation rules for quadrature amplitudes:
  \begin{eqnarray}
\label{eq:SQZ_quad_transform}
 \hat{a}_c^{\sqz} &=& \hat{a}_c\left(\cosh\rho t+\cos2\phi\sinh\rho t\right)+\hat{a}_s\sin2\phi\sinh\rho t\,,\nonumber\\
 \hat{a}_s^{\sqz} &=& \hat{a}_c\sin2\phi\sinh\rho t+\hat{a}_s\left(\cosh\rho t-\cos2\phi\sinh\rho t\right)\,,
\end{eqnarray}
where $\hat{a}_c=(\hat{a}_++\hat{a}^\dag_-)/\sqrt{2}$ and $\hat{a}_s=(\hat{a}_+-\hat{a}^\dag_-)/(i\sqrt{2})$ stand for initial values of spectral quadrature amplitudes of the two-mode light wave created in the PDC process. A close look at these transformations written in the matrix form reveals that it can be represented as the following sequence:
  \begin{equation}
\label{eq:SQZ_matrix_transform}
  \hat{\boldsymbol{a}}^{\sqz} = \begin{pmatrix}
    \hat{a}_c^{\sqz}\\
    \hat{a}_s^{\sqz}
  \end{pmatrix} = \mathbb{S}_{\sqz}[\rho t,\phi]\hat{\boldsymbol{a}} = \mathbb{P}[\phi]\mathbb{S}_{\sqz}[\rho t,0]\mathbb{P}[-\phi]\hat{\boldsymbol{a}}
\end{equation}
where
\begin{equation}
\label{eq:SQZ_matrix_transform_def}
  \mathbb{S}_{\sqz}[\rho t,\phi]\equiv
  \begin{bmatrix}
    \cosh\rho t+\cos2\phi\sinh\rho t & \sin2\phi\sinh\rho t\\
    \sin2\phi\sinh\rho t & \cosh\rho t-\cos2\phi\sinh\rho t
  \end{bmatrix}\ \Longrightarrow\ \mathbb{S}_{\sqz}[\rho t,0] =
  \begin{bmatrix}
    e^{\rho t} & 0\\
    0 & e^{-\rho t}
  \end{bmatrix}
\end{equation}
are squeezing matrices in general and in special ($\phi=0$) case, while $\mathbb{P}[\phi]$ stands for a counterclockwise 2D-rotation matrix by angle $\phi$ defined by~\eqref{eq:CCW_rotation_matrix}.
Therefore, the evolution of a two-mode light quadrature amplitude vector $\hat{\boldsymbol{a}}$ in a PDC process described by the Hamiltonian~\eqref{eq:PDC_hamiltonian} consists of a clockwise rotation by an angle $\phi$ followed by a deformation along the main axes (stretching along the $a_c$-axis and proportional squeezing along the $a_s$-axis) and rotation back by the same angle. It is straightforward to show that vector $\hat{\boldsymbol{X}}^{\sqz} = \left\{\hat{X}(t),\,\hat{Y}(t)\right\}^{\mathsf{T}} = \hat{\boldsymbol{a}}^{\sqz}e^{-i\Omega t}+\hat{\boldsymbol{a}}^{\sqz*}e^{i\Omega t}$ transforms similarly (here $\hat{\boldsymbol{a}}^{\sqz*}=\left\{\hat{a}_c^{\sqz\dag},\,\hat{a}_s^{\sqz\dag}\right\}^{\mathsf{T}}$ ).

This geometric representation is rather useful, particularly for the characterization of a squeezed state. If the initial state of the two-mode field is a vacuum state then the outgoing field will be in a squeezed vacuum state. One can define it as a result of action of a special squeezing operator $\hat{S}[\rho t,\phi]$ on the vacuum state
\begin{equation}
\label{eq:SQZ_operator_action}
  \ket{\sqz_0(\rho t,\phi)} = \hat{S}[\rho t,\phi]\ket{\vac}\,.
\end{equation}
This operator is no more and no less than the evolution operator for the PDC process in the interaction picture, i.e.,
\begin{equation}
\label{eq:SQZ_operator_def}
  \hat{S}[\rho t,\phi] \equiv \exp\left\{\rho t(\hat{a}_+\hat{a}_-e^{-2i\phi}-\hat{a}_+^\dag\hat{a}_-^\dag e^{2i\phi})\right\}\,.
\end{equation}
Action of this operator on the two-photon quadrature amplitudes is fully described by Eqs.~\eqref{eq:SQZ_operator_action}:
\begin{equation*}
  \hat{\boldsymbol{a}}^{\sqz}=\hat{S}^\dag[\rho t,\phi]\hat{\boldsymbol{a}}\hat{S}[\rho t,\phi] = \mathbb{P}[\phi]\mathbb{S}_{\sqz}[\rho t,0]\mathbb{P}[-\phi]\hat{\boldsymbol{a}}\,,
\end{equation*}
while annihilation operators of the corresponding modes $\hat{a}_\pm$
are transformed in accordance with Eqs.~\eqref{eq:PDC_HEqs_solution}.

\epubtkImage{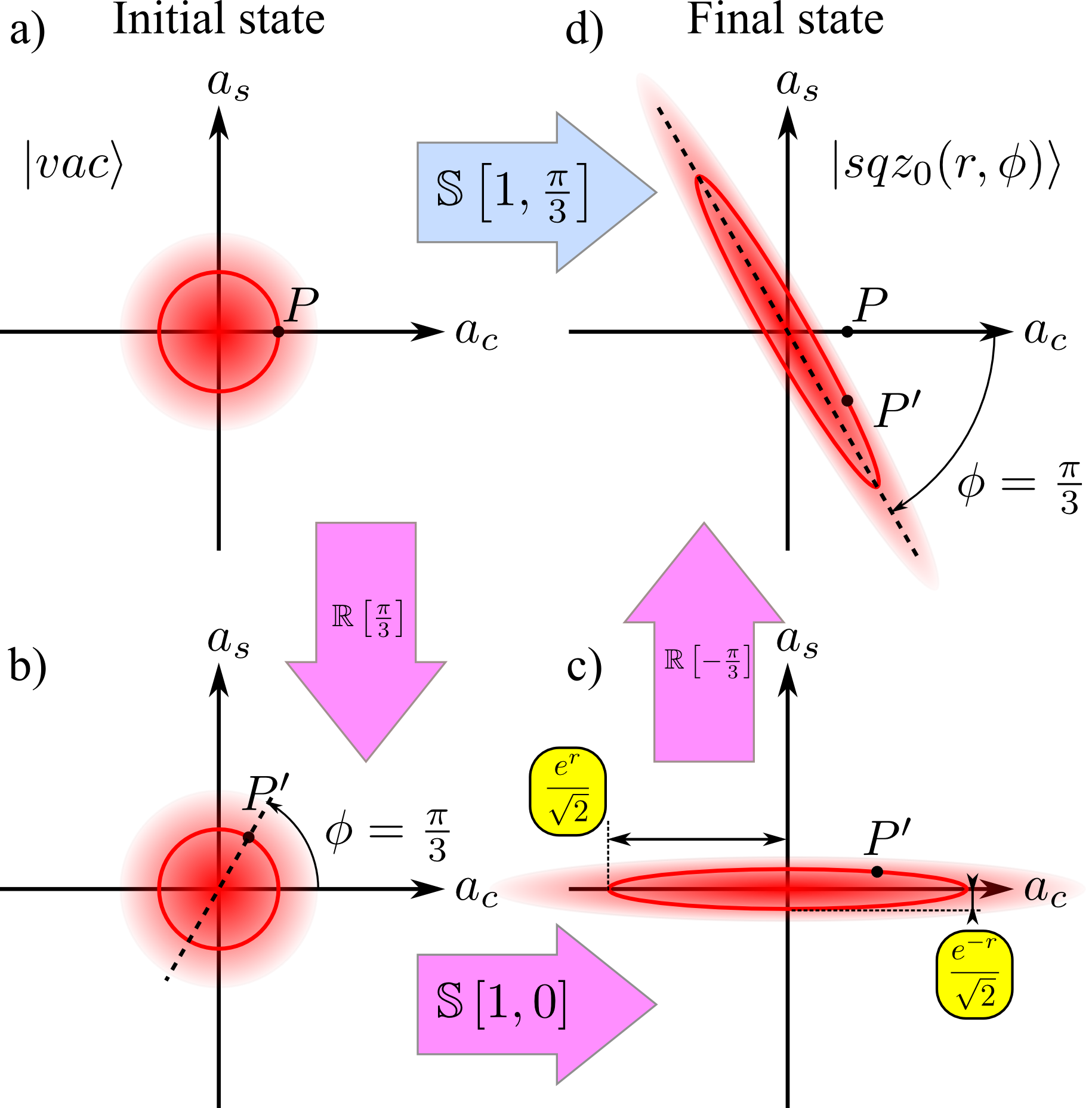}{%
\begin{figure}[htb]
  \centerline{\includegraphics[width=.6\textwidth]{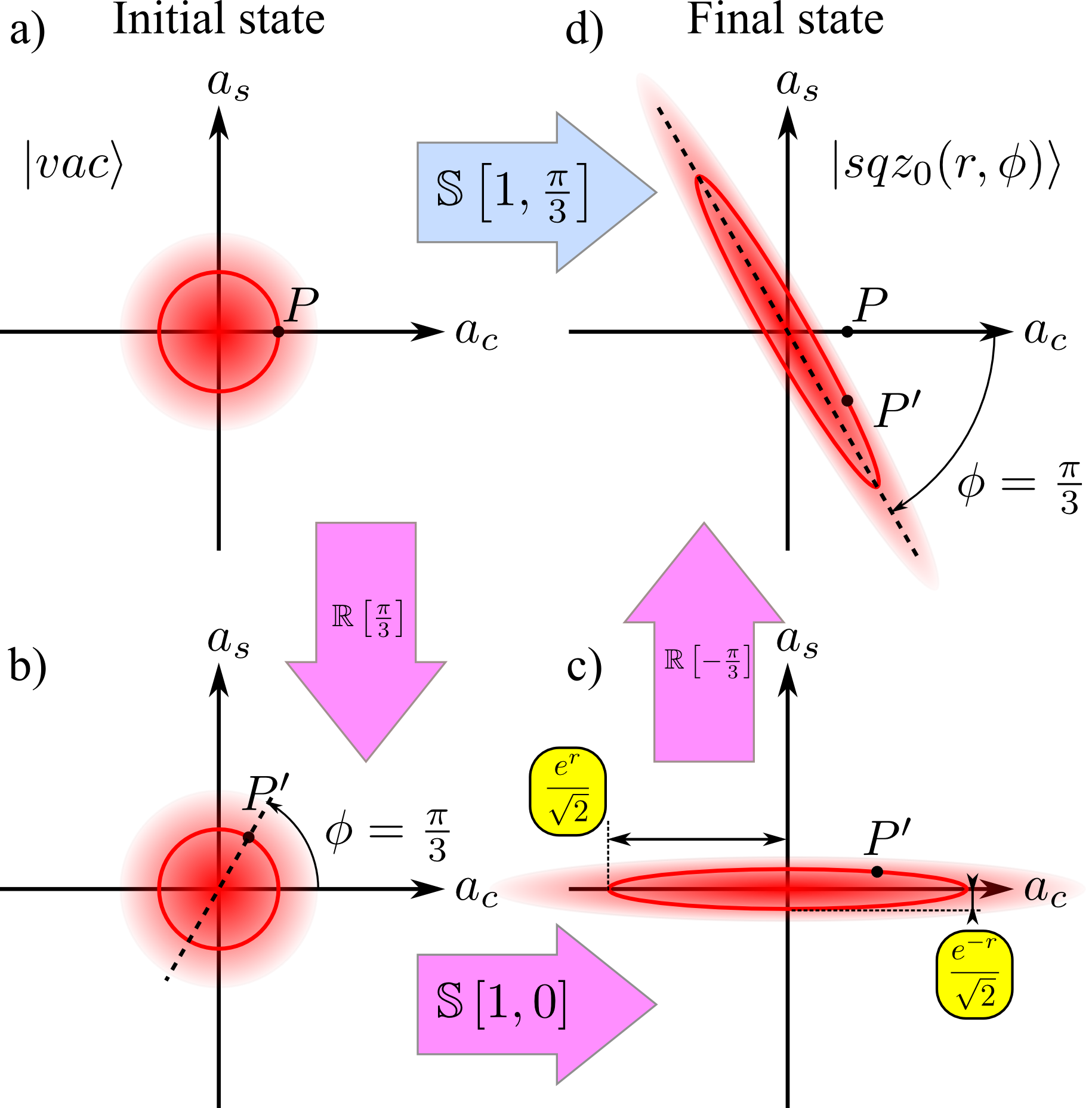}}
  \caption{Schematic plot of a vacuum state transformation under the action of the squeezing operator $\hat{S}[r,\phi]$. Eqs.~\eqref{eq:SQZ_matrix_transform} demonstrate the equivalence of the general squeezing operator $\hat{S}[r,\phi]$ to a sequence of phase plane counterclockwise rotation by an angle $\phi$ (transition from a) to b)), phase plane squeezing and stretching by a factor $e^r$ (transition from b) to c)) and rotation back by the same angle $\phi$ (transition from c) to d)). Point $P'$ tracks how transformations change the initial state marked with point $P$.}
  \label{fig:sqz_op_action_on_vac}
\end{figure}}

The linearity of the squeezing transformations implies that the squeezed vacuum state is Gaussian since it is obtained from the Gaussian vacuum state and therefore can be fully characterized by the expectation values of operators $\hat{X}$ and $\hat{Y}$ and their covariance matrix $\mathbb{V}_{\sqz}$. Let us calculate these values:
\begin{equation*}
  \mean{\hat{X}}_{\sqz} = \bra{\sqz_0(r,\phi)}\hat{X}\ket{\sqz_0(r,\phi)} = \bra{\vac}\hat{S}^\dag[r,\phi]\hat{X}\hat{S}[r,\phi]\ket{\vac} =\bra{\vac}\hat{X}(t)\ket{\vac} = 0\,,\quad \mean{\hat{Y}}_{\sqz} = 0\,,
\end{equation*}
and for a covariance matrix one can get the following expression:
\begin{eqnarray}
\label{eq:SQZ_V}
  \mathbb{V}_{\sqz} = \bra{\sqz_0(r,\phi)}\hat{\boldsymbol{X}}\circ\hat{\boldsymbol{X}}^{\mathsf{T}}\ket{\sqz_0(r,\phi)} = \mathbb{P}[-\phi]\mathbb{S}_{\sqz}[r,0]\mathbb{V}_{\vac}\mathbb{S}_{\sqz}[r,0]\mathbb{P}[\phi] =\nonumber\\ \frac12\mathbb{P}[-\phi]\mathbb{S}_{\sqz}[2r,0]\mathbb{P}[\phi]=
  \frac12\begin{bmatrix}
    \cos\phi & \sin\phi\\
    -\sin\phi & \cos\phi
  \end{bmatrix}
  \begin{bmatrix}
    e^{2r} & 0\\
    0 & e^{-2r}
  \end{bmatrix}
\begin{bmatrix}
    \cos\phi & -\sin\phi\\
    \sin\phi & \cos\phi
  \end{bmatrix}\,,
\end{eqnarray}
where we introduced squeezing parameter $r\equiv\rho t$ and used a short notation for the symmetrized outer product of vector $\hat{\boldsymbol{X}}$ with itself:
$$
\hat{\boldsymbol{X}}\circ\hat{\boldsymbol{X}}^{\mathsf{T}}\equiv
\begin{bmatrix}
  \hat{X}\circ\hat{X} & \hat{X}\circ\hat{Y}\\
  \hat{Y}\circ\hat{X} & \hat{Y}\circ\hat{Y}
\end{bmatrix}\,.
$$

The squeezing parameter $r$ is the quantity reflecting the strength of the squeezing. This way of characterizing the squeezing strength, though convenient enough for calculations, is not very ostensive. Conventionally, squeezing strength is measured in decibels (dB) that are related to the squeezing parameter $r$ through the following simple formula:
\begin{equation}
\label{eq:SQZ_factor_dB_def}
  r_{\mathrm{dB}} = 10\log_{10}e^{2r}=20r\log_{10}e\ \Longleftrightarrow r = r_{\mathrm{dB}}/(20\log_{10}e)\,.
\end{equation}
For example, 10~dB squeezing corresponds to $r\simeq1.15$.

The covariance matrix~\eqref{eq:SQZ_V} refers to a unique error ellipse on a phase plane with semi-major axis  $e^r/\sqrt{2}$ and semi-minor axis $e^{-r}/\sqrt{2}$ rotated by angle $\phi$ clockwise as featured in Figure~\ref{fig:sqz_op_action_on_vac}.

It would be a wise guess to make, that a squeezed vacuum Wigner function can be obtained from that of a vacuum state, using these simple geometric considerations. Indeed, for a squeezed vacuum state it reads:
\begin{equation}
\label{eq:SQZ_WignerFcn}
W_{\ket{\sqz}}(X,\,Y) = \dfrac{1}{2\pi\sqrt{\det\mathbb{V}_{\sqz}}}\exp\left\{-\frac12\{X,\,Y\}^{\mathsf{T}}\mathbb{V}_{\sqz}^{-1}\{X,\,Y\}\right\}\,,
\end{equation}
where the error ellipse refers to the level where the Wigner function
value falls to $1/\sqrt{e}$ of the maximum. The corresponding plot and
phase plane picture of the squeezed vacuum Wigner function are
featured in Figure~\ref{fig:sqz_Wigner_fcn}.

\epubtkImage{fig17.png}{%
\begin{figure}[htbp]
  \centerline{\includegraphics[width=.47\textwidth]{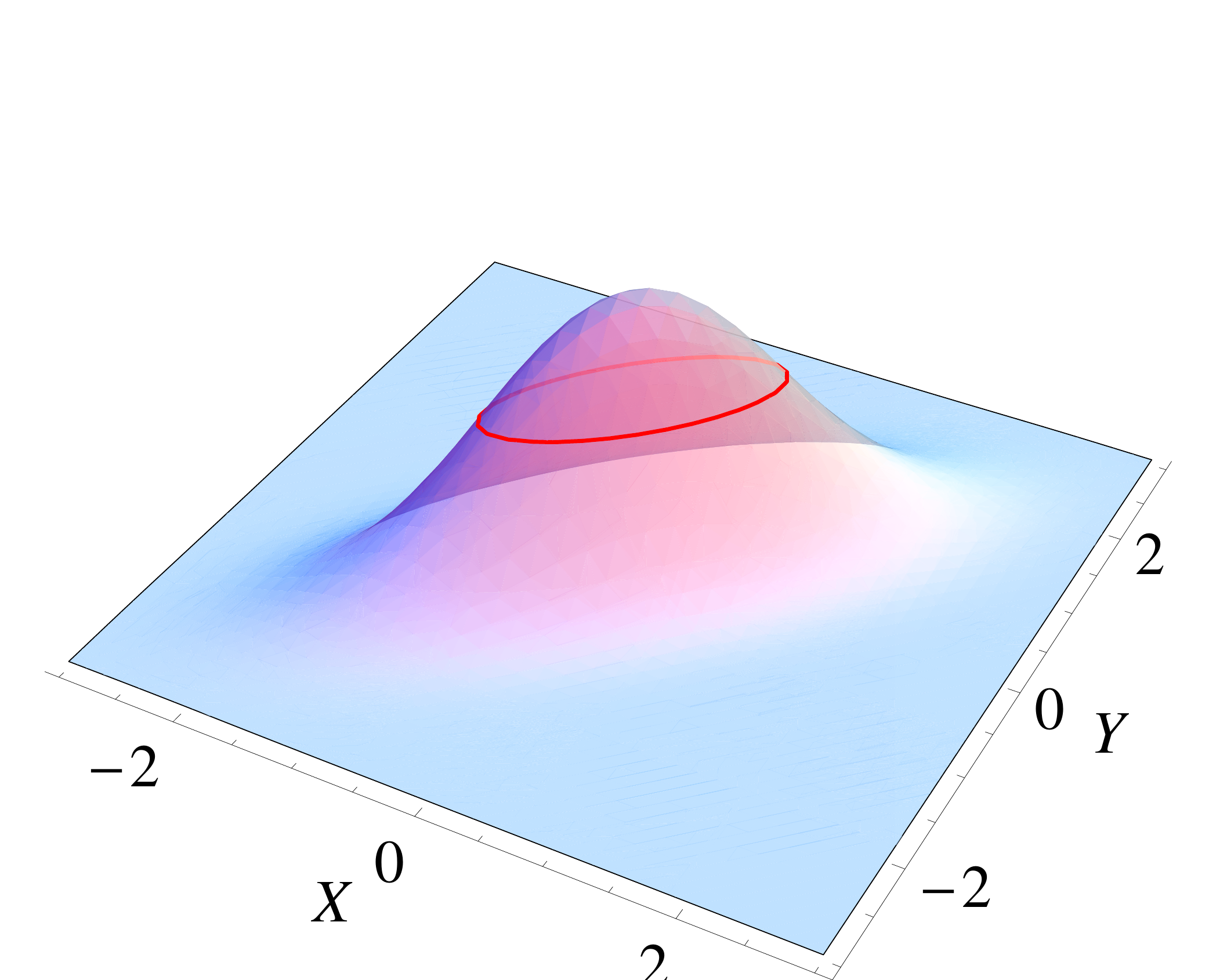}\hfill\includegraphics[width=.33\textwidth]{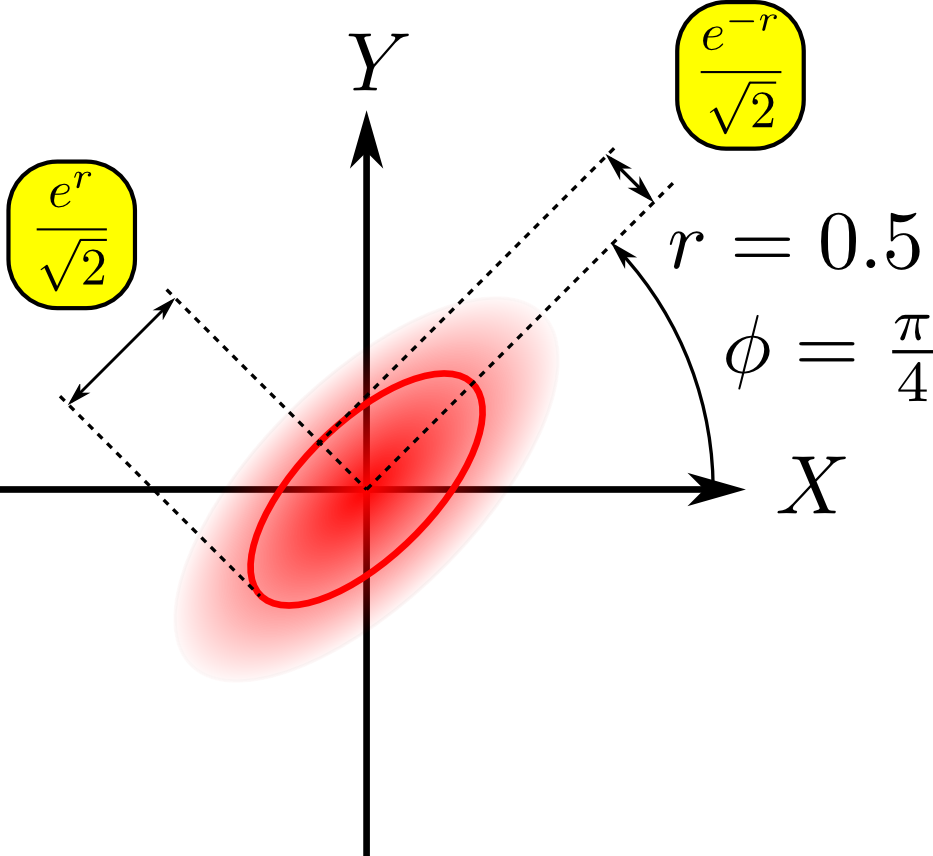}}
  \caption{\emph{Left panel}: Wigner function of a squeezed vacuum state with squeeze parameter $r=0.5$ (5~dB) and rotation angle $\phi=-\pi/4$. \emph{Right panel:} Error ellipse corresponding to that Wigner function.}
  \label{fig:sqz_Wigner_fcn}
\end{figure}}

Another important state that arises in GW detectors is the displaced squeezed state $\ket{\sqz_\alpha(r,\phi)}$ that is obtained from the squeezed vacuum state in the same manner as the coherent state yields from the vacuum state, i.e., by the application of the displacement operator (equivalent to the action of a classical force):
\begin{equation}
  \ket{\sqz_\alpha(r,\phi)} = \hat{D}[\alpha]\ket{\sqz_0(r,\phi)} = \hat{D}[\alpha]\hat{S}[r,\phi]\ket{\vac}\,.
\end{equation}
The light leaving a GW interferometer from the signal port finds itself in such a state, if a classical GW-like force changes the difference of the arm lengths, thus displacing a ponderomotively squeezed vacuum state in phase quadrature $Y$ by an amount proportional to the magnitude of the signal force. Such a displacement has no other consequence than simply to shift the mean values of $\hat{X}$ and $\hat{Y}$ by some constant values dependent on shift complex amplitude $\alpha$:
\begin{equation*}
\mean{\hat{X}}_{\sqz}=\sqrt{2}\mathrm{Re}[\alpha]\,, \qquad \mean{\hat{Y}}_{\sqz}=\sqrt{2}\mathrm{Im}[\alpha]\,.
\end{equation*}

\epubtkImage{fig18.png}{%
\begin{figure}[htb]
  \centerline{\includegraphics[width=\textwidth]{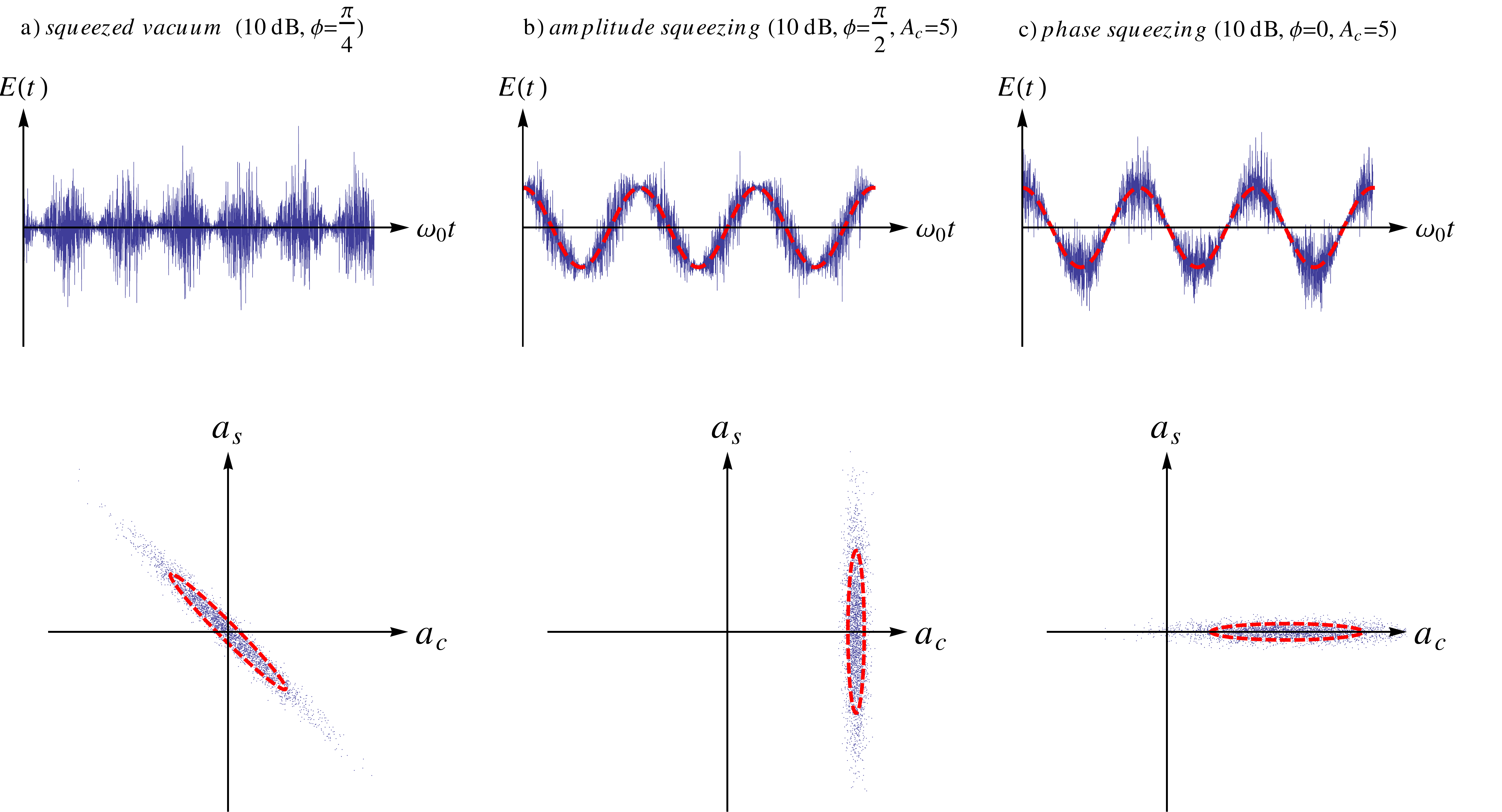}}
  \caption{Light field in a squeezed state $\ket{\sqz_\alpha(r,\phi)}$. \emph{Upper row} features time dependence of the electric field strength $E(t)$ in three different squeezed states (10~dB squeezing assumed for all): a) squeezed vacuum state with squeezing angle $\phi=\pi/4$; b) displaced squeezed state with classical amplitude $A_c=5$ (mean field strength oscillations $\mean{\hat{E}(t)}_{\sqz}$ are given by red dashed line) and amplitude squeezing ($\phi=\pi/2$); c) displaced squeezed state with classical amplitude $A_c=5$ and phase squeezing ($\phi=0$). \emph{Lower row} features error ellipses (red dashed lines) for the corresponding plots in the upper row.}
  \label{fig:sqz_meas_result}
\end{figure}}

Let us now generalize the results of a two-mode consideration to a continuous spectrum case. Apparently, quadrature operators $\hat{X}(t)$ and $\hat{Y}(t)$ are similar to $\hat{a}_c(t)$ and $\hat{a}_s(t)$ for the traveling wave case. Utilizing this similarity, let us define a squeezing operator for the continuum of modes as:
\begin{equation}
  \hat{S}[r(\Omega),\phi(\Omega)]\equiv\exp\left\{\intinfty\frac{d\Omega}{2\pi}r(\Omega)\left[\hat{a}_+\hat{a}_-e^{-2i\phi(\Omega)}-\hat{a}^\dag_+\hat{a}^\dag_-e^{2i\phi(\Omega)}\right]\right\}\,,
\end{equation}
where $r(\Omega)$ and $\phi(\Omega)$ are frequency-dependent squeezing factor and angle, respectively. Acting with this operator on a vacuum state of the travelling wave yields a squeezed vacuum state of a continuum of modes in the very same manner as in Eq.~\eqref{eq:SQZ_operator_action}. The result one could get in the measurement of the electric field amplitude of light in a squeezed state as a function of time is presented in Figure~\ref{fig:sqz_meas_result}. Quadrature amplitudes for each frequency $\Omega$ transform in accordance with Eqs.~\eqref{eq:SQZ_quad_transform}. Thus, we are free to use these formulas for calculation of the power spectral density matrix for a traveling wave squeezed vacuum state. Indeed, substituting $\hat{a}_{c,s}(\Omega)\to\hat{a}^{\sqz}_{c,s}(\Omega)$ in Eq.~\eqref{eq:SpDens_matrix_def} and using Eq.~\eqref{eq:SQZ_matrix_transform} one immediately gets:
\begin{equation}
\label{eq:SQZ_SpDens_matrix}
  \mathbb{S}_{\sqz}(\Omega) = \mathbb{P}[-\phi(\Omega)]\mathbb{S}_{\sqz}[r(\Omega),0]\mathbb{S}_{\vac}(\Omega)\mathbb{S}_{\sqz}[r(\Omega),0]\mathbb{P}[\phi(\Omega)] = \mathbb{S}_{\sqz}(r(\Omega),\phi(\Omega))\,.
\end{equation}
Note that entries of $\mathbb{S}_{\sqz}(\Omega)$ might be frequency dependent if squeezing parameter $r(\Omega)$ and squeezing angle $\phi(\Omega)$ are frequency dependent as is the case in all physical situations. This indicates that quantum noise in a squeezed state of light is not Markovian and this can easily be shown by calculating the the covariance matrix, which is simply a Fourier transform of $\mathbb{S}(\Omega)$ according to the Wiener--Khinchin theorem:
\begin{equation}
  \mathbb{V}_{\sqz}(t-t') = \intinfty\frac{d\Omega}{2\pi}\mathbb{S}_{\sqz}(\Omega)e^{-i\Omega(t-t')}\,.
\end{equation}
Of course, the exact shape of $\mathbb{V}_{\sqz}(t-t')$ could be obtained only if we specify $r(\Omega)$ and $\phi(\Omega)$. Note that the noise described by $\mathbb{V}_{\sqz}(t-t')$ is stationary since all the entries of the covariance matrix (correlation functions) depend on the difference of times $t-t'$.

The spectral density matrix allows for pictorial representation of a
multimode squeezed state where an error ellipse is assigned to each
sideband frequency $\Omega$. This effectively adds one more dimension
to a phase plane picture already used by us for the characterization
of a two-mode squeezed states. Figure~\ref{fig:sqz_cont_3D_diagram}
exemplifies the state of a ponderomotively squeezed light that would
leave the speedmeter type of the interferometer (see Section~\ref{sec:speedmeter}).

\epubtkImage{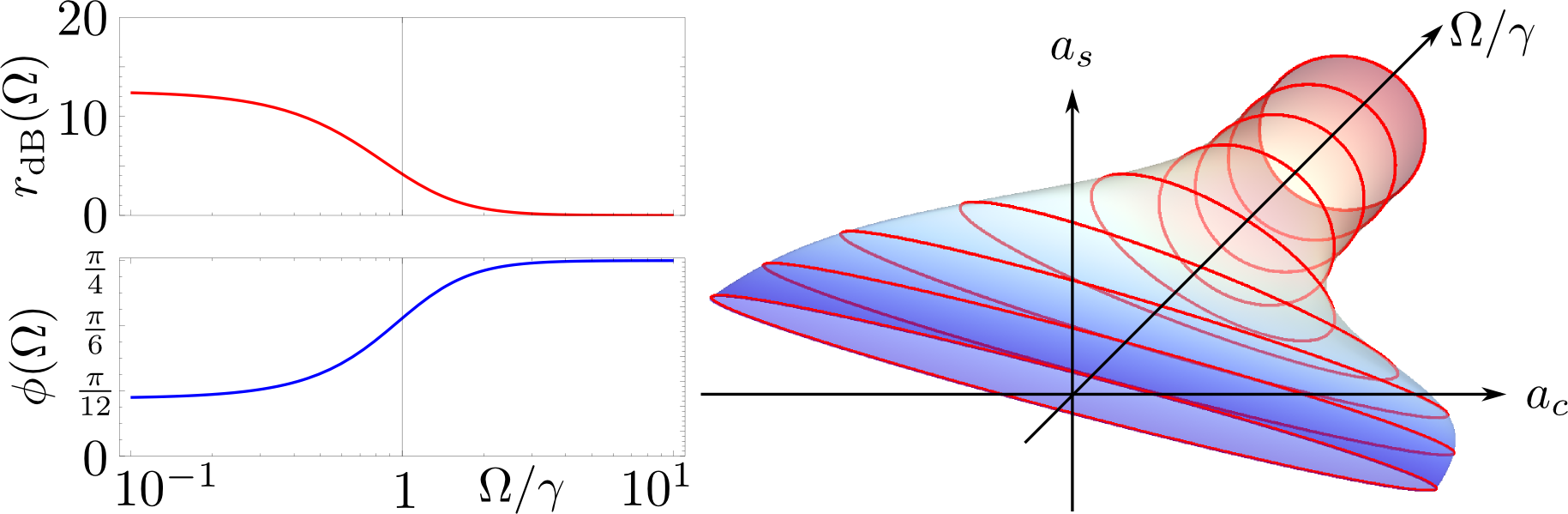}{%
\begin{figure}[htbp]
  \centerline{\includegraphics[width=\textwidth]{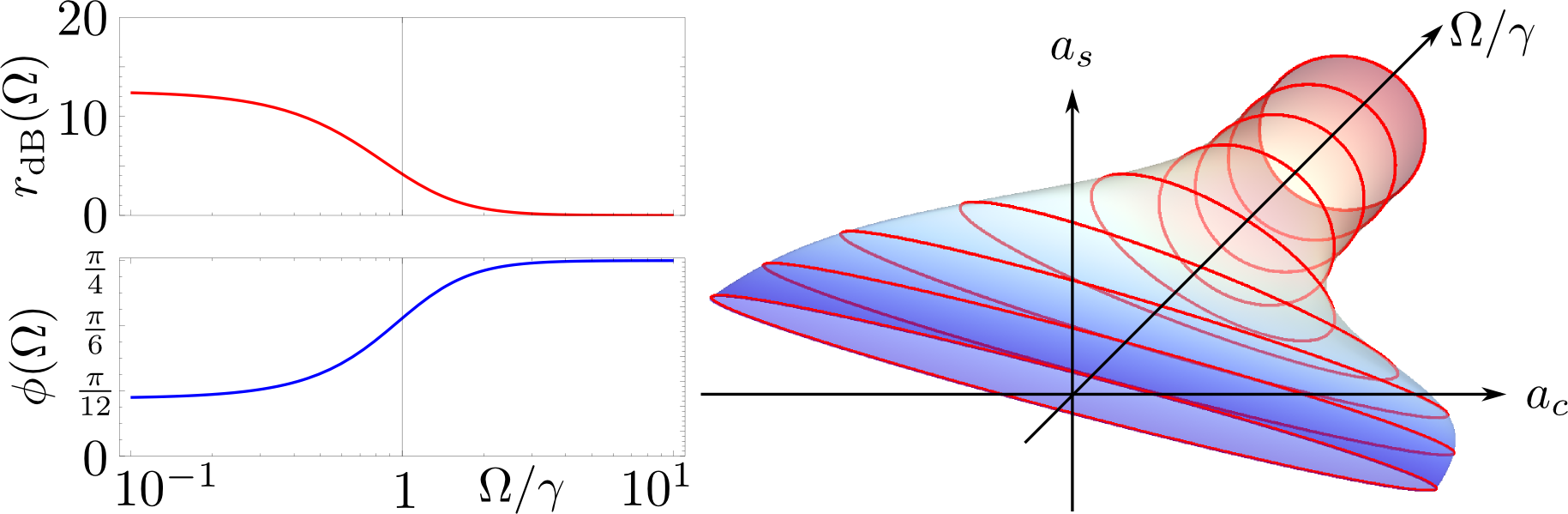}}
  \caption{Example of a squeezed state of the continuum of modes: output state of a speedmeter interferometer. \emph{Left panel} shows the plots of squeezing parameter $r_{\mathrm{dB}}(\Omega)$ and squeezing angle $\phi(\Omega)$ versus normalized sideband frequency $\Omega/\gamma$ (here $\gamma$ is the interferometer half-bandwidth). \emph{Right panel} features a family of error ellipses for different sideband frequencies $\Omega$ that illustrates the squeezed state defined by $r_{\mathrm{dB}}(\Omega)$ and $\phi(\Omega)$ drawn in the left panel.}
  \label{fig:sqz_cont_3D_diagram}
\end{figure}}

\subsection{How to calculate spectral densities of quantum noise in linear optical measurement?}

In this section, we give a brief introduction to calculation of the power spectral densities of quantum noise one usually encounters in linear optical measurement. In optomechanical sensors, as we have discussed earlier, the outgoing light carries the information about the measured quantity (e.g., the displacement due to GW tidal forces) in its phase and (sometimes) amplitude quadratures. The general transformation from the input light characterized by a vector of quadrature amplitudes $\hat{\boldsymbol{a}}(\Omega) = \left\{\hat{a}_c(\Omega),\,\hat{a}_s(\Omega)\right\}^{\mathsf{T}}$ to the readout quantity of a meter is linear and can be written in spectral form as:
\begin{equation}
\label{eq:generic_Y}
  \hat{Y}(\Omega) = \vb{\mathcal{Y}}^\dag(\Omega)\hat{\boldsymbol{a}}(\Omega) + G(\Omega) = \mathcal{Y}_c^*(\Omega)\hat{a}_c(\Omega) + \mathcal{Y}_s^*(\Omega)\hat{a}_s(\Omega) + G(\Omega),
\end{equation}
where $G(\Omega)$ is the spectrum of the measured quantity, $\mathcal{Y}_{c,s}(\Omega)$ are some complex-valued functions of $\Omega$ that characterize how the light is transformed by the device. Quantum noise is represented by the terms of the above expression not dependent on the measured quantity $G$, i.e.,
\begin{equation}
\label{eq:generic_N_Y}
  \hat{N}_Y(\Omega) = \vb{\mathcal{Y}}^\dag(\Omega)\hat{\boldsymbol{a}}(\Omega) =
  \begin{pmatrix}
    \mathcal{Y}^*_{c}(\Omega) & \mathcal{Y}^*_{s}(\Omega)
  \end{pmatrix}
  \cdot
  \begin{pmatrix}
    \hat{a}_c(\Omega)\\
    \hat{a}_s(\Omega)
  \end{pmatrix}.
\end{equation}

The measure of quantum noise is the power spectral density $S_Y(\Omega)$ that is defined by the following expression:
\begin{equation} 
\label{eq:generic_S_Y}
  2\pi S_Y(\Omega)\delta(\Omega-\Omega') = \bra{\psi}\hat{N}_Y(\Omega)\circ\hat{N}^\dag_Y(\Omega')\ket{\psi} = \mean{\hat{N}_Y(\Omega)\circ\hat{N}_Y^\dag(\Omega')}\,.
\end{equation}
Here $\ket{\psi}$ is the quantum state of the light wave.

In our review, we will encounter two types of quantum states that we have described above, i.e., vacuum $\ket{\vac}$ and squeezed vacuum $\ket{\sqz_0(r,\phi)}$ states. Let us show how to calculate the power (double-sided) spectral density of a generic quantity $\hat{Y}(\Omega)$ in a vacuum state. To do so, one should substitute Eq.~\eqref{eq:generic_N_Y} into Eq.~\eqref{eq:generic_S_Y} and obtain that:
\begin{eqnarray}
\label{eq:S_Y_vac}
  S^{\vac}_Y(\Omega) &=& \vb{\mathcal{Y}}^\dag(\Omega)\mean{\hat{\boldsymbol{a}}(\Omega)\circ\hat{\boldsymbol{a}}^\dag(\Omega)}_{\vac}\vb{\mathcal{Y}}(\Omega) = \vb{\mathcal{Y}}^\dag(\Omega)\mathbb{S}_{\vac}(\Omega)\vb{\mathcal{Y}}(\Omega) \nonumber\\
 &=& \frac{\vb{\mathcal{Y}}^\dag(\Omega)\vb{\mathcal{Y}}(\Omega)}{2} = \frac{|\mathcal{Y}_c(\Omega)|^2+|\mathcal{Y}_s(\Omega)|^2}{2}\,,
\end{eqnarray}
where we used the definition of the power spectral density matrix of light in a vacuum state~\eqref{eq:SpDens_matrix_vac}\epubtkFootnote{Hereafter we will omit, for the sake of brevity, the factor $2\pi\delta(\Omega-\Omega')$ in equations that define the power (double-sided) spectral densities of relevant quantum observables, as well as assume $\Omega=\Omega'$, though keeping in mind that a mathematically rigorous definition should be written in the form of Eq.~\eqref{eq:generic_S_Y}.}.
Similarly, one can calculate the spectral density of quantum noise if the light is in a squeezed state $\ket{\sqz_0(r,\phi)}$, utilizing the definition of the squeezed state density matrix given in Eq.~\eqref{eq:SQZ_SpDens_matrix}:
\begin{eqnarray}
\label{eq:S_Y_sqz}
  S^{\sqz}_Y(\Omega) &=&
  \vb{\mathcal{Y}}^\dag(\Omega)\mean{\hat{\boldsymbol{a}}(\Omega)\circ\hat{\boldsymbol{a}}^\dag(\Omega)}_{\sqz}\vb{\mathcal{Y}}(\Omega) = \vb{\mathcal{Y}}^\dag(\Omega)\mathbb{S}_{\sqz}(\Omega)\vb{\mathcal{Y}}(\Omega) = \frac12\vb{\mathcal{Y}}^\dag(\Omega)\mathbb{P}[-\phi]\mathbb{S}_{\sqz}[2r,0]\mathbb{P}[\phi]\vb{\mathcal{Y}}(\Omega) \nonumber\\ &=&
  \frac{|\mathcal{Y}_c(\Omega)|^2}{2}(\cosh2r+\sinh2r\cos2\phi)+\frac{|\mathcal{Y}_s(\Omega)|^2}{2}(\cosh2r-\sinh2r\cos2\phi)\nonumber\\ && -\mathrm{Re}\left[\mathcal{Y}_c(\Omega)\mathcal{Y}^*_s(\Omega)\right]\sinh2r\sin2\phi
\,.
\end{eqnarray}

It might also be necessary to calculate also cross-correlation spectral density $S_{YZ}(\Omega)$ of $\hat{Y}(\Omega)$ with some other quantity $\hat{Z}(\Omega)$ with quantum noise defined as:
\begin{equation*}
  \hat{N}_Z(\Omega) = \vb{\mathcal{Z}}(\Omega)^\dag\hat{\boldsymbol{a}}(\Omega) = \mathcal{Z}_c^*(\Omega)\hat{a}_c(\Omega) + \mathcal{Z}_s^*(\Omega)\hat{a}_s(\Omega)\,.
\end{equation*}
Using the definition of cross-spectral density $S_{YZ}(\Omega)$ similar to~\eqref{eq:generic_S_Y}:
\begin{equation}
  2\pi S_{YZ}(\Omega)\delta(\Omega-\Omega') = \bra{\psi}\hat{N}_Y(\Omega)\circ\hat{N}^\dag_Z(\Omega')\ket{\psi} = \mean{\hat{N}_Y(\Omega)\circ\hat{N}_Z^\dag(\Omega')}\,,
\end{equation}
one can get the following expressions for spectral densities in both cases of the vacuum state:
\begin{eqnarray}
\label{eq:S_YZ_vac}
  S^{\vac}_{YZ}(\Omega) &=& \vb{\mathcal{Y}}^\dag(\Omega)\mean{\hat{\boldsymbol{a}}(\Omega)\circ\hat{\boldsymbol{a}}^\dag(\Omega)}_{\vac}\vb{\mathcal{Z}}(\Omega) = \vb{\mathcal{Y}}^\dag(\Omega)\mathbb{S}_{\vac}(\Omega)\vb{\mathcal{Z}}(\Omega) \nonumber\\
 &=& \frac{\vb{\mathcal{Y}}^\dag(\Omega)\vb{\mathcal{Z}}(\Omega)}{2} = \frac{\mathcal{Y}^*_c(\Omega)\mathcal{Z}_c(\Omega)+\mathcal{Y}^*_s(\Omega)\mathcal{Z}_s(\Omega)}{2}\,,
\end{eqnarray}
and the squeezed state:
\begin{eqnarray}
\label{eq:S_YZ_sqz}
  S^{\sqz}_{YZ}(\Omega) &=& \vb{\mathcal{Y}}^\dag(\Omega)\mean{\hat{\boldsymbol{a}}(\Omega)\circ\hat{\boldsymbol{a}}^\dag(\Omega)}_{\sqz}\vb{\mathcal{Z}}(\Omega) = \vb{\mathcal{Y}}^\dag(\Omega)\mathbb{S}_{\sqz}(\Omega)\vb{\mathcal{Z}}(\Omega)\nonumber\\ &=& \frac12\vb{\mathcal{Y}}^\dag(\Omega)\mathbb{P}[-\phi]\mathbb{S}_{\sqz}[2r,0]\mathbb{P}[\phi]\vb{\mathcal{Z}}(\Omega) \nonumber\\ &=& \frac{\mathcal{Y}^*_c(\Omega)\mathcal{Z}_c(\Omega)}{2}(\cosh2r+\sinh2r\cos2\phi)+\frac{\mathcal{Y}^*_s(\Omega)\mathcal{Z}_s(\Omega)}{2}(\cosh2r-\sinh2r\cos2\phi) \nonumber\\&&- \frac{\mathcal{Y}_s^*(\Omega)\mathcal{Z}_c(\Omega)+\mathcal{Y}^*_c(\Omega)\mathcal{Z}_s(\Omega)}{2}\sinh2r\sin2\phi
\,.
\end{eqnarray}

Note that since the observables $\hat{Y}(t)$ and $\hat{Z}(t)$ that one calculates spectral densities for are Hermitian, it is compulsory, as is well known, for any operator to represent a physical quantity, then the following relation holds for their spectral coefficients $\mathcal{Y}_{c,s}(\Omega)$ and $\mathcal{Z}_{c,s}(\Omega)$:
\begin{equation}
\label{eq:Y_Z_hermiticity_cont}
  \mathcal{Y}^*_{c,s}(\Omega) = \mathcal{Y}_{c,s}(-\Omega)\,,\qquad\mbox{and}\qquad\mathcal{Z}^*_{c,s}(\Omega) = \mathcal{Z}_{c,s}(-\Omega)\,.
\end{equation}
This leads to an interesting observation that the coefficients $\mathcal{Y}_{c,s}(\Omega)$ and $\mathcal{Z}_{c,s}(\Omega)$ should be real-valued functions of variable $s = i\Omega$.

Now we can make further generalizations and consider multiple light and vacuum fields comprising the quantity of interest:
\begin{equation}
  \hat{Y}(\Omega)\ \to\ \hat{N}_Y(\Omega) = \sum\limits_{i=1}^N \vb{\mathcal{Y}}^\dag_i(\Omega)\hat{\boldsymbol{a}}_i(\Omega)\,,
\end{equation}
where $\hat{\boldsymbol{a}}_i(\Omega)$ stand for quadrature amplitude vectors of $N$ independent electromagnetic fields, and $\vb{\mathcal{Y}}^\dag_i(\Omega)$ are the corresponding complex-valued coefficient functions indicating how these fields are transmitted to the output. In reality, the readout observable of a GW detector is always a combination of the input light field and vacuum fields that mix into the output optical train as a result of optical loss of various origin. This statement can be exemplified by a single lossy mirror I/O-relations given by Eq.~\eqref{eq:IO_lossy_mirror_relations_first} of Section~\ref{sec:losses_in_OE}.

Thus, to calculate the spectral density for such an observable one needs to know the initial state of all light fields under consideration. Since we assume $\hat{\boldsymbol{a}}_i(\Omega)$ independent from each other, the initial state will simply be a direct product of the initial states for each of the fields:
\begin{equation*}
  \ket{\Psi} = \bigotimes_{i=1}^N\ket{\psi_i}\,,
\end{equation*}
and the formula for the corresponding power (double-sided) spectral density reads:
\begin{equation}
  S_Y(\Omega) = \sum\limits_{i=1}^N \vb{\mathcal{Y}}^\dag_i(\Omega)\bra{\psi_i}\hat{\boldsymbol{a}}_i(\Omega)\circ\hat{\boldsymbol{a}}^\dag_i(\Omega)\ket{\psi_i}\vb{\mathcal{Y}}_i(\Omega) = \sum\limits_{i=1}^N\vb{\mathcal{Y}}^\dag_i(\Omega)\mathbb{S}_i(\Omega)\vb{\mathcal{Y}}_i(\Omega) = \sum\limits_{i=1}^N S_{Y_i}(\Omega)\,,
\end{equation}
with $\mathbb{S}_i(\Omega)$ standing for the i-th input field spectral density matrix.
Hence, the total spectral density is just a sum of spectral densities of each of the fields. The cross-spectral density for two observables $\hat{Y}(\Omega)$ and $\hat{Z}(\Omega)$ can be built by analogy and we leave this task to the reader.

\newpage
\section{Linear Quantum Measurement}
\label{sec:linear_quantum_measurement}

In Section~\ref{sec:quantum_light}, we discussed the quantum nature of light and fluctuations of the light field observables like phase and amplitude that stem thereof and yield what is usually called the quantum noise of optical measurement. In GW detection applications, where a sensitivity of the phase measurement is essential, as discussed in Section~\ref{sec:GW_interaction_with_IF}, the natural question arises: is there a limit to the measurement precision imposed by quantum mechanics? A seemingly simple answer would be that such a limit is set by the quantum fluctuations of the outgoing light phase quadrature, which are, in turn, governed by the quantum state the outgoing light finds itself in. The difficult part is that on its way through the interferometer, the light wave inflicts an additional back-action noise that adds up to the phase fluctuations of the incident wave and contaminates the output of the interferometer. The origin of this back action is in amplitude fluctuations of the incident light, giving rise to a random radiation pressure force that acts on the interferometer mirrors along with the signal GW force, thus effectively mimicking it. And it is the fundamental principle of quantum mechanics, the Heisenberg uncertainty principle, that sets a limit on the product of the phase and amplitude uncertainties (since these are complementary observables), thus leading up to the lower bound of the achievable precision of phase measurement. This limit appears to be a general feature for a very broad class of measurement known as linear measurement and is referred to as the SQL~\cite{Sov.Phys.JETP_26.831_1968, 92BookBrKh}.

In this section, we try to give a brief introduction to quantum measurement theory, starting from rather basic examples with discrete measurement and then passing to a general theory of continuous linear measurement. We introduce the concept of the SQL and derive it for special cases of probe bodies. We also discuss briefly possible ways to overcome this limit by contriving smarter ways of weak force measurement then direct coordinate monitoring.

\subsection{Quantum measurement of a classical force}

\subsubsection{Discrete position measurement}
\label{sec:linear_toy}

Let us consider a very simple measurement scheme, which, nevertheless,
embodies all key features of a general position measurement. In the
scheme shown in Figure~\ref{fig:toy_0}, a sequence of very short light
pulses are used to monitor the displacement of a probe body $M$. The
position $x$ of $M$ is probed periodically with time interval
$\vartheta$. In order to make our model more realistic, we suppose
that each pulse reflects from the test mass $\digamma>1$ times, thus
increasing the optomechanical coupling and thereby the information of
the measured quantity contained in each reflected pulse. We also
assume mass $M$ large enough to neglect the displacement inflicted by
the pulses radiation pressure in the course of the measurement
process.

\epubtkImage{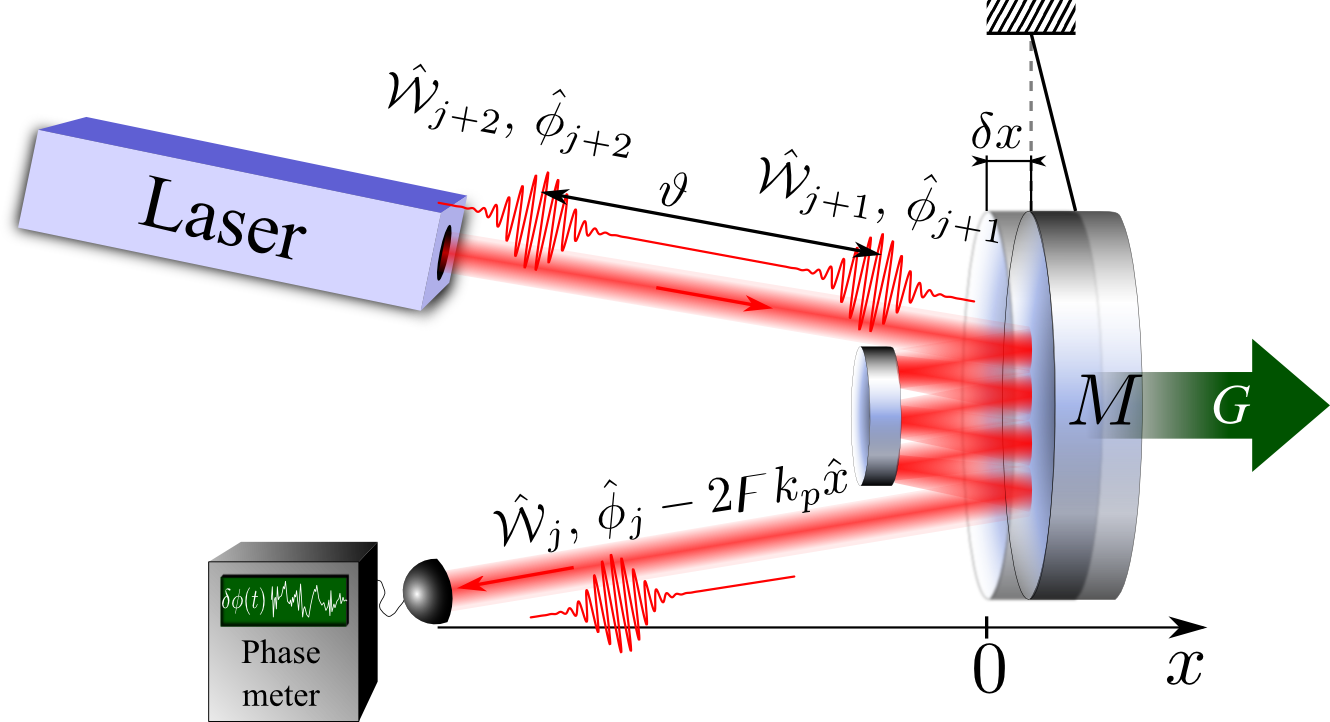}{%
\begin{figure}[htbp]
  \centerline{\includegraphics[width=.85\textwidth]{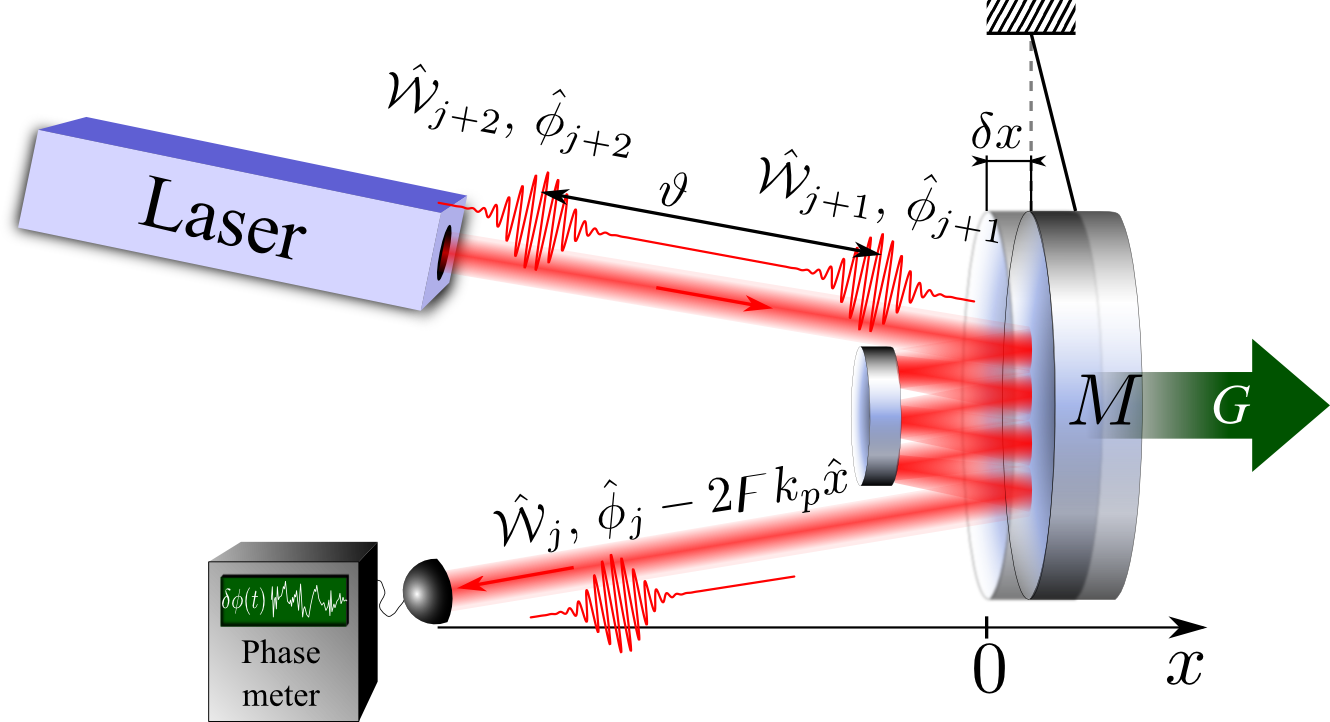}}
  \caption{Toy example of a linear optical position measurement.}
\label{fig:toy_0}
\end{figure}}

Then each $j$-th pulse, when reflected, carries a phase shift proportional to the value of the test-mass position $x(t_j)$ at the moment of reflection:
\begin{equation}
\label{eq:phi_refl}
  \hat{\phi}_j^{\mathrm{refl}} = \hat{\phi}_j - 2\digamma k_p\hat{x}(t_j) \,,
\end{equation}
where $k_p=\omega_p/c$, $\omega_p$ is the light frequency, $j=\dots,-1,0,1,\dots$ is the pulse number and $\hat{\phi}_j$ is the initial (random) phase of the $j$-th pulse. We assume that the mean value of all these phases is equal to zero, $\mean{\hat{\phi}_j}=0$, and their root mean square (RMS) uncertainty $\mean{(\hat{\phi^2}}-\mean{\hat{\phi}}^2)^{1/2}$ is equal to $\Delta\phi$.

The reflected pulses are detected by a phase-sensitive device (the phase detector). The implementation of an optical phase detector is considered in detail in Section~\ref{sec:homodyne}. Here we suppose only that the phase $\hat{\phi}_j^{\mathrm{refl}}$ measurement error introduced by the detector is much smaller than the initial uncertainty of the phases $\Delta\phi$. In this case, the initial uncertainty will be the only source of the position measurement error:
\begin{equation}
\label{eq:Delta_x_meas}
  \Delta x_{\mathrm{meas}} = \frac{\Delta\phi}{2\digamma k_p} \,.
\end{equation}
For convenience, we renormalize Eq.~\eqref{eq:phi_refl} as the equivalent test-mass displacement:
\begin{equation}
\label{eq:tilde_x_j}
  \tilde{x}_j \equiv -\frac{\hat{\phi}_j^{\mathrm{refl}}}{2\digamma k_p}
    = \hat{x}(t_j) + \hat{x}_{\mathrm{fl}}(t_j) \,,
\end{equation}
where
\begin{equation}
\label{eq:x_j_fl}
  \hat{x}_{\mathrm{fl}}(t_j) = -\frac{\hat{\phi}_j}{2\digamma k_p}
\end{equation}
are the independent random values with the RMS uncertainties given by Eq.~\eqref{eq:Delta_x_meas}.

Upon reflection, each light pulse kicks the test mass, transferring to it a back-action momentum equal to
\begin{equation}
\label{eq:p_j}
  \hat{p}_j^{\mathrm{after}} - \hat{p}_j^{\mathrm{before}} = \hat{p}_j^{\mathrm{b.a.}}
  = \frac{2\digamma}{c}\hat{\mathcal{W}}_j \,,
\end{equation}
where $\hat{p}_j^{\mathrm{before}}$ and $\hat{p}_j^{\mathrm{after}}$ are the test-mass momentum values just before and just after the light pulse reflection, and $\mathcal{W}_j$ is the energy of the $j$-th pulse. The major part of this perturbation is contributed by classical radiation pressure:
\begin{equation}
  \mean{\hat{p}_j^{\mathrm{b.a.}}} = \frac{2\digamma}{c}\mathcal{W} \,,
\end{equation}
with $\mathcal{W}$ the mean energy of the pulses. Therefore, one could neglect its effect, for it could be either subtracted from the measurement result or compensated by an actuator. The random part, which cannot be compensated, is proportional to the deviation of the pulse energy:
\begin{equation}
\label{eq:p_j_fl}
  \hat{p}^{\mathrm{b.a.}}(t_j) = \hat{p}_j^{\mathrm{b.a.}} - \mean{\hat{p}_j^{\mathrm{b.a.}}}
    = \frac{2\digamma}{c}\bigl(\hat{\mathcal{W}}_j - \mathcal{W}\bigr) \,,
\end{equation}
and its RMS uncertainly is equal to
\begin{equation}
\label{eq:Delta_p_pert}
  \Delta p_{\mathrm{b.a.}} = \frac{2\digamma\Delta\mathcal{W}}{c} \,,
\end{equation}
with $\Delta\mathcal{W}$ the RMS uncertainty of the pulse energy.

The energy and the phase of each pulse are canonically conjugate observables and thus obey the following uncertainty relation:
\begin{equation}
\label{eq:dE_dphi}
  \Delta\mathcal{W}\Delta\phi \ge \frac{\hbar\omega_p}{2} \,.
\end{equation}
Therefore, it follows from Eqs.~(\ref{eq:Delta_x_meas} and \ref{eq:Delta_p_pert}) that the position measurement error $\Delta x_{\mathrm{meas}}$ and the momentum perturbation $\Delta p_{\mathrm{b.a.}}$ due to back action also satisfy the uncertainty relation:
\begin{equation}
\label{eq:Delta_x_Delta_p}
  \Delta x_{\mathrm{meas}}\Delta p_{\mathrm{b.a.}} \ge \frac{\hbar}{2} \,.
\end{equation}

This example represents a simple particular case of a \emph{linear measurement}. This class of measurement schemes can be fully described by two linear equations of the form~\eqref{eq:tilde_x_j} and \eqref{eq:p_j}, provided that both the measurement uncertainty and the object back-action perturbation ($\hat{x}_{\mathrm{fl}}(t_j)$ and $\hat{p}^{\mathrm{b.a.}}(t_j)$ in this case) are statistically independent of the test object initial quantum state and satisfy the same uncertainty relation as the measured observable and its canonically conjugate counterpart (the object position and momentum in this case).

 \subsubsection{From discrete to continuous measurement}
\label{sec:disc2cont}

 Suppose the test mass to be heavy enough for a single pulse to either perturb its momentum noticeably, or measure its position with the required precision (which is a perfectly realistic assumption for the kilogram-scale test masses of GW interferometers). In this case, many pulses should be used to accumulate the measurement precisions; at the same time, the test-mass momentum perturbation will be accumulated as well. Choose now such a time interval $T$, which, on the one hand, is long enough to comprise a large number of individual pulses:
\begin{equation}
  N = \frac{T}{\vartheta} \gg 1 \,,
\end{equation}
and, on the other hand, is sufficiently short for the test-mass position $x$ not to change considerably during this time due to the test-mass self-evolution. Then one can use all the $N$ measurement results to refine the precision of the test-mass position $x$ estimate, thus getting $\sqrt{N}$ times smaller uncertainty
  \begin{equation}
\label{eq:Dx_T}
    \Delta x_T = \frac{\Delta x_{\mathrm{meas}}}{\sqrt{N}}
      = \Delta x_{\mathrm{meas}}\sqrt{\frac{\vartheta}{T}} \,.
  \end{equation}
At the same time, the accumulated random kicks the object received from each of the pulses random kicks, see Eq.~\eqref{eq:p_j_fl}, result in random change of the object's momentum similar to that of Brownian motion, and thus increasing in the same diffusive manner:
  \begin{equation}
\label{eq:Dp_T}
    \Delta p_T = \Delta p_{\mathrm{b.a.}}\sqrt{N} = \Delta p_{\mathrm{b.a.}}\sqrt{\frac{T}{\vartheta}} \,.
  \end{equation}

If we now assume the interval between the measurements to be infinitesimally small ($\vartheta\to0$), keeping at the same time each single measurement strength infinitesimally weak:
\begin{equation*}
  \Delta x_{\mathrm{meas}} \to \infty \Leftrightarrow \Delta p_{\mathrm{b.a.}} \to 0 \,,
\end{equation*}
then we get a \emph{continuous} measurement of the test-mass position $\hat{x}(t)$ as a result.
We need more adequate parameters to characterize its `strength' than $\Delta x_{\mathrm{meas}}$ and $\Delta p_{\mathrm{b.a.}}$. For continuous measurement we introduce the following parameters instead:
\begin{equation}
\label{eq:S_xS_F_lim}
  S_x = \lim_{\vartheta\to0}(\Delta x_{\mathrm{meas}})^2\vartheta = \frac{S_\phi}{4\digamma^2k_p^2}\,,\qquad
  S_F = \lim_{\vartheta\to0}\frac{(\Delta p_{\mathrm{b.a.}})^2}{\vartheta} = \frac{4\digamma^2S_\mathcal{I}}{c^2}
    \,,
\end{equation}
with
\begin{equation}
\label{S_phiS_I_lim}
  S_\phi = \lim_{\vartheta\to0}(\Delta \phi)^2\vartheta \,,\qquad
  S_\mathcal{I} = \lim_{\vartheta\to0}\frac{(\Delta\mathcal{W})^2}{\vartheta} \,.
\end{equation}
This allows us to rewrite Eqs.~\eqref{eq:Dx_T} and \eqref{eq:Dp_T} in a form that does not contain the time interval $\vartheta$:
\begin{equation}
\label{dx_dp_cont}
  \Delta x_T = \sqrt{\frac{S_x}{T}} \,,\qquad \Delta p_T = \sqrt{S_FT} \,.
\end{equation}
To clarify the physical meaning of the quantities $S_x$ and $S_\phi$ let us rewrite Eq.~\eqref{eq:tilde_x_j} in the continuous limit:
\begin{equation}
\label{eq:tilde_x}
  \tilde{x}(t) = \hat{x}(t) + \hat{x}_{\mathrm{fl}}  (t) \,,
\end{equation}
where
\begin{equation}
\label{eq:x_fl_cont}
  \hat{x}_{\mathrm{fl}}(t) = -\frac{\hat{\phi}(t)}{2\digamma k_p}
\end{equation}
stands for \emph{measurement noise}, proportional to the phase $\hat{\phi}(t)$ of the light beam (in the continuous limit the sequence of individual pulses transforms into a continuous beam). Then there is no difficulty in seeing that $S_x$ is a power (double-sided) spectral density of this noise, and $S_\phi$ is a power double-sided spectral density of $\hat{\phi}(t)$.

If we turn to Eq.~\eqref{eq:p_j}, which describes the meter back action, and rewrite it in a continuous limit we will get the following differential equation for the object momentum:
\begin{equation}
\label{eq:F_fl_def}
  \frac{d\hat{p}(t)}{dt} = \hat{F}_{\mathrm{fl}}(t) + \dots
\end{equation}
where $\hat{F}_{\mathrm{fl}}(t)$ is a continuous Markovian random force, defined as a limiting case of the following discrete Markov process:
\begin{equation}
\label{eq:F_fl_cont}
  \hat{F}_{\mathrm{fl}}(t_j) = \lim_{\vartheta\to0}\frac{\hat{p}^{\mathrm{b.a.}}(t_j)}{\vartheta}
  = \frac{2\digamma}{c}
      \lim_{\vartheta\to0}\frac{\hat{\mathcal{W}}_j-\mathcal{W}}{\vartheta}
  = \frac{2\digamma}{c}[\hat{\mathcal{I}}(t_j)-\mathcal{I}_0] \,,
\end{equation}
with $\hat{\mathcal{I}}(t)$ the optical power, $\mathcal{I}_0$ its mean value, and `$\dots$' here meaning all forces (if any), acting on the object but having nothing to do with the meter (light, in our case). Double-sided power spectral density of $\hat{F}_{\mathrm{b.a.}}$ is equal to $S_F$, and double-sided power spectral density of $\hat{\mathcal{I}}$ is $S_\mathcal{I}$.

We have just built a simple model of a \emph{continuous linear measurement}, which nevertheless comprises the main features of a more general theory, i.e., it contains equations for the calculation of measurement noise~\eqref{eq:tilde_x} and also for back action~\eqref{eq:F_fl_def}. The precision of this measurement and the object back action in this case are described by the spectral densities $S_x$ and $S_F$ of the two meter noise sources, which are assumed to not be correlated in our simple model, and thus satisfy the following relation (cf.\ Eqs.~\eqref{eq:S_xS_F_lim}):
\begin{equation}
\label{eq:S_xS_F_simple}
  S_xS_F = \frac{S_\phi S_\mathcal{I}}{\omega_p^2} \ge \frac{\hbar^2}{4} \,.
\end{equation}
This relation (as well as its more general version to be discussed later) for continuous linear measurements plays the same role as the uncertainty relation~\eqref{eq:Delta_x_Delta_p} for discrete measurements, establishing a universal connection between the accuracy of the monitoring and the perturbation of the monitored object.

\paragraph*{Simple case: light in a coherent state.}
Recall now that scheme of representing the quantized light wave as a sequence of short statistically-independent pulses with duration $\varepsilon\equiv\vartheta$ we referred to in Section~\ref{sec:light_quantum_states}. It is the very concept we used here, and thus we can use it to calculate the spectral densities of the measurement and back-action noise sources for our simple device featured in Figure~\ref{fig:toy_0} assuming the light to be in a coherent state with classical amplitude $A_c=\sqrt{2\mathcal{I}_0/(\hbar\omega_p)}$ (we chose $A_s=0$ thus making the mean phase of light $\mean{\hat{\phi}}=0$). To do so we need to express phase $\hat{\phi}$ and energy $\hat{\mathcal{W}}$ in the pulse in terms of the quadrature amplitudes $\hat{a}_{c,s}(t)$. This can be done if we refer to Eq.~\eqref{eq:2photon_E_strain} and make use of the following definition of the mean electromagnetic energy of the light wave contained in the volume $v_\vartheta\equiv\mathcal{A}c\vartheta$ (here, $\mathcal{A}$ is the effective cross-sectional area of the light beam):
\begin{equation}
  \hat{\mathcal{W}} = \frac{v_\vartheta}{4\pi}\overline{\hat{E}^2(t)} = \frac{v_\vartheta}{4\pi \vartheta}\int_{-\vartheta/2}^{\vartheta/2}d\tau\,\hat{E}^2(\tau) = \mathcal{W}+\delta\hat{\mathcal{W}}\,,
\end{equation}
where $\mathcal{W}=v_\vartheta \mathcal{C}_0^2 A_c^2/(8\pi) = \mathcal{I}_0\vartheta$ is the mean pulse energy, and
 \begin{equation}
   \delta\hat{\mathcal{W}} \simeq \frac{\mathcal{A}c\mathcal{C}_0^2}{4\pi}2A_c\int_{-\vartheta/2}^{\vartheta/2}d\tau\,\hat{a}_c(\tau) = \sqrt{2\hbar\omega_p\mathcal{I}_0\vartheta}\hat{X}_\vartheta(t) = \sqrt{2\hbar\omega_p\mathcal{W}}\hat{X}_\vartheta(t)\,
 \end{equation}
is a fluctuating part of the pulse energy\epubtkFootnote{Here, we
  omitted the terms of $\delta\hat{\mathcal{W}}$ proportional to
  $\hat{a}^2_{c,s}(t)$ since their contribution to the integral is of
  the second order of smallness in $\hat{a}_{c,s}/A_c$ compared to the
  one for the first order term.}. We used here the definition of the
mean pulse quadrature amplitude operators introduced in
Eqs.~\eqref{eq:X_and_Y_quadrature_def}. In the same manner, one can
define a phase for each pulse using Eqs.~\eqref{eq:EMW_quadrature_def}
and with the assumption of small phase fluctuations ($\Delta\phi\ll1$) one
can get:
\begin{equation}
  \hat{\phi} \simeq \frac{1}{A_c \vartheta}\int_{-\vartheta/2}^{\vartheta/2}d\tau \hat{a}_s(\tau) = \sqrt{\frac{\hbar\omega_p}{2\mathcal{I}_0\vartheta}}\hat{Y}_\vartheta= \sqrt{\frac{\hbar\omega_p}{2\mathcal{W}}}\hat{Y}_\vartheta\,.
\end{equation}

Thus, since in a coherent state $\Delta\hat{X}^2_\vartheta = \Delta\hat{Y}^2_\vartheta = 1/2$ the phase and energy uncertainties are equal to
\begin{equation}
\label{DphiDE_coh}
  \Delta\phi = \frac{1}{2}\sqrt{\frac{\hbar\omega_p}{\mathcal{W}}} \,, \qquad
  \Delta\mathcal{W} = \sqrt{\hbar\omega_p\mathcal{W}} \,,
\end{equation}
and hence
\begin{equation}
\label{DxDp_coh}
  \Delta x_{\mathrm{meas}} = \frac{c}{4\digamma}\sqrt{\frac{\hbar}{\omega_p\mathcal{W}}} \,, \qquad
  \Delta p_{\mathrm{b.a.}} = \frac{2\digamma}{c}\sqrt{\hbar\omega_p\mathcal{W}} \,.
\end{equation}
Substituting these expressions into Eqs.~(\ref{S_phiS_I_lim}, \ref{eq:S_xS_F_lim}), we get the following expressions for the power (double-sided) spectral densities of the measurement and back-action noise sources:
\begin{equation}
\label{eq:S_phiS_I_toy}
  S_\phi = \frac{\hbar\omega_p}{4\mathcal{I}_0} \,, \qquad S_\mathcal{I} = \hbar\omega_p\mathcal{I}_0 \,,
\end{equation}
and
\begin{equation}
\label{S_xS_F_toy}
  S_x = \frac{\hbar c^2}{16\omega_p \mathcal{I}_0\digamma^2} \,, \qquad
  S_F = \frac{4\hbar\omega_p\mathcal{I}_0\digamma^2}{c^2} \,.
\end{equation}

We should emphasize that this simple measurement model and the corresponding uncertainty relation~\eqref{eq:S_xS_F_simple} are by no means general. We have made several rather strong assumptions in the course of derivation, i.e., we assumed:
\begin{enumerate}
  \item energy and phase fluctuations in each of the light pulses uncorrelated: $\mean{\hat{\mathcal{W}}(t_j)\hat{\phi}(t_j)}=0$;
  \item all pulses to have the same energy and phase uncertainties $\Delta\mathcal{W}$ and $\Delta\phi$, respectively;
  \item the pulses statistically independent from each other, particularly taking $\mean{\hat{\mathcal{W}}(t_i)\hat{\mathcal{W}}(t_j)}=\mean{\hat{\phi}(t_i)\hat{\phi}(t_j)}=\mean{\hat{\mathcal{W}}(t_i)\hat{\phi}(t_j)}=0$ with $t_i\neq t_j$.
\end{enumerate}

These assumptions can be mapped to the following features of the fluctuations $\hat{x}_{\mathrm{fl}}(t)$ and $\hat{F}_{\mathrm{b.a.}}(t)$ in the continuous case:
\begin{enumerate}
  \item these noise sources are mutually not correlated;
  \item they are stationary (invariant to the time shift) and, therefore, can be described by spectral densities $S_x$ and $S_F$;
  \item they are Markovian (white) with constant (frequency-independent) spectral densities.
\end{enumerate}

The features 1 and 2, in turn, lead to characteristic fundamentally-looking sensitivity limitations, the SQL. We will call linear quantum meters, which obey these limitations (that is, with mutually non-correlated and stationary noises $\hat{x}_{\mathrm{fl}}$ and $\hat{F}_{\mathrm{b.a.}}$), \emph{Simple Quantum Meters} (SQM).

\subsection{General linear measurement}
\label{sec:gen_linear_measurement}

In this Section, we generalize the concept of linear quantum
measurement discussed above and give a comprehensive overview of the
formalism introduced in~\cite{92BookBrKh} and further elaborated
in~\cite{PhysRevD.65.042001, 03pth1Ch}. This formalism can be applied
to any system that performs a transformation from some unknown
classical observable (e.g., GW tidal force in GW interferometers) into
another classical observable of a measurement device that can be
measured with (ideally) arbitrarily high precision (e.g., in GW
detectors, the readout photocurrent serves such an observable) and its
value depends on the value of unknown observable linearly. For
definiteness, let us keep closer to GW detectors and assume the
continuous measurement of a classical force.

\epubtkImage{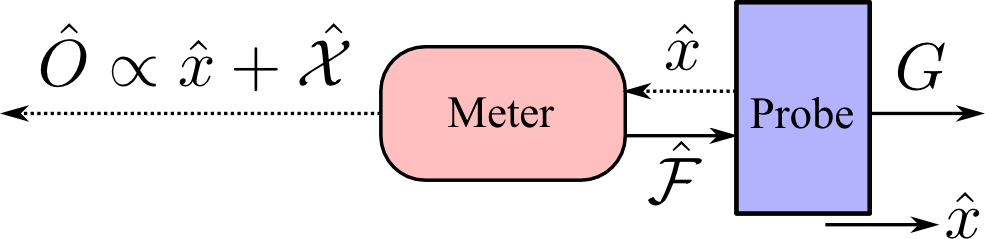}{%
\begin{figure}[htbp]
  \centerline{\includegraphics[width=0.5\textwidth]{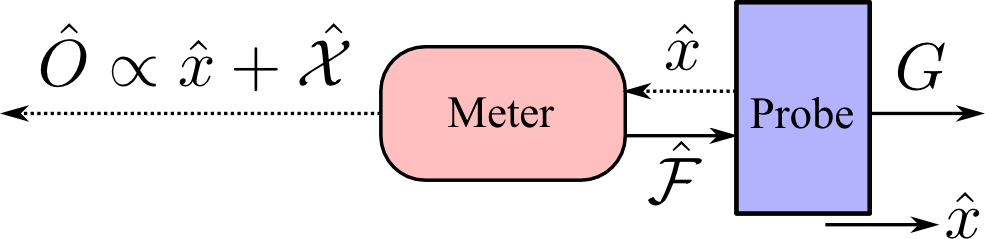}}
  \caption{General scheme of the continuous linear measurement with $G$ standing for measured classical force, $\hat{\mathcal{X}}$ the measurement noise, $\hat{\mathcal{F}}$ the back-action noise, $\hat{O}$ the meter readout observable, $\hat{x}$ the actual probe's displacement.}
\label{fig:gen_pos}
\end{figure}}

The abstract scheme of such a device is drawn in Figure~\ref{fig:gen_pos}. It consists of a probe $\mathcal{P}$ that is exposed to the action of a classical force $G(t)$, and the meter. The action of this force on the probe causes its displacement $\hat{x}$ that is monitored by the meter (e.g., light, circulating in the interferometer). The output observable of the meter $\hat{O}$ is monitored by some arbitrary classical device that makes a measurement record $o(t)$. The quantum nature of the probe--meter interaction is reflected by the back-action force $\hat{F}$ that randomly kicks the probe on the part of the meter (e.g., radiation pressure fluctuations). At the same time, the meter itself is the source of additional quantum noise $\hat{O}_{\mathrm{fl}}(t)$ in the readout signal. Quantum mechanically, this system can be described by the following Hamiltonian:
\begin{equation}
\label{eq:linear_system_H}
  \hat{\mathcal{H}} = \hat{\mathcal{H}}^{(0)}_{\mathrm{probe}} + \hat{\mathcal{H}}^{(0)}_{\mathrm{meter}} + \hat{V}(t) \,,
\end{equation}
where $\hat{\mathcal{H}}^{(0)}_{\mathrm{probe}}$ and $\hat{\mathcal{H}}^{(0)}_{\mathrm{meter}}$ are the Hamiltonians describing the free evolution of the probe and the meter, respectively, i.e., when there is no coupling between these systems, and $\hat{V}(t) = - \hat{x}(G(t)+\hat{F})$ is the interaction Hamiltonian. The evolution of this system can be found, in general, by solving Heisenberg equations for all of the system observables. However, it is convenient to rewrite it first in the interaction picture factoring out the free evolution of the probe and the meter (see Appendix 3.7 of~\cite{03pth1Ch} for detailed derivation):
\begin{equation}
\label{eq:linear_system_H_IP}
  \hat{\mathcal{H}}_I(t)
  = \exp\left\{\frac{i}{\hbar}\hat{\mathcal{H}}^{(0)}(t-t_0)\right\}
      \hat{V}(t)\exp\left\{-\frac{i}{\hbar}\hat{\mathcal{H}}^{(0)}(t-t_0)\right\}
  = -\hat{x}^{(0)}(t)(G(t)+\hat{F}^{(0)}(t))\,,
\end{equation}
where $\hat{\mathcal{H}}^{(0)} = \hat{\mathcal{H}}^{(0)}_{\mathrm{probe}} + \hat{\mathcal{H}}^{(0)}_{\mathrm{meter}}$, and $\hat{x}^{(0)}(t)$ and $\hat{F}^{(0)}(t)$ are the Heisenberg operators of the probe's displacement and the meter back-action force, respectively, in the case of no coupling between these systems, i.e., the solution to the following system of independent Heisenberg equations:
\begin{equation*}
 \frac{d\hat{x}^{(0)}(t)}{dt} = \frac{i}{\hbar}\left[\hat{\mathcal{H}}^{(0)}_{\mathrm{probe}},\hat{x}^{(0)}(t)\right]\,, \qquad \frac{d\hat{F}^{(0)}(t)}{dt} = \frac{i}{\hbar}\left[\hat{\mathcal{H}}^{(0)}_{\mathrm{meter}},\hat{F}^{(0)}(t)\right]\,,
\end{equation*}
and $t_0$ is the arbitrary initial moment of time that can be set to $-\infty$ without loss of generality.

The following statement can be proven (see~\cite{Kubo1956}, Section~VI of~\cite{92BookBrKh}, and Theorems~3 and 4 in Appendix~3.7 of~\cite{03pth1Ch} for proof):
\textit{For a linear system with Hamiltonian~\eqref{eq:linear_system_H}, for any linear observable $\hat{A}$ of the probe and for any linear observable $\hat{B}$ of the meter, their full Heisenberg evolutions are given by:
\begin{eqnarray}
\label{eq:linear_system_AB_evolution}
  \hat{A}(t) &=& \hat{A}^{(0)}(t) + \int_{t_0}^t dt'\,\chi_{Ax}(t,t')[\hat{F}(t')+G(t')]\,,\nonumber\\
  \hat{B}(t) &=& \hat{B}^{(0)}(t) + \int_{t_0}^t dt'\,\chi_{BF}(t,t')\hat{x}(t')\,,
\end{eqnarray}
where $\hat{A}^{(0)}(t)$ and $\hat{B}^{(0)}(t)$ stand for the free Heisenberg evolutions in the case of no coupling, and the functions $\chi_{Ax}(t,t')$ and $\chi_{BF}(t,t')$ are called (time-domain) susceptibilities and defined as:}
\begin{eqnarray}
\label{eq:linear_system_AB_chi}
  \chi_{Ax}(t,t') &\equiv& \begin{cases}
      \dfrac{i}{\hbar}\left[\hat{A}^{(0)}(t),\hat{x}^{(0)}(t')\right]\,, &
        t\geqslant t' \\
      0 \,, \quad t<t'
    \end{cases}
  \nonumber \\
  \chi_{BF}(t,t') &\equiv& \begin{cases}
      \dfrac{i}{\hbar}\left[\hat{B}^{(0)}(t),\hat{F}^{(0)}(t')\right]\,, &
        t\geqslant t' \\
      0 \,, \quad t<t'
   \end{cases}\,.
\end{eqnarray}
The second clauses in these equations maintain the causality principle.

For time independent Hamiltonian $\hat{\mathcal{H}}^{(0)}$ and operator $\hat{F}$ (in the Schr\"odinger picture), the susceptibilities are invariant to time shifts, i.e., $\chi(t,t')=\chi(t+\tau,t'+\tau)$, therefore they depend only on the difference of times: $\chi(t,t')\to\chi(t-t')$. In this case, one can rewrite Eqs.~\eqref{eq:linear_system_AB_evolution} in frequency domain as:
\begin{equation}
\label{eq:linear_system_AB_Fourier}
  \hat{A}(\Omega) = \hat{A}^{(0)}(\Omega) + \chi_{Ax}(\Omega)[\hat{F}(\Omega)+G(\Omega)]\,, \qquad \hat{B}(\Omega) = \hat{B}^{(0)}(\Omega) + \chi_{BF}(\Omega)\hat{x}(\Omega)\,,
\end{equation}
where the Fourier transforms of all of the observables are defined in accordance with Eq.~\eqref{eq:Fourier_transform}.

Let us now use these theorems to find the full set of equations of motion for the system of linear observables $\hat{x}$, $\hat{F}$ and $\hat{O}$ that fully characterize our linear measurement process in the scheme featured in Figure~\ref{fig:gen_pos}:
\begin{eqnarray}
\label{eq:gen_linear_meas_EOM}
 \hat{O}(t) &=& \hat{O}^{(0)}(t) + \int_{t_0}^t dt'\,\chi_{OF}(t-t')\hat{x}(t')\,,\nonumber \\
 \hat{F}(t) &=& \hat{F}^{(0)}(t) + \int_{t_0}^t dt'\,\chi_{FF}(t-t')\hat{x}(t')\,,\nonumber\\
 \hat{x}(t) &=& \hat{x}^{(0)}(t) + \int_{t_0}^t dt'\,\chi_{xx}(t-t')\left[G(t')+\hat{F}(t')\right]\,,
\end{eqnarray}
where time-domain susceptibilities are defined as
\begin{eqnarray}
\label{eq:gen_linear_meas_succep}
  \chi_{OF}(t-t') &=& \frac{i}{\hbar}\left[\hat{O}^{(0)}(t),\hat{F}^{(0)}(t')\right]\,, \nonumber\\
  \chi_{FF}(t-t') &=& \frac{i}{\hbar}\left[\hat{F}^{(0)}(t),\hat{F}^{(0)}(t')\right]\,, \nonumber\\
  \chi_{xx}(t-t') &=& \frac{i}{\hbar}\left[\hat{x}^{(0)}(t),\hat{x}^{(0)}(t')\right]\,.
\end{eqnarray}

The meaning of the above equations is worth discussing. The first of Eqs.~\eqref{eq:gen_linear_meas_EOM} describes how the readout observable $\hat{O}(t)$ of the meter, say the particular quadrature of the outgoing light field measured by the homodyne detector (cf.\ Eq.~\eqref{eq:homodyne_photocurrent}), depends on the actual displacement $\hat{x}(t)$ of the probe, and the corresponding susceptibility $\chi_{OF}(t-t')$ is the transfer function for the meter from $\hat{x}$ to $\hat{O}$. The term $\hat{O}^{(0)}(t)$ stands for the free evolution of the readout observable, provided that there was no coupling between the probe and the meter. In the case of the GW detector, this is just a pure quantum noise of the outgoing light that would have come out were all of the interferometer test masses fixed. It was shown explicitly in~\cite{02a1KiLeMaThVy} and we will demonstrate below that this noise is fully equivalent to that of the input light except for the insignificant phase shift acquired by the light in the course of propagation through the interferometer.

The following important remark should be made concerning the meter's output observable $\hat{O}(t)$. As we have mentioned already, the output observable in the linear measurement process should be precisely measurable at any instance of time. This implies a \emph{simultaneous measurability condition}~\cite{Sci.209.4456.547_1980_BrVoTh, 80a1CaThDrSaZi, PhysRevD.19.2888, 92BookBrKh, 03pth1Ch, PhysRevD.65.042001} on the observable $\hat{O}(t)$ requiring that it should commute with itself at any moment of time:
\begin{equation}
\label{eq:measurability_cond}
  \left[\hat{O}(t),\,\hat{O}(t')\right]=0\,,\quad \forall t,\,t'\,.
\end{equation}
Initially, this condition was introduced as the definition of the quantum non-demolition (QND) observables by Braginsky et al.~\cite{Sov.Phys.Uspekhi.17.5.644_1975_BrVo, 77a1eBrKhVo}. In our case it means that the measurement of $\hat{O}(t_1)$ at some moment of time $t_1$ shall not disturb the measurement result at any other moments of time and therefore the sample data $\left\{\hat{O}(t_1),\hat{O}(t_1),\ldots,\hat{O}(t_n)\right\}$ can be stored directly as bits of classical data in a classical storage medium, and any noise from subsequent processing of the signal can be made arbitrarily small. It means that all noise sources of quantum origin are already included in the quantum fluctuations of $\hat{O}(t)$~\cite{03pth1Ch, PhysRevD.65.042001}. And the fact that due to~\eqref{eq:measurability_cond} this susceptibility turns out to be zero reflects the fact that $\hat{O}(t)$ should be a classical observable.

The second equation in~\eqref{eq:gen_linear_meas_EOM} describes how the back-action force exerted by the meter on the probe system evolves in time and how it depends on the probe's displacement. The first term, $\hat{F}^{(0)}(t)$, meaning is rather obvious. In GW interferometer, it is the radiation pressure force that the light exerts on the mirrors while reflecting off them. It depends only on the mean value and quantum fluctuations of the amplitude of the incident light and does not depend on the mirror motion. The second term here stands for a \emph{dynamical back-action} of the meter and since, by construction, it is the part of the back-action force that depends, in a linear way, from the probe's displacement, the meaning of the susceptibility $\chi_{FF}(t-t')$ becomes apparent: it is the generalized rigidity that the meter introduces, effectively modifying the dynamics of the probe. We will see later how this effective rigidity can be used to improve the sensitivity of the GW interferometers without introducing additional noise and thus enhancing the SNR of the GW detection process.

The third equation of~\eqref{eq:gen_linear_meas_EOM} concerns the evolution of the probe's displacement in time. Three distinct parts comprise this evolution. Let us start with the second and the third ones:
\begin{equation}
  x_s(t) = \int_{t_0}^t dt'\,\chi_{xx}(t-t')G(t')\,,\quad\mbox{and}\quad \hat{x}_{\mathrm{b.a.}}(t) = \int_{t_0}^t dt'\,\chi_{xx}(t-t')\hat{F}(t')\,.
\end{equation}
Here $x_s(t)$ is the probe's response on the signal force $G(t)$ and is, actually, the part we are mostly interested in. This expression also unravels the role of susceptibility $\chi_{xx}(t-t')$: it is just the Green's function of the equation of motion governing the probe's bare dynamics (also known as impulse response function) that can be shown to be a solution of the following initial value problem:
\begin{eqnarray*}
  \mathbf{D}\chi_{xx}(t-t') = \delta(t-t')\,,\quad \chi_{xx}(t-t')|_{t\to t'} = 0\,,\quad \ldots\\ \ldots\ \frac{\partial^{n-2}\chi_{xx}(t-t')}{\partial t^{n-2}}\bigr|_{t\to t'+0} = 0\,,\quad \frac{\partial^{n-1}\chi_{xx}(t-t')}{\partial t^{n-1}}\bigr|_{t\to t'+0} = \frac{1}{a_n}\,,
\end{eqnarray*}
where $\mathbf{D} = \sum_{k=0}^n a_k\frac{d^k}{dt^k}$ is the linear differential operator that is governed by the dynamics of the probe, e.g., it is equal to $\mathbf{D}_{\mathrm{f.m.}} =M\frac{d^2}{dt^2}$ for a free mass $M$ and to $\mathbf{D}_{\mathrm{osc}} =M\frac{d^2}{dt^2}+M\Omega^2_m$
for a harmonic oscillator with eigenfrequency $\Omega_m$. Apparently, operator $\mathbf{D}$ is an inverse of the integral operator $\vb{\chi}_{xx}$ whose kernel is $\chi_{xx}(t-t')$:
\begin{equation*}
  x_s(t) = \mathbf{D}^{-1}G(t) = \vb{\chi}_{xx}G(t) = \int_{t_0}^t dt'\,\chi_{xx}(t-t')G(t')\,.
\end{equation*}

The second value, $\hat{x}_{\mathrm{b.a.}}(t)$, is the displacement of the probe due to the back-action force exerted by the meter on the probe. Since it enters the probe's response in the very same way the signal does, it is the most problematic part of the quantum noise that, as we demonstrate later, imposes the SQL~\cite{Sov.Phys.JETP_26.831_1968, 92BookBrKh}.

And finally, $\hat{x}^{(0)}(t)$ simply features a free evolution of the probe in accordance with its equations of motion and thus depends on the initial values of the probe's displacement $\hat{x}^{(0)}(t_0)$, momentum $\hat{p}^{(0)}(t_0)$, and, possibly, on higher order time derivatives of $\hat{x}^{(0)}(t)$ taken at $t_0$, as per the structure of the operator $\mathbf{D}$ governing the probe's dynamics. It is this part of the actual displacement that bears quantum uncertainties imposed by the initial quantum state of the probe. One could argue that these uncertainties might become a source of additional quantum noise obstructing the detection of GWs, augmenting the noise of the meter. This is not the case as was shown explicitly in~\cite{03a1BrGoKhMaThVy}, since our primary interest is in the detection of a classical force rather than the probe's displacement. Therefore, performing over the measured data record $o(t)$ the linear transformation corresponding to first applying the operator $\mathrm{\vb{\chi}}_{OF}^{-1}$ on the readout quantity that results in expressing $o(t)$ in terms of the probe's displacement:
\begin{equation*}
  \tilde{x}(t)=\mathrm{\vb{\chi}}_{OF}^{-1} O(t) = x_{\mathrm{fl}}(t)+x^{(0)}(t)+x_s(t)+x_{\mathrm{b.a.}}(t)\,,
\end{equation*}
with $x_{\mathrm{fl}}(t) = \mathrm{\vb{\chi}}_{OF}^{-1} O^{(0)}(t)$ standing for the meter's own quantum noise (measurement uncertainty), and then applying a probe dynamics operator $\mathbf{D}$ that yields a force signal equivalent to the readout quantity $o(t)$:
\begin{equation*}
  \tilde{F}(t) = \mathbf{D}\tilde{x}(t) = \mathbf{D}x_{\mathrm{fl}}(t)+F_{\mathrm{b.a.}}(t)+G(t)\,.
\end{equation*}
The term $\mathbf{D}x^{(0)}(t)$ vanishes since $x^{(0)}(t)$ is the solution of a free-evolution equation of motion. Thus, we see that the result of measurement contains two noise sources, $\hat{x}_{\mathrm{fl}}(t)$ and $\hat{F}_{\mathrm{b.a.}}(t)$, which comprise the sum noise masking the signal force $G(t)$.

Since we can remove initial quantum uncertainties associated with the state of the probe, it would be beneficial to turn to the Fourier domain and rewrite Eqs.~\eqref{eq:gen_linear_meas_EOM} in the spectral form:
\begin{eqnarray}
\label{eq:gen_linear_meas_EOM_Fourier}
  \hat{O}(\Omega) &=& \hat{O}^{(0)}(\Omega) + \chi_{OF}(\Omega)\hat{x}(\Omega)\,,\nonumber\\
  \hat{F}(\Omega) &=& \hat{F}^{(0)}(\Omega) + \chi_{FF}(\Omega)\hat{x}(\Omega)\,,\nonumber\\
  \hat{x}(\Omega) &=& \chi_{xx}(\Omega)[\hat{F}(\Omega)+G(\Omega)]\,,
\end{eqnarray}
where spectral susceptibilities are defined as:
\begin{equation}
  \chi_{AB}(\Omega) = \int_0^{\infty}d\tau\,\chi_{AB}(\tau)e^{i\Omega\tau},
\end{equation}
with $(A,B) \Rightarrow (O,F,x)$, and we omit the term $\hat{x}^{(0)}(\Omega)$ in the last equation for the reasons discussed above.
The solution of these equations is straightforward to get and reads:
\begin{eqnarray}
  \hat{O}(\Omega) &=& \hat{O}^{(0)}(\Omega) +\frac{\chi_{xx}(\Omega)\chi_{OF}(\Omega)}{1-\chi_{xx}(\Omega)\chi_{FF}(\Omega)}\left[G(\Omega)+\hat{F}^{(0)}(\Omega)\right]\,,\label{eq:gen_lin_meas_output_spectral}\\
  \hat{F}(\Omega) &=& \frac{1}{1-\chi_{xx}(\Omega)\chi_{FF}(\Omega)}\left[\hat{F}^{(0)}(\Omega)+\chi_{FF}(\Omega)\chi_{xx}(\Omega)G(\Omega)\right]\,,\label{eq:gen_lin_meas_F_ba_spectral}\\
  \hat{x}(\Omega) &=& \frac{\chi_{xx}(\Omega)}{1-\chi_{xx}(\Omega)\chi_{FF}(\Omega)}\left[G(\Omega)+\hat{F}^{(0)}(\Omega)\right]\label{eq:gen_lin_meas_x_spectral}\,.
\end{eqnarray}

It is common to normalize the output quantity of the meter $\hat{O}(\Omega)$ to unit signal. In GW interferometers, two such normalizations are popular. The first one tends to consider the tidal force $G$ as a signal and thus set to 1 the coefficient in front of $G(\Omega)$ in Eq.~\eqref{eq:gen_lin_meas_output_spectral}. The other one takes GW spectral amplitude $h(\Omega)$ as a signal and sets the corresponding coefficient in $\hat{O}(\Omega)$ to unity. Basically, these normalizations are equivalent by virtue of Eq.~\eqref{eq:GW_force_to_h_rel} as:
\begin{equation}
\label{eq:gen_lin_F_to_h_transform}
  -M\Omega^2 x_h(\Omega)\equiv-M\Omega^2 \frac{Lh(\Omega)}{2} = G(\Omega)\quad\Rightarrow\quad h(\Omega) = -2 G(\Omega)/(ML\Omega^2)\,.
\end{equation}
In both cases, the renormalized output quantities can be considered as a sum of the noise and signal constituents:
\begin{equation}
\label{eq:gen_out_noise_def}
  \hat{O}^{F} = \hat{\mathcal{N}}^F+G \qquad\mbox{or}\qquad \hat{O}^{h} = \hat{\mathcal{N}}^h+h(\Omega)\,.
\end{equation}
And it is the noise term in both cases that we are seeking to calculate to determine the sensitivity of the GW detector. Let us rewrite $\hat{O}(\Omega)$ in force normalization:
\begin{eqnarray}
\label{eq:gen_force_noise_def}
  \hat{O}^F(\Omega) &=& \frac{1-\chi_{FF}(\Omega)\chi_{xx}(\Omega)}{\chi_{OF}(\Omega)\chi_{xx}(\Omega)} \hat{O}(\Omega) =
  \frac{\hat{O}^{(0)}(\Omega)}{\chi_{OF}(\Omega)\chi_{xx}(\Omega)}+\left(\hat{F}^{(0)}(\Omega)-\frac{\chi_{FF}(\Omega)}{\chi_{OF}(\Omega)}\hat{O}^{(0)}(\Omega)\right)+G(\Omega)\nonumber\\&\equiv&
  \frac{\hat{\mathcal{X}}(\Omega)}{\chi_{xx}(\Omega)}+\hat{\mathcal{F}}(\Omega)+G(\Omega)\,,
\end{eqnarray}
where we introduce two new linear observables $\hat{\mathcal{X}}$ and $\hat{\mathcal{F}}$ of the meter defined as:
\begin{equation}
\label{eq:gen_noise_def}
  \hat{\mathcal{X}}(\Omega) \equiv \frac{\hat{O}^{(0)}(\Omega)}{\chi_{OF}(\Omega)}\,,\qquad \hat{\mathcal{F}}(\Omega) \equiv \hat{F}^{(0)}(\Omega)-\frac{\chi_{FF}(\Omega)}{\chi_{OF}(\Omega)}\hat{O}^{(0)}(\Omega)\,,
\end{equation}
that have the following meaning:
\begin{itemize}
  \item $\hat{\mathcal{X}}$ is the effective output fluctuation of the meter not dependent on the probe. Henceforth, we will refer to it as the \emph{effective measurement noise} (shot noise, in the GW interferometer common terminology);
  \item $\hat{\mathcal{F}}$ is the effective response of the output at time $t$ to the meter's back-action force at earlier times $t<t'$. In the following we will refer to $\hat{\mathcal{F}}$ as the \emph{effective back-action noise} (radiation-pressure noise, in the GW interferometer common terminology).
\end{itemize}
These two new observables that embody the two types of noise inherent in any linear measurement satisfy the following commutation relations:
\begin{eqnarray}
\label{eq:gen_lin_commutator}
  \left[\hat{\mathcal{X}}(\Omega),\hat{\mathcal{X}}^\dag(\Omega')\right]=\left[\hat{\mathcal{F}}(\Omega),\hat{\mathcal{F}}^\dag(\Omega')\right] = 0\,,\quad &\Longleftrightarrow\quad &\left[\hat{\mathcal{X}}(t),\hat{\mathcal{X}}(t')\right]=\left[\hat{\mathcal{F}}(t),\hat{\mathcal{F}}(t')\right] = 0\,,\\
  \left[\hat{\mathcal{X}}(\Omega),\hat{\mathcal{F}}^\dag(\Omega')\right]=-2\pi i\hbar\delta(\Omega-\Omega')\,,\quad &\Longleftrightarrow\quad &\left[\hat{\mathcal{X}}(t),\hat{\mathcal{F}}^\dag(t')\right]=-i\hbar\delta(t-t')\,,
\end{eqnarray}
that can be interpreted in the way that $\hat{\mathcal{X}}(t)$ and $\hat{\mathcal{F}}(t)$ can be seen at each instance of time as the canonical momentum and the coordinate of different effective linear measuring devices (meter+probe), thus defining an infinite set of subsequent measurements similar to the successive independent monitors of von Neumann's model~\cite{1996Book_vonNeumann}. In this case, however, the monitors described by $\hat{\mathcal{X}}(t)$ and $\hat{\mathcal{F}}(t)$ are not, generally speaking, independent. In GW detectors, these monitors appear correlated due to the internal dynamics of the detector, i.e., the noise processes they describe are non-Markovian.

In particular, this can be seen when one calculates the power (double-sided) spectral density of the sum noise $\hat{\mathcal{N}}^{F}(t)$:
\begin{equation}
\label{eq:gen_spdens}
  S^F(\Omega) = \intinfty dt\,\mean{\hat{\mathcal{N}}^{F}(t)\circ\hat{\mathcal{N}}^{F}(t')}e^{i\Omega(t-t')} = \frac{S_{\mathcal{X}\mathcal{X}}(\Omega)}{|\chi_{xx}(\Omega)|^2}+S_{\mathcal{F}\mathcal{F}}(\Omega)+2\mathrm{Re}\left[\frac{S_{\mathcal{X}\mathcal{F}}(\Omega)}{\chi_{xx}(\Omega)}\right]\,,
\end{equation}
where spectral densities:
\begin{eqnarray}
  S_{\mathcal{X}\mathcal{X}}(\Omega) &=& \intinfty dt\,\mean{\hat{\mathcal{X}}(t)\circ\hat{\mathcal{X}}(t')}e^{i\Omega(t-t')}\,,\\
  S_{\mathcal{F}\mathcal{F}}(\Omega) &=& \intinfty dt\,\mean{\hat{\mathcal{F}}(t)\circ\hat{\mathcal{F}}(t')}e^{i\Omega(t-t')}\,,\\
  S_{\mathcal{X}\mathcal{F}}(\Omega) &=& \intinfty dt\,\mean{\hat{\mathcal{X}}(t)\circ\hat{\mathcal{F}}(t')}e^{i\Omega(t-t')}\,,
\end{eqnarray}
are not necessarily constant and, thus, describe non-Markovian random processes. It can also be shown that since $\hat{\mathcal{X}}(t)$ and $\hat{\mathcal{F}}(t)$ satisfy commutation relations~\eqref{eq:gen_lin_commutator}, their spectral densities shall satisfy the \emph{Schr{\"o}dinger--Robertson} uncertainty relation:
\begin{equation}
\label{eq:gen_spdens_uncert_rel}
  S_{\mathcal{X}\mathcal{X}}(\Omega)S_{\mathcal{F}\mathcal{F}}(\Omega)-|S_{\mathcal{X}\mathcal{F}}(\Omega)|^2\geqslant \frac{\hbar^2}{4}\,
\end{equation}
that is the generalization of a Heisenberg uncertainty relation in the case of correlated observables.

The general structure of quantum noise in the linear measurement process, comprising two types of noise sources whose spectral densities are bound by the uncertainty relation~\eqref{eq:gen_spdens_uncert_rel}, gives a clue to several rather important corollaries. One of the most important is the emergence of the SQL, which we consider in detail below.

\subsection{Standard Quantum Limit}
\label{sec:SQL}

Recall the SQM in Section~\ref{sec:disc2cont}.

The SQM has non-correlated effective measurement and back-action noises that results in $S_{\mathcal{X}\mathcal{F}}(\Omega)=0$. Apparently, under these conditions $\hat{\mathcal{X}}$ and $\hat{\mathcal{F}}$ turn into $\hat{x}_{\mathrm{fl}}$ and $\hat{F}_{\mathrm{fl}}$ of Eqs.~\eqref{eq:tilde_x} and \eqref{eq:F_fl_def}, respectively. Hence, we will use $S_x(\Omega)$ instead of $S_{\mathcal{X}\mathcal{X}}(\Omega)$ and $S_F(\Omega)$ instead of $S_{\mathcal{F}\mathcal{F}}(\Omega)$ Then the uncertainty relation~\eqref{eq:gen_spdens_uncert_rel} transforms into:
\begin{equation}
\label{eq:simple_spdens_uncert_rel}
   S_x(\Omega)S_F(\Omega)\geqslant\frac{\hbar^2}{4}\,.
\end{equation}

The SQL is the name for an ultimate lower bound of a sum noise spectral density the SQM can, in principle, have \emph{at any given frequency $\Omega$}. To derive this limit we assume noise sources $x_{\mathrm{fl}}$ and $F_{\mathrm{b.a.}}$ to have minimal values allowed by quantum mechanics, i.e.
\begin{equation}
\label{eq:minimal_qnoise_req}
  S_x(\Omega)S_F(\Omega)=\frac{\hbar^2}{4}\,.
\end{equation}
Then, using this condition, one can minimize SQM's sum noise:
\begin{equation*}
  S^F(\Omega) = \frac{S_x(\Omega)}{|\chi_{xx}(\Omega)|^2}+S_F(\Omega) = \frac{S_x(\Omega)}{|\chi_{xx}(\Omega)|^2}+\frac{\hbar^2}{4S_x(\Omega)}
\end{equation*}
to yield:
\begin{equation}
\label{eq:SQL_force}
  S^F_{\mathrm{SQL}}(\Omega) = \frac{\hbar}{|\chi_{xx}(\Omega)|}
\end{equation},
that is \emph{achieved when contributions of measurement noise and back-action noise to the sum noise are equal to each other}, i.e., when
\begin{equation}
\label{eq:SQL_optimization}
 S_x(\Omega) = \frac{\hbar}{2}|\chi_{xx}(\Omega)|\,,\qquad\Longleftrightarrow\qquad S_F(\Omega) = \frac{\hbar}{2|\chi_{xx}(\Omega)|}\,.
\end{equation}

It is instructive to cite the forms of the SQL in other normalizations, i.e., for $h$-normalization and for $x$-normalization. The former is obtained from~\eqref{eq:SQL_force} via multiplication by $4/(M^2L^2\Omega^4)$:
\begin{equation}
\label{eq:SQL_h}
  S^h_{\mathrm{SQL}}(\Omega) = \frac{4 S^F_{\mathrm{SQL}}(\Omega)}{M^2L^2\Omega^4} = \frac{4\hbar}{M^2L^2\Omega^4|\chi_{xx}(\Omega)|}\,.
\end{equation}
The latter is obtained fromEq.~\eqref{eq:SQL_force} using the obvious connection between force and displacement $x(\Omega) = \chi_{xx}(\Omega)F(\Omega)$:
\begin{equation}
\label{eq:SQL_x}
  S^x_{\mathrm{SQL}}(\Omega) = |\chi_{xx}(\Omega)|^2 S^F_{\mathrm{SQL}}(\Omega) = \hbar|\chi_{xx}(\Omega)|\,.
\end{equation}

These limits look fundamental. There are no parameters of the meter (only $\hbar$ as a reminder of the uncertainty relation~\eqref{eq:S_xS_F_simple}), and only the probe's dynamics is in there. Nevertheless, this is not the case and, actually, this limit can be beaten by more sophisticated, but still linear, position meters. 

At the same time, the SQL represents an important landmark beyond which the ordinary brute-force methods of sensitivity improving cease working, and methods that allow one to blot out the back-action noise $\hat{\mathcal{F}}(t)$ from the meter output signal have to be used instead. Due to this reason, the SQL, and especially the SQL for the simplest test object -- free mass -- is usually considered as a borderline between the classical and the quantum domains.

\subsubsection{Free mass SQL}
\label{sec:SQL_fm}

In the rest of this section, we consider in more detail the SQLs for a free mass and for a harmonic oscillator. We also assume the minimal quantum noise requirement~\eqref{eq:minimal_qnoise_req} to hold.

The free mass is not only the simplest model for the probe's dynamics,
but also the most important class of test objects for GW detection. Test
masses of GW detectors must be isolated as much as possible from
the noisy environment. To this end, the design of GW interferometers
implies suspension of the test masses on thin fibers. The real
suspensions are rather sophisticated and comprise several stages slung
one over another, with mechanical eigenfrequencies $f_m$ in $\lesssim
1\mathrm{Hz}$ range. The sufficient degree of isolation is provided at
frequencies much higher than $f_m$, where the dynamics of test masses
can be approximated with good precision by that of a free mass.

Let us introduce the following convenient measure of measurement strength (precision) in the first place:
\begin{equation}
\label{eq:Omega_q_def}
  \Omega_q = \left(\frac{S_F}{M^2S_x}\right)^{1/4} .
\end{equation}
Using the uncertainty relation~\eqref{eq:minimal_qnoise_req}, the noise spectral densities $S_x$ and $S_F$ can be expressed through $\Omega_q$ as follows:
\begin{equation}
\label{eq:Omega_q}
  S_x = \frac{\hbar}{2M\Omega_q^2} \,, \qquad S_F = \frac{\hbar M\Omega_q^2}{2} \,.
\end{equation}
Therefore, the larger $\Omega_q$ is, the smaller $S_x$ is (the higher is the measurement precision), and the larger $S_F$ is (the stronger the meter back action is).

In the case of interferometers, $\Omega_q^2$ is proportional to the circulating optical power. For example, for the toy optical meter considered above,
\begin{equation}
\label{eq:Omega_q_toy}
  \Omega_q = \sqrt{\frac{8\omega_p\mathcal{I}_0\digamma^2}{Mc^2}} \,,
\end{equation}
see Eqs.~\eqref{S_xS_F_toy}.
Using this notation, and taking into account that for a free mass $M$,
\begin{equation}
\label{D_fm}
  \chi_{xx}^{\mathrm{f.m.}}(\Omega) = -\frac{1}{M\Omega^2} \,,
\end{equation}
the sum quantum noise power (double-sided) spectral density can be written as follows:
\begin{equation}
\label{S_F_sum_fm}
  S^F_{\mathrm{f.m.}}(\Omega) = M^2\Omega^4S_x + S_F
  = \frac{\hbar M\Omega_q^2}{2}\left(\frac{\Omega^4}{\Omega_q^4} + 1\right) \,.
\end{equation}
The SQL optimization~\eqref{eq:SQL_optimization} takes the following simple form in this case:
\begin{equation}
\label{eq:SQL_fm_cond}
  \Omega_q = \Omega \,,
\end{equation}
giving:
\begin{equation}
\label{eq:S_F_SQL_fm}
  S^F_{\mathrm{SQL\,f.m.}}(\Omega) = \hbar M\Omega^2 \,.
\end{equation}

This consideration is illustrated by Figure~\ref{fig:S_F_SQL} (left),
where power (double-sided) spectral density~\eqref{S_F_sum_fm} is plotted for three
different values of $\Omega_q$. It is easy to see that these plots
never dive under the SQL line~\eqref{eq:S_F_SQL_fm}, which embodies a
common envelope for them. Due to this reason, the sensitivities area
above this line is typically considered as the `classical domain', and
below it -- as the `quantum domain'.

\epubtkImage{fig22.png}{%
\begin{figure}[htbp]
  \centerline{
    \includegraphics[width=.45\textwidth]{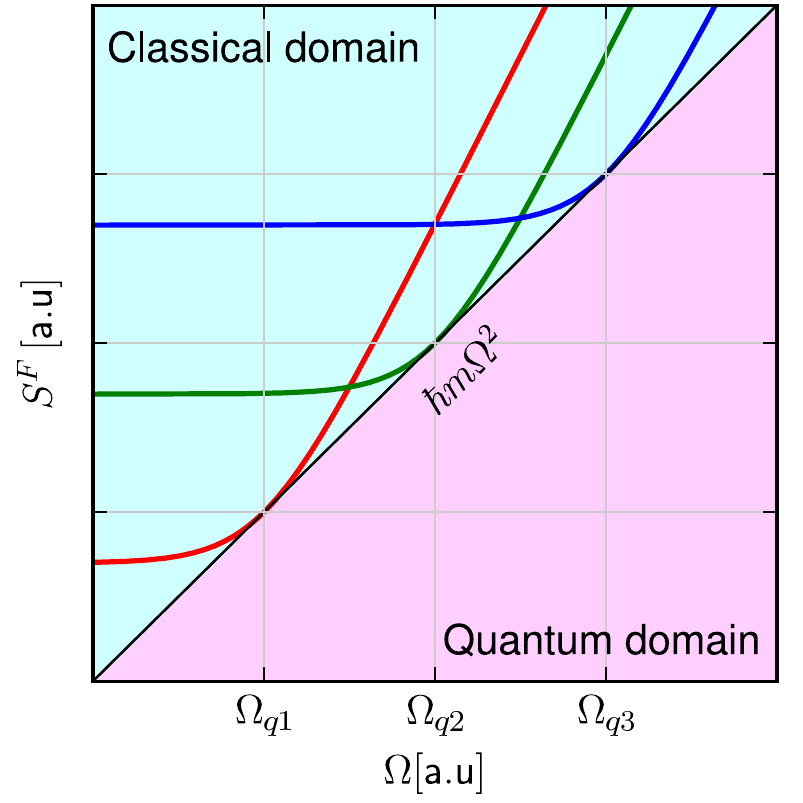}\hfill
    \includegraphics[width=.45\textwidth]{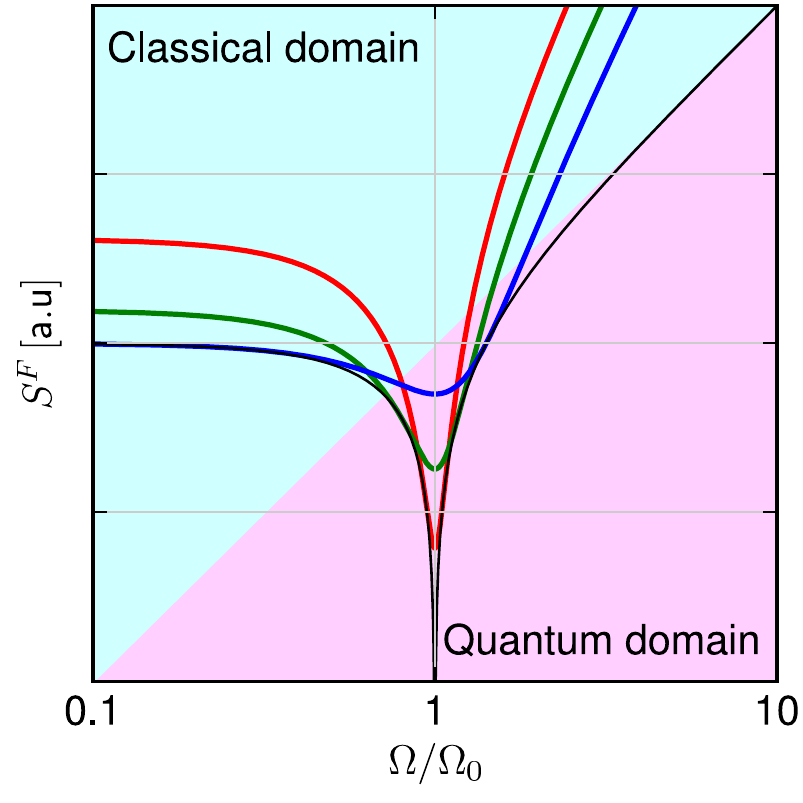}
  }
  \caption{Sum quantum noise power (double-sided) spectral densities of the Simple Quantum Meter for different values of measurement strength $\Omega_q(\mathrm{red})<\Omega_q(\mathrm{green})<\Omega_q(\mathrm{blue})$. Thin black line: SQL. \emph{Left:} free mass. \emph{Right:} harmonic oscillator}
\label{fig:S_F_SQL}
\end{figure}}

\subsubsection{Harmonic oscillator SQL}
\label{sec:SQL_osc}

The simplest way to overcome the limit~\eqref{eq:S_F_SQL_fm}, which does not require any quantum tricks with the meter, is to use a harmonic oscillator as a test object, instead of the free mass. It is easy to see from Eq.\eqref{eq:SQL_force} that the more responsive the test object is at some given frequency $\Omega$ (that is, the bigger $\chi_{xx}(\Omega)$) is, the smaller its force SQL at this frequency is. In the harmonic oscillator case,
\begin{equation}
  \chi_{xx}^{\mathrm{osc}}(\Omega) = \frac{1}{M(\Omega_0^2-\Omega^2)} \,,
\end{equation}
with $\Omega_0$ standing for the oscillator mechanical eigenfrequency, and the sum quantum noise power (double-sided) spectral density equal to
\begin{equation}
\label{S_F_sum_osc}
  S^F_{\mathrm{osc}}(\Omega) = M^2(\Omega_0^2-\Omega^2)^2S_x + S_F
  = \frac{\hbar M\Omega_q^2}{2}\left[\frac{(\Omega_0^2-\Omega^2)^2}{\Omega_q^4} + 1\right].
\end{equation}
Due to a strong response of the harmonic oscillator near resonance, the first (measurement noise) term in Eq.~\eqref{S_F_sum_fm} goes to zero in the vicinity of $\Omega_0$. Therefore, reducing the value of $\Omega_q$, that is, using weaker measurement, it is possible to increase the sensitivity in a narrow band around $\Omega_0$. At the same time, the smaller $\Omega_q$ is, the more narrow the bandwidth is  where this sensitivity is achieved, as can be seen from the plots drawn in Figure~\ref{fig:S_F_SQL} (right).

Consider, in particular, the following minimax optimization of the narrow-band sensitivity. Let $\nu=\Omega-\Omega_0$ be the detuning from the resonance frequency. Suppose also
\begin{equation}
  |\nu| \ll \Omega_0 \,.
\end{equation}
In this case, the sum noise power (double-sided) spectral density~\eqref{S_F_sum_osc} can be approximated as follows:
\begin{equation}
\label{eq:S_F_osc_nb}
  S^F_{\mathrm{osc}}(\Omega_0+\nu)
    = \frac{\hbar M\Omega_q^2}{2}\left(\frac{4\Omega_0^2\nu^2}{\Omega_q^4} + 1\right) .
\end{equation}
Then require the maximum of $S^F_{\mathrm{osc}}$ in a given frequency range $\Delta\Omega$ be as small as possible.

It is evident that this frequency range has to be centered around the resonance frequency $\Omega_0$, with the maximums at its edges, $\nu=\pm\Delta\Omega/2$. The sum noise power (double-sided) spectral density at these points is equal to
\begin{equation}
  S^F_{\mathrm{osc}}(\Omega_0\pm\Delta\Omega/2)
    = \frac{\hbar M\Omega_q^2}{2}\left(\frac{\Omega_0^2\Delta\Omega^2}{\Omega_q^4} + 1\right) .
\end{equation}
The minimum of this expression is provided by
\begin{equation}
\label{eq:Omega_q_osc}
  \Omega_q = \sqrt{\Omega_0\Delta\Omega} \,.
\end{equation}
Substitution of this value back into Eq.~\eqref{eq:S_F_osc_nb} gives the following optimized power (double-sided) spectral density:
\begin{equation}
\label{eq:S_F_osc_nb_opt}
  S^F_{\mathrm{osc}}(\Omega_0+\nu)
  = \frac{\hbar M\Omega_0}{2}\left(\frac{4\nu^2}{\Delta\Omega} + \Delta\Omega\right) ,
\end{equation}
with
\begin{equation}
  S^F_{\mathrm{osc}}(\Omega_0\pm\Delta\Omega/2)= \hbar M\Omega_0\Delta\Omega \,.
\end{equation}
Therefore, the harmonic oscillator can provide a narrow-band sensitivity gain, compared to the \emph{free mass} SQL~\eqref{eq:S_F_SQL_fm}, which reads

\begin{equation}
\label{eq:xi2_osc}
  \xi^2_{\mathrm{osc}} = \frac{S^F_{\mathrm{osc}}(\Omega_0+\nu)}{S^F_{\mathrm{SQL\,f.m.}}(\Omega_0)} \simeq \frac12\left(\frac{4 \nu^2}{\Omega_q^2}+\frac{\Omega_q^2}{\Omega_0^2}\right)\,,
\end{equation}
and can be further written accounting for the above optimization as:
\begin{equation}
\label{eq:osc_fm_SQL}
  \frac{S^F_{\mathrm{osc}}(\Omega_0+\nu)}
    {S^F_{\mathrm{SQL\,f.m.}}(\Omega_0)}\biggr|_{|\nu|\le\Delta\Omega/2}
  \le \frac{S^F_{\mathrm{osc}}(\Omega_0\pm\Delta\Omega/2)}{S^F_{\mathrm{SQL\,f.m.}}(\Omega_0)}
  = \frac{\Delta\Omega}{\Omega_0} \,.
\end{equation}
Of course, the \emph{oscillator} SQL, equal to
\begin{equation}
\label{eq:S_F_SQL_osc}
  S^F_{\mathrm{SQL\,osc}} = \hbar M|\Omega_0^2-\Omega^2| \approx 2\hbar M\Omega_0|\nu|
\end{equation}
cannot be beaten is this way, and the question of whether the sensitivity~\eqref{eq:osc_fm_SQL} is the `true' beating of the SQL or not, is the question to answer (and the subject of many discussions).

\subsubsection{Sensitivity in different normalizations. Free mass and harmonic oscillator}

Above, we have discussed, in brief, different normalizations of the sum noise spectral density and derived the general expressions for the SQL in these normalizations (cf.\ Eqs.~\eqref{eq:SQL_h} and \eqref{eq:SQL_x}). Let us consider how these expressions look for the free mass and harmonic oscillator and how the sensitivity curves transform when changing to different normalizations.

\paragraph*{$h$-normalization:}
The noise spectral density in $h$-normalization can be obtained using Eq.~\eqref{eq:GW_force_to_h_rel}. Where the SQM is concerned, the sum noise in $h$-normalization reads
\begin{equation*}
  h_{\mathrm{sum}}(\Omega)\equiv\hat{\mathcal{N}}^h(\Omega) = -\frac{2}{ML\Omega^2}\left[\frac{\hat{x}_{\mathrm{fl}}(\Omega)}{\chi_{xx}(\Omega)}+\hat{F}_{\mathrm{fl}}(\Omega)\right]\,.
\end{equation*}
In the case of a free mass with $\chi_{xx}(\Omega) = -1/(M\Omega^2)$ the above expression transforms as:
\begin{equation*}
  h^{\mathrm{f.m.}}_{\mathrm{sum}}(\Omega) = \frac{2\hat{x}_{\mathrm{fl}}(\Omega)}{L} - \frac{2\hat{F}_{\mathrm{fl}}}{ML\Omega^2}
\end{equation*}
and that results in the following power (double-sided) spectral density formula:
\begin{equation}\label{eq:S_h_fm}
  S^h_{\mathrm{f.m.}}(\Omega) = \frac{4}{L^2}\left[S_x+\frac{S_F}{M^2\Omega^4}\right] = \frac{2\hbar}{ML^2\Omega_q^2}\left(1+\frac{\Omega_q^4}{\Omega^4}\right)
\end{equation}
and results in the following formula for free mass SQL in $h$-normalization:
\begin{equation}\label{eq:S_h_SQL_fm}
 S^h_{\mathrm{SQL\,f.m.}}(\Omega) = \frac{4\hbar}{ML^2\Omega^2}\,.
\end{equation}

The plots of these spectral densities at different values of
$\Omega_q$ are given in the left panel of Figure~\ref{fig:S_X_SQL}.

\epubtkImage{fig23.png}{%
\begin{figure}[htbp]
  \centerline{
    \includegraphics[width=.45\textwidth]{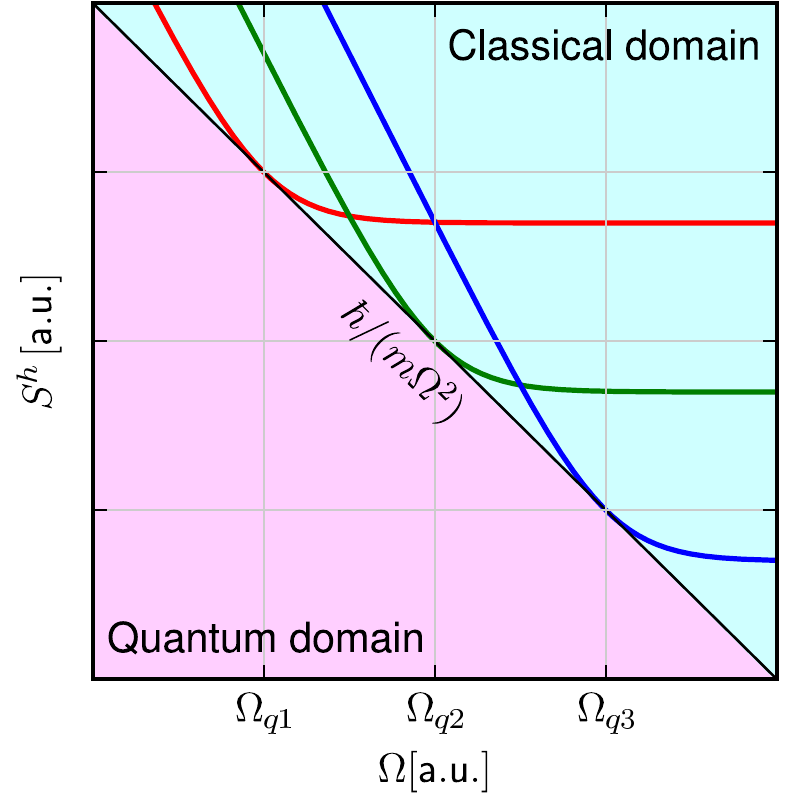}\hfill
    \includegraphics[width=.45\textwidth]{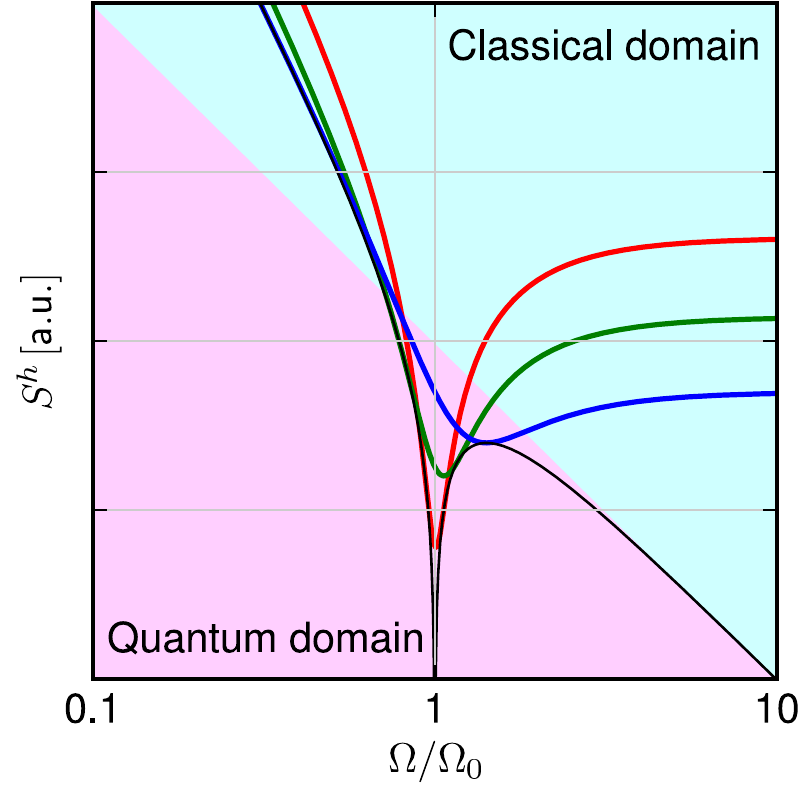}
  }
  \caption{Sum quantum noise power (double-sided) spectral densities of the Simple Quantum Meter in the $h$-normalization for different values of measurement strength: $\Omega_q(\mathrm{red})<\Omega_q(\mathrm{green})<\Omega_q(\mathrm{blue})$. Thin black line: SQL. \emph{Left:} free mass. \emph{Right:} harmonic oscillator.}
\label{fig:S_X_SQL}
\end{figure}}

As for the harmonic oscillator, similar formulas can be obtained taking into account that $\chi_{xx}^{\mathrm{osc}}(\Omega) = 1/(M(\Omega_0^2-\Omega^2))$. Thus, one has:
\begin{equation*}
  h^{\mathrm{osc}}_{\mathrm{sum}}(\Omega) = -\frac{2(\Omega_0^2-\Omega^2)\hat{x}_{\mathrm{fl}}(\Omega)}{L\Omega^2} - \frac{2\hat{F}_{\mathrm{fl}}}{ML\Omega^2}
\end{equation*}
that results in the following power (double-sided) spectral density formula:
\begin{equation}
\label{eq:S_h_sum_osc}
  S^h_{\mathrm{osc}}(\Omega) = \frac{4}{L^2}\left[\left(1-\frac{\Omega_0^2}{\Omega^2}\right)^2S_x+\frac{S_F}{M^2\Omega^4}\right] = \frac{2\hbar}{ML^2\Omega_q^2}\left[\frac{\Omega_q^4}{\Omega^4}\left(1+\frac{(\Omega_0^2-\Omega^2)^2}{\Omega_q^4}\right)\right]
\end{equation}
and results in the following formula for free mass SQL in $h$-normalization:
\begin{equation}\label{eq:S_h__SQL_osc}
  S^h_{\mathrm{SQL\,osc}}(\Omega) = \frac{4\hbar|\Omega_0^2-\Omega^2|}{ML^2\Omega^4}\,.
\end{equation}

The corresponding plots are drawn in the right panel of Figure~\ref{fig:S_X_SQL}. Despite a quite different look, in essence, these spectral densities are the same \emph{force} spectral densities as those drawn in Figure~\ref{fig:S_F_SQL}, yet tilted rightwards by virtue of factor $1/\Omega^4$. In particular, they are characterized by the same minimum at the resonance frequency, created by the strong response of the harmonic oscillator on a near-resonance force, as the corresponding force-normalized spectral densities (\ref{S_F_sum_osc}, \ref{eq:S_F_SQL_osc}).

\paragraph*{$x$-normalization:}
Another normalization that is worth considering is the actual probe displacement, or $x$-normalization. In this normalization, the sum noise spectrum is obtained by multiplying noise term $\hat{\mathcal{N}}^F(\Omega)$ in Eq.~\eqref{eq:gen_out_noise_def} by the probe's susceptibility
\begin{equation}
\label{x_sum}
  \hat{x}_{\mathrm{sum}}(\Omega) = \hat{x}_{\mathrm{fl}}(\Omega) + \chi_{xx}(\Omega)\hat{F}_{\mathrm{fl}}(\Omega)\,.
\end{equation}

It looks rather natural at a first glance; however, as we have shown below, it is less heuristic than the force normalization and could even be misleading. Nevertheless, for completeness, we consider this normalization here.

Spectral density of $\hat{x}_{\mathrm{sum}}(\Omega)$ and the corresponding SQL are equal to
\begin{eqnarray}
  S^x(\Omega) &=& S_x(\Omega) + |\chi_{xx}(\Omega)|^2S_F(\Omega) \,, \\
  S^x_{\mathrm{SQL}}(\Omega) &=& \hbar|\chi_{xx}(\Omega)| \,.
\end{eqnarray}

In the free mass case, the formulas are the same as in $h$-normalization except for the multiplication by $4/L^2$:
\begin{equation}
\label{eq:S_x_fm}
   S^x_{\mathrm{f.m.}}(\Omega) = \left[S_x+\frac{S_F}{M^2\Omega^4}\right] = \frac{\hbar}{2M\Omega_q^2}\left(1+\frac{\Omega_q^4}{\Omega^2}\right)
\end{equation}
with SQL equal to:
\begin{equation}
\label{eq:SQL_x_fm}
  S^x_{\mathrm{SQL\, f.m.}}(\Omega) = \frac{\hbar}{M\Omega^2}\,.
\end{equation}

In the harmonic oscillator case, these equations have the following form:
\begin{eqnarray}
  S^x_{\mathrm{osc}}(\Omega) = S_x + \frac{S_F}{M^2(\Omega_0^2-\Omega^2)^2}
    &=& \frac{\hbar}{2M\Omega_q^2}\left[\frac{\Omega_q^4}{\Omega^4}\left(1+\frac{(\Omega_0^2-\Omega^2)^2}{\Omega_q^4}\right)\right]
      \,, \\
  S^x_{\mathrm{SQL\,osc}}(\Omega) &=& \frac{\hbar}{m|\Omega_0^2-\Omega^2|} \,.
\end{eqnarray}

The corresponding plot of the harmonic oscillator power (double-sided) spectral density in
$x$-normalization is given in Figure~\ref{fig:S_x_SQL}.

\epubtkImage{fig24.png}{%
\begin{figure}[htbp]
  \centerline{\includegraphics{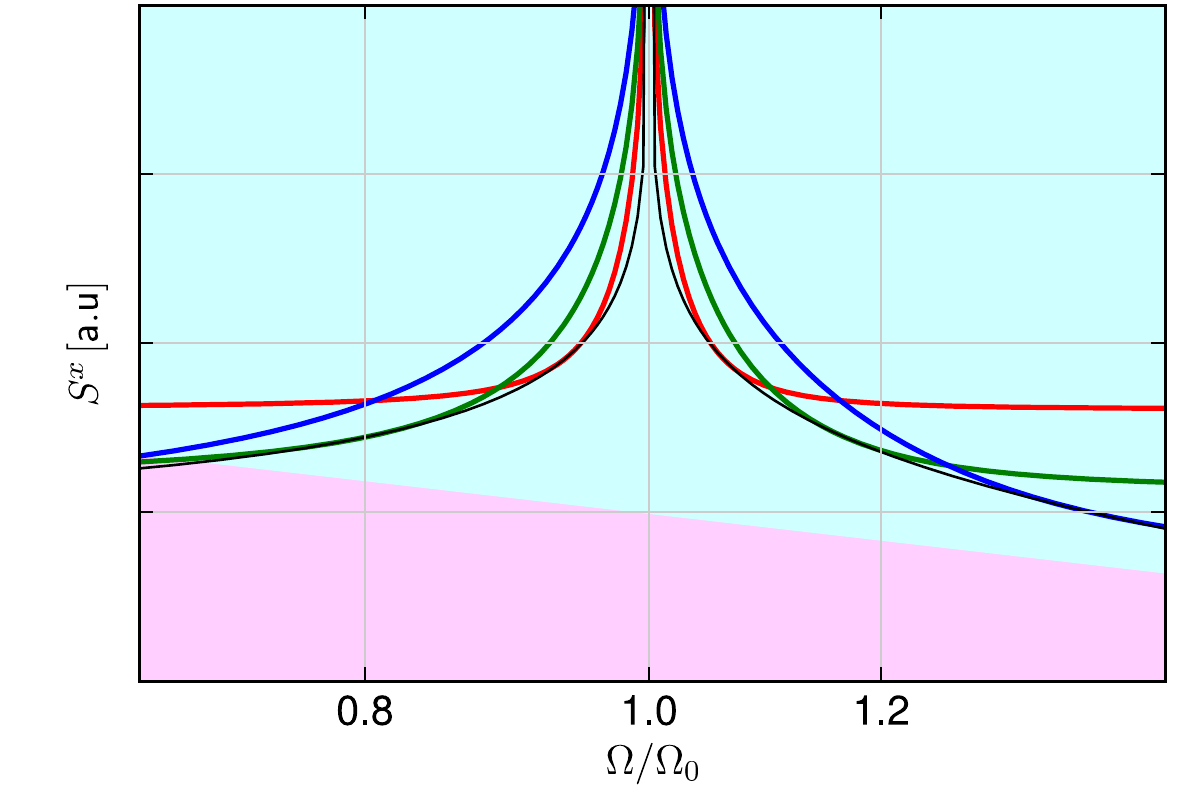}}
  \caption{Sum quantum noise power (double-sided) spectral densities of Simple Quantum Meter and harmonic oscillator in displacement normalization for different values of measurement strength: $\Omega_q(\mathrm{red})<\Omega_q(\mathrm{green})<\Omega_q(\mathrm{blue})$. Thin black line: SQL.}
\label{fig:S_x_SQL}
\end{figure}}

Note that the curves display a sharp upsurge of noise around the resonance frequencies. However, the resonance growth of the displacement due to signal force $G$ have a long start over this seeming noise outburst, as we have shown already, leads to the substantial sensitivity gain for a near-resonance force. This sensitivity increase is clearly visible in the force and equivalent displacement normalization, see Figures~\ref{fig:S_F_SQL} and \ref{fig:S_X_SQL}, but completely masked in Figure~\ref{fig:S_x_SQL}.

\subsection{Beating the SQL by means of noise cancellation}
\label{sec:toy_FI_correlation}

The SQL is not a fundamental limitation as we have mentioned already, and the clue to how to overcome it can be devised from the expression for the general linear measurement sum noise spectral density~\eqref{eq:gen_spdens}. One can see that a properly constructed cross-correlation between the measurement noise $\hat{\mathcal{X}} = \hat{O}^{(0)}/\chi_{OF}(\Omega)$ and back-action noise $\hat{F}^{(0)}$, i.e., the right choice of $\chi_{FF}(\Omega)$ that should be at any frequency equal to:
\begin{equation}
\label{eq:sub_SQL_corr}
  \chi_{FF}(\Omega) = S_{\mathcal{X}F}(\Omega)/S_{\mathcal{X}\mathcal{X}}(\Omega)
\end{equation}
 can compensate the back-action term and leave only the measurement noise-related contribution to the final sum quantum noise:
 \begin{equation*}
   S^F(\Omega) = S_{\mathcal{X}\mathcal{X}}(\Omega)/|\chi_{xx}(\Omega)|^2\,.
 \end{equation*}
This spectral density could be arbitrarily small, providing the unbound measurement strength. However, there is a significant obstacle on the way towards back-action free measurement: the optimal correlation should be frequency dependent in the right and rather peculiar way. Another drawback of such back-action evasion via noise cancellation resides in the dissipation that is always present in real measurement setups and, according to Fluctuation-Dissipation Theorem~\cite{PhysRev.83.34, Landau_Lifshitz_v5} is a source of additional noise that undermines any quantum correlations that might be built in the ideal system.

The simplest way is to make the relation~\eqref{eq:sub_SQL_corr} hold at some fixed frequency, which can always be done either (i) by preparing the meter in some special initial quantum state that has measurement and back-action fluctuations correlated (Unruh~\cite{Unruh1982, PhysRevD.19.2888} proposed to prepare input light in a squeezed state to achieve such correlations), or (ii) by monitoring a linear combination of the probe's displacement and momentum~\cite{94a1VyZuMa, 96a2eVyMa, 96a1eVyMa, 94a1VyZu, 95a1VyZu, Phys.Lett.A.300.547_2002_Danilishin, Phys.Lett.A.278.3.123_2000_Danilishin} that can be accomplished, e.g., via homodyne detection, as we demonstrate below.

We consider the basic principles of the schemes, utilizing the noise cancellation via building cross-correlations between the measurement and back-action noise. We start from the very toy example discussed in Section~\ref{sec:linear_toy}.

\epubtkImage{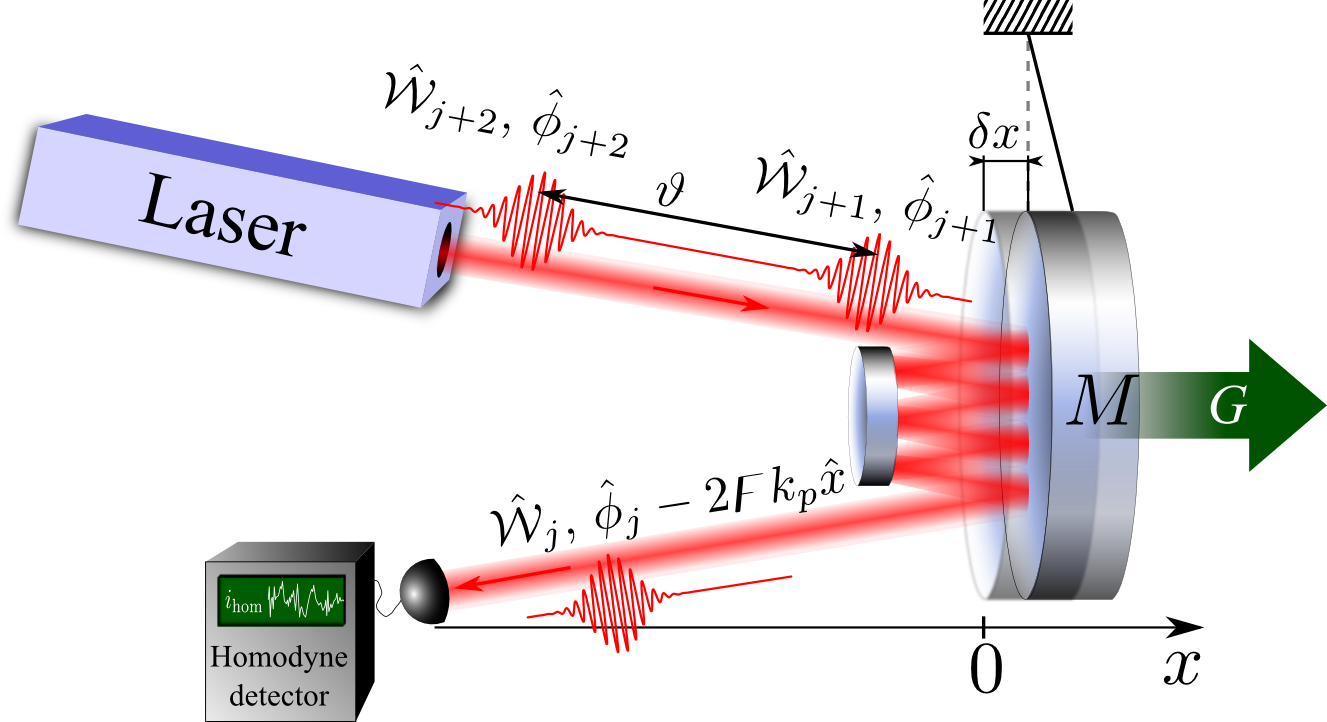}{%
\begin{figure}[htbp]
  \centerline{\includegraphics[width=.85\textwidth]{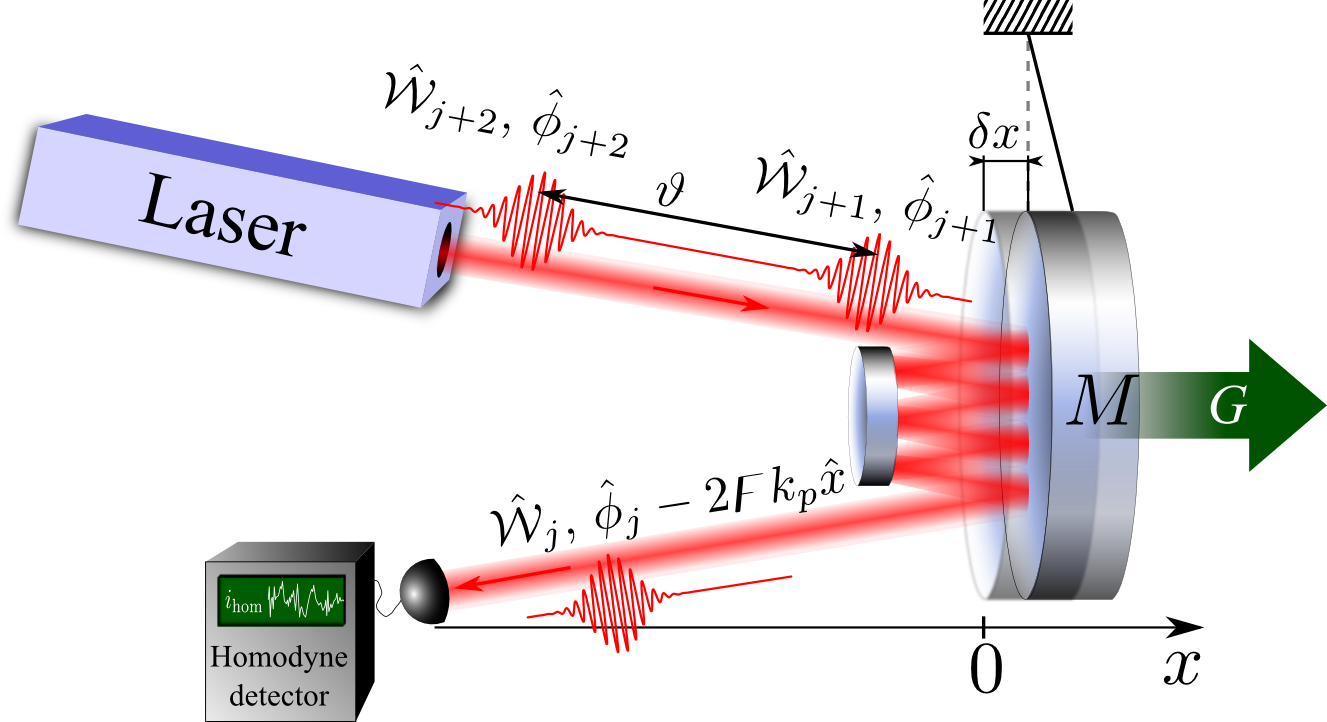}}
  \caption{Toy example of a linear optical position measurement.}
\label{fig:toy_1}
\end{figure}}

The advanced version of that example is shown in Figure~\ref{fig:toy_1}. The only difference between this scheme and the initial one (see Figure~\ref{fig:toy_0}) is that here the detector measures not the phase of light pulses, but linear combination of the phase and energy, parametrized by the \emph{homodyne angle} $\phi_{\mathrm{LO}}$ (cf.\ Eq.~\eqref{eq:homodyne_photocurrent}):
\begin{equation}
\label{i_j}
  \hat{O}(t_j) = -\hat{\phi}_j^{\mathrm{refl}}\sin\phi_{\mathrm{LO}}
    + \frac{\hat{\mathcal{W}}_j - \mathcal{W}}{2\mathcal{W}}\cos\phi_{\mathrm{LO}} \,,
\end{equation}
(we subtracted the regular term proportional to the mean energy $\mathcal{W}$ from the output signal $\hat{O}(t_j)$).

Similar to Eq.~\eqref{eq:tilde_x_j}, renormalize this output signal as the equivalent test object displacement:
\begin{equation}
\label{tilde_x_j_corr}
  \tilde{x}(t_j) \equiv \frac{\hat{O}_j}{2\digamma k_p\sin\phi_{\mathrm{LO}}} 
    = \hat{x}(t_j) + \hat{x}_{\mathrm{fl}}(t_j) \,,
\end{equation}
where the noise term has in this case the following form:
\begin{equation}
\label{x_j_corr}
  \hat{x}_{\mathrm{fl}}(t_j) = \frac{1}{2\digamma k_p}\biggl(
      -\hat{\phi}(t_j) + \frac{\hat{\mathcal{W}}_j - \mathcal{W}}{2\mathcal{W}}\cot\phi_{\mathrm{LO}}
    \biggr).
\end{equation}
RMS uncertainty of this value (the measurement error) is equal to
\begin{equation}
\label{Delta_x_corr}
  \Delta x_{\mathrm{meas}} = \frac{1}{2\digamma k_p}
    \sqrt{(\Delta\phi)^2 + \frac{(\Delta\mathcal{W})^2}{4\mathcal{W}^2}\cot^2\phi_{\mathrm{LO}}} \,.
\end{equation}
At first glance, it seems like we just obtained the increased measurement error, for the same value of the test object perturbation, which is still described by Eq.~\eqref{eq:Delta_p_pert}. However, the additional term in Eq.~\eqref{i_j} can be viewed not only as the additional noise, but as the source of information about the test object perturbation. It can be used to subtract, at least in part, the terms induced by this perturbation from the subsequent measurement results. Quantitatively, this information is characterized by the cross-correlation of the measurement error and the back action:
\begin{equation}
\label{eq:Delta_xp}
  \Delta(xp) = \mean{\hat{x}_{\mathrm{fl}}(t_j)\circ\hat{p}^{\mathrm{b.a.}}(t_j)}
    = \frac{(\Delta\mathcal{W})^2}{2\omega_p\mathcal{W}}\cot\phi_{\mathrm{LO}} \,.
\end{equation}
It is easy to see that the uncertainties (\ref{eq:Delta_p_pert}, \ref{Delta_x_corr}, \ref{eq:Delta_xp}) satisfy the following Schr\"odinger--Robertson uncertainty relation:
\begin{equation}
\label{eq:DxDp_corr}
  (\Delta x_{\mathrm{meas}})^2(\Delta p_{\mathrm{b.a.}})^2 - [\Delta(xp)]^2
    = \frac{(\Delta\phi)^2(\Delta\mathcal{W})^2}{\omega_p^2} \ge \frac{\hbar^2}{4} \,.
\end{equation}

Now we can perform the transition to the continuous measurement limit as we did in Section~\ref{sec:disc2cont}:
  \begin{eqnarray}
    S_x &=& \lim_{\vartheta\to0}(\Delta x_{\mathrm{meas}})^2\vartheta
      = \frac{1}{4\digamma^2k_p^2}\left(S_\phi + \frac{S_\mathcal{I}}{4\mathcal{I}^2}\cot^2\phi_{\mathrm{LO}}\right)\,,\nonumber \\
    S_F &=& \lim_{\vartheta\to0}\frac{(\Delta p_{\mathrm{b.a.}})^2}{\vartheta}
      = \frac{4\digamma^2 S_\mathcal{I}}{c^2}\,, \nonumber\\
    S_{xF} &=& \lim_{\vartheta\to0}\Delta(xp) = \frac{S_\mathcal{I}}{2\omega_p\mathcal{I}}\cot\phi_{\mathrm{LO}} \,.\label{eq:S_xS_F_corr_lim},
  \end{eqnarray}
which transform inequality~\eqref{eq:DxDp_corr} to the Sch\"odinger--Robertson uncertainty relation for continuous measurements:
\begin{equation}
  S_xS_F - S_{xF}^2 = \frac{S_\phi S_\mathcal{I}}{\omega_p^2} \ge \frac{\hbar^2}{4} \,.
\end{equation}

In the particular case of the light pulses in a coherent quantum state~\eqref{DphiDE_coh}, the measurement error~\eqref{Delta_x_corr}, the momentum perturbation~\eqref{eq:Delta_p_pert}, and the cross-correlation term~\eqref{eq:Delta_xp} are equal to:
\begin{equation}
\label{eq:DxDp_corr_coh}
  \Delta x_{\mathrm{meas}}
    = \frac{c}{4\digamma}\sqrt{\frac{\hbar}{\omega_p\mathcal{W}\sin^2\phi_{\mathrm{LO}}}}\,, \qquad
  \Delta p_{\mathrm{b.a.}} = \frac{2\digamma}{c}\sqrt{\hbar\omega_p\mathcal{W}} \,, \qquad
  \Delta(xp) = \frac{\hbar}{2}\cot\phi_{\mathrm{LO}}
\end{equation}
(the momentum perturbation $\Delta p_{\mathrm{b.a.}}$ evidently remains the same as in the uncorrelated case, and we provided its value here only for convenience), which gives the exact equality in the Schr\"odinger--Robertson uncertainty relation:
\begin{equation}
  (\Delta x_{\mathrm{meas}})^2(\Delta p_{\mathrm{b.a.}})^2 - [\Delta(xp)]^2 = \frac{\hbar^2}{4} \,.
\end{equation}
Correspondingly, substituting the coherent quantum state power (double-sided) spectral densities~\eqref{eq:S_phiS_I_toy} into Eqs.~\eqref{eq:S_xS_F_corr_lim}, we obtain:
\begin{equation}
\label{eq:S_x_S_F_corr}
  S_x = \frac{\hbar c^2}{16\omega_p \mathcal{I}_0\digamma^2\sin^2\phi_{\mathrm{LO}}} \,, \qquad
  S_F = \frac{4\hbar\omega_p\mathcal{I}_0\digamma^2}{c^2} \qquad,
  S_{xF} = \frac{\hbar}{2}\cot\phi_{\mathrm{LO}} \,,
\end{equation}
with
\begin{equation}
  S_xS_F - S_{xF}^2 = \frac{\hbar^2}{4}
\end{equation}
[compare with Eqs.~\eqref{S_xS_F_toy}].
  
 The cross-correlation between the measurement and back-action fluctuations is equivalent to the \emph{virtual rigidity} $\chi^{\mathrm{virt}}_{FF}(\Omega)\equiv-K_{\mathrm{virt}}=\mathrm{const.}$ as one can conclude looking at Eqs.~\eqref{eq:gen_noise_def}. Indeed,
\begin{equation*}
  \hat{x}_{\mathrm{fl}}(t) = \hat{\mathcal{X}}(t)\,,\qquad\mbox{and}\qquad\hat{F}^{(0)} = \hat{\mathcal{F}}(t)+K_{\mathrm{virt}}\hat{x}_{\mathrm{fl}}(t)\,.
\end{equation*}
The sum noise does not change under the above transformation and can be written as:
\begin{equation*}
  \hat{F}_{\mathrm{sum}}(t) = \mathbf{D}\hat{x}_{\mathrm{fl}}(t) + \hat{F}^{(0)}_{\mathrm{fl}}(t)
  = \mathbf{D}_{\mathrm{eff}}\hat{x}_{\mathrm{fl}}(t) + \hat{\mathcal{F}}(t) \,,
\end{equation*}
where the new effective dynamics that correspond to the new noise are governed by the following differential operator
\begin{equation}
\label{D_new}
  \mathbf{D}_{\mathrm{eff}} = \mathbf{D} + K_{\mathrm{virt}} \,.
\end{equation}
The above explains why we refer to $K_{\mathrm{virt}}$ as `virtual rigidity'.

To see how virtual rigidity created by cross-correlation of noise sources can help beat the SQL consider a free mass as a probe body in the above considered toy example. The modified dynamics:
\begin{equation}
  \mathbf{D}_{\mathrm{eff}} = M\frac{d^2}{dt^2} + K_{\mathrm{virt}} \,, \qquad S_{xF} = K_{\mathrm{virt}} S_x \ne 0 \,,
\end{equation}
correspond to a harmonic oscillator with eigenfrequency $\Omega_0^2=K_{\mathrm{virt}}/M$, and as we have demonstrated in Eq.~\eqref{eq:osc_fm_SQL} provide a narrow-band sensitivity gain versus a free mass SQL near the resonance frequency $\Omega_0$ .

However, there is a drawback of virtual rigidity compared to the real one: it requires higher measurement strength, and therefore higher power, to reach the same gain in sensitivity as provided by a harmonic oscillator. This becomes evident if one weighs the back-action spectral density $S_F$, which is a good measure of measurement strength according to Eqs.~\eqref{eq:Omega_q}, for the virtual rigidity against the real one.
For the latter, to overcome the free mass SQL by a factor
\begin{equation}
\label{xi2_nb}
  \xi^2 = \frac{\Delta\Omega}{\Omega_0}
\end{equation}
(see Eq.~\eqref{eq:osc_fm_SQL}) at a given frequency $\Omega_0$, the back-action noise spectral density has to be \emph{reduced} by this factor:
\begin{equation}
  S_F = \frac{\hbar M \Omega_q^2}{2} = \xi^2S_F^{\mathrm{opt\,f.m.}} \,,
\end{equation}
see Eqs.~(\ref{eq:Omega_q}, \ref{eq:SQL_fm_cond}, \ref{eq:Omega_q_osc}). Here, $S_F^{\mathrm{opt\,f.m.}} = \hbar M \Omega_0/2$ is the back-action noise spectral density, which allows one to reach the free mass SQL~\eqref{eq:S_F_SQL_fm} at frequency $\Omega_0$. Such a sensitivity gain is achieved at the expense of proportionally reduced bandwidth:
\begin{equation}
  \Delta\Omega = \xi^2\Omega_0 \,.
\end{equation}
For the virtual rigidity, the optimal value of $S_F$ results from Eq.~\eqref{eq:Omega_q_osc}:
\begin{equation}
  S_F = \frac{\hbar^2/4 + S_{xF}^2}{S_x}
    = \frac{\hbar M}{2}\left(\Omega_q^2 + \frac{\Omega_0^4}{\Omega_q^2}\right)
    = S_F^{\mathrm{opt\,f.m.}}\left(\xi^2 + \frac{1}{\xi^2}\right) .
\end{equation}
Hence, the better the sensitivity (the smaller $\xi^2$), the larger $S_F$ must be and, therefore, measurement strength.

Another evident flaw of the virtual rigidity, which it shares with the real one, is the narrow-band character of the sensitivity gain it provides around $\Omega_0$ and that this band shrinks as the sensitivity gain rises (cf.\ Eq.~\eqref{xi2_nb}). In order to provide a broadband enhancement in sensitivity, either the real rigidity $K=M\Omega_0^2$, or the virtual one $K_{\mathrm{virt}}=S_{xF}/S_F$ should depend on frequency in such a way as to be proportional (if only approximately) to $\Omega^2$ in the frequency band of interest. Of all the proposed solutions providing frequency dependent virtual rigidity, the most well known are the \emph{quantum speedmeter}~\cite{90a1BrKh} and the \emph{filter cavities}~\cite{02a1KiLeMaThVy} schemes. Section~\ref{sec:toy_speedmeter}, we consider the basic principles of the former scheme. Then, in Section~\ref{sec:sub-SQL_schemes} we provide a detailed treatment of both of them.

\subsection{Quantum speedmeter}
\label{sec:toy_speedmeter}

\subsubsection{The idea of the quantum speedmeter}

The toy scheme that demonstrates a bare idea of the quantum speedmeter is shown in Figure~\ref{fig:speedmeter}. The main difference of this scheme from the position meters considered above (see Figures~\ref{fig:toy_0}, \ref{fig:toy_1}) is that each light pulse reflects from the test mass twice: first from the front and then from the rear face after passing the delay line with delay time $\tau$. An outgoing pulse acquires a phase shift proportional to the difference of the test-object positions at time moments separated by $\tau$, which is proportional to the test-mass average velocity $\hat{\bar{v}}(t_j)$ in this time interval ($t_j$ indicates the time moment after the second reflection):
\begin{equation}
\label{phi_sm_refl}
  \hat{\phi}^{\mathrm{refl}}(t_j) = \hat{\phi}(t_j) + 2\digamma k_p\tau\hat{\bar{v}}(t_j) \,,
\end{equation}
where
\begin{equation}
\label{V_j}
  \hat{\bar{v}}(t_j) = \frac{\hat{x}(t_j)-\hat{x}(t_j-\tau)}{\tau} \,.
\end{equation}

\epubtkImage{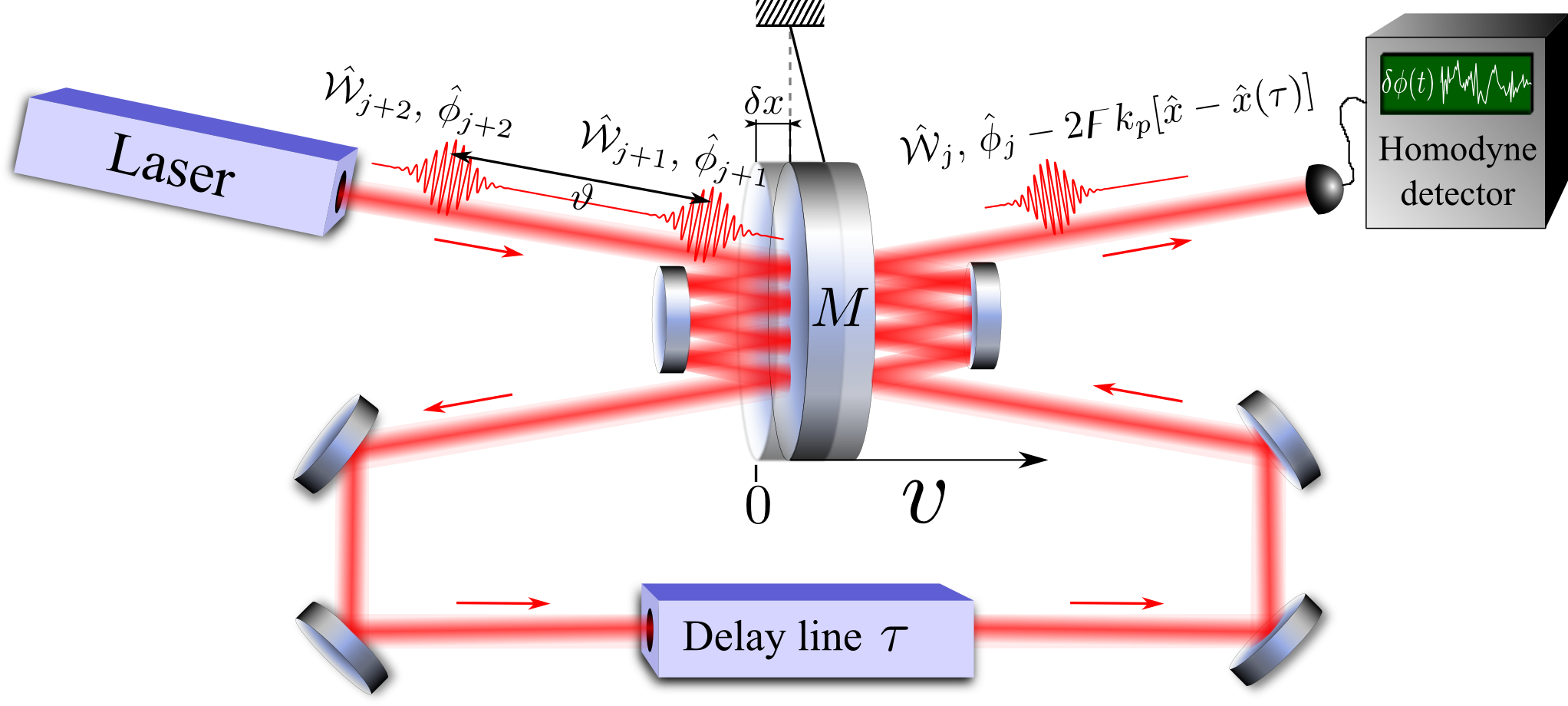}{
\begin{figure}
  \centerline{\includegraphics[width=\textwidth]{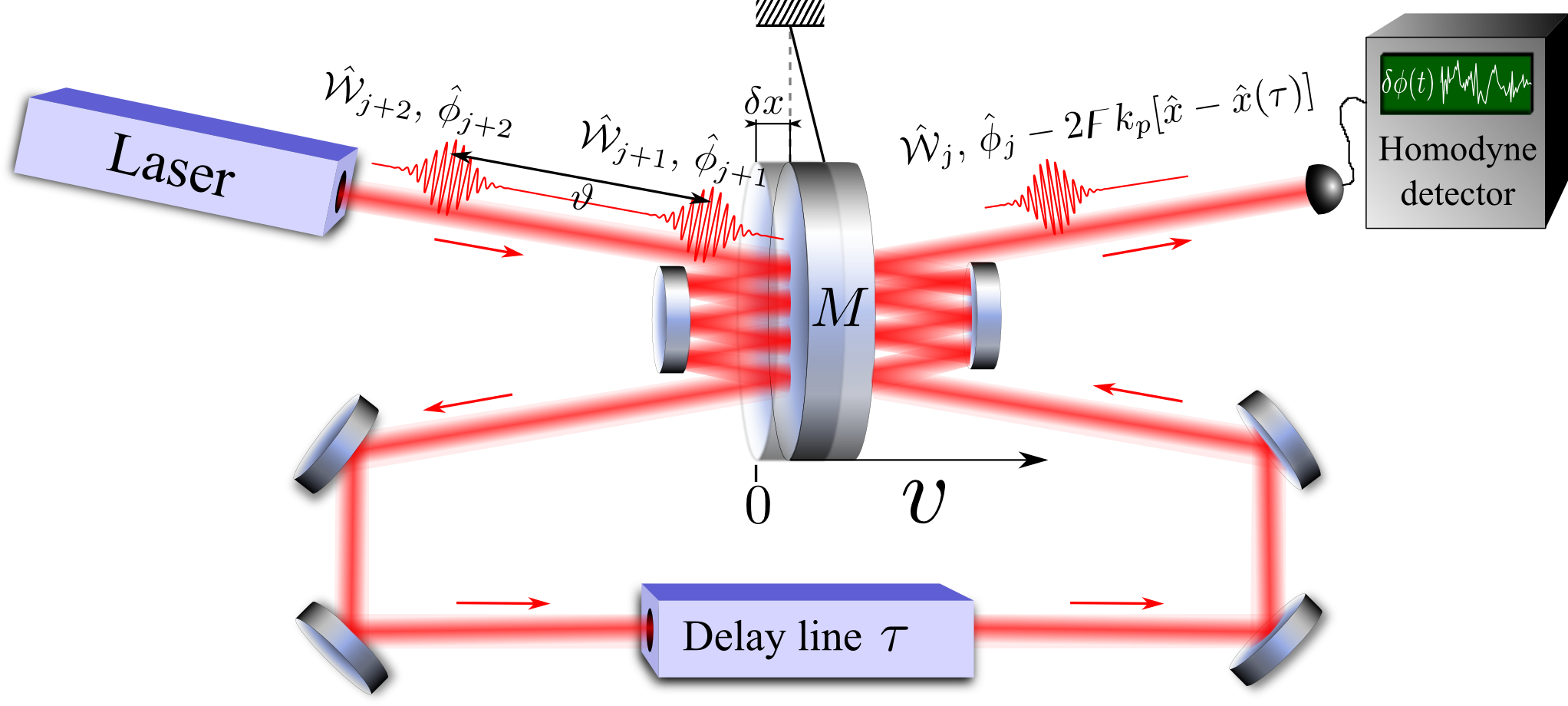}}
  \caption{Toy example of the quantum speedmeter scheme.}
\label{fig:speedmeter}
\end{figure}}

We omit here mathematical details of the transition to the continuous measurement limit as they are  essentially the same as in the position measurement case, see Section~\ref{sec:disc2cont}, and start directly with the continuous time equations. The output signal of the homodyne detector in the speedmeter case is described by the following equations:
  \begin{eqnarray}
    \hat{O}(t) &=& -\hat{\phi}^{\mathrm{refl}}(t)\sin\phi_{\mathrm{LO}} + \frac{\hat{\mathcal{I}}(t) - \mathcal{I}_0}{2\mathcal{I}_0}\,\cos\phi_{\mathrm{LO}}
       \,,\\
    \hat{\phi}^{\mathrm{refl}}(t)
      &=& \hat{\phi}(t) + 2\digamma k_p[\hat{x}(t)-\hat{x}(t-\tau)] \,.
  \end{eqnarray}
In spectral representation these equations yield:
\begin{equation}
  \tilde x(\Omega) \equiv -\frac{\hat{O}(\Omega)}{2\digamma k_p\sin\phi_{\mathrm{LO}}}
    = \hat{x}(\Omega) + \hat{x}_{\mathrm{fl}}(\Omega) \,,
\end{equation}
where
\begin{equation}
\label{eq:sm_x_fl_raw}
  \hat{x}_{\mathrm{fl}}(\Omega) = \frac{1}{2\digamma k_p(1-e^{i\Omega\tau})}
    \left[\hat{\phi}(\Omega) - \frac{\hat{\mathcal{I}}(\Omega) - \mathcal{I}_0}{2\mathcal{I}_0}\,\cot\phi_{\mathrm{LO}}\right]
\end{equation}
is the equivalent displacement measurement noise.

The back-action force with account for the two subsequent light reflections off the faces of the probe, can be written as:
\begin{equation}
\label{sm_F_pert_raw}
  \hat{F}_{\mathrm{b.a.}}(t) = \frac{2\digamma}{c}[\hat{\mathcal{I}}(t+\tau)-\hat{\mathcal{I}}(t)] \,
\end{equation}
and in spectral form as:
\begin{equation}
  \hat{F}_{\mathrm{b.a.}}(\Omega)
    = \frac{2\digamma}{c}\hat{\mathcal{I}}(\Omega)(e^{-i\Omega\tau}-1) .
\end{equation}

Then one can make a reasonable assumption that the time between the two reflections $\tau$ is much smaller than the signal force variation characteristic time ($\sim\Omega^{-1}$) that spills over into the following condition:
\begin{equation}
  \Omega\tau \ll 1 \,,
\end{equation}
and allows one to expand the exponents in Eqs.~(\ref{eq:sm_x_fl_raw}, \ref{sm_F_pert_raw}) into a Taylor series:
\begin{equation}
  \hat{x}_{\mathrm{fl}}(\Omega) = \frac{\hat{v}_{\mathrm{fl}}(\Omega)}{-i\Omega} \,, \qquad
  \hat{F}_{\mathrm{b.a.}}(t) = -i\Omega\hat{p}_{\mathrm{b.a.}}(\Omega) \,,
\end{equation}
where
\begin{eqnarray}
  \hat{v}_{\mathrm{fl}}(\Omega) &=& \frac{1}{2\digamma k_p\tau}
    \left[\hat{\phi}(\Omega) - \frac{\hat{\mathcal{I}}(\Omega) - \mathcal{I}}{2\mathcal{I}}\,\cot\phi_{\mathrm{LO}}\right] , \\
  \hat{p}_{\mathrm{b.a.}}(\Omega) &=& \frac{\hat{F}_{\mathrm{b.a.}}(\Omega)}{-i\Omega}= \frac{2\digamma\tau\hat{\mathcal{I}}(\Omega)}{c} \,.
\end{eqnarray}
Spectral densities of theses noises are equal to
\begin{equation}
\label{eq:sm_S_xS_F}
  S_x(\Omega) = \frac{S_v}{\Omega^2} \,, \qquad S_F(\Omega) = \Omega^2 S_p(\Omega) \,, \qquad
  S_{xF}(\Omega) = -S_{vp}(\Omega) \,,
\end{equation}
where
  \begin{eqnarray}
\label{eq:SM_toy_spdens}
    S_v &=& \frac{1}{4\digamma^2k_p^2\tau^2}
      \left(S_\phi + \frac{S_\mathcal{I}}{4\mathcal{I}^2}\cot^2\phi_{\mathrm{LO}}\right)\,,\nonumber \\
    S_p &=& \frac{4\digamma^2\tau^2S_\mathcal{I}}{c^2}\,, \nonumber\\
    S_{vp} &=& -\frac{S_\mathcal{I}}{2\omega_p\mathcal{I}}\cot\phi_{\mathrm{LO}}
  \end{eqnarray}
Note also that
\begin{equation}
\label{eq:SvSp-Svp2_uncert_rel}
  S_x(\Omega)S_F(\Omega) - S_{xF}^2(\Omega) = S_vS_p-S_{vp}^2
    = \frac{S_\phi S_\mathcal{I}}{\omega_p^2} \ge \frac{\hbar^2}{4} \,.
\end{equation}

The apparent difference of the spectral densities presented in
Eq.~\eqref{eq:sm_S_xS_F} from the ones describing the `ordinary'
position meter (see Eqs.~\eqref{eq:S_xS_F_corr_lim}) is that they now
have rather special frequency dependence. It is this frequency
dependence that together with the cross-correlation of the measurement
and back-action fluctuations, $S_{xF}\ne0$, allows the reduction of
the sum noise spectral density to arbitrarily small values. One can
easily see it after the substitution of Eq.~\eqref{eq:sm_S_xS_F} into
Eq.~\eqref{eq:gen_spdens} with a free mass
$\chi_{xx}(\Omega)=-1/(M\Omega^2)$ in mind:
\begin{equation}
\label{eq:SM_F_sum_spdens}
  S^F = M^2\Omega^4S_x(\Omega) + S_F(\Omega) - 2M\Omega^2S_{xF}(\Omega) = \Omega^2(M^2 S_v + 2MS_{vp} + S_p) \,.
\end{equation}

If there was no correlation between the back-action and measurement fluctuations, i.e., $S_{vp} = 0$, then by virtue of the uncertainty relation, the sum sensitivity appeared limited by the SQL~\eqref{eq:S_F_SQL_fm}:
\begin{equation}
\label{eq:SM_F_sum_spdens_uncorr}
  S^F = \Omega^2\left(\frac{\hbar^2M^2}{4S_p} + S_p\right)\geqslant \hbar M\Omega^2\,.
\end{equation}
One might wonder, what is the reason to implement such a complicated measurement strategy to find ourselves at the same point as in the case of a simple coordinate measurement? However, recall that in the position measurement case, a constant cross-correlation $S_{xF}\propto\cot\phi_{\mathrm{LO}}$ allows one to get only a narrow-band sub-SQL sensitivity akin to that of a harmonic oscillator. This effect we called \emph{virtual rigidity}, and showed that for position measurement this rigidity $K_{\mathrm{virt}} = S_{xF}/S_x$ is constant. In the speedmeter case, the situation is totally different; it is clearly seen if one calculates virtual rigidity for a speedmeter:
\begin{equation}
   K_{\mathrm{virt}}^{\mathrm{SM}} = \frac{S_{xF}}{S_x(\Omega)} = -\Omega^2\frac{S_{vp}}{S_v}\,.
 \end{equation}
It turns out to be frequency dependent exactly in the way that is necessary to compensate the free mass dynamical response to the back-action fluctuations. Indeed, in order to minimize the sum noise spectral density~\eqref{eq:SM_F_sum_spdens} conditioned on uncertainty relation~\eqref{eq:SvSp-Svp2_uncert_rel}, one needs to set
\begin{equation}
  S_{xF} = -S_{vp} = \frac{S_{F}}{M\Omega^2} = \frac{S_{p}}{M} = \mathrm{ const.},
\end{equation}
which allows one to overcome the SQL simply by choosing the right fixed homodyne angle:
\begin{equation}
  \cot\phi_{\mathrm{LO}} = \frac{8\digamma^2\tau^2\omega_p\mathcal{I}_0}{Mc^2} \,.
\end{equation}
Then the sum noise is equal to
\begin{equation}
\label{cont_opt}
  S^F(\Omega) = \frac{S_\phi S_\mathcal{I}}{\omega_p^2}\,\frac{M^2\Omega^2}{S_p}
  = \frac{M^2\Omega^2}{4\digamma^2k_p^2\tau^2}S_\phi \,.
\end{equation} 
and, in principle, can be made arbitrarily small, if a sufficient value of $S_p$ is provided; that is, ifthe optomechanical coupling is sufficiently strong.

\paragraph*{Simple case: light in a coherent state.}
Let us consider how the spectral density of a speedmeter will appear if the light field is in a coherent state. The spectral densities of phase and power fluctuations are given in Eqs.~\eqref{eq:S_phiS_I_toy}, hence the sum noise power (double-sided) spectral density for the speedmeter takes the following form:
\begin{equation}
S^F(\Omega) = \frac{\hbar M^2c^2\Omega^2}{16\omega_p\mathcal{I}_0\digamma^2\tau^2} = \frac{\hbar M}{2\tau^2}\left(\frac{\Omega}{\Omega_q}\right)^2 = \frac{S^F_{\mathrm{SQL\,f.m.}}}{2\Omega_q^2\tau^2}\,,
\end{equation}
where $\Omega_q$ for our scheme is defined in Eq.~\eqref{eq:Omega_q_toy}. This formula indicates the ability of a speedmeter to have a sub-SQL sensitivity in all frequencies provided high enough optical power and no optical loss.

\subsubsection{QND measurement of a free mass velocity}

The initial motivation to consider speed measurement rested on the assumption that a velocity $\hat{v}$ of a free mass is directly proportional to its momentum $\hat{p}=M\hat{v}$. And the momentum in turn is, as an integral of motion, a QND-observable, i.e., it satisfies the simultaneous measurability condition~\eqref{eq:measurability_cond}:
\begin{equation*}
  \left[\hat{p}(t),\,\hat{p}(t')\right] = 0\quad \forall t,\,t'.
\end{equation*}
But this connection between $\hat{p}$ and $\hat{v}$ holds only if one considers an isolated free mass not coupled to a meter. As the measurement starts, the velocity value gets perturbed by the meter and it is not proportional to the momentum anymore. Let us illustrate this statement by our simple velocity measurement scheme. The distinctive feature of this example is that the meter probes the test position $\hat{x}$ twice, with opposite signs of the coupling factor. Therefore, the Lagrangian of this scheme can be written as:
\begin{equation}
\label{sm_L_orig}
  \hat{\mathcal{L}} = \frac{M\hat{v}^2}{2} + \beta(t)\hat{x}\hat{\mathcal{N}} + \hat{\mathcal{L}}_{\mathrm{meter}} \,,
\end{equation}
where
\begin{equation}
\label{sm_v}
  \hat{v} = \frac{d\hat{x}}{dt}
\end{equation}
is the test-mass velocity, $\mathcal{N}$ stands for the meter's observable, which provides coupling to the test mass, $\mathcal{L}_{\mathrm{meter}}$ is the self-Lagrangian of the meter, and $\beta(t)$ is the coupling factor, which has the form of two short pulses with the opposite signs, separated by the time $\tau$, see Figure~\ref{fig:sm_alpha_beta}. We suppose for simplicity that the evolution of the meter observable $\mathcal{N}$ can be neglected during the measurement (this is a reasonable assumption, for in real schemes of the speedmeter and in the gedanken experiment considered above, this observable is proportional to the number of optical quanta, which does not change during the measurement). This assumption allows one to omit the term $\mathcal{L}_{\mathrm{meter}}$ from consideration.

\epubtkImage{fig27.png}{
\begin{figure}
  \centerline{\includegraphics{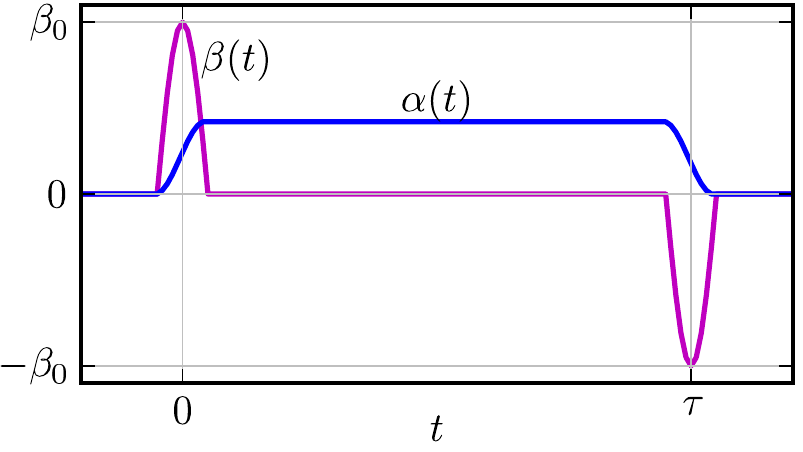}}
  \caption{Real ($\beta(t)$) and effective ($\alpha(t)$) coupling constants in the speedmeter scheme.}
\label{fig:sm_alpha_beta}
\end{figure}}

This Lagrangian does not satisfy the most well-known sufficient (but not necessary!) condition of the QND measurement, namely the commutator of the interaction Hamiltonian $\hat{\mathcal{H}}_{\mathrm{int}}=-\beta(t)\hat{x}\hat{\mathcal{N}}$ with the operator of measured observable $\hat{x}$ does not vanish~\cite{77a1eBrKhVo}. However, it can be shown that a more general condition is satisfied:
\begin{equation*}
   [\hat{U}(0,\tau),\hat{x}]=0
\end{equation*}
where $\hat{U}(0,\tau)$ is the evolution operator of probe-meter dynamics from the initial moment $t_{\mathrm{start}}=0$ when the measurement starts till the final moment $t_{\mathrm{end}}=\tau$ when it ends. Basically, the latter condition guarantees that the value of $\hat{x}$ before the measurement will be equal to that after the measurement, but does not say what it should be in between (see Section~4.4 of~\cite{92BookBrKh} for details).

Moreover, using the following nice trick~\epubtkFootnote{Personal communication with Yanbei Chen.}, the Lagrangian~\eqref{sm_L_orig} can be converted to the form, satisfying the simple condition of~\cite{77a1eBrKhVo}:
\begin{equation}
  \hat{\mathcal{L}}' = \hat{\mathcal{L}} - \frac{d\alpha(t)\hat{x}}{dt}\hat{\mathcal{N}} \,,
\end{equation}
where
\begin{equation}
  \alpha(t) = \int_{-\infty}^t\beta(t')\,dt' \,,
\end{equation}
see Figure~\ref{fig:sm_alpha_beta}.
It is known that two Lagrangians are equivalent if they differ only by a full time derivative of and arbitrary function of the generalized coordinates. Lagrangian equations of motion for the coordinates of the system are invariant to such a transformation.

The new Lagrangian has the required form with the interaction term proportional to the test-mass velocity:
\begin{equation}
  \hat{\mathcal{L}} = \frac{M\hat{v}^2}{2} - \alpha(t)\hat{v}\hat{\mathcal{N}} \,.
\end{equation}
Note that the antisymmetric shape of the function $\beta(t)$ guarantees that the coupling factor $\alpha(t)$ becomes equal to zero when the measurement ends. The canonical momentum of the mass $M$ is equal to
\begin{equation}
  p = \partdiff{\mathcal{L}}{v} = Mv - \alpha(t)\mathcal{N} \,,
\end{equation}
and the Hamiltonian of the system reads
\begin{equation}
\label{sm_gen_H}
  \hat{\mathcal{H}} = pv - \mathcal{L} = \frac{[p+\alpha(t)\mathcal{N}]^2}{2M}\,.
\end{equation}

The complete set of observables describing our system includes in addition apart to $\hat{x}$, $\hat{p}$, and $\hat{\mathcal{N}}$, the observable $\hat{\Phi}$ canonically conjugated to $\hat{\mathcal{N}}$:
\begin{equation}
\label{sm_gen_Phi_N}
  [\hat{\Phi},\hat{\mathcal{N}}] = i\hbar
\end{equation}
(if $\hat{\mathcal{N}}$ is proportional to the number of quanta in the light pulse then $\hat{\Phi}$ is proportional to its phase), which represents the output signal $\hat{O}$ of the meter. The Heisenberg equations of motion for these observables are the following:
  \begin{equation}
  \begin{array}{ll}
\label{eq:sm_gen_eqs}
\displaystyle
    \frac{d\hat{x}(t)}{dt} = \hat{v}(t) = \frac{\hat{p} + \alpha(t)\hat{\mathcal{N}}}{M} \,, & \qquad
\displaystyle
    \frac{d\hat{p}(t)}{dt} = 0 \,,\\[1em]
\displaystyle
    \frac{d\hat\Phi(t)}{dt} = \frac{\alpha(t)[\hat{p} + \alpha(t)\hat{\mathcal{N}}]}{M}
      = \alpha(t)\hat{v}(t) \,, & \qquad
\displaystyle
    \frac{d\hat{\mathcal{N}}(t)}{dt} = 0 \,.
  \end{array}
  \end{equation}
These equations show clearly that (i) a canonical momentum $\hat{p}$ is preserved by this measurement scheme while (ii) the test-mass velocity $\hat{v}$, as well as its kinematic momentum $M\hat{v}$, are perturbed during the measurement (that is, while $\alpha\ne0$), yet restore their initial values after the measurement, and (iii) the output signal of the meter $\hat{\Phi}$ carries the information about this perturbed value of the velocity.

Specify $\alpha(t)$ to be of a simple rectangular shape:
\begin{equation*}
  \alpha(t) = \begin{cases}
      1 \,, & \mbox{if}\ 0\le t<\tau \\
      0 \,, & \mbox{if}\ t<0\ \mbox{or}\ t\ge\tau \,.
    \end{cases}
\end{equation*}
This assumption does not affect our main results, but simplifies the calculation. In this case, the solution of Eqs.~\eqref{eq:sm_gen_eqs} reads:
\begin{subequations}
  \begin{eqnarray}
    \hat{x}(\tau) &=& \hat{x}(0) + \hat{\bar{v}}\tau \,, \\
    \hat{\Phi}(\tau) &=& \hat{\Phi}(0) + \hat{\bar{v}}\tau \,,
  \end{eqnarray}
\end{subequations}
with
\begin{equation}
\label{sm_gen_bar_v}
  \hat{\bar{v}} = \frac{\hat{p}+\hat{\mathcal{N}}}{M}
\end{equation}
the test-mass velocity during the measurement [compare with Eqs.~(\ref{phi_sm_refl}, \ref{V_j})].

Therefore, by detecting the variable $\Phi(\tau)$, the \emph{perturbed} value of velocity $\bar v$ is measured with an imprecision
\begin{equation}
\label{sm_gen_v_meas}
  \Delta v_{\mathrm{meas}} = \frac{\Delta\Phi}{\tau} \,,
\end{equation}
where $\Delta\Phi$ is the initial uncertainty of $\Phi$. The test-mass position perturbation after this measurement is proportional to the initial uncertainty of $\mathcal{N}$:
\begin{equation}
\label{sm_gen_pert}
  \Delta x_{\mathrm{b.a.}} = \Delta v_{\mathrm{b.a.}}\tau = \frac{\Delta\mathcal{N}\tau}{M} \,.
\end{equation}
It follows from the last two equations that
\begin{equation}
  \Delta v_{\mathrm{meas}}\Delta x_{\mathrm{b.a.}}
    = \frac{\Delta\Phi\Delta\mathcal{N}}{M} \ge \frac{\hbar}{2M} \,.
\end{equation}
The sum error of the initial velocity estimate $v_{\mathrm{init}} = p/m$ yielding from this measurement is thus equal to:
\begin{equation}
  \Delta v_{\mathrm{sum}}
    = \sqrt{\left(\frac{\Delta\Phi}{\tau}\right)^2 + \left(\frac{\Delta\mathcal{N}}{M}\right)^2}
    \ge \sqrt{\frac{\hbar}{M\tau}} \,.
\end{equation}
As we see, it is limited by a value of the \emph{velocity measurement} SQL:
\begin{equation}
\label{eq:v_SQL}
  \Delta v_{\mathrm{SQL}} = \sqrt{\frac{\hbar}{M\tau}}\,.
\end{equation}

To overcome this SQL one has to use cross-correlation between the measurement error and back-action. Then it becomes possible to measure $v_{\mathrm{init}}$ with arbitrarily high precision. Such a cross-correlation can be achieved by measuring the following combination of the meter observables
\begin{equation}
  \hat{\Phi}(\tau)-\hat{\mathcal{N}}\tau/m = \hat{\Phi} (0) + \frac{\hat{p}\tau}{M}
\end{equation}
instead of $\hat{\Phi}(\tau)$, which gives a sum measurement uncertainty for the initial velocity $\hat{p}/m$, proportional to the initial uncertainty of $\hat{\Phi}$ only:
\begin{equation}
  \Delta v_{\mathrm{sum}} = \frac{\Delta\Phi}{\tau} \,,
\end{equation}
and hence not limited by the SQL.

\newpage
\section{Quantum Noise in Conventional GW Interferometers}
\label{sec:QN_in_GW_interferometers}

FY{In Section~\ref{sec:linear_quantum_measurement}}, we have talked about the quantum measurement, the general structure of quantum noise implied by the quantum mechanics and the restrictions on the achievable sensitivity it imposes. In this section, we turn to the application of these general and lofty principles to real life, i.e., we are going to calculate quantum noise for several types of the schemes of GW interferometers and consider the advantages and drawbacks they possess.

To grasp the main features of quantum noise in an advanced GW interferometer it would be elucidating to consider first two elementary examples: (i) a single movable mirror coupled to a free optical field, reflecting from it, and (ii) a Fabry--P{\'e}rot cavity comprising two movable mirrors and pumped from both sides. These two systems embody all the main features and phenomena that also mold the advanced and more complicated interferometers' quantum noise. Should one encounter these phenomena in real-life GW detectors, knowledge of how they manifest themselves in these simple situations would be of much help in successfully discerning them.

\subsection{Movable mirror}

The scheme of the mirror is drawn in
Figure~\ref{fig:Single_Mirror}. It is illuminated from both sides by
the two independent laser sources with frequency $\omega_p$, and mean
power values $\mathcal{I}_1$ and $\mathcal{I}_2$. In
terms of the general linear measurement theory of
Section~\ref{sec:gen_linear_measurement} we have two meters
represented by these two incident light waves. The two arbitrary
quadratures of the reflected waves are deemed as measured quantities
$\hat{O}_1$ and $\hat{O}_2$. Measurement can be performed, e.g., by
means of two independent homodyne detectors. Let us analyze quantum
noise in such a model keeping to the scheme given by
Eqs.~\eqref{eq:gen_linear_meas_EOM_Fourier}.

\epubtkImage{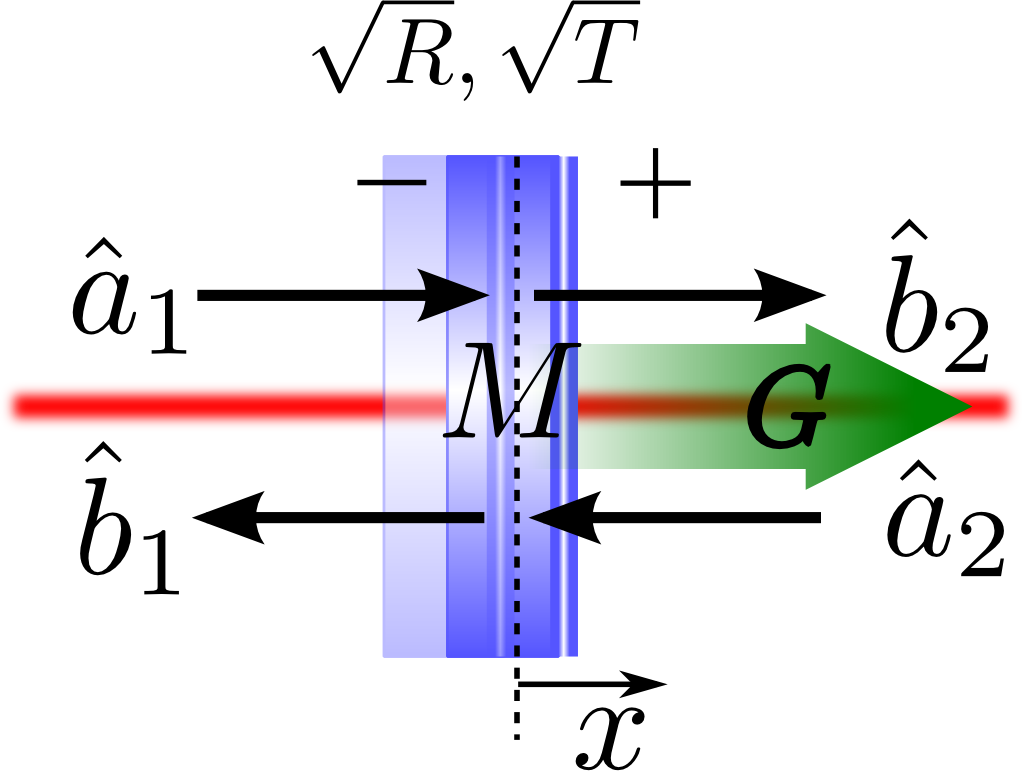}{
\begin{figure}[htbp]
  \centerline{\includegraphics[width=.3\textwidth]{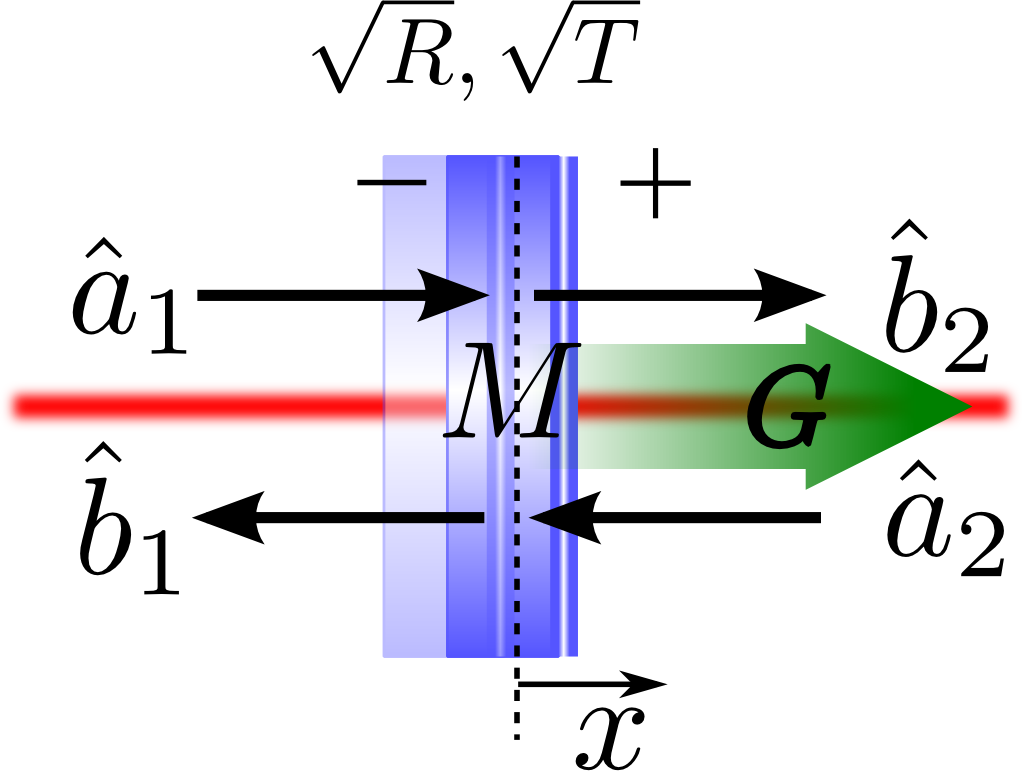}}
  \caption{Scheme of light reflection off the single movable mirror of mass $M$ pulled by an external force $G$.}
\label{fig:Single_Mirror}
\end{figure}}

\subsubsection{Optical transfer matrix of the movable mirror}

The first one of Eqs.~\eqref{eq:gen_linear_meas_EOM_Fourier} in our simple scheme is represented by the input-output coupling equations (\ref{eq:IO_mirror_matrix_4x4_zero}, \ref{eq:IO_mirror_relations_first+x}) of light on a movable mirror derived in Section~\ref{sec:light_reflection}. We choose the transfer matrix of the mirror to be real:
\begin{equation}
\label{eq:mirror_Mreal}
  \mathbb{M}_{\mathrm{real}} = 
  \begin{bmatrix}
  -\sqrt{R} & \sqrt{T}\\
   \sqrt{T} & \sqrt{R}
  \end{bmatrix}
\end{equation}
according to Eq.~\eqref{eq:IO_mirror_matrix}. Then we can write down the coupling equations, substituting electric field strength amplitudes $\vb{\mathcal{E}}_{1,2}$ and $\hat{\boldsymbol{e}}_{1,2}(\Omega)$ by their dimensionless counterparts as introduced by Eq.~\eqref{eq:2photon_E_strain} of Section~\ref{sec:2photon_formalism}:
\begin{equation}
\label{eq:IO_mirror_relations+x_quant}
\begin{bmatrix}
  \hat{\mathbf{B}}_{1}(\Omega)\\
  \hat{\mathbf{B}}_{2}(\Omega)
\end{bmatrix}=
\mathbb{M}_{\mathrm{real}}
\cdot
\begin{bmatrix}
  \hat{\mathbf{A}}_{1}(\Omega)\\
  \hat{\mathbf{A}}_{2}(\Omega)
\end{bmatrix}\,,\qquad\mbox{and}\qquad
\begin{bmatrix}
  \hat{\boldsymbol{b}}_{1}(\Omega)\\
  \hat{\boldsymbol{b}}_{2}(\Omega)
\end{bmatrix}=
\mathbb{M}_{\mathrm{real}}
\cdot
\begin{bmatrix}
  \hat{\boldsymbol{a}}_{1}(\Omega)\\
  \hat{\boldsymbol{a}}_{2}(\Omega)
\end{bmatrix} +
\begin{bmatrix}
\boldsymbol{R}_1\\
\boldsymbol{R}_2
\end{bmatrix}\hat{x}(\Omega)\,,
\end{equation}
where
\begin{equation}
  \boldsymbol{R}_1 = 2\sqrt{R}\frac{\omega_p}{c}
  \begin{bmatrix}
    \mathrm{A}_{1s}\\
    -\mathrm{A}_{1c}
  \end{bmatrix}\,,\qquad\mbox{and}\qquad
\boldsymbol{R}_2 = 2\sqrt{R}\frac{\omega_p}{c}
  \begin{bmatrix}
    \mathrm{A}_{2s}\\
    -\mathrm{A}_{2c}
  \end{bmatrix}\,.
\end{equation}
Without any loss of generality one can choose the phase of the light incident from the left to be such that $\mathrm{A}_{1s} = 0$ and $\mathrm{A}_{1c} = \sqrt{2\mathcal{I}_1/(\hbar\omega_p)}$. Then factoring in the constant phase difference between the left and the right beams equal to $\Phi_0$, one would obtain for the left light $\{A_{2c},A_{2s}\} = \sqrt{2\mathcal{I}_2/(\hbar\omega_p)}\{\cos\Phi_0,\sin\Phi_0\}$.

The two output measured quantities will then be given by the two homodyne photocurrents:
\begin{eqnarray}
\label{eq:mirror_readout_quant}
  \hat{O}_1(\Omega) &=& \boldsymbol{H}^{\mathsf{T}}[\phi_1]\hat{\boldsymbol{b}}_1 = \hat{b}_{1c}(\Omega)\cos\phi_1 + \hat{b}_{1s}(\Omega)\sin\phi_1 \,,\\
  \hat{O}_2(\Omega) &=& \boldsymbol{H}^{\mathsf{T}}[\phi_2]\hat{\boldsymbol{b}}_2 = \hat{b}_{2c}(\Omega)\cos\phi_2 + \hat{b}_{2s}(\Omega)\sin\phi_2\,,
\end{eqnarray}
where vector $\boldsymbol{H}[\phi]$ was first introduced in Section~\ref{sec:heterodyne} after Eq.~\eqref{eq:heterodyne_photocurrent} as:
\begin{equation}
  \boldsymbol{H}[\phi] \equiv \begin{bmatrix}
                   \cos\phi\\
                   \sin\phi
                 \end{bmatrix}\,.
\end{equation}
Then, after substitution of~\eqref{eq:mirror_Mreal} into~\eqref{eq:IO_mirror_relations+x_quant}, and then into~\eqref{eq:mirror_readout_quant}, one gets
\begin{eqnarray}
  \hat{O}^{(0)}_1(\Omega) &=& \boldsymbol{H}^{\mathsf{T}}[\phi_1](-\sqrt{R}\hat{\boldsymbol{a}_1}+\sqrt{T}\hat{\boldsymbol{a}_2})\,,\\
  \hat{O}^{(0)}_2(\Omega) &=& \boldsymbol{H}^{\mathsf{T}}[\phi_2](\sqrt{T}\hat{\boldsymbol{a}_1}+\sqrt{R}\hat{\boldsymbol{a}_2})\,,
\end{eqnarray}
and
\begin{eqnarray}
  \chi_{O_1F}(\Omega) &=& \boldsymbol{H}^{\mathsf{T}}[\phi_1]\boldsymbol{R}_1 = -\frac{2\sqrt{2R\hbar\omega_p\mathcal{I}_1}}{\hbar c}\sin\phi_1\,,\\
  \chi_{O_2F}(\Omega) &=& \boldsymbol{H}^{\mathsf{T}}[\phi_2]\boldsymbol{R}_2 = -\frac{2\sqrt{2R\hbar\omega_p\mathcal{I}_2}}{\hbar c}\sin(\phi_2-\Phi_0)\,.
\end{eqnarray}

\subsubsection{Probe's dynamics: radiation pressure force and ponderomotive rigidity}

Now we can write down the equation of motion for the mirror assuming it is pulled by a GW tidal force $G$:
\begin{equation}
\label{eq:Mirror_EOMs}
  m\ddot{\hat{x}}(t) = \hat{F}_{\mathrm{r.p.}}(t) + G(t)\qquad\Longrightarrow\qquad -M\Omega^2\hat{x}(\Omega) = \hat{F}_{\mathrm{r.p.}}(\Omega) + G(\Omega)\,,
\end{equation}
that gives us the probe's dynamics equation (the third one) in~\eqref{eq:gen_linear_meas_EOM_Fourier}:
\begin{equation}
  \hat{x}(\Omega) = \chi_{xx}(\Omega)[\hat{F}_{\mathrm{b.a.}}(\Omega) + G(\Omega)] = -\frac{1}{M\Omega^2}[\hat{F}_{\mathrm{r.p.}}(\Omega) + G(\Omega)]\,,
\end{equation}
where the free mirror mechanical susceptibility $\chi_{xx}(\Omega) = -1/(M\Omega^2)$.
The term $\hat{F}_{\mathrm{b.a.}}(t)$ stands for the radiation pressure force imposed by the light that can be calculated as
\begin{eqnarray}
 &&\hat{F}_{\mathrm{r.p.}}(\Omega) = F_0 + \hat{F}_{\mathrm{b.a.}}(\Omega) =  \frac{\hat{\mathcal{I}}_{a_1}(\Omega)+\hat{\mathcal{I}}_{b_1}(\Omega)-\hat{\mathcal{I}}_{a_2}(\Omega)-\hat{\mathcal{I}}_{b_2}(\Omega)}{c} \simeq\nonumber\\ &&\frac{\hbar k_p}{2}\left[({\mathbf{A}}_1^{\mathsf{T}}{\mathbf{A}}_1+{\mathbf{B}}_1^{\mathsf{T}}{\mathbf{B}}_1-{\mathbf{A}}_2^{\mathsf{T}}{\mathbf{A}}_2-{\mathbf{B}}_2^{\mathsf{T}}{\mathbf{B}}_2)+ 2({\mathbf{A}}_1^{\mathsf{T}}\hat{\boldsymbol{a}}_1(t)+{\mathbf{B}}_1^{\mathsf{T}}\hat{\boldsymbol{b}}_1(t)-{\mathbf{A}}_2^{\mathsf{T}}\hat{\boldsymbol{a}}_2(t)-{\mathbf{B}}_2^{\mathsf{T}}\hat{\boldsymbol{b}}_2(t))\right]
\end{eqnarray}
where $k_p\equiv\omega_p/c$
\begin{equation}
F_0 =  \frac{\hbar k_p}{2}({\mathbf{A}}_1^{\mathsf{T}}{\mathbf{A}}_1+{\mathbf{B}}_1^{\mathsf{T}}{\mathbf{B}}_1-{\mathbf{A}}_2^{\mathsf{T}}{\mathbf{A}}_2-{\mathbf{B}}_2^{\mathsf{T}}{\mathbf{B}}_2) = \frac{2 R}{c}(\mathcal{I}_1-\mathcal{I}_2)-\frac{4\sqrt{RT}}{c}\sqrt{\mathcal{I}_1\mathcal{I}_2}\cos\Phi_0\,,
\end{equation}
is the regular part of the radiation pressure force\epubtkFootnote{Note the second term proportional to $\sqrt{\mathcal{I}_1\mathcal{I}_2}\cos\Phi_0$, which owes its existence to the interference of the two traveling waves running in opposite directions. An interesting consequence of this is that the radiation pressure does not vanish even if the two waves have equal powers, i.e., $\mathcal{I}_1=\mathcal{I}_2$, that is, in order to compensate for the radiation pressure force of one field on the semi-transparent mirror, the other one should not only have the right intensity but also the right phase with respect to the former one:
$$\cos\Phi_0 = \frac{1}{2}\sqrt{\frac{R}{T}}\frac{\mathcal{I}_1-\mathcal{I}_2}{\sqrt{\mathcal{I}_1\mathcal{I}_2}}\,.$$
}. It is constant and thus can be compensated by applying a fixed restoring force of the same magnitude but with opposite direction, which can be done either by employing an actuator, or by turning the mirror into a low-frequency pendulum with $\omega_m\ll\Omega_{\mathrm{GW}}$ by suspending it on thin fibers, as is the case for the GW interferometers, that provides a necessary gravity restoring force in a natural way. However, it does not change the quantum noise and thus can be omitted from further consideration.
The latter term represents a quantum correction to the former one
\begin{equation}
\label{eq:Mirror_F_ba}
  \hat{F}_{\mathrm{b.a.}}(\Omega) \simeq \hbar k_p({\mathbf{A}}_1^{\mathsf{T}}\hat{\boldsymbol{a}}_1(\Omega)+{\mathbf{B}}_1^{\mathsf{T}}\hat{\boldsymbol{b}}_1(\Omega)-{\mathbf{A}}_2^{\mathsf{T}}\hat{\boldsymbol{a}}_2(\Omega)-{\mathbf{B}}_2^{\mathsf{T}}\hat{\boldsymbol{b}}_2(\Omega)) = \boldsymbol{F}_1^{\mathsf{T}}\hat{\boldsymbol{a}}_1(\Omega)+\boldsymbol{F}_2^{\mathsf{T}}\hat{\boldsymbol{a}}_2(\Omega)-K\hat{x}(\Omega)
\end{equation}
where $\hat{F}_{\mathrm{b.a.}}^{(0)} \equiv \boldsymbol{F}_1^{\mathsf{T}}\hat{\boldsymbol{a}}_1(\Omega)+\boldsymbol{F}_2^{\mathsf{T}}\hat{\boldsymbol{a}}_2(\Omega)$ is the random part of the radiation pressure that depends on the input light quantum fluctuations described by quantum quadrature amplitudes vectors $\hat{\boldsymbol{a}}_1(\Omega)$ and $\hat{\boldsymbol{a}}_2(\Omega)$ with coefficients given by vectors:
\begin{equation*}
  \boldsymbol{F}_1 = \frac{2\sqrt{2\hbar\omega_p R}}{c}
  \begin{bmatrix}
    \sqrt{R \mathcal{I}_1}-\sqrt{T\mathcal{I}_2}\cos \Phi_0\\
    -\sqrt{T\mathcal{I}_2}\sin\Phi_0
  \end{bmatrix}\,,\quad\mbox{and}\quad
\boldsymbol{F}_2 = -\frac{2\sqrt{2\hbar\omega_p R}}{c}
  \begin{bmatrix}
    \sqrt{T \mathcal{I}_1}+\sqrt{R\mathcal{I}_2}\cos \Phi_0\\
    \sqrt{R\mathcal{I}_2}\sin\Phi_0
  \end{bmatrix}\,,
\end{equation*}
and the term $-K\hat{x}(\Omega)$ represents the dynamical back action with
\begin{equation}
  K = \frac{8\omega_p\sqrt{RT\mathcal{I}_1\mathcal{I}_2}\sin\Phi_0}{c^2}
\end{equation}
being a ponderomotive rigidity that arises in the potential created by the two counter propagating light waves. Eq.~\eqref{eq:Mirror_F_ba} gives us the second of Eqs.~\eqref{eq:gen_linear_meas_EOM_Fourier}. Here $\chi_{FF}(\Omega) = -K$.

\subsubsection{Spectral densities}

We can reduce both our readout quantities to the units of the signal force $G$ according to Eq.~\eqref{eq:gen_force_noise_def}:
\begin{equation}
\label{eq:mirror_sum_noises}
  \hat{O}^F_1 = \frac{\hat{\mathcal{X}_1}(\Omega)}{\chi_{xx}^{\mathrm{eff}}(\Omega)} + \hat{\mathcal{F}}(\Omega) +G\,,
  \qquad
  \hat{O}^F_2 = \frac{\hat{\mathcal{X}_2}(\Omega)}{\chi_{xx}^{\mathrm{eff}}(\Omega)} + \hat{\mathcal{F}}(\Omega) +G
\end{equation}
and define the two effective measurement noise sources as
\begin{equation}
  \hat{\mathcal{X}}_1(\Omega) = \frac{\hat{O}_1^{(0)}}{\chi_{O_1F}(\Omega)}\,,\quad\mbox{and}\quad\hat{\mathcal{X}}_2(\Omega) = \frac{\hat{O}_2^{(0)}}{\chi_{O_2F}(\Omega)}
\end{equation}
and an effective force noise as
\begin{equation}
  \hat{\mathcal{F}}(\Omega) = \hat{F}_{\mathrm{b.a.}}^{(0)}\,,
\end{equation}
absorbing optical rigidity $K$ into the effective mechanical susceptibility:
\begin{equation}
  \chi_{xx}^{\mathrm{eff}}(\Omega) = \frac{\chi_{xx}(\Omega)}{1+K\chi_{xx}(\Omega)} = \frac{1}{K-M\Omega^2}\,.
\end{equation}
One can then easily calculate their power (double-sided) spectral densities according to Eq.~\eqref{eq:gen_spdens}:
\begin{eqnarray}
\label{eq:mirror_spdens_sum}
  S^F_{11} = \frac{S_{\mathcal{X}_1\mathcal{X}_1}(\Omega)}{|\chi^{\mathrm{eff}}_{xx}(\Omega)|^2}&+&S_{\mathcal{F}\mathcal{F}}(\Omega)+2\mathrm{Re}\left[\frac{S_{\mathcal{X}_1\mathcal{F}}(\Omega)}{\chi^{\mathrm{eff}}_{xx}(\Omega)}\right]\,,\nonumber\\
  S^F_{22} = \frac{S_{\mathcal{X}_2\mathcal{X}_2}(\Omega)}{|\chi^{\mathrm{eff}}_{xx}(\Omega)|^2}&+&S_{\mathcal{F}\mathcal{F}}(\Omega)+2\mathrm{Re}\left[\frac{S_{\mathcal{X}_2\mathcal{F}}(\Omega)}{\chi^{\mathrm{eff}}_{xx}(\Omega)}\right]\,,\nonumber\\
  S^F_{12} =  S^{F*}_{21} = \frac{S_{\mathcal{X}_1\mathcal{X}_2}(\Omega)}{|\chi^{\mathrm{eff}}_{xx}(\Omega)|^2}&+&S_{\mathcal{F}\mathcal{F}}(\Omega)+\left[\frac{S_{\mathcal{X}_1\mathcal{F}}(\Omega)}{\chi^{\mathrm{eff}}_{xx}(\Omega)}+\frac{S^*_{\mathcal{X}_2\mathcal{F}}(\Omega)}{\chi^{\mathrm{eff*}}_{xx}(\Omega)}\right],
\end{eqnarray}
where, if both lights are in coherent states that implies
$$\mean{\hat{\boldsymbol{a}}_i(\Omega)\circ\hat{\boldsymbol{a}}^\dag_j(\Omega')} = 2\pi\mathbb{S}_{\vac}\delta_{ij}\delta(\Omega-\Omega')\,,\qquad (i,j)=(1,2)$$,
one can get:
\begin{eqnarray}
\label{eq:mirror_spdens_partial}
  &&S_{\mathcal{X}_1\mathcal{X}_1}(\Omega) = \frac{\hbar c^2}{16\omega_p\mathcal{I}_1R\sin^2\phi_1}\,,\qquad S_{\mathcal{X}_2\mathcal{X}_2}(\Omega) = \frac{\hbar c^2}{16\omega_p\mathcal{I}_2R\sin^2(\phi_2-\Phi_0)}\,,\qquad S_{\mathcal{X}_1\mathcal{X}_2}(\Omega) = 0\,,\nonumber\\
  &&S_{\mathcal{F}\mathcal{F}}(\Omega) = \frac{4\hbar\omega_p R(\mathcal{I}_1+\mathcal{I}_2)}{c^2}\,,\qquad S_{\mathcal{X}_1\mathcal{F}}(\Omega) = \frac{\hbar}{2}\cot\phi_1\,,\qquad S_{\mathcal{X}_2\mathcal{F}}(\Omega) = \frac{\hbar}{2}\cot(\phi_2-\Phi_0)\,.
\end{eqnarray}
Comparison of these expressions with the Eqs.~\eqref{eq:S_x_S_F_corr} shows that we have obtained the results similar to that of the toy example in Section~\ref{sec:toy_FI_correlation}. If we switched one of the pumping carriers off, say the right one, the resulting spectral densities for $\hat{O}^F_1(\Omega)$ would be exactly the same as in the simple case of Eqs.~\eqref{eq:S_x_S_F_corr}, except for the substitution of $\digamma^2\mathcal{I}_0\to R\mathcal{I}_1$ and $\phi_1\to\phi_{\mathrm{LO}}$.

\subsubsection{Full transfer matrix approach to the calculation of quantum noise spectral densities}
\label{sec:mirror_full_transfer_mat}

It was easy to calculate the above spectral densities by parts, distinguishing the effective measurement and back-action noise sources and making separate calculations for them. Had we considered a bit more complicated situation with the incident fields in the squeezed states with arbitrary squeezing angles, the calculation of all six of the above individual spectral densities~\eqref{eq:mirror_spdens_partial} and subsequent substitution to the sum spectral densities expressions~\eqref{eq:mirror_spdens_sum} would be more difficult. Thus, it would be beneficial to have a tool to do all these operations at once numerically.

It is achievable if we build a \emph{full transfer matrix} of our system. To do so, let us first consider the readout observables separately. We start with $\hat{O}_1$ and rewrite it as follows:
\begin{eqnarray}
  \hat{O}_1 &=&
  \boldsymbol{H}^{\mathsf{T}}[\phi_1]\left(-\sqrt{R}\hat{\boldsymbol{a}}_1+\chi_{xx}^{\mathrm{eff}}\boldsymbol{R}_1\boldsymbol{F}^{\mathsf{T}}_1\hat{\boldsymbol{a}}_1+\sqrt{T}\hat{\boldsymbol{a}}_2+\chi_{xx}^{\mathrm{eff}}\boldsymbol{R}_1\boldsymbol{F}^{\mathsf{T}}_2\hat{\boldsymbol{a}}_2\right)+\chi_{xx}^{\mathrm{eff}}\boldsymbol{H}^{\mathsf{T}}[\phi_1]\boldsymbol{R}_1G \nonumber\\ &=& \boldsymbol{H}^{\mathsf{T}}[\phi_1]\cdot\mathbb{M}_1\cdot
  \begin{bmatrix}
    \hat{\boldsymbol{a}}_1\\
    \hat{\boldsymbol{a}}_2
  \end{bmatrix}
+\chi_{xx}^{\mathrm{eff}}\boldsymbol{H}^{\mathsf{T}}[\phi_1]\boldsymbol{R}_1G
\,,
\end{eqnarray}
where we omitted the frequency dependence of the constituents for the sake of brevity and introduced a $2\times4$ full transfer matrix $\mathbb{M}_1$ for the first readout observable defined as
\begin{equation}
 \mathbb{M}_1 =
 \begin{bmatrix}
   -\sqrt{R}\mathbb{I} + \chi_{xx}^{\mathrm{eff}}\boldsymbol{R}_1\boldsymbol{F}^{\mathsf{T}}_1 & \sqrt{T}\mathbb{I} + \chi_{xx}^{\mathrm{eff}}\boldsymbol{R}_1\boldsymbol{F}^{\mathsf{T}}_2
 \end{bmatrix}
\end{equation}
with outer product of two arbitrary vectors $\vb{\alpha} = \{\alpha_1,\alpha_2\}^{\mathsf{T}}$ and $\vb{\beta} = \{\beta_1,\beta_2\}^{\mathsf{T}}$ written in short notation as:
\begin{equation}
  \vb{\alpha}\vb{\beta}^{\mathsf{T}} \equiv
  \begin{bmatrix}
    \alpha_1\beta_1 & \alpha_1\beta_2\\
    \alpha_2\beta_1 & \alpha_2\beta_2
  \end{bmatrix}\,.
\end{equation}
In a similar manner, the full transfer matrix for the second readout can be defined as:
\begin{equation}
 \mathbb{M}_2 =
 \begin{bmatrix}
   \sqrt{T}\mathbb{I} + \chi_{xx}^{\mathrm{eff}}\boldsymbol{R}_2\boldsymbol{F}^{\mathsf{T}}_1 & \sqrt{R}\mathbb{I} + \chi_{xx}^{\mathrm{eff}}\boldsymbol{R}_2\boldsymbol{F}^{\mathsf{T}}_2
 \end{bmatrix}\,.
\end{equation}
Having accomplished this, one is prepared to calculate all the spectral densities~\eqref{eq:mirror_spdens_sum} at once using the following matrix formulas:
\begin{equation}
\label{eq:mirror_gen_spdens_lossless}
  S^F_{ij}(\Omega) =\frac{1}{2|\chi_{xx}^{\mathrm{eff}}|^2} \frac{\boldsymbol{H}^{\mathsf{T}}[\phi_i]\cdot\mathbb{M}_i\mathbb{S}_{\in}\mathbb{M}^\dag_j\cdot\boldsymbol{H}[\phi_j]+\boldsymbol{H}^{\mathsf{T}}[\phi_i]\cdot\mathbb{M}^*_j\mathbb{S}_{\in}\mathbb{M}^{\mathsf{T}}_i\cdot\boldsymbol{H}[\phi_j]}{\boldsymbol{H}^{\mathsf{T}}[\phi_i]\boldsymbol{R}_i\boldsymbol{R}^\dag_j\boldsymbol{H}[\phi_j]}\,,\qquad (i,j)=(1,2)
\end{equation}

where $\mathbb{M}^*\equiv (\mathbb{M}^\dag)^{\mathsf{T}}$ and $\mathbb{S}_{\in}$ is the $4\times4$-matrix of spectral densities for the two input fields:
\begin{equation}
  \mathbb{S}_{\in} =
  \begin{bmatrix}
    \mathbb{S}_{\sqz}[r_1,\theta_1] & 0\\
    0 & \mathbb{S}_{\sqz}[r_2,\theta_2]
  \end{bmatrix}
\end{equation}
with $\mathbb{S}_{\sqz}[r_i,\theta_i]$ defined by Eq.~\eqref{eq:SQZ_SpDens_matrix}.

Thus, we obtain the formula that can be (and, actually, is) used for the calculation of quantum noise spectral densities of any, however complicated, interferometer given the full transfer matrix of this interferometer.

\subsubsection{Losses in a readout train}

Thus far we have assumed that there is no dissipation in the transition from the outgoing light to the readout photocurrent of the homodyne detector. This is, unfortunately, not the case since any real photodetector has the finite quantum efficiency $\eta_d<1$ that indicates how many photons absorbed by the detector give birth to photoelectrons, i.e., it is the measure of the probability for the photon to be transformed into the photoelectron. As with any other dissipation, this loss of photons gives rise to an additional noise according to the FDT that we should factor in. We have shown in Section~\ref{sec:losses_in_OE} that this kind of loss can be taken into account by means of a virtual asymmetric beamsplitter with transmission coefficients $\sqrt{\eta_d}$ and $\sqrt{1-\eta_d}$ for the signal light and for the additional noise, respectively. This beamsplitter is set into the readout optical train as shown in Figure~\ref{fig:mov_mirr_refl} and the $i$-th readout quantity needs to be modified in the following way:
\begin{equation}
  \hat{O}_i^{\mathrm{loss}}(\Omega) = \sqrt{\eta_d}\hat{O}_i(\Omega)+\sqrt{1-\eta_d}\hat{n}_i(\Omega)\,,
\end{equation}
 where
 $$\hat{n}_i(\Omega) = \boldsymbol{H}^{\mathsf{T}}[\phi_i]\hat{\boldsymbol{n}}_i(\Omega) = \hat{n}_{i,c}(\Omega)\cos\phi_i+\hat{n}_{i,s}(\Omega)\sin\phi_i$$
 is the additional noise that is assumed to be in a vacuum state. Since the overall factor in front of the readout quantity does not matter for the final noise spectral density, one can redefine $\hat{O}_i^{\mathrm{loss}}(\Omega)$ in the following way:
 \begin{equation}
\label{eq:homodyne_lossy}
  \hat{O}_i^{\mathrm{loss}}(\Omega) = \hat{O}_i(\Omega)+\epsilon_d\hat{n}_i(\Omega)\,,\qquad\mbox{where}\qquad\epsilon_d\equiv\sqrt{\frac{1}{\eta_d}-1}\,.
\end{equation}

The influence of this loss on the final sum spectral densities~\eqref{eq:mirror_gen_spdens_lossless} is straightforward to calculate if one assumes the additional noise sources in different readout trains to be uncorrelated. If it is so, then Eq.~\eqref{eq:mirror_gen_spdens_lossless} modifies as follows:
\begin{eqnarray}
\label{eq:mirror_gen_spdens_lossy}
  S^{F,\mathrm{loss}}_{ij}(\Omega) =\frac{1}{2|\chi_{xx}^{\mathrm{eff}}|^2} \left\{\frac{\boldsymbol{H}^{\mathsf{T}}[\phi_i]\cdot\mathbb{M}_i\mathbb{S}_{\in}\mathbb{M}^\dag_j\cdot\boldsymbol{H}[\phi_j]+\boldsymbol{H}^{\mathsf{T}}[\phi_i]\cdot\mathbb{M}^*_j\mathbb{S}_{\in}\mathbb{M}^{\mathsf{T}}_i\cdot\boldsymbol{H}[\phi_j]}{\boldsymbol{H}^{\mathsf{T}}[\phi_i]\boldsymbol{R}_i\boldsymbol{R}^\dag_j\boldsymbol{H}[\phi_j]} +\right.\nonumber\\\left. \frac{\epsilon_d^2\delta_{ij}}{\boldsymbol{H}^{\mathsf{T}}[\phi_i]\boldsymbol{R}_i\boldsymbol{R}^\dag_i\boldsymbol{H}[\phi_i]}\right\}\,,\qquad (i,j)=(1,2)\,.
\end{eqnarray}

Now, when we have considered all the stages of the quantum noise spectral densities calculation on a simple example of a single movable mirror, we are ready to consider more complicated systems. Our next target is a Fabry--P{\'e}rot cavity.

\subsection{Fabry--P{\'e}rot cavity}
\label{sec:Fabry-Perot}

The schematic view of a Fabry--P{\'e}rot cavity with two movable mirrors is drawn in Figure~\ref{fig:Fabry-Perot}. This simple scheme is important for at least two reasons: (i) it is the most common element for more sophisticated interferometer configurations, which are considered below; and (ii) the analysis of real-life high-sensitivity interferometers devoted, in particular, to detection of GWs, can be reduced to a single Fabry--P{\'e}rot cavity by virtue of the `scaling law' theorem~\cite{Buonanno2003}, see Section~\ref{sec:advligo}.

\epubtkImage{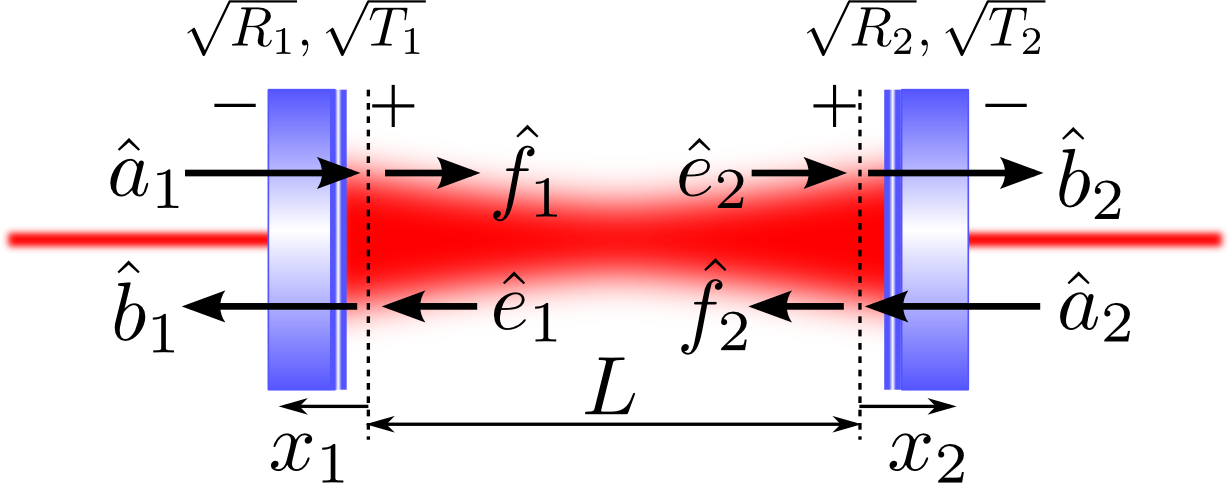}{
\begin{figure}
  \centerline{\includegraphics[width=.5\textwidth]{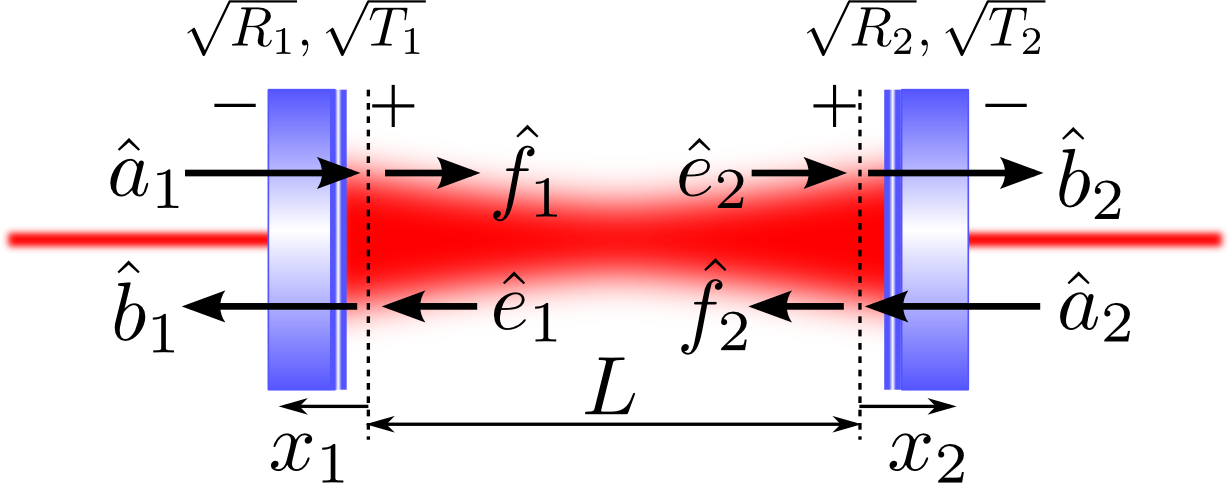}}
  \caption{Fabry--P{\'e}rot cavity}
\label{fig:Fabry-Perot}
\end{figure}}

A Fabry--P{\'e}rot cavity consists of two movable mirrors that are separated by a distance $L+x_1+x_2$, where $L=c\tau$ is the distance at rest with $\tau$ standing for a single pass light travel time, and $x_1$ and $x_2$ are the small deviations of the mirrors' position from the equilibrium. Each of the mirrors is described by the transfer matrix $\mathbb{M}_{1,2}$ with real coefficients of reflection $\sqrt{R_{1,2}}$ and transmission $\sqrt{T_{1,2}}$ according to Eq.~\eqref{eq:mirror_Mreal}.
As indicated on the scheme, the outer faces of the mirrors are assumed to have negative reflectivities. While the intermediate equations depend on this choice, the final results are invariant to it. The cavity is pumped from both sides by two laser sources with the same optical frequency $\omega_p$.

The coupling equations for the ingoing and outgoing fields at each of the mirrors are absolutely the same as in Section~\ref{sec:Mirror_motion}.  The only new thing is the free propagation of light between the mirrors that adds two more field continuity equations to the full set, describing the transformation of light in the Fabry--P{\'e}rot cavity. It is illuminating to write down input-output relations first in the time domain:
\begin{equation}
\begin{array}{ll}
\label{eq:FP_I/O_time}
  \hat{E}_{b_1}(t) = -\sqrt{R_1}\hat{E}_{a_1}(t+2x_1/c)+\sqrt{T_1}\hat{E}_{e_1}(t)\,, & \qquad \hat{E}_{b_2}(t) = -\sqrt{R_2}\hat{E}_{a_2}(t+2x_2/c)+\sqrt{T_2}\hat{E}_{e_2}(t)\,,\\
  \hat{E}_{f_1}(t) = \sqrt{R_1}\hat{E}_{e_1}(t-2x_1/c)+\sqrt{T_1}\hat{E}_{a_1}(t)\,, & \qquad \hat{E}_{f_2}(t) = \sqrt{R_2}\hat{E}_{a_2}(t-2x_2/c)+\sqrt{T_2}\hat{E}_{e_2}(t)\,,\\
  \hat{E}_{e_1}(t) = \hat{E}_{f_2}(t-L/c)\,, & \qquad \hat{E}_{e_2}(t) =
  \hat{E}_{f_1}(t-L/c)\,.
\end{array}
\end{equation}
Further we use notation $\tau=L/c$ for the light travel time between the mirrors.

The frequency domain version of the above equations and their solutions are derived in Appendix~\ref{app:FP_I/O-relations}. We write these I/O-relations given in Eqs.~\eqref{eq:FP_1} in terms of complex amplitudes instead of 2 photon amplitudes, for the expressions look much more compact in that representation. However, one can simplify them even more using the \emph{single-mode approximation}.

\paragraph*{Single-mode approximation.}

We note that:

(i) in GW detection, rather high-finesse cavities are used, which implies low transmittance coefficients for the mirrors
\begin{equation}
    T_{1,2}\ll 1 \,;
  \end{equation}
(ii) the cavities are relatively short, so their Free Spectral Range (FSR) $f_{\mathrm{FSR}}=(2\tau)^{-1}$ is much larger than the characteristic frequencies of the mirrors' motion:
\begin{equation}
   |\Omega|\tau\ll 1 \,;
\end{equation}
and (iii) the detuning of the pump frequency from one of the cavity eigenfrequencies:
\begin{equation}
  \delta = \omega_p - \frac{\pi n}{\tau} \quad (n\ \mbox{is an integer})
\end{equation}
is also small in comparison with the FSR:
\begin{equation}
    |\delta|\tau \ll 1 \,.
\end{equation}
In this case, only this mode of the cavity can be taken into account, and the cavity can be treated as a single-mode lumped resonator.
Note also that, while our intermediate equations below depend on whether $n$ is even or odd, the final results do not. Therefore, we assume for simplicity that $n$ is even.

Expanding the numerators and denominators of Eqs.~(\ref{eq:FP_0}, \ref{eq:FP_1}) into Taylor series in $\tau$ and keeping only the first non-vanishing terms, we obtain that
  \begin{eqnarray}
\label{eq:FP_0_short}
    \mathrm{B}_{1,2} &=& \mathcal{R}_{1,2}(0)\mathrm{A}_{1,2} + \mathcal{T}(0)\mathrm{A}_{2,1}\,,\nonumber\\
    \mathrm{E}_{1,2} &=& \mathrm{F}_{1,2} = \mathrm{E}
      = \frac{\sqrt{\gamma_1}\mathrm{A}_1 + \sqrt{\gamma_2}\mathrm{A}_2}{\ell(0)\sqrt{\tau}} \,,
  \end{eqnarray}
and
  \begin{eqnarray}
    \hat{\mathrm{b}}_{1,2}(\omega) &=& \mathcal{R}_{1,2}(\Omega)\hat{\mathrm{a}}_{1,2}(\omega)
      + \mathcal{T}(\Omega)\hat{\mathrm{a}}_{2,1}(\omega)
      + \frac{2\sqrt{\gamma_{1,2}}X(\Omega)}{\ell(\Omega)} \,,\label{eq:FP_1_short}\\
    \hat{\mathrm{e}}_{1,2}(\omega) &=& \hat{\mathrm{f}}_{1,2}(\omega) = \hat{\mathrm{e}}(\omega)
      = \frac{
          \sqrt{\gamma_1}\hat{\mathrm{a}}_1(\omega) + \sqrt{\gamma_2}\hat{\mathrm{a}}_2(\omega)
          + \hat{X}(\Omega)
        }{\ell(\Omega)\sqrt{\tau}} \,, \label{FP_1_short_e}
  \end{eqnarray}
where
\begin{eqnarray}
  \gamma_{1,2} = \frac{T_{1,2}}{4\tau} \,, \\
  \gamma = \gamma_1 + \gamma_2
\end{eqnarray}
is the cavity half-bandwidth,
\begin{equation}
  \ell(\Omega) = \gamma - i(\delta+\Omega) \,,
\end{equation}
\begin{equation}
\label{eq:FP_RT}
  \mathcal{R}_{1,2}(\Omega) = \frac{2\gamma_{1,2}}{\ell(\Omega)} - 1 \,, \qquad
  \mathcal{T}(\Omega) = \frac{2\sqrt{\gamma_1\gamma_2}}{\ell(\Omega)}
\end{equation}
are the cavity left and right reflectivities and its transmittance,
\begin{equation}
  \hat{X}(\Omega) = \frac{ik_p\mathrm{E}\hat{x}(\Omega)}{\sqrt{\tau}} \,,
\end{equation}
and
\begin{equation}
  \hat{x} = \hat{x}_1 + \hat{x}_2
\end{equation}
is the sum variation of the cavity length.

\subsubsection{Optical transfer matrix for a Fabry--P{\'e}rot cavity}
\label{sec:FP_cav_OTMat}

The above optical I/O-relations are obtained in terms of the complex amplitudes. In order to transform them to two-photon quadrature notations, one needs to employ the following linear transformations:
\begin{enumerate}
  \item change frequency $\omega\to\omega_p\pm\Omega$ and rewrite the relations between the input $\hat{\alpha}(\omega)$ and output operators $\hat{\beta}(\omega)$ in the form:
  \begin{eqnarray}
    \hat{\beta}(\omega) =
    f(\Omega)\hat{\alpha}(\omega)\ \to\ \hat{\beta}_+&\equiv&\hat{\beta}(\omega_p+\Omega)
    = f(\omega_p+\Omega)\hat{\alpha}(\omega_p+\Omega) \equiv
    f_+\hat{\alpha}_+ \ \mbox{and} \nonumber\\
    \hat{\beta}_-^\dag&\equiv&\hat{\beta}^\dag(\omega_p-\Omega) = f^*(\omega_p-\Omega)\hat{\alpha}^\dag(\omega_p-\Omega) \equiv f^*_-\hat{\alpha}^\dag_-\,;
  \end{eqnarray}
  where $f(\Omega)$ is an arbitrary complex-valued function of sideband frequency $\Omega$;
  \item use the definition~\eqref{eq:2photon_quads_def} to get the following relations for two-photon quadrature operators:
  \begin{equation}
\label{eq:sideband_to_2photon}
    \begin{bmatrix}
      \hat{\beta}_c(\Omega)\\
      \hat{\beta}_s(\Omega)
    \end{bmatrix} = \frac12
    \begin{bmatrix}
      (f_++f^*_-) & i(f_+-f^*_-)\\
      -i(f_+-f^*_-) & (f_++f^*_-)
    \end{bmatrix}\cdot
    \begin{bmatrix}
      \hat{\alpha}_c(\Omega)\\
      \hat{\alpha}_s(\Omega)\,
    \end{bmatrix}.
  \end{equation}
\end{enumerate}

Applying transformations~\eqref{eq:sideband_to_2photon} to Eqs.~\eqref{eq:FP_1_short}, we rewrite the I/O-relations for a Fabry--P{\'e}rot cavity in the two-photon quadratures notations:
  \begin{eqnarray}
    \hat{\boldsymbol{b}}_{1,2}(\Omega) &=& \mathbb{R}_{1,2}(\Omega)\hat{\boldsymbol{a}}_{1,2}(\Omega)
        + \mathbb{T}(\Omega)\hat{\boldsymbol{a}}_{2,1}(\Omega)
        + 2\sqrt{\gamma_{1,2}}\mathbb{L}(\Omega)\hat{\mathbf{X}}(\Omega) \,,
        \label{FP_1_quad(a)} \\
    \hat{\boldsymbol{e}}(\Omega) &=& \frac{1}{\sqrt{\tau}}\mathbb{L}(\Omega)\left[
        \sqrt{\gamma_{1,2}}\hat{\boldsymbol{a}}_{1,2}(\Omega)
        + \sqrt{\gamma_{2,1}}\hat{\boldsymbol{a}}_{2,1}(\Omega)
        + \hat{\mathbf{X}}(\Omega)
      \right] ,  \label{FP_1_quad(b)}
  \end{eqnarray}
where
\begin{eqnarray}
  \hat{\mathbf{X}}(\Omega)
    = \svector{-\mathrm{E}_s}{\mathrm{E}_c}\frac{k_p\hat{x}(\Omega)}{\sqrt{\tau}} \,, \\
  \mathrm{E}_c = \sqrt{2}\Re\mathrm{E}\,,\qquad \mathrm{E}_s = \sqrt{2}\Im\mathrm{E}\,,
\end{eqnarray}
\begin{eqnarray}
  \mathbb{L}(\Omega) = \frac{1}{\mathcal{D}(\Omega)}
    \smatrix{\gamma-i\Omega}{-\delta}{\delta}{\gamma-i\Omega} , \label{FP_bbL} \\
  \mathcal{D}(\Omega) = \ell(\Omega)\ell^*(-\Omega) = (\gamma-i\Omega)^2 + \delta^2 \,,
\end{eqnarray}
\begin{equation}
\label{FP_bbRT}
  \mathbb{R}_{1,2}(\Omega) = 2\gamma_{1,2}\mathbb{L}(\Omega) - \mathbb{I} \,, \qquad
  \mathbb{T}(\Omega) = 2\sqrt{\gamma_1\gamma_2}\mathbb{L}(\Omega) \,.
\end{equation}

Therefore, the I/O-relations in standard form read:
\begin{equation}
  \begin{bmatrix}
    \hat{\boldsymbol{b}}_1(\Omega)\\
    \hat{\boldsymbol{b}}_2(\Omega)
  \end{bmatrix}=
  \mathbb{M}^{(0)}_{\mathrm{FP}}\cdot
  \begin{bmatrix}
    \hat{\boldsymbol{a}}_1(\Omega)\\
    \hat{\boldsymbol{a}}_2(\Omega)
  \end{bmatrix}+
  \begin{bmatrix}
    \boldsymbol{R}^{\mathrm{FP}}_1(\Omega)\\
    \boldsymbol{R}^{\mathrm{FP}}_2(\Omega)
  \end{bmatrix}\hat{x}(\Omega)
\end{equation}
 with optical transfer matrix defined as:
\begin{equation}
\label{eq:FP_OTMat}
  \mathbb{M}^{(0)}_{\mathrm{FP}}(\Omega) =
  \begin{bmatrix}
    \mathbb{R}_1(\Omega) & \mathbb{T}(\Omega)\\
    \mathbb{T}(\Omega) & \mathbb{R}_2(\Omega)
  \end{bmatrix}
\end{equation}
and the response to the cavity elongation $\hat{x}(\Omega)$ defined as:
\begin{equation}
    \boldsymbol{R}^{\mathrm{FP}}_1(\Omega)= 2k_p\sqrt{\frac{\gamma_1}{\tau}}\mathbb{L}(\Omega)\cdot
    \begin{bmatrix}
      -\mathrm{E}_s\\
      \mathrm{E}_c
    \end{bmatrix}\quad\mbox{and}\quad
\boldsymbol{R}^{\mathrm{FP}}_2(\Omega)=2k_p\sqrt{\frac{\gamma_2}{\tau}}\mathbb{L}(\Omega)\cdot
    \begin{bmatrix}
      -\mathrm{E}_s\\
      \mathrm{E}_c
    \end{bmatrix}\,.
\end{equation}

Note that due to the fact that $(\mathbb{M}^{(0)}_{\mathrm{FP}})^\dag = (\mathbb{M}^{(0)}_{\mathrm{FP}})^{-1}$ the reflectivity and the transmission matrices $\mathbb{R}_{1,2}$ and $\mathbb{T}$ satisfy the following unitarity relations:
\begin{equation}
\label{eq:FP_bbRT_u}
  \mathbb{R}_1\mathbb{R}_1^\dagger + \mathbb{T}\mathbb{T}^\dagger
    = \mathbb{R}_2\mathbb{R}_2^\dagger + \mathbb{T}\mathbb{T}^\dagger
    = \mathbb{I} \,, \qquad
  \mathbb{R}_1\mathbb{T}^\dagger + \mathbb{T}\mathbb{R}_2^\dagger = 0 \,.
\end{equation}

\subsubsection{Mirror dynamics, radiation pressure forces and ponderomotive rigidity}

The mechanical equations of motion of the Fabry--P{\'e}rot cavity mirrors, in spectral representation, are the following:
\begin{equation}
\label{eq:x_12}\mathrm{f}
  \chi_{xx,i}^{-1}(\Omega)\hat{x}_{i}(\Omega)
    = \hat{F}_{i}(\Omega) + G_i(\Omega)\qquad i=1,2\,,
\end{equation}
where $\chi_{xx,i}$ are the mechanical susceptibilities of the mirrors, $G_{i}$ stand for any external classical forces that could act on the mirrors (for example, a signal force to be detected),
\begin{equation}
  \hat{F}_{i} = \frac{
    \hat{\mathcal{I}}_{\mathrm{e}\,i} + \hat{\mathcal{I}}_{\mathrm{f}\,i}
    - \hat{\mathcal{I}}_{\mathrm{a}\,i} - \hat{\mathcal{I}}_{\mathrm{b}\,i}
  }{c}
\end{equation}
are the radiation pressure forces acting on the mirrors, and $\hat{\mathcal{I}}_{\mathrm{e}\,i}$, $\hat{\mathcal{I}}_{\mathrm{f}\,i}$, $\hat{\mathcal{I}}_{\mathrm{a}\,i}$, $\hat{\mathcal{I}}_{\mathrm{b}\,i}$ are the powers of the waves $\mathrm{e}_{i}$, $\mathrm{f}_{i}$, etc. The signs for all forces are chosen in such a way that the positive forces are oriented outwards from the cavity, increasing the corresponding mirror displacement $x_{1,2}$.

In the spectral representation, using the quadrature amplitudes notation, the radiation pressure forces read:
\begin{eqnarray}
  \hat{F}_{1,2}(\Omega) &=& \frac{\hbar k_p}{2}\left(
      \mathbf{E}_{1,2}^{\mathsf{T}}\mathbf{E}_{1,2} + \mathbf{F}_{1,2}^{\mathsf{T}}\mathbf{F}_{1,2}
      - \mathbf{A}_{1,2}^{\mathsf{T}}\mathbf{A}_{1,2} - \mathbf{B}_{1,2}^{\mathsf{T}}\mathbf{B}_{1,2}
    \right) \nonumber\\
    && + \hbar k_p\left[
        \mathbf{E}_{1,2}^{\mathsf{T}}\hat{\boldsymbol{e}}_{1,2}(\Omega)
        + \mathbf{F}_{1,2}^{\mathsf{T}}\hat{\boldsymbol{f}}_{1,2}(\Omega)
        - \mathbf{A}_{1,2}^{\mathsf{T}}\hat{\boldsymbol{a}}_{1,2}(\Omega)
        - \mathbf{B}_{1,2}^{\mathsf{T}}\hat{\boldsymbol{b}}_{1,2}(\Omega)
      \right] .
\end{eqnarray}
The first group, as we have already seen, describes the regular constant force; therefore, we omit it henceforth.

In the single-mode approximation, the radiation pressure forces acting on both mirrors are equal to each other:
\begin{equation}
\label{eq:FP_F_RP}
  \hat{F}_{1,2}(\Omega) \equiv \hat{F}_{\mathrm{b.a.}}(\Omega)
    = 2\hbar k_p\mathbf{E}^{\mathsf{T}}\hat{\boldsymbol{e}}(\Omega) \,,
\end{equation}
and the optical field in the cavity is sensitive only to the elongation mechanical mode described by the coordinate $x$. Therefore, combining Eqs.~\eqref{eq:x_12}, we obtain for this mode:
\begin{equation}
\label{eq_x_raw}
  \chi_{xx}^{-1}(\Omega)\hat{x}(\Omega) = \hat{F}_{\mathrm{b.a.}}(\Omega) + G(\Omega) \,,
\end{equation}
where
\begin{equation}
  \chi_{xx}(\Omega) = \left[\chi_{xx,1}(\Omega)+\chi_{xx,2}(\Omega)\right]
\end{equation}
is the reduced mechanical susceptibility and
\begin{equation}
  G(\Omega) = \frac{
      \chi_{xx,1}(\Omega)G_1(\Omega) + \chi_{xx,2}(\Omega)G_2(\Omega)
    }{\chi_{xx}(\Omega)}
\end{equation}
is the effective external force.

In the simplest and at the same time the most important particular case of free mirrors:
\begin{equation}
  \chi_{xx,i}(\Omega) = -\frac{1}{m_{i}\Omega^2}\qquad i=1,2\,,
\end{equation}
the reduced mechanical susceptibility and the effective external force are equal to
\begin{equation}
  \chi_{xx}(\Omega) = -\frac{1}{\mu\Omega^2} \,,
\end{equation}
and
\begin{equation}
  G(\Omega)
    = \mu\left[\frac{G_1(\Omega)}{m_1} + \frac{G_2(\Omega)}{m_2}\right] ,
\end{equation}
where
\begin{equation}
  \mu = \left(\frac{1}{m_1}+\frac{1}{m_2}\right)^{-1} 
\end{equation}
is the effective mass of the elongation mechanical mode.

It follows from Eqs.~(\ref{FP_1_quad(b)}) and~(\ref{eq:FP_F_RP}) that the radiation pressure force can be written as a sum of the random and dynamical back-action terms, similarly to the single mirror case:
\begin{equation}
\label{eq:FP_F_fl_K}
  \hat{F}_{\mathrm{b.a.}}(\Omega) = \hat{F}^{(0)}_{\mathrm{b.a.}}(\Omega) - K(\Omega)\hat{x}(\Omega) \,,
\end{equation}
with the random component equal to
\begin{equation}
\label{eq:FP_F_fl}
  \hat{F}_{\mathrm{b.a.}}^{(0)}(\Omega) = \frac{2\hbar k_p\mathbf{E}^{\mathsf{T}}}{\sqrt{\tau}}
    \mathbb{L}(\Omega)\left[
      \sqrt{\gamma_1}\hat{\boldsymbol{a}}_1(\Omega) + \sqrt{\gamma_2}\hat{\boldsymbol{a}}_2(\Omega)
    \right]
\end{equation}
and the ponderomotive rigidity that reads
\begin{equation}
\label{eq:FP_K}
  K(\Omega) = \frac{MJ\delta}{\mathcal{D}(\Omega)}\,.
\end{equation}

We introduced here the normalized optical power
\begin{equation}
\label{eq:FP_J}
  J = \frac{4\hbar k_p^2|E|^2}{M\tau} = \frac{4\omega_p\mathcal{I}_c}{McL}
\end{equation}
with
\begin{equation}
  \mathcal{I}_c = \hbar\omega_p|E|^2
\end{equation}
standing for the mean optical power circulating inside the cavity, and $M$ is some (in general, arbitrary) mass. Typically, it is convenient to make it equal to the reduced mass $\mu$.

Substitution of the force~\eqref{eq:FP_F_fl_K} into Eq.~\eqref{eq_x_raw} gives the following final equation of motion:
\begin{equation}
\label{eq:FP_mech_eq}
  [\chi_{xx}^{-1}(\Omega)+K(\Omega)]\hat{x}(\Omega)
  = \hat{F}_{\mathrm{b.a.}}^{(0)}(\Omega) + G(\Omega)  \,.
\end{equation}
Thus, the effective mechanical susceptibility $\chi_{xx}^{\mathrm{eff,\,FP}}(\Omega)$ for a Fabry--P{\'e}rot cavity reads:
\begin{equation}
\label{eq:chi_xx_eff_FP}
  \chi_{xx}^{\mathrm{eff,\,FP}}(\Omega) = \bigl(\chi_{xx}^{-1}(\Omega)+K(\Omega)\bigr)^{-1} = \frac{1}{K(\Omega)-\mu\Omega^2}\,.
\end{equation}

\subsection{Fabry--P{\'e}rot--Michelson interferometer}
\label{sec:advligo}


In real-life high-precision experiments with mechanical test objects,
interferometer schemes that are much more sophisticated than the
simple Fabry--P{\'e}rot cavity are used. In particular, the best
sensitivity in mechanical displacement measurements is achieved by the
laser GW detectors. The typical scheme of such a
detector is shown in Figure~\ref{fig:advligo}. It is this scheme
which is planned to be used for the next generation Advanced
LIGO~\cite{Thorne2000, Fritschel2002, Smith2009}, Advanced
VIRGO~\cite{Acernese2006-2},  and LCGT~\cite{LCGTsite} GW detectors, and its simplified versions are (or were) in use in
the contemporary first generation detectors: Initial
LIGO~\cite{Abramovici1992, LIGOsite}, VIRGO~\cite{VIRGOsite},
GEO\,600~\cite{Willke2002, GEOsite}, and TAMA~\cite{Ando2001,
  TAMAsite}.

\epubtkImage{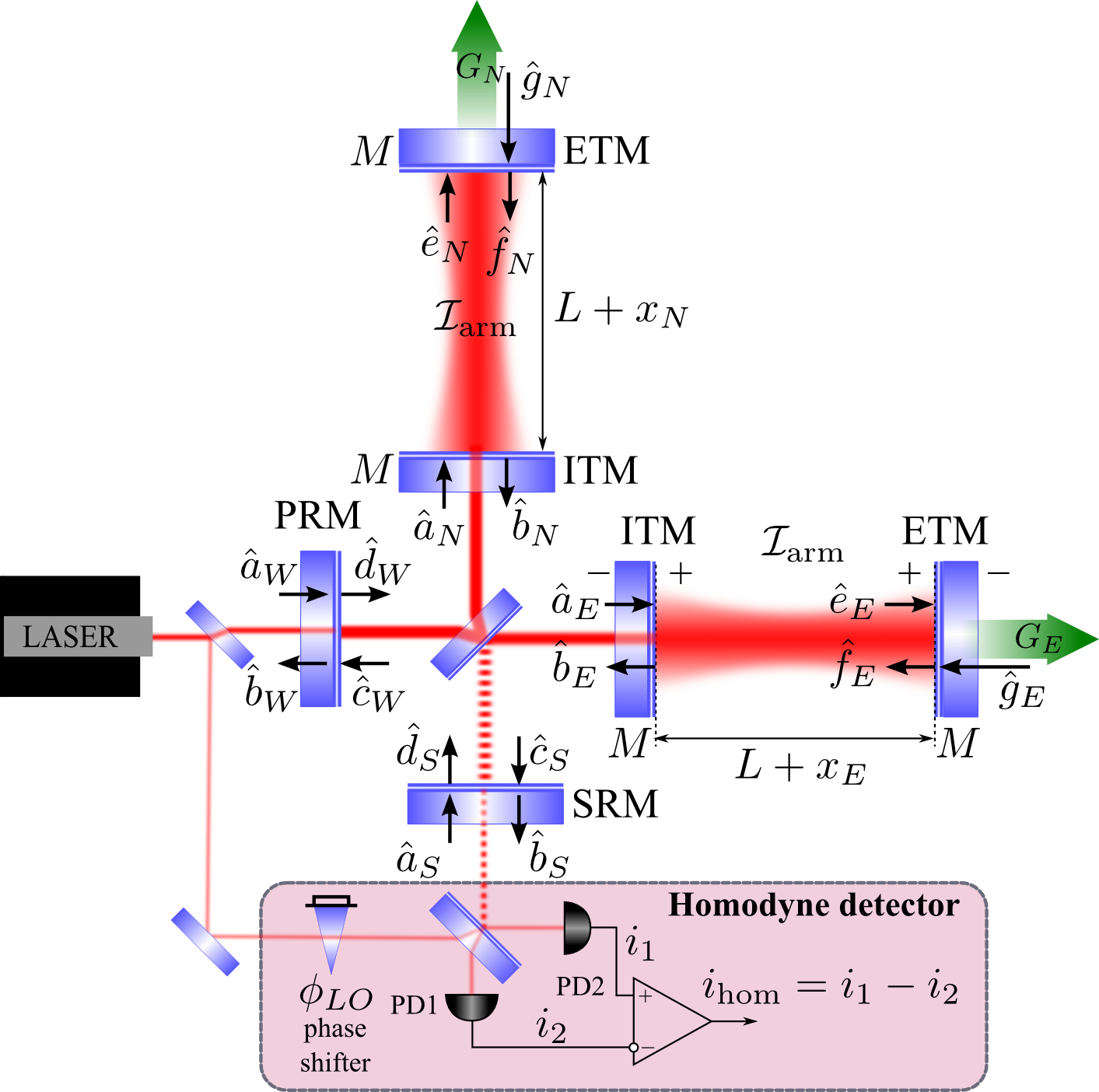}{
\begin{figure}[htbp]
  \centerline{\includegraphics[width=.8\textwidth]{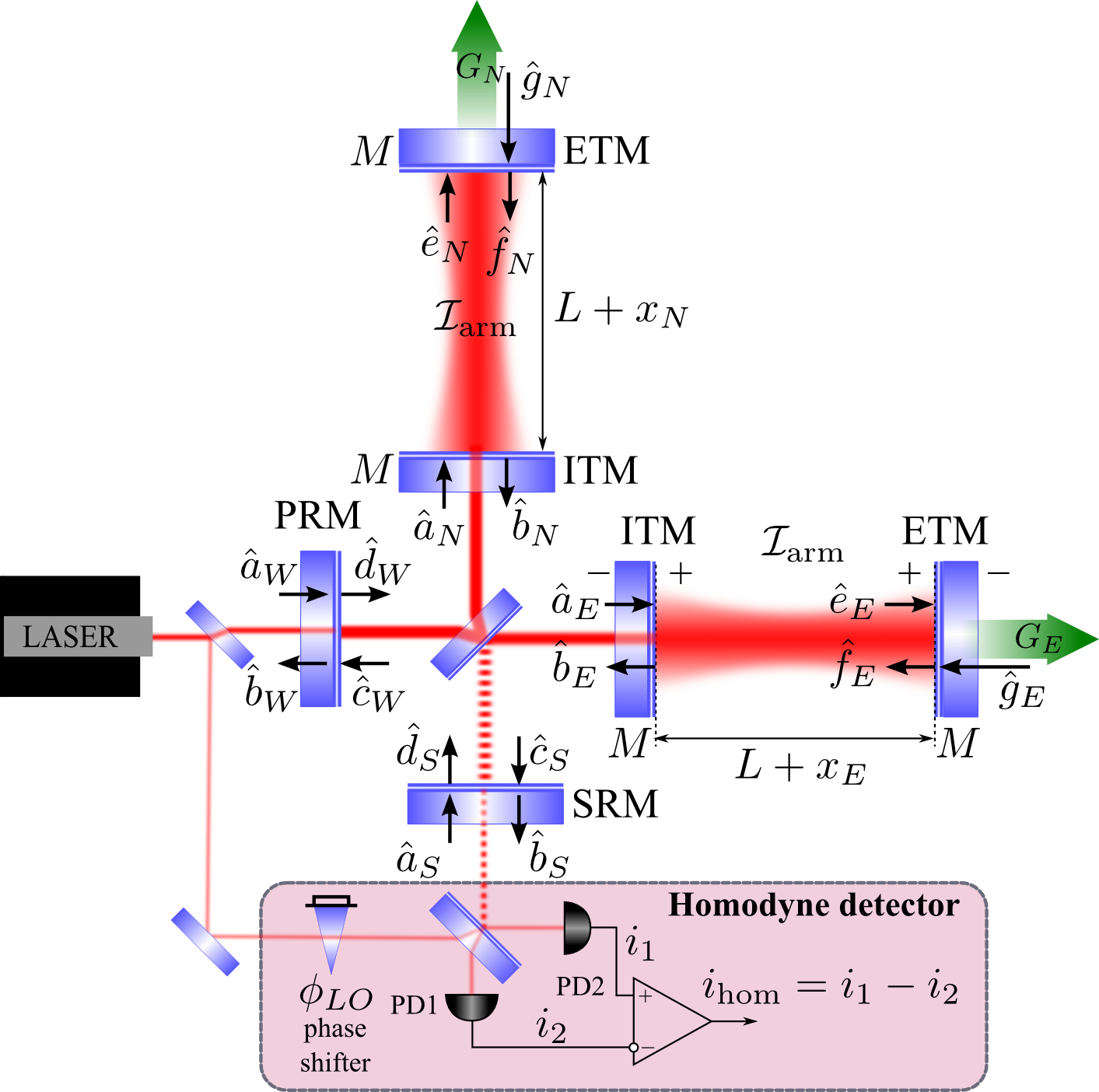}}
  \caption{Power- and signal-recycled Fabry--P{\'e}rot--Michelson interferometer.}
\label{fig:advligo}
\end{figure}}

This scheme works similar to the ordinary Michelson interferometer considered briefly in Section~\ref{sec:MI}. The beamsplitter \textsf{BS} distributes the pump power from the laser evenly between the arms. The beams, reflected off the Fabry--P{\'e}rot cavities are recombined on the beamsplitter in such a way that, in the ideal case of perfect symmetry of the arms, all the light goes back to the laser, i.e., keeping the signal (`south') port dark.  Any imbalance of the interferometer arms, caused by signal forces acting on the end test masses (ETMs) makes part of the pumping light leak into the dark port where it is monitored by a photodetector.

The Fabry--P{\'e}rot cavities in the arms, formed by the input test masses (ITMs) and the end test masses, provide the increase of the optomechanical coupling, thus making photons bounce many times in the cavity and therefore carry away a proportionally-amplified mirror displacement signal in their phase (cf.\ with the $\digamma$ factor in the toy systems considered in Section~\ref{sec:linear_quantum_measurement}). The two auxiliary \emph{recycling} mirrors: the PRM and the signal recycling (SRM) allow one to increase the power, circulating inside the Fabry--P{\'e}rot cavities, for a given laser power, and provide the means for fine-tuning of the quantum noise spectral density~\cite{PhysRevD.38.2317_1988_Meers, PhysRevD.38.433_1988_Vinet}, respectively.

It was shown in~\cite{Buonanno2003} that quantum noise of this dual (power and signal) recycled interferometer is equivalent to that of a single Fabry--P{\'e}rot cavity with some effective parameters (the analysis in that paper was based on earlier works~\cite{Mizuno1995, 00pth1Ra}, where the classical regime had been considered). Here we reproduce this \emph{scaling law} theorem, extending it in two aspects: (i) we factor in optical losses in the arm cavities by virtue of modeling it by the finite transmissivity of the ETMs, and (ii) we do not assume the arm cavities tuned in resonance (the detuned arm cavities could be used, in particular, to create optical rigidity in non-signal-recycled configurations).

\subsubsection{Optical I/O-relations}

We start with Eqs.~(\eqref{eq:FP_0_short}) and~(\eqref{eq:FP_1_short}) for the arm cavities. The notation for the field amplitudes is shown in Figure~\ref{fig:advligo}. The fields referring to the interferometer arms are marked with the subscripts $N$ (`northern') and $E$ (`eastern') following the convention of labeling the GW interferometer parts in accordance with the cardinal directions they are located at with respect to the drawing (up-direction coincides with north). In order to avoid subscripts, we rename some of the field amplitudes as follows:
\begin{equation}
  \mathrm{a}_{1\,N,E} \to \mathrm{a}_{N,E} \,, \qquad \mathrm{b}_{1\,N,E} \to \mathrm{b}_{N,E} \,, \qquad \mathrm{a}_{2\,N,E} \to \mathrm{g}_{N,E} \,;
\end{equation}
see also Figure~\ref{fig:advligo}. Note that the fields $\mathrm{g}_{N,E}$ now describe the noise sources due to optical losses in the arm cavities.

Rewrite those Eqs.~(\ref{eq:FP_0_short}), (\ref{eq:FP_1_short}), (\ref{FP_1_short_e}) that are relevant to our consideration, in these notations:
  \begin{eqnarray}
\label{eq:sl_cav_0}
    \mathrm{B}_{N,E} &=& \mathcal{R}_{\mathrm{arm}}(0)\mathrm{A}_{N,E}  \,,\nonumber \\
    \mathrm{E}_{N,E} &=& \frac{1}{\ell_{\mathrm{arm}}(0)}\sqrt{\frac{\gamma_{1\mathrm{arm}}}{\tau}}
      \mathrm{A}_{N,E} \,,
  \end{eqnarray}
  \begin{eqnarray}
\label{eq:sl_cav_1}
    \hat{\mathrm{b}}_{N,E}(\omega) &=& \mathcal{R}_{\mathrm{arm}}(\Omega)\hat{\mathrm{a}}_{N,E}(\omega)
      + \mathcal{T}_{\mathrm{arm}}(\Omega)\hat{\mathrm{g}}_{N,E}(\omega)
      + \frac{2\sqrt{\gamma_{1\mathrm{arm}}}\hat{X}_{N,E}(\Omega)}{\ell_{\mathrm{arm}}(\Omega)} \,,\nonumber \\
    \hat{\mathrm{e}}_{N,E}(\omega) &=& \frac{
        \sqrt{\gamma_{1\mathrm{arm}}}\hat{\mathrm{a}}_{N,E}(\omega)
        + \sqrt{\gamma_2}\hat{\mathrm{g}}_{N,E}(\omega)
        + \hat{X}_{N,E}(\Omega)
      }{\ell_{\mathrm{arm}}(\Omega)\sqrt{\tau}} \,,
  \end{eqnarray}
where
\begin{equation}
\label{sl_gammas}
  \gamma_{1\mathrm{arm}} = \frac{T_{\mathrm{arm}}}{4\tau} \,, \qquad
  \gamma_2 = \frac{A_{\mathrm{arm}}}{4\tau} \,,
\end{equation}
$T_{\mathrm{arm}}$ is the input mirrors power transmittance, $A_{\mathrm{arm}}$ is the arm cavities power losses per bounce,
\begin{equation}
  \gamma_{\mathrm{arm}} = \gamma_{1\mathrm{arm}} + \gamma_2
\end{equation}
is the arm cavities half-bandwidth,
\begin{equation}
  \ell_{\mathrm{arm}}(\Omega) = \gamma_{\mathrm{arm}} - i(\delta_{\mathrm{arm}}+\Omega) \,,
\end{equation}
$\delta_{\mathrm{arm}}$ is the arm cavities detuning,
\begin{equation}
\label{eq:FP_rt}
  \mathcal{R}_{\mathrm{arm}}(\Omega)
    = \frac{2\gamma_{1\mathrm{arm}}}{\ell_{\mathrm{arm}}(\Omega)} - 1 \,, \qquad
  \mathcal{T}_{\mathrm{arm}}(\Omega)
    = \frac{2\sqrt{\gamma_{1\mathrm{arm}}\gamma_2}}{\ell_{\mathrm{arm}}(\Omega)}
\end{equation}
and
\begin{equation}
  \hat{X}_{N,E} = \frac{ik_p\mathrm{E}_{N,E}\hat{x}_{N,E}(\Omega)}{\sqrt{\tau}} \,.
\end{equation}

Assume then that the beamsplitter is described by the matrix~\eqref{eq:IO_beamsplitter_matrix}, with $R=T=1/2$. Let $l_W=c\tau_W$ be the power recycling cavity length (the optical distance between the power recycling mirror and the input test masses) and $l_S=c\tau_S$ -- power recycling cavity length (the optical distance between the signal recycling mirror and the input test masses). In this case, the light propagation between the recycling mirrors and the input test masses is described by the following equations for the classical field amplitudes:
  \begin{eqnarray}
\label{eq:sl_bs_0}
    \mathrm{A}_{N,E} &=& \frac{\mathrm{D}_We^{i\phi_W} \pm \mathrm{D}_Se^{i\phi_S}}{\sqrt{2}} \,,\nonumber \\
    \mathrm{C}_{W,S} &=& \frac{\mathrm{B}_N \pm \mathrm{B}_E}{\sqrt{2}}e^{i\phi_{W,S}} \,,
  \end{eqnarray}
where
\begin{equation}
  \phi_{W,S} = \omega_p\tau_{W,S} \,,
\end{equation}
are the phase shifts gained by the carrier light with frequency $\omega_p$ passing through the power and signal recycling cavities, and the similar equations:
\begin{eqnarray}\label{eq:sl_bs_1}
    \hat{\mathrm{a}}_{N,E}(\omega) &=& \frac{
        \hat{\mathrm{d}}_W(\omega)e^{i\omega\tau_W} \pm \hat{\mathrm{d}}_S(\omega)e^{i\omega\tau_S}
      }{\sqrt{2}} \,, \nonumber\\
    \hat{\mathrm{c}}_{W,S}(\omega)
      &=& \frac{\hat{\mathrm{b}}_N(\omega) \pm \hat{\mathrm{b}}_E(\omega)}{\sqrt{2}}e^{i\omega\tau_{W,S}}
  \end{eqnarray}
appling to the quantum fields' amplitudes.

The last group of equations that completes our equations set is for the coupling of the light fields at the recycling mirrors:
\begin{equation}
\begin{array}{ll}
\label{eq:sl_rec}
    \hat{\mathrm{b}}_W = -\sqrt{R_W}\hat{\mathrm{a}}_W + \sqrt{T_{\mathrm{W}}}\hat{\mathrm{c}}_W \,, & \qquad
    \hat{\mathrm{d}}_W = \sqrt{T_W}\hat{\mathrm{a}}_W + \sqrt{R_{\mathrm{W}}}\hat{\mathrm{c}}_W \,, \\
    \hat{\mathrm{b}}_S = -\sqrt{R_S}\hat{\mathrm{a}}_S + \sqrt{T_{\mathrm{S}}}\hat{\mathrm{c}}_S \,, & \qquad
    \hat{\mathrm{d}}_S = \sqrt{T_S}\hat{\mathrm{a}}_S + \sqrt{R_{\mathrm{W}}}\hat{\mathrm{c}}_S \,,
\end{array}
\end{equation}
where $R_W$, $T_W$ and $R_S$, $T_S$ are the reflectivities and transmissivities of the power and signal recycling mirrors, respectively. These equations, being linear and frequency independent, are valid both for the zeroth-order classical amplitudes and for the first-order quantum ones.

\subsubsection{Common and differential optical modes}

The striking symmetry of the above equations suggests that the convenient way to describe this system is to decompose all the optical fields in the interferometer arms into the superposition of the symmetric (common) and antisymmetric (differential) modes, which we shall denote by the subscripts $+$ and $-$, respectively:
\begin{equation}
\label{eq:ne2pm}
  \hat{\mathrm{a}}_\pm = \frac{\hat{\mathrm{a}}_N\pm\hat{\mathrm{a}}_E}{\sqrt{2}} \,, \qquad
  \hat{\mathrm{b}}_\pm = \frac{\hat{\mathrm{b}}_N\pm\hat{\mathrm{b}}_E}{\sqrt{2}} \,, \qquad
  \hat{\mathrm{e}}_\pm = \frac{\hat{\mathrm{e}}_N\pm\hat{\mathrm{e}}_E}{\sqrt{2}} \,, \qquad
  \hat{\mathrm{g}}_\pm = \frac{\hat{\mathrm{g}}_N\pm\hat{\mathrm{g}}_E}{\sqrt{2}} \,.
\end{equation}
It follows from Eqs.~\eqref{eq:sl_bs_1} that the symmetric mode is coupled solely to the `western' (bright) port, while the antisymmetric one couples exclusively to the `southern' (dark) port of the interferometer.

It is easy to see that the classical field amplitudes of the antisymmetric mode are equal to zero. For the common mode, combining Eqs.~(\ref{eq:sl_cav_0}), (\ref{eq:sl_bs_0}), (\ref{eq:sl_rec}), (\ref{eq:ne2pm}), it is easy to obtain the following set of equation:
  \begin{eqnarray}
\label{eq:sl_pm_eq_0}
    \mathrm{B}_+ &=& \mathcal{R}_{\mathrm{arm}}(0)\mathrm{A}_+ \,,\nonumber \\
    \mathrm{E}_+ &=& \frac{1}{\ell_{\mathrm{arm}}(0)}\sqrt{\frac{\gamma_{1\mathrm{arm}}}{\tau}}\mathrm{A}_+ \,,\nonumber \\
    \mathrm{A}_+ &=& \mathrm{D}_We^{i\phi_W} \,,\nonumber \\
    \mathrm{C}_W &=& \mathrm{B}_+e^{i\phi_W} \,, \nonumber\\
    \mathrm{B}_W &=& -\sqrt{R_W}\mathrm{A}_W + \sqrt{T_W}\mathrm{C}_W \,, \nonumber\\
    \mathrm{D}_W &=& \sqrt{T_W}\mathrm{A}_W + \sqrt{R_W}\mathrm{C}_W \,.
  \end{eqnarray}
Its solution is equal to (only those amplitudes are provided that we shall need below):
\begin{subequations}
\label{sl_pm_raw_0}
  \begin{eqnarray}
    \mathrm{B}_W &=& \frac{\mathcal{R}_{\mathrm{arm}}(0)e^{2i\phi_W} - \sqrt{R_W}}
      {1 - \mathcal{R}_{\mathrm{arm}}(0)\sqrt{R_W}e^{2i\phi_W}}\,\mathrm{A}_W  \,, \\
    \mathrm{E}_+ &=& \frac{1}{\ell_{\mathrm{arm}}(0)}\sqrt{\frac{\gamma_{1\mathrm{arm}}}{\tau}}
      \frac{\sqrt{T_W}e^{i\phi_W}}{1 - \mathcal{R}_{\mathrm{arm}}(0)\sqrt{R_W}e^{2i\phi_W}}\,\mathrm{A}_W \,.
  \end{eqnarray}
\end{subequations}

In the differential mode, the first non-vanishing terms are the first-order quantum-field amplitudes. In this case, using Eqs.~(\ref{eq:sl_cav_1}), (\ref{eq:sl_bs_1}), (\ref{eq:sl_rec}), (\ref{eq:ne2pm}), and taking into account that
\begin{equation}
\label{E_plus}
  \mathrm{E}_N = \mathrm{E}_E = \frac{\mathrm{E}_+}{\sqrt{2}} \,,
\end{equation}
we obtain:
  \begin{eqnarray}
\label{eq:sl_pm_eq_1}
    \hat{\mathrm{b}}_-(\omega) &=& \mathcal{R}_{1\mathrm{arm}}(\Omega)\hat{\mathrm{a}}_-(\omega)
      + \mathcal{T}_{\mathrm{arm}}(\Omega)\hat{\mathrm{g}}_-(\omega)
      + \frac{2\sqrt{\gamma_{1\mathrm{arm}}}\hat{X}_-(\Omega)}{\ell_{\mathrm{arm}}(\Omega)} \,, \nonumber\\
    \hat{\mathrm{e}}_-(\omega) &=& \frac{
        \sqrt{\gamma_{1\mathrm{arm}}}\hat{\mathrm{a}}_-(\omega)
        + \sqrt{\gamma_2}\hat{\mathrm{g}}_-(\omega)
        + \hat{X}_-(\Omega)
      }{\ell_{\mathrm{arm}}(\Omega)\sqrt{\tau}} \,, \\
    \hat{\mathrm{a}}_-(\omega) &=& \hat{\mathrm{d}}_S(\omega)e^{i\omega\tau_S} \,, \nonumber\\
    \hat{\mathrm{c}}_S(\omega) &=& \hat{\mathrm{b}}_-(\omega)e^{i\omega\tau_S} \,, \nonumber\\
    \hat{\mathrm{b}}_S(\omega) &=& -\sqrt{R_S}\hat{\mathrm{a}}_S + \sqrt{T_S}\hat{\mathrm{c}}_S(\omega) \,, \nonumber\\
    \hat{\mathrm{d}}_S(\omega) &=& \sqrt{T_S}\hat{\mathrm{a}}_S + \sqrt{R_S}\hat{\mathrm{c}}_S(\omega) \,,
  \end{eqnarray}
where
\begin{eqnarray}
  \hat{X}_-(\Omega) &=& \frac{ik_p\mathrm{E}_+\hat{x}_-(\Omega)}{\sqrt{\tau}} \,, \\
  \hat{x}_- &=& \frac{\hat{x}_N - \hat{x}_E}{2} \,. \label{eq:x_minus}
\end{eqnarray}
The solution of this equation set is the following:
\begin{eqnarray}
\label{eq:sl_pm_raw_1}
    \hat{\mathrm{b}}_S(\omega) &=&
  \frac{
        [\mathcal{R}_{\mathrm{arm}}(\Omega)e^{2i\omega\tau_S} - \sqrt{R_S}]\hat{\mathrm{a}}_S(\Omega)
        + \sqrt{T_S}e^{i\omega\tau_S}\biggl[
            \mathcal{T}_{\mathrm{arm}}(\Omega)\hat{\mathrm{g}}_-(\omega)
            + \dfrac{2\sqrt{\gamma_{1\mathrm{arm}}}\hat{X}_-(\Omega)}{\ell_{\mathrm{arm}}(\Omega)}
          \biggr]
      }{1 - \mathcal{R}_{\mathrm{arm}}(\Omega)\sqrt{R_S}e^{2i\omega\tau_S}} \,,\nonumber \\
    \hat{\mathrm{e}}_-(\omega) &=& \frac{
        \sqrt{T_S}e^{i\omega\tau_S}\sqrt{\gamma_{1\mathrm{arm}}}\hat{\mathrm{a}}_S(\Omega)
        + [1+\sqrt{R_S}e^{2i\omega\tau_S}]
            [\sqrt{\gamma_2}\hat{\mathrm{g}}_-(\omega) +
              \hat{X}_-(\Omega)]}
        {\ell_{\mathrm{arm}}(\Omega)\sqrt{\tau}[1 - \mathcal{R}_{\mathrm{arm}}(\Omega)\sqrt{R_S}e^{2i\omega\tau_S}]}
      \,.
  \end{eqnarray}

Eqs.~\eqref{eq:sl_pm_eq_0} and \eqref{sl_pm_raw_0}, on the one hand,
and \eqref{eq:sl_pm_eq_1} and \eqref{eq:sl_pm_raw_1}, on the other,
describe two almost independent optical configurations each consisting
of the two coupled Fabry--P{\'e}rot cavities as featured in
Figure~\ref{fig:pm_modes_1}. `Almost independent' means that they do
not couple in an ordinary linear way (and, thus, indeed represent
two optical modes). However, any variation of the differential
mechanical coordinate $x_-$ makes part of the pumping carrier energy
stored in the common mode pour into the differential mode, which
means a non-linear parametric coupling between these modes.

\epubtkImage{fig34.png}{
\begin{figure}
  \centerline{\includegraphics[width=.6\textwidth]{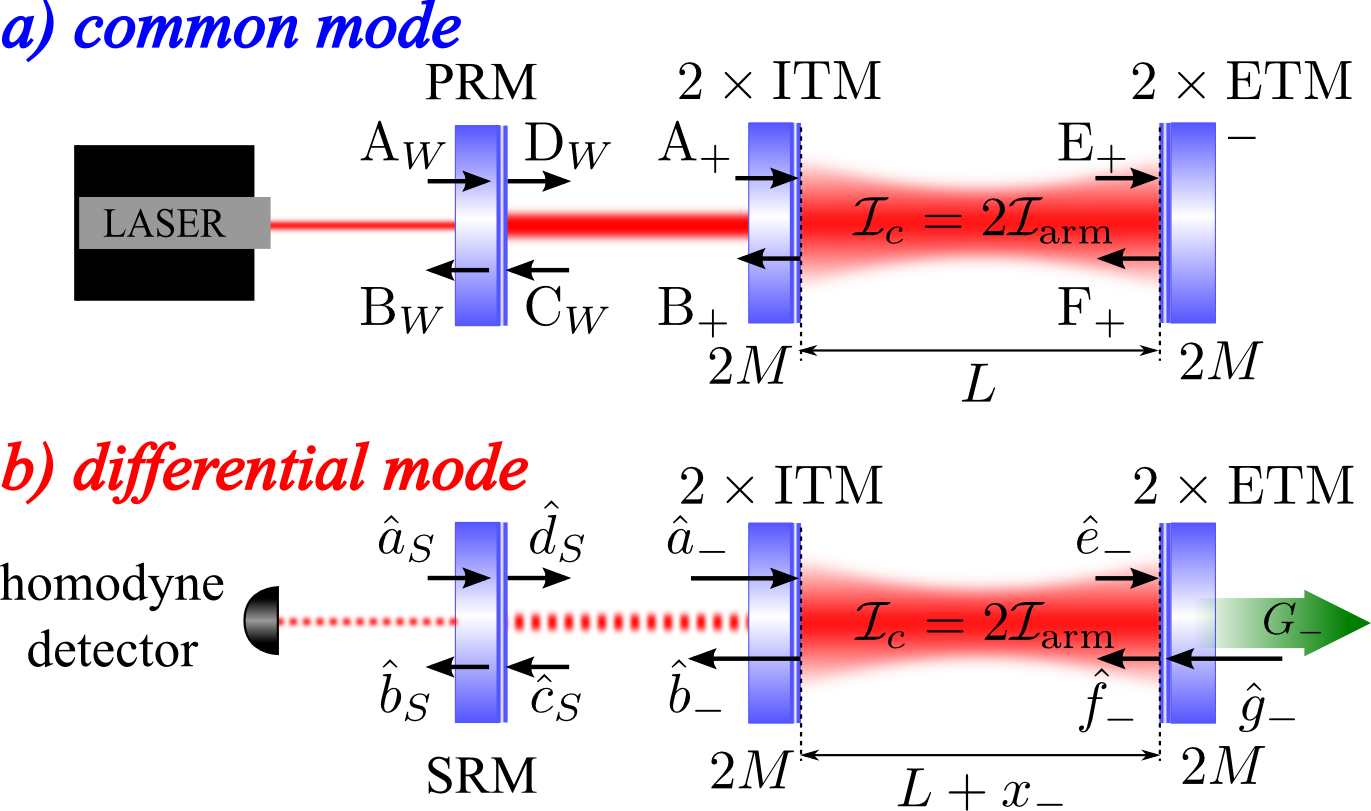}}
  \caption{Effective model of the dual-recycled Fabry--P{\'e}rot--Michelson  interferometer, consisting of the common (a) and the differential (b) modes, coupled only through the mirrors displacements.}
\label{fig:pm_modes_1}
\end{figure}}

\subsubsection{Interferometer dynamics: mechanical equations of motion, radiation pressure forces and ponderomotive rigidity}

The mechanical elongation modes of the two Fabry--P{\'e}rot cavities are described by the following equations of motion [see Eq.~\eqref{eq_x_raw}]:
\begin{equation}
\label{eq:x_NE_eq}
  -\mu\Omega^2\hat{x}_{N,E}(\Omega)
    = 2\hbar k_p\mathbf{E}_{N,E}^{\mathsf{T}}\hat{\boldsymbol{e}}_{N,E}(\Omega)
    + \frac{G_{N,E}(\Omega)}{2} \,,
\end{equation}
where $\mu=M/2$ is the effective mass of these modes and $G_{N,E}$ are the external classical forces acting on the cavities end mirrors. The differential mechanical mode  equation of motion~\eqref{eq:x_minus} taking into account Eq.~\eqref{E_plus} reads:
\begin{equation}
\label{eq:x_minus_eq_raw}
  -2\mu\Omega^2\hat{x}_-(\Omega) = \hat{F}_-^{\mathrm{r.p.}}(\Omega) + \frac{G_-(\Omega)}{2}
    \,,
\end{equation}
where
\begin{equation}
\label{F_minus_RP}
  F_-^{\mathrm{r.p.}}(\Omega) = 2\hbar k_p\mathbf{E}_+^{\mathsf{T}}\hat{\boldsymbol{e}}_-(\Omega)
\end{equation}
is the differential radiation-pressure force and
\begin{equation}
  G_- = G_N-G_E
\end{equation}
is the differential external force.

Equations~\eqref{eq:sl_pm_raw_1} and \eqref{eq:x_minus_eq_raw} together form a complete set of equations describing the differential optomechanical mode of the interferometer featured in Figure~\ref{fig:pm_modes_1}(b). Eq.~\eqref{eq:x_minus_eq_raw} implies that the effective mass of the differential mechanical degree of freedom coincides with the single mirror mass:
\begin{equation}
\label{sl_M_eff}
  2\mu=M,
\end{equation}
which prescribes the mirrors of the effective cavity to be twice as heavy as the real mirrors, i.e., $2M$. For the same reason Eq.~\eqref{E_plus} implies for the effective optical power a value twice as high as the power of light, circulating in the arm cavities:
\begin{equation}
  \mathcal{I}_c = \hbar\omega_p\mathrm{E}_+^2 = 2\mathcal{I}_{\mathrm{arm}} = 2\hbar\omega_p\mathrm{E}_{N,E}^2 \,.
\end{equation}

\subsubsection{Scaling law theorem}
\label{sec:scaling_low}

Return to Eqs.~\eqref{sl_pm_raw_0} for the common mode. Introduce the following notations:
  \begin{eqnarray}
    \gamma_{1W} = \gamma_{1\mathrm{arm}}\Re\frac{1-\sqrt{R_W}e^{2i\phi_W}}{1+\sqrt{R_W}e^{2i\phi_W}}
      &=& \frac{\gamma_{1\mathrm{arm}}T_W}{1 + 2\sqrt{R_W}\cos2\phi_W + R_W} \,, \\
    \delta_W = \delta_{\mathrm{arm}} - \gamma_{1\mathrm{arm}}\Im\frac{1-\sqrt{R_W}e^{2i\phi_W}}{1+\sqrt{R_W}e^{2i\phi_W}}
      &=& \delta_{\mathrm{arm}}
        + \frac{2\gamma_{1\mathrm{arm}}\sqrt{R_W}\sin2\phi_W}{1 + 2\sqrt{R_W}\cos2\phi_W + R_W} \,, \\
    \gamma_W &=& \gamma_{1W} + \gamma_2 \,, \\
    \ell_W(0) &=& \gamma_W - i\delta_W \,.
  \end{eqnarray}
In these notation, Eqs.~\eqref{sl_pm_raw_0} have the following form:
  \begin{eqnarray}
    \mathrm{B}_W &=& \mathcal{R}_W\mathrm{A}_We^{2i\alpha_W}  \,, \nonumber \\
    \mathrm{E}_+ &=& \frac{\sqrt{\gamma_{1W}}}{\ell_W(0)\sqrt{\tau}}\mathrm{A}_We^{i\alpha_W} \,,
      \label{eq:BE_SL_raw}
  \end{eqnarray}
where
\begin{eqnarray}
  \mathcal{R}_W &=& \frac{2\gamma_{1W}}{\ell_W(0)} - 1 \,, \\
  \alpha_W &=& \arg\frac{e^{i\phi_W}}{1+\sqrt{R_W}e^{2i\phi_W}} \,.
\end{eqnarray}
It is easy to see that these equations have the same form as Eqs.~\eqref{eq:FP_0_short} for the single Fabry--P{\'e}rot cavity, with the evident replacement of $\gamma_1$ and $\delta$ with the effective parameters $\gamma_{1W}$ and $\delta_W$. The only difference is an additional phase shift $\alpha_W$. Thus, we have shown that the power recycling cavity formed by the PRM and the ITMs can be treated as a single mirror with some effective parameters defined implicitly by Eqs.~\eqref{eq:BE_SL_raw}, complemented by light propagation over length $\alpha_W/k_p$. Note also that the phase shift $\alpha_W$ can be absorbed into the field amplitudes simply by renaming
\begin{equation}
\label{eq:sl_noalpha_0}
  \mathrm{A}_We^{i\alpha_W} \to \mathrm{A}_W \,, \qquad \mathrm{B}_We^{-i\alpha_W} \to \mathrm{B}_W \,,
\end{equation}
which yields:
  \begin{eqnarray}
\label{eq:sl_pm_0}
    \mathrm{B}_W &=& \mathcal{R}_W\mathrm{A}_W \,,\nonumber \\
    \mathrm{E}_+ &=& \frac{\sqrt{\gamma_{1W}}}{\ell_W(0)\sqrt{\tau}}\mathrm{A}_W \,.
  \end{eqnarray}
The corresponding equivalent model of the common mode is shown in
Figure~\ref{fig:pm_modes_2}(a).

\epubtkImage{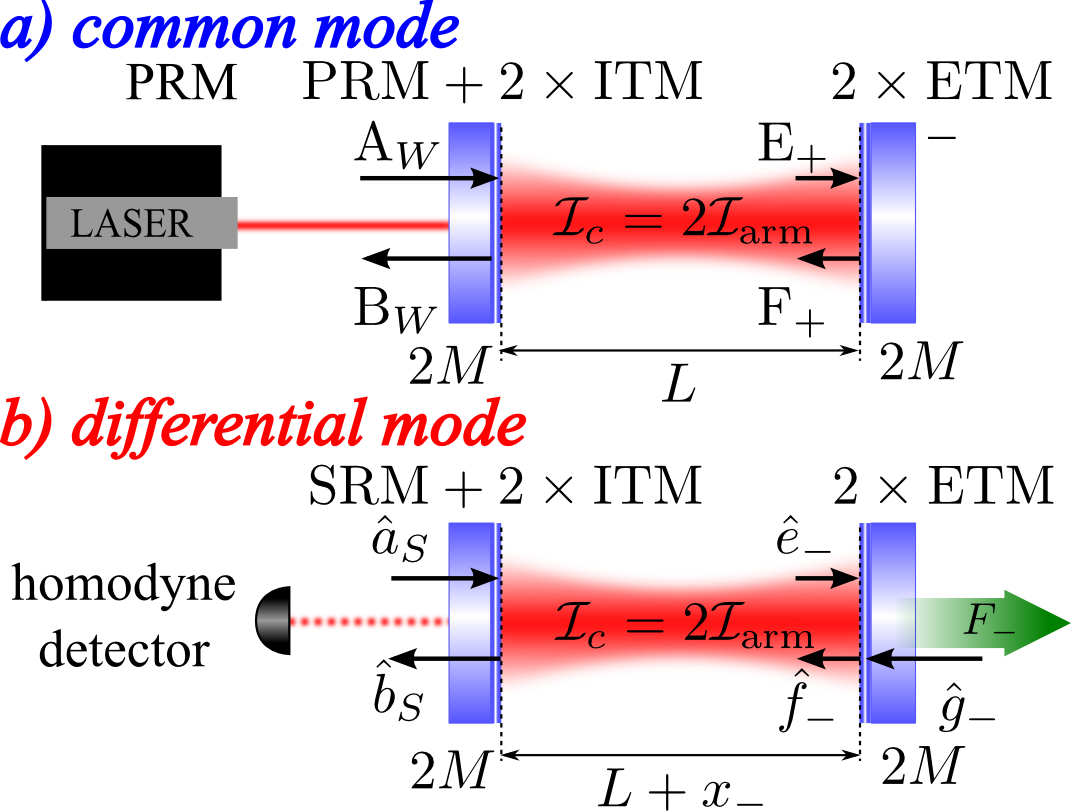}{
\begin{figure}
  \centerline{\includegraphics[width=.5\textwidth]{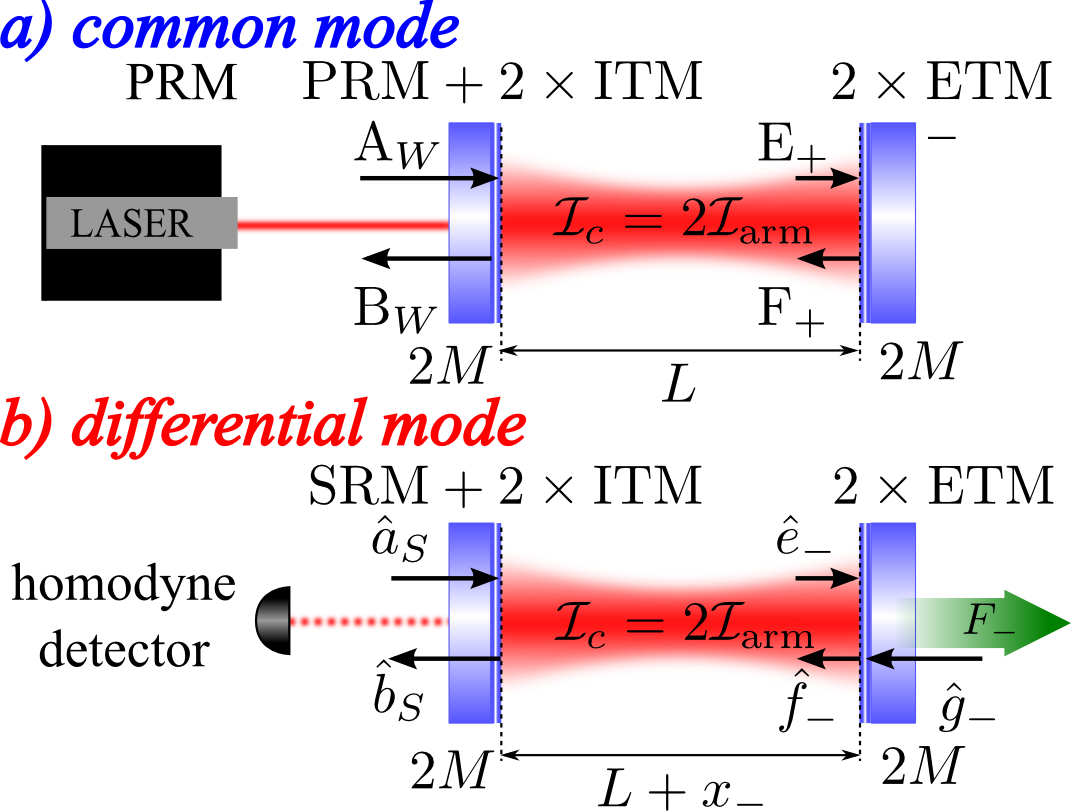}}
  \caption{Common (top) and differential (bottom) modes of the dual-recycled Fabry--P{\'e}rot--Michelson  interferometer, reduced to the single cavities using the scaling law model.}
\label{fig:pm_modes_2}
\end{figure}}

Taking into account that the main goal of power recycling is the increase of the power $\mathcal{I}_c=\hbar\omega_p|E_+|^2$ circulating in the arm cavities, for a given laser power $\mathcal{I}_0=\hbar\omega_p|A_W|^2$, the optimal tuning of the power recycling cavity corresponds to the critical coupling of the common mode with the laser:
\begin{equation}
  \gamma_{1W} = \gamma_2 \,, \qquad \delta_W = 0 \,.
\end{equation}
In this case,
\begin{equation}
  \mathrm{E}_+ = \frac{\mathrm{A}_W}{2\sqrt{\gamma_2\tau}} \Longrightarrow
  \mathcal{I}_c = 2\mathcal{I}_{\mathrm{arm}} = \frac{\mathcal{I}_0}{4\gamma_2\tau} \,.
\end{equation}
Note that this regime can be achieved even with the detuned arm cavities, $\delta\ne0$.

Consider now the differential mode quantum field amplitudes as given in Eqs.~\eqref{eq:sl_pm_raw_1}. Note a factor $e^{i\omega\tau_S}$ that describes a frequency-dependent phase shift the sideband fields acquire on their pass through the signal recycling cavity. It is due to this frequency-dependent phase shift that the differential mode cannot be reduced, strictly speaking, to a single effective cavity mode, and a more complicated two-cavity model of Figure~\ref{fig:pm_modes_1} should be used instead. The reduction to a single mode is nevertheless possible in the special case of a short signal-recycling cavity, i.e., such that:
\begin{equation}
  |\Omega|\tau_S \ll 1 \,.
\end{equation}
The above condition is satisfied in a vast majority of the proposed schemes of advanced GW interferometers and in all current interferometers that make use of the recycling techniques~\cite{VIRGOsite,GEOsite}.
In this case, the phase shift $\phi_S$ can be approximated by the frequency-independent value:
\begin{equation}
  \omega\tau_S \approx \omega_p\tau_S \equiv \phi_S \,.
\end{equation}
It allows one to introduce the following effective parameters:
  \begin{eqnarray}
\label{sl_gd}
    \gamma_1 = \gamma_{1\mathrm{arm}}\Re\frac{1-\sqrt{R_S}e^{2i\phi_S}}{1+\sqrt{R_S}e^{2i\phi_S}}
      &=& \frac{\gamma_{1\mathrm{arm}}T_S}{1 + 2\sqrt{R_S}\cos2\phi_S + R_S} \,,\nonumber \\
    \delta = \delta_{\mathrm{arm}}
      - \gamma_{1\mathrm{arm}}\Im\frac{1-\sqrt{R_S}e^{2i\phi_S}}{1+\sqrt{R_S}e^{2i\phi_S}}
      &=& \delta_{\mathrm{arm}}
        + \frac{2\gamma_{1\mathrm{arm}}\sqrt{R_S}\sin2\phi_S}{1 + 2\sqrt{R_S}\cos2\phi_S + R_S}  \,, \nonumber \\
    \gamma &=& \gamma_1 + \gamma_2 \,, \nonumber \\
    \ell(\Omega) &=& \gamma - i(\delta+\Omega) \,
  \end{eqnarray}
and to rewrite Eqs.~\eqref{eq:sl_pm_raw_1} as follows:
\begin{eqnarray}
  \hat{\mathrm{b}}_S(\omega) &=&
    \left[\mathcal{R}_1(\Omega)\hat{\mathrm{a}}_S(\omega)e^{i\alpha_S}
    + \mathcal{T}(\Omega)\hat{\mathrm{g}}_-(\omega)
    + \frac{2\sqrt{\gamma_1}\hat{X}_-(\omega)}{\ell(\Omega)}\right]e^{i\alpha_S} \,,
      \label{eq:sl_pm_1_b} \\
  \hat{\mathrm{e}}_-(\omega) &=& \frac{
      \sqrt{\gamma_1}\hat{\mathrm{a}}_S(\omega)e^{i\alpha_S}
      + \sqrt{\gamma_2}\hat{\mathrm{g}}_-(\omega)
      + \hat{X}_-(\omega)
    }{\ell(\Omega)\sqrt{\tau}} \,, \label{eq:sl_pm_1_e}
\end{eqnarray}
where the reflectivity and transmittance of the equivalent Fabry--P{\'e}rot cavity are still defined by the Eqs.~\eqref{eq:FP_RT}, but with new effective parameters~\eqref{sl_gd}, and
\begin{equation}
  \alpha_S = \arg\frac{e^{i\phi_S}}{1+\sqrt{R_S}e^{2i\phi_S}}
\end{equation}
is the phase shift introduced by the signal recycled cavity. Along similar lines as in the common mode case, we make the following change of variables
\begin{equation}
  \hat{\mathrm{a}}_Se^{i\alpha_S} \to \hat{\mathrm{a}}_S \,, \qquad
  \hat{\mathrm{b}}_Se^{-i\alpha_S} \to \hat{\mathrm{b}}_S \,.
\end{equation}
Eqs.~\eqref{eq:sl_pm_1_b} and \eqref{eq:sl_pm_1_e} have exactly the same form as the corresponding equations for the Fabry--P{\'e}rot cavity~\eqref{eq:FP_1_short}. Thus, we have successfully built a single cavity model for the differential mode, see Figure~\ref{fig:pm_modes_2}(b).

The mechanical equations of motion for the effective cavity are absolutely the same as for an ordinary Fabry--P{\'e}rot cavity considered in Section~\ref{sec:Fabry-Perot} except for the new values of the effective mirrors' mass $2M$ and effective circulating power $\mathcal{I}_c=2\mathcal{I}_{\mathrm{arm}}$. Bearing this in mind, we can procede to the quantum noise spectral density calculation for this interferometer.

\subsubsection{Spectral densities for the Fabry--P{\'e}rot--Michelson interferometer}
\label{sec:sl_noises}

The scaling law we have derived above enables us to calculate spectral
densities of quantum noise for a dual-recycled
Fabry--P{\'e}rot--Michelson featured in Figure~\ref{fig:advligo} as if
it were a bare Fabry--P{\'e}rot cavity with movable mirrors pumped
from one side, similar to that shown in Figure~\ref{fig:pm_modes_2}.

\epubtkImage{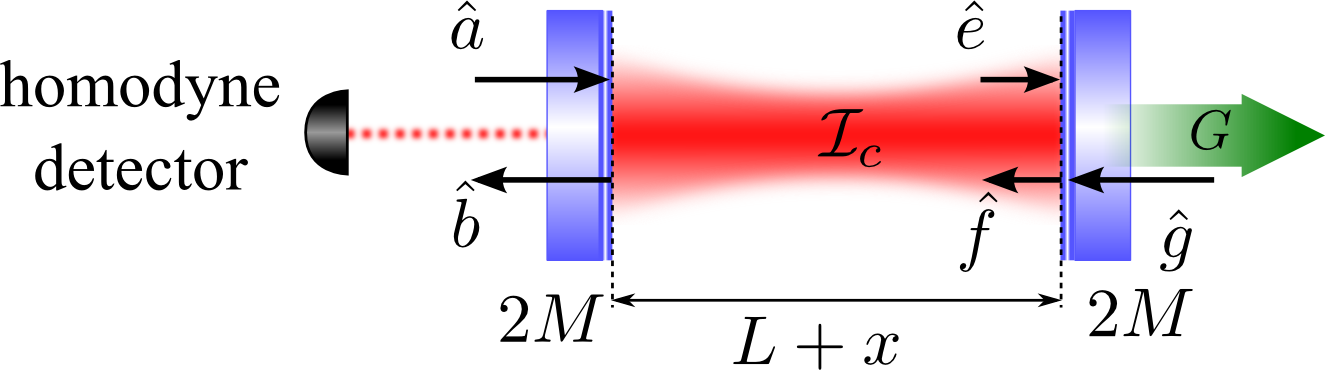}{
\begin{figure}
  \centerline{\includegraphics[width=.55\textwidth]{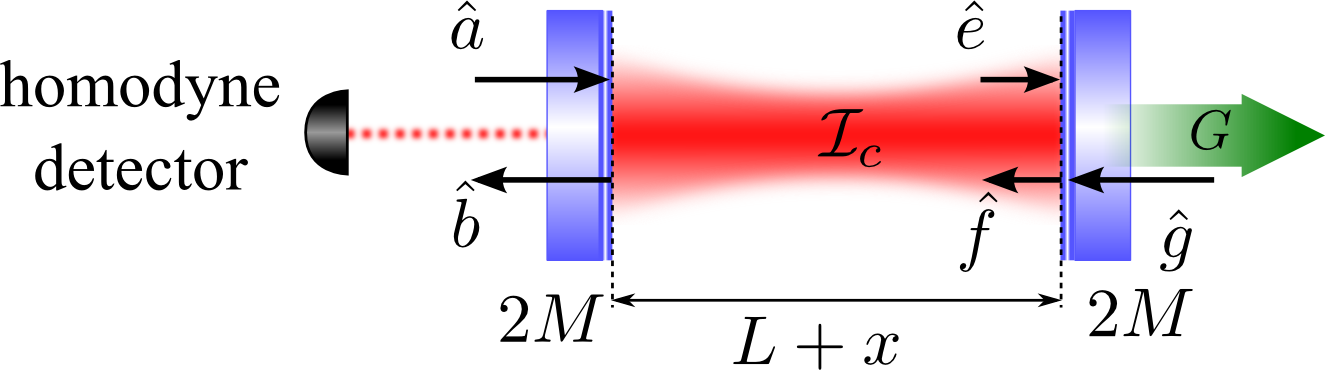}}
  \caption{The differential mode of the dual-recycled Fabry--P{\'e}rot--Michelson interferometer in simplified notation~\eqref{eq:simple_notations}.}
\label{fig:model}
\end{figure}}

We remove some of the subscripts in our notations, for the sake of notational brevity:
\begin{equation}
    \label{eq:simple_notations}
  \begin{array}{llll}
    \hat{\mathrm{a}}_S \to \hat{\mathrm{a}} \,, & \qquad \hat{\mathrm{b}}_S \to \hat{\mathrm{b}} \,, & \qquad \hat{\mathrm{e}}_- \to \hat{\mathrm{e}} \,, & \qquad \mathrm{E}_- \to \mathrm{E} \,,\\
    x_- \to x \,, & \qquad F_-^{\mathrm{r.p.}} \to F_{\mathrm{r.p.}} \,, & \qquad  
    G_- \to G \,, & \qquad F_-^{\mathrm{b.a.}} \to F_{\mathrm{b.a.}},
  \end{array}
\end{equation}
(compare Figures~\ref{fig:pm_modes_2} and \ref{fig:model}). We also choose the phase of the classical field $\mathrm{E}$ amplitude inside the arm cavities to be zero:
\begin{equation}
  \Im\mathrm{E} = 0 \Longrightarrow \mathbf{E} = \sqrt{2}\mathrm{E}\svup
\end{equation}
that obviously does not limit the generality of our consideration, yet sets the reference point for all the classical and quantum fields' phases.

With this in mind, we rewrite I/O-relations~\eqref{eq:sl_pm_1_b} and \eqref{eq:sl_pm_1_e} in the two-photon quadratures notation:
  \begin{eqnarray}
    \hat{\boldsymbol{b}}(\Omega) &=& \mathbb{R}_1(\Omega)\hat{\boldsymbol{a}}(\Omega)
      + \mathbb{T}(\Omega)\hat{\boldsymbol{g}}(\Omega)
      + \sqrt{\frac{2MJ\gamma_1}{\hbar}}\,
          \frac{\mathbf{D}(\Omega)\hat{x}(\Omega)}{\mathcal{D}(\Omega)}
    \,, \label{eq:sl_quad_b} \\
    \hat{\boldsymbol{e}}(\Omega) &=& \frac{1}{\sqrt{\tau}}\biggl\{
        \mathbb{L}(\Omega)
          [\sqrt{\gamma_1}\hat{\boldsymbol{a}}(\Omega) + \sqrt{\gamma_2}\hat{\boldsymbol{g}}(\Omega)]
        + \sqrt{\frac{MJ}{2\hbar}}\,
            \frac{\mathbf{D}(\Omega)\hat{x}(\Omega)}{\mathcal{D}(\Omega)}
      \biggr\} , \label{eq:sl_quad_e}
  \end{eqnarray}
where the matrices $\mathbb{L}$, $\mathbb{R}_1$, $\mathbb{T}$ are defined by Eqs.~\eqref{FP_bbL} and \eqref{FP_bbRT},
\begin{equation}
  \mathbf{D}(\Omega) = \mathcal{D}(\Omega)\mathbb{L}(\Omega)\svdown
  = \svector{-\delta}{\gamma-i\Omega}
\end{equation}
and
\begin{equation}
\label{eq:FPMI_J_def}
  J = \frac{4\hbar k_p^2\mathrm{E}^2}{M\tau} = \frac{4\omega_p\mathcal{I}_c}{McL}
  = \frac{8\omega_p\mathcal{I}_{\mathrm{arm}}}{McL}
\end{equation}
is the normalized optical power, circulating in the interferometer arms.

Suppose that the output beam is registered by the homodyne detector; see Section~\ref{sec:homodyne}. Combining Eqs.~\eqref{eq:sl_quad_b} and \eqref{eq:homodyne_lossy}, we obtain for the homodyne detector a readout expressed in units of signal force:
\begin{equation}
\label{ord_i_meas}
  \hat{O}^{F,\mathrm{loss}}(\Omega) = \frac{\hat{\mathcal{X}}^{\mathrm{loss}}}{\chi_{xx}^{\mathrm{eff,\,FP}}(\Omega)} + \hat{\mathcal{F}}(\Omega)+\frac{G(\Omega)}{2} \,,
\end{equation}
where
\begin{equation}
\label{eq:ord_x_meas}
  \hat{\mathcal{X}}^{\mathrm{loss}}(\Omega) = \frac{\hat{O}^{(0),\,\mathrm{loss}}(\Omega)}{\chi_{OF}(\Omega)} = \sqrt{\frac{\hbar}{2MJ\gamma_1}}\,
    \frac{\mathcal{D}(\Omega)}{\boldsymbol{H}^{\mathsf{T}}[\phi_{\mathrm{LO}}]\mathbf{D}(\Omega)}
    \boldsymbol{H}^{\mathsf{T}}[\phi_{\mathrm{LO}}][
      \mathbb{R}_1(\Omega)\hat{\boldsymbol{a}}(\Omega) + \mathbb{T}(\Omega)\hat{\boldsymbol{g}}(\Omega)
      + \epsilon_d\hat{\boldsymbol{n}}(\Omega)
    ]
\end{equation}
stands for the measurement noise, which is typically referred to as shot noise in optomechanical measurement with the interferometer opto-mechanical response function defined as
 \begin{equation}
  \chi_{OF}(\Omega) =  \sqrt{\frac{2MJ\gamma_1}{\hbar}}\frac{\boldsymbol{H}^{\mathsf{T}}[\phi_{\mathrm{LO}}]\mathbf{D}(\Omega)}{\mathcal{D}(\Omega)}
 \end{equation}
 and the back-action noise caused by the radiation pressure fluctuations equal to
\begin{equation}
\label{eq:ord_F_pert}
  \hat{\mathcal{F}}(\Omega)\equiv\hat{F}_{\mathrm{b.a.}}(\Omega) = \sqrt{2\hbar MJ}\begin{bmatrix}1\\0\end{bmatrix}^{\mathsf{T}}\mathbb{L}(\Omega)
    [\sqrt{\gamma_1}\hat{\boldsymbol{a}}(\Omega) + \sqrt{\gamma_2}\hat{\boldsymbol{g}}(\Omega)]
\end{equation}

The dynamics of the interferometer is described by the effective susceptibility $\chi_{xx}^{\mathrm{eff,\,FP}}(\Omega)$ that is the same as the one given by Eq.~\eqref{eq:chi_xx_eff_FP} where
\begin{equation}
\label{eq:sl_K}
  K(\Omega) = \frac{MJ\delta}{\mathcal{D}(\Omega)}
\end{equation}
is the frequency-dependent optical rigidity that has absolutely the same form as that of a single Fabry--P{\'e}rot cavity given by Eq.~\eqref{eq:FP_K}.

Suppose then that the input field of the interferometer is in the squeezed quantum state that is equivalent to the following transformation of the input fields:
\begin{equation}
\label{a_sqz}
  \hat{\boldsymbol{a}} = \mathbb{S}_{\sqz}[r,\theta]\hat{\boldsymbol{a}}^{\vac} \,,
\end{equation}
where the squeezing matrix is defined by Eq.~\eqref{eq:SQZ_matrix_transform}, and the quadrature vector $\hat{\boldsymbol{a}}^{\vac}$ corresponds to the vacuum state.

Using the rules of spectral densities computation given in Eqs.~\eqref{eq:S_Y_sqz} and \eqref{eq:S_YZ_sqz}, taking into account unitarity conditions~\eqref{eq:FP_bbRT_u}, one can get the following expressions for the power (double-sided) spectral densities of the dual-recycled Fabry--P{\'e}rot--Michelson interferometer measurement and back-action noise sources as well as their cross-correlation spectral density:
  \begin{eqnarray}
    S_{\mathcal{X}\mathcal{X}}(\Omega) &=& \frac{\hbar}{4MJ\gamma_1}
      \frac{|\mathcal{D}(\Omega)|^2}{\left|\boldsymbol{H}^{\mathsf{T}}[\phi_{\mathrm{LO}}]\mathbf{D}(\Omega)\right|^2}\nonumber\\
        &&\times \bigl\{
          \boldsymbol{H}^{\mathsf{T}}(\phi_{\mathrm{LO}})\mathbb{R}_1(\Omega)
            [\mathbb{S}_{\sqz}(2r,\theta) - \mathbb{I}]
            \mathbb{R}_1^\dagger(\Omega)\boldsymbol{H}[\phi_{\mathrm{LO}}]
          + 1 + \epsilon_d^2
        \bigr\} ,\label{eq:FPMI_Sx} \\
    S_{\mathcal{F}\mathcal{F}}(\Omega) &=& \hbar MJ\,\begin{bmatrix}1\\0\end{bmatrix}^{\mathsf{T}}\mathbb{L}(\Omega)
        [\gamma_1\mathbb{S}_{\sqz}(2r,\theta) + \gamma_2]\mathbb{L}^\dagger(\Omega)\svup \,,\label{eq:FPMI_SF} \\
    S_{\mathcal{X}\mathcal{F}}(\Omega) &=& \frac{\hbar}{2}
      \frac{\mathcal{D}(\Omega)}{\boldsymbol{H}^{\mathsf{T}}[\phi_{\mathrm{LO}}]\mathbf{D}(\Omega)}
        \boldsymbol{H}^{\mathsf{T}}(\phi_{\mathrm{LO}})\bigl[
          \mathbb{R}_1(\Omega)\mathbb{S}_{\sqz}(2r,\theta)
          + \sqrt{\gamma_2/\gamma_1}\mathbb{T}(\Omega)
        \bigr]\mathbb{L}^\dagger(\Omega)\svup \,. \label{eq:FPMI_SxF}
  \end{eqnarray}
These spectral densities satisfy the Sch\"rodinger--Robertson uncertainty relation:
\begin{equation}
\label{SxSFSxf}
  S_{\mathcal{X}\mathcal{X}}(\Omega)S_{\mathcal{F}\mathcal{F}}(\Omega) - |S_{\mathcal{X}\mathcal{F}}(\Omega)|^2 \ge \frac{\hbar^2}{4}
\end{equation}
of the same form as in the general linear measurement case considered in Section~\ref{sec:gen_linear_measurement}, see Eq.~\eqref{eq:gen_spdens_uncert_rel}, with the exact equality in the ideal lossless case:
\begin{equation}
\label{no_loss}
  \gamma_2 = 0 \,, \qquad \eta_d = 1 \,.
\end{equation}
see Appendix~\ref{app:SxSFSxf}

\subsubsection{Full transfer matrix approach to calculation of the Fabry--P{\'e}rot--Michelson interferometer quantum noise}

In order to compute the sum quantum noise spectral density one has to first calculate $S_{\mathcal{X}\mathcal{X}}(\Omega)$, $S_{\mathcal{F}\mathcal{F}}(\Omega)$ and $S_{\mathcal{X}\mathcal{F}}(\Omega)$ using Eqs.~\eqref{eq:FPMI_Sx}, \eqref{eq:FPMI_SF}, and \eqref{eq:FPMI_SxF} and then insert them into the general formula~\eqref{eq:gen_spdens}.

However, there is another option that is more convenient from the computational point of view. One can compute the full quantum noise transfer matrix of the Fabry--P{\'e}rot--Michelson interferometer in the same manner as for a single mirror in Section~\ref{sec:mirror_full_transfer_mat}. The procedure is rather straightforward. Write down the readout observable of the homodyne detector in units of signal force:
\begin{eqnarray}
\label{eq:X_sum_BC}
  \hat{O}^F(\Omega) =
  \frac{\hat{\mathcal{X}}^{\mathrm{loss}}(\Omega)}{\chi_{xx}^{\mathrm{eff,\,FP}}(\Omega)}+\hat{\mathcal{F}}(\Omega)+\frac{G(\Omega)}{2} = \frac{G(\Omega)}{2}+ 
    \sqrt{\frac{\hbar}{2MJ\gamma_1}}
    \frac{-M\Omega^2}{\boldsymbol{H}^{\mathsf{T}}[\phi_{\mathrm{LO}}]\mathbf{D}(\Omega)}
    \boldsymbol{H}^{\mathsf{T}}[\phi_{\mathrm{LO}}] \nonumber\\ \times\biggl\{
      \mathbb{C}_1(\Omega)\mathbb{S}_{\sqz}[r,\theta]\hat{\boldsymbol{a}}^{\vac}(\Omega)
      + \mathbb{C}_2(\Omega)\hat{\boldsymbol{g}}(\Omega)
      + \left[\mathcal{D}(\Omega) - \frac{J\delta}{\Omega^2}\right]
          \epsilon_d\hat{\boldsymbol{n}}(\Omega)
    \biggr\}\,,
\end{eqnarray}
where matrices $\mathbb{C}_{1,2}(\Omega)$ can be computed using the fact that
\begin{equation*}
  [\delta + \mathbf{D}(\Omega)\begin{bmatrix}1\\0\end{bmatrix}^{\mathsf{T}}]\mathbb{L}(\Omega) = \smatrix{0}{0}{1}{0} ,
\end{equation*}
which yields:
\begin{eqnarray}
    \mathbb{C}_1(\Omega)
      &=& \smatrix{2\gamma_1(\gamma-i\Omega) - \mathcal{D}(\Omega) + J\delta/\Omega^2}
          {-2\gamma_1\delta}{2\gamma_1\delta-2J\gamma_1/\Omega^2}
          {2\gamma_1(\gamma-i\Omega) - \mathcal{D}(\Omega) + J\delta/\Omega^2} , \\
    \mathbb{C}_2(\Omega) &=& 2\sqrt{\gamma_1\gamma_2}
      \smatrix{\gamma-i\Omega}{-\delta}{\delta-J/\Omega^2}{\gamma-i\Omega} .
  \end{eqnarray}
In the GW community, it is more common to normalize the signal of the interferometer in units of GW amplitude spectrum $h(\Omega)$. This can easily be done using the simple rule given in Eq.~\eqref{eq:gen_lin_F_to_h_transform} and taking into account that in our case $G(\Omega)\to G(\Omega)/2$:
\begin{eqnarray}
  \hat{O}^h(\Omega) = h_{\mathrm{GW}}(\Omega)+\hat{h}_n(\Omega) = h_{\mathrm{GW}}(\Omega)+
    \frac{2}{L}\sqrt{\frac{\hbar}{2MJ\gamma_1}}
    \frac{1}{\boldsymbol{H}^{\mathsf{T}}[\phi_{\mathrm{LO}}]\mathbf{D}(\Omega)}
    \boldsymbol{H}^{\mathsf{T}}[\phi_{\mathrm{LO}}] \nonumber\\ \times\biggl\{
      \mathbb{C}_1(\Omega)\mathbb{S}_{\sqz}[r,\theta]\hat{\boldsymbol{a}}^{\vac}(\Omega)
      + \mathbb{C}_2(\Omega)\hat{\boldsymbol{g}}(\Omega)
      + \left[\mathcal{D}(\Omega) - \frac{J\delta}{\Omega^2}\right]
          \epsilon_d\hat{\boldsymbol{n}}(\Omega)
    \biggr\}\,,
\end{eqnarray}
where $h_{\mathrm{GW}}(\Omega)$ is the spectrum of the GW signal and the second term $\hat{h}_n(\Omega)$ stands for the sum quantum noise expressed in terms of metrics variation spectrum units, i.e., in $\mathrm{Hz}^{-1/2}$.

The power (double-sided) spectral density of the sum quantum noise then reads:
\begin{eqnarray}
\label{eq:S_sum_BC}
  S^h(\Omega) = \frac{4  S^F(\Omega)}{M^2L^2\Omega^4} &=& \frac{\hbar}{MJ\gamma_1L^2}
    \frac{1}{|\boldsymbol{H}^{\mathsf{T}}[\phi_{\mathrm{LO}}]\mathbf{D}(\Omega)|^2} \nonumber\\ \times\biggl\{
      \boldsymbol{H}^{\mathsf{T}}[\phi_{\mathrm{LO}}]\bigl[
        \mathbb{C}_1(\Omega)\mathbb{S}_{\sqz}[2r,\theta]\mathbb{C}_1^\dagger(\Omega)
        &+& \mathbb{C}_2(\Omega)\mathbb{C}_2^\dagger(\Omega)
      \bigr]\boldsymbol{H}[\phi_{\mathrm{LO}}]
      + \left|\mathcal{D}(\Omega) - \frac{J\delta}{\Omega^2}\right|^2\epsilon_d^2
    \biggr\} .
\end{eqnarray}

In conclusion, we should say that the quantum noise of the Fabry--P{\'e}rot--Michelson  interferometer has been calculated in many papers, starting from the seminal work by Kimble et al.~\cite{02a1KiLeMaThVy} where a resonance-tuned case with $\delta=0$ was analyzed, and then by Buonanno and Chen in~\cite{Buonanno2001, Buonanno2003}, who considered a more general detuned case.
Thus, treading their steps, we have shown that the quantum noise of the Fabry--P{\'e}rot--Michelson  interferometer (as well as the single cavity Fabry--P{\'e}rot one) has the following distinctive features:
 \begin{itemize}
   \item It comprises two effective noise sources as in any quantum
     linear measurement device. These are measurement noise
     $\hat{\mathcal{X}}^{\mathrm{loss}}$, more frequently called \emph{quantum
       shot noise} in the GW community, and the back-action noise $\hat{\mathcal{F}}$, often referred to as \emph{quantum radiation-pressure noise}.
   \item These noise sources are correlated and this correlation depends not only on the homodyne angle $\phi_{\mathrm{LO}}$ or the correlations in the input light (e.g., squeezing angle $\theta$ in case of squeezed input), but also on the interferometer effective detuning $\delta$, which, according to the scaling law theorem, can be changed by varying signal-recycling cavity parameters.
   \item The scaling law theorem also shows that changing the arm cavities' detuning is equivalent to the modification of the signal recycling cavity parameters in terms of effective detuning and bandwidth of the interferometer.
   \item Another important corollary of the scaling law is that the effective bandwidths and detunings for the common and differential optical modes can be chosen independently, thus making it possible to tune the former in resonance with the pumping laser to keep as high a value of the circulating optical power in the arms as possible, and to detune the latter one to modify the test masses dynamical response by virtue of the introduction of optical rigidity that arises in the detuned cavity as we have shown.
 \end{itemize}

All of these features can be used to decrease the quantum noise of the interferometer and reach a sensitivity below the SQL in a decent range of frequencies as we show in Section~\ref{sec:sub-SQL_schemes}.

\newpage
\section{Schemes of GW Interferometers with Sub-SQL Sensitivity}
\label{sec:sub-SQL_schemes}

\subsection{Noise cancellation by means of cross-correlation}
\label{sec:noise_cancellation}

\subsubsection{Introduction}

In this section, we consider the interferometer configurations that use the idea of the cross-correlation of the shot and the radiation pressure noise sources discussed in Section~\ref{sec:toy_FI_correlation}. This cross-correlation allows the measurement and the back-action noise to partially cancel each other out and thus effectively reduce the sum quantum noise to below the SQL.

As we noted above, Eq.~\eqref{eq:FPMI_SxF} tells us that this cross-correlation can be created by tuning either the homodyne angle $\phi_{\mathrm{LO}}$, the squeezing angle $\theta$, or the detuning $\delta$. In Section~\ref{sec:toy_FI_correlation}, the simplest particular case of the frequency-independent correlation created by means of measurement of linear combination of the phase and amplitude quadratures, that is, by using the homodyne angle $\phi_{\mathrm{LO}}\ne\pi/2$, has been considered. We were able to obtain a narrow-band sensitivity gain at some given frequency that was similar to the one achievable by introducing a constant rigidity to the system, therefore such correlation was called effective rigidity.

However, the broadband gain requires a frequency-dependent correlation, as it was first demonstrated for optical interferometric position meters~\cite{Unruh1982}, and then for general position measurement case~\cite{87a1eKh}. Later, this idea was developed in different contexts by several authors~\cite{JaekelReynaud1990, Pace1993, 96a2eVyMa, 02a1KiLeMaThVy, PhysRevD.68.042001_2003_Harms, 06pth1Ha, PhysRevLett.95.211102_2005_Vahlbruch, Arcizet2006}. In particular, in~\cite{02a1KiLeMaThVy}, a practical method of creation of the frequency-dependent correlation was proposed, based on the use of additional \emph{filter cavities}, which were proposed to be placed either between the squeeze light source and the main interferometer, creating the frequency-dependent squeezing angle (called \emph{pre-filtering}), or between the main interferometer and the homodyne detector, creating the effective frequency-dependent squeezing angle (\emph{post-filtering}). As we show below, in principle, both pre- and post-filtering can be used together, providing some additional sensitivity gain.

It is necessary to note also an interesting method of noise cancellation proposed by Tsang and Caves recently~\cite{PhysRevLett.105.123601_2010_Tsang}. The idea was to use \emph{matched squeezing}; that is, to place an additional cavity inside the main interferometer and couple the light inside this additional cavity with the differential mode of the interferometer by means of an optical parametric amplifier (OPA). The squeezed light created by the OPA should compensate for the ponderomotive squeezing created by back-action at all frequencies and thus decrease the quantum noise below the SQL at a very broad frequency band. However, the thorough analysis of the optical losses influence, that as we show later, are ruinous for the subtle quantum correlations this scheme is based on, was not performed.

Coming back to the filter-cavities--based interferometer topologies, we limit ourselves here by the case of the resonance-tuned interferometer, $\delta=0$. This assumption simplifies all the equations considerably, and allows one to clearly separate the sensitivity gain provided by the quantum noise cancellation due to cross-correlation from the one provided by the optical rigidity, which will be considered in Section~\ref{sec:optical_rigidity}.

We also neglect optical losses \emph{inside} the interferometer, assuming that $\gamma_2=0$. In broadband interferometer configurations considered here, with typical values of $\gamma\gtrsim10^3\mathrm{\ s}^{-1}$, the influence of these losses is negligible compared to those of the photodetector inefficiency and the losses in the filter cavities. Indeed, taking into account the fact that with modern high-reflectivity mirrors, the losses per bounce do not exceed $A_{\mathrm{arm}}\lesssim10^{-4}$, and the arms lengths of the large-scale GW detectors are equal to several kilometers, the values of $\gamma_2\lesssim1\mathrm{\ s}^{-1}$, and, correspondingly, $\gamma_2/\gamma\lesssim10^{-3}$, are feasible. At the same time, the value of photodetector quantum inefficiency $\epsilon_d^2\approx1-\eta_d\approx0.05$ (factoring in the losses in the interferometer output optical elements as well) is considered quite optimistic. Note, however, that in narrow-band regimes considered in Section~\ref{sec:optical_rigidity}, the bandwidth $\gamma$ can be much smaller and influence of $\gamma_2$ could be significant; therefore, we take these losses into account in Section~\ref{sec:optical_rigidity}.

Using these assumptions, the quantum noises power (double-sided) spectral densities~\eqref{eq:FPMI_Sx}, \eqref{eq:FPMI_SF} and \eqref{eq:FPMI_SxF} can be rewritten in the following explicit form:
  \begin{eqnarray}
    S_{\mathcal{X}\mathcal{X}}(\Omega) &=& \frac{\hbar}{2M\Omega^2\mathcal{K}(\Omega)\sin^2\phi_{\mathrm{LO}}}
      \bigl[\cosh2r + \sinh2r\cos2(\theta-\phi_{\mathrm{LO}}) + \epsilon_d^2\bigr] , \label{eq:vr_noises_x} \\
    S_{\mathcal{F}\mathcal{F}}(\Omega)
      &=& \frac{\hbar M\Omega^2\mathcal{K}(\Omega)}{2}(\cosh2r + \sinh2r\cos2\theta) , \label{eq:vr_noises_F}\\
    S_{\mathcal{X}\mathcal{F}}(\Omega)
      &=& \frac{\hbar}{2\sin\phi_{\mathrm{LO}}}[\cosh2r\cos\phi_{\mathrm{LO}} + \sinh2r\cos(2\theta-\phi_{\mathrm{LO}})]
      \label{eq:vr_noises_xF} ,
  \end{eqnarray}
where
\begin{equation}
\label{eq:varK}
  \mathcal{K}(\Omega) = \frac{2J\gamma}{\Omega^2(\gamma^2+\Omega^2)}
\end{equation}
is the convenient optomechanical coupling factor introduced in~\cite{02a1KiLeMaThVy}.

Eq.~\eqref{eq:X_sum_BC} and \eqref{eq:S_sum_BC} for the sum quantum noise and its power (double-sided) spectral density in this case can also be simplified significantly:
\begin{eqnarray}
  \hat{h}_{n}(\Omega) &=& \frac{2}{L}\sqrt{\frac{\hbar}{2MJ\gamma}}\,
    \frac{1}{\sin\phi_{\mathrm{LO}}}
    \boldsymbol{H}^{\mathsf{T}}[\phi_{\mathrm{LO}}]\bigl[
      (\gamma+i\Omega)\mathbb{K}(\Omega)\mathbb{S}_{\sqz}[r,\theta]\hat{\boldsymbol{a}}^{\vac}(\Omega)
      + (\gamma-i\Omega)\epsilon_d\hat{\boldsymbol{n}}(\Omega)
    \bigr] , \label{eq:vr_X_sum} \\
  S^h(\Omega) &=& \frac{2\hbar}{ML^2\Omega^2\mathcal{K}(\Omega)\sin^2\phi_{\mathrm{LO}}}
    \bigl[
      \boldsymbol{H}^{\mathsf{T}}[\phi_{\mathrm{LO}}]\mathbb{K}(\Omega)\mathbb{S}_{\sqz}[2r,\theta]
        \mathbb{K}^\dagger(\Omega)\boldsymbol{H}[\phi_{\mathrm{LO}}]
      + \epsilon_d^2
    \bigr] , \label{eq:vr_sum_noise}
\end{eqnarray}

where
\begin{equation}
  \mathbb{K}(\Omega) = \smatrix{1}{0}{-\mathcal{K}(\Omega)}{1} .
\end{equation}

In Section~\ref{sec:corr_ideal} we consider the optimization of the spectral density~\eqref{eq:vr_sum_noise}, assuming that the arbitrary frequency dependence of the homodyne and/or squeezing angles can be implemented. As we see below, this case corresponds to the ideal lossless filter cavities. In Section~\ref{sec:corr_real}, we consider two realistic schemes, taking into account the losses in the filter cavities.

\subsubsection{Frequency-dependent homodyne and/or squeezing angles}
\label{sec:corr_ideal}

\paragraph*{Classical optimization.}

As a reference point, consider first the simplest case of frequency \emph{in}dependent homodyne and squeezing angles. We choose the specific values of these parameters following the \emph{classical optimization}, which minimizes the shot noise~\eqref{eq:vr_noises_x} without taking into account the back action. Because, the shot noise dominates at high frequencies, therefore, this optimization gives a smooth broadband shape of the sum noise spectral density.

It is evident that this minimum is provided by
\begin{equation}
   \phi_{\mathrm{LO}} = \frac{\pi}{2} \,, \qquad \theta = 0 \,.
\end{equation}
In this case, the sum quantum noise power (double-sided) spectral density is equal to
\begin{equation}
\label{eq:S_X_Caves}
  S^h(\Omega) = \frac{2\hbar}{ML^2\Omega^2}\left[
    \frac{e^{-2r} + \epsilon_d^2}{\mathcal{K}(\Omega)} + \mathcal{K}(\Omega)e^{2r}
  \right] .
\end{equation}
It is easy to note the similarity of this spectral density with the ones of the toy position meter considered above, see Eq.~\eqref{eq:S_h_fm}. The only significant differences introduced here are the optical losses and the decrease of the optomechanical coupling at high frequencies due to the finite bandwidth $\gamma$ of the interferometer. If $\epsilon_d=0$ and $\gamma\gg\Omega$, then Eqs.~\eqref{eq:S_h_fm} and \eqref{eq:S_X_Caves} become identical, with the evident correspondence
\begin{equation}
  \mathcal{I}_0\digamma^2 \leftrightarrow \frac{\mathcal{I}_c}{\gamma\tau} \,.
\end{equation}
In particular, the spectral density~\eqref{eq:S_X_Caves} can never be smaller than the free mass SQL $S^h_{\mathrm{SQL\,f.m.}}(\Omega)$ (see~\eqref{eq:S_h_SQL_fm}). Indeed, it can be minimized at any given frequency $\Omega$ by setting
\begin{equation}
  \mathcal{K}(\Omega) = e^{-r}\sqrt{e^{-2r} + \epsilon_d^2} \,,
\end{equation}
and in this case,
\begin{equation}
  \xi^2(\Omega) \equiv \frac{S^h(\Omega)}{S^h_{\mathrm{SQL\,f.m.}}(\Omega)}
   = \sqrt{1 + \epsilon_d^2e^{2r}}  \ge 1 \,.
\end{equation}

The spectral density~\eqref{eq:S_X_Caves} was first calculated in the pioneering work of~\cite{81a1Ca}, where the existence of two kinds of quantum noise in optical interferometric devices, namely the measurement (shot) noise and the back action (radiation pressure) noise, were identified for the first time, and it was shown that the injection of squeezed light with $\theta=0$ into the interferometer dark port is equivalent to the increase of the optical pumping power. However, it should be noted that in the presence of optical losses this equivalence holds unless squeezing is not too strong, $e^{-r}>\epsilon_d$.

\epubtkImage{fig34.png}{
\begin{figure}[htb]
  \centerline{\includegraphics[width=.8\textwidth]{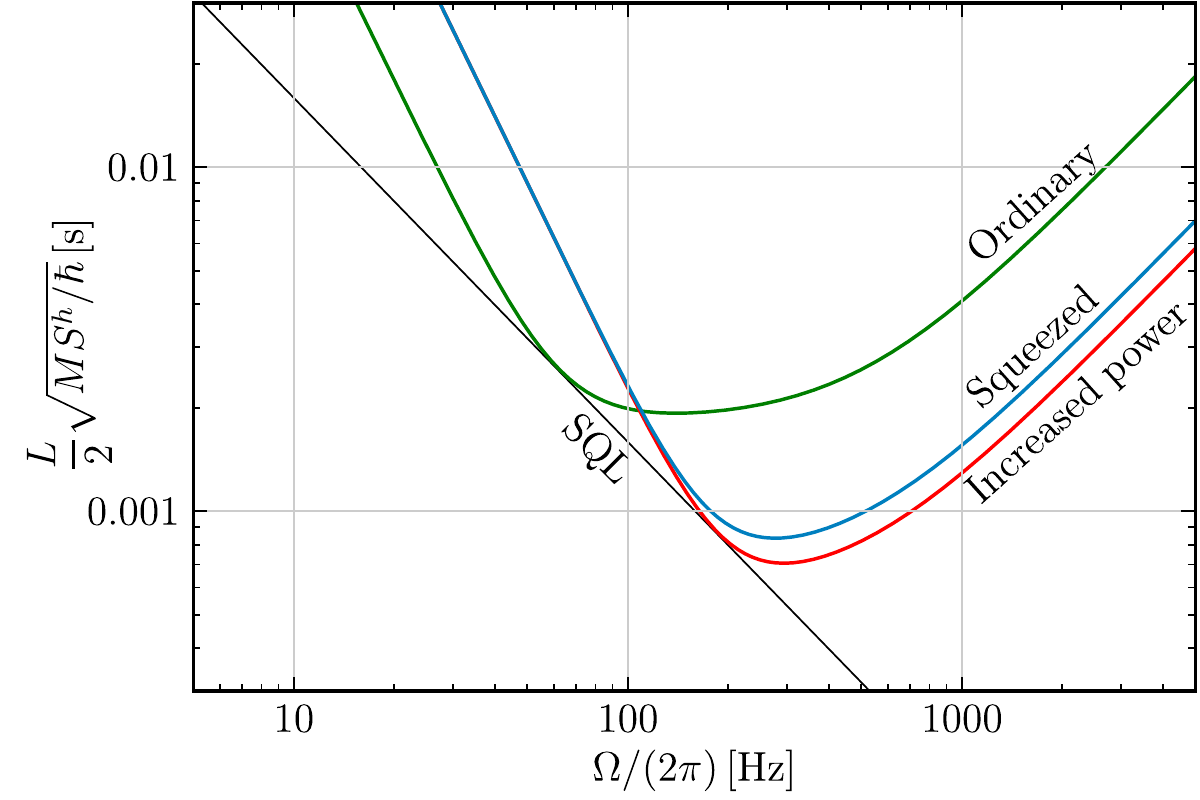}}
  \caption{Examples of the sum quantum noise spectral densities of the classically-optimized ($\phi_{\mathrm{LO}}=\pi/2$, $\theta=0$) resonance-tuned interferometer. `Ordinary': $J=J_{\mathrm{aLIGO}}$, no squeezing. `Increased power': $J=10J_{\mathrm{aLIGO}}$, no squeezing. `Squeezed': $J=J_{\mathrm{aLIGO}}$, 10~dB squeezing. For all plots, $\gamma=2\pi\times500\mathrm{\ s}^{-1}$ and $\eta_d=0.95$.}
\label{fig:LIGO_res}
\end{figure}}

The noise spectral density curves for the resonance-tuned interferometer are drawn in Figure~\ref{fig:LIGO_res}. The default parameters for this and all subsequent similar plots are chosen to be close to those planned for the Advanced LIGO interferometer: the value of $J = J_{\mathrm{aLIGO}} \equiv (2\pi\times100)^3\mathrm{\ s}^{-3}$ corresponds to the circulating power of  $\mathcal{I}_{\mathrm{arm}} = 840\mathrm{\ kW}$, $L=4\mathrm{\ km}$, and $M=40\mathrm{\ kg}$; the interferometer bandwidth $\gamma=2\pi\times500\mathrm{\ s}^{-1}$ is close to the one providing the best sensitivity for Advanced LIGO in the presence of technical noise~\cite{08a1KoSiKhDa}; 10~dB squeezing ($e^{2r}=10$), which corresponds to the best squeezing available at the moment (2011) in the low-frequency band~\cite{McKenzie2004, Vahlbruch2006, Vahlbruch_CQG_27_084027_2010}); $\eta_d=0.95$ can be considered a reasonably optimistic estimate for the real interferometer quantum efficiency.

Noteworthy is the proximity of the plots for the interferometer with 10~dB input squeezing and the one with 10-fold increased optical power. The noticeable gap at higher frequencies is due to optical loss.

\paragraph*{Frequency dependent squeezing angle.}
\label{sec:var_sqz}

Now suppose that the homodyne angle can be frequency dependent, and calculate the corresponding minimum of the sum noise spectral density~\eqref{eq:vr_sum_noise}. The first term in square brackets in this equation can be rewritten as:
\begin{equation}
\label{vr_sqz_1}
  \boldsymbol{V}^{\mathsf{T}}\mathbb{P}[\theta]\mathbb{S}_{\sqz}[2r,0]\mathbb{P}^\dagger[\theta]\boldsymbol{V}
  = e^{2r}(V_c\cos\theta + V_s\sin\theta)^2 + e^{-2r}(-V_c\sin\theta + V_s\cos\theta)^2
  \,,
\end{equation}
where
\begin{equation}
  \mathbf{V} \equiv \svector{V_c}{V_s} = \mathbb{K}^\dagger(\Omega)\boldsymbol{H}[\phi_{\mathrm{LO}}] \,.
\end{equation}
It is evident that the minimum of~\eqref{vr_sqz_1} is provided by
\begin{equation}
\label{eq:vr_preopt_theta}
  \tan\theta = -\frac{V_c}{V_s} = -\cot\phi_{\mathrm{LO}} + \mathcal{K}(\Omega)
\end{equation}
and is equal to $\boldsymbol{V}^{\mathsf{T}}\boldsymbol{V}e^{-2r}$. Therefore,
\begin{eqnarray}
\label{eq:vr_opt_2}
  S^h(\Omega) &=& \frac{2\hbar}{ML^2\Omega^2\mathcal{K}(\Omega)\sin^2\phi_{\mathrm{LO}}}
    \Bigl[
      \boldsymbol{H}^{\mathsf{T}}[\phi_{\mathrm{LO}}]\mathbb{K}(\Omega)
        \mathbb{K}^\dagger(\Omega)\boldsymbol{H}[\phi_{\mathrm{LO}}]e^{-2r}
      + \epsilon_d^2
    \Bigr] \nonumber\\
   &=& \frac{2\hbar}{ML^2\Omega^2}\left\{
      \biggl[
        \frac{1}{\mathcal{K}(\Omega)\sin^2\phi_{\mathrm{LO}}} - 2\cot\phi_{\mathrm{LO}} + \mathcal{K}(\Omega)
      \biggr]e^{-2r}
      + \frac{\epsilon_d^2}{\mathcal{K}(\Omega)\sin^2\phi_{\mathrm{LO}}}
    \right\} \,.
\end{eqnarray}
Thus, we obtaine a well-known result~\cite{02a1KiLeMaThVy} that, using an optimal squeezing angle, the quantum noise spectral density can be reduced by the squeezing factor $e^{-2r}$ in comparison with the vacuum input case. Note, however, that the noise contribution due to optical losses remains unchanged. Concerning the homodyne angle $\phi_{\mathrm{LO}}$, we use again the classical optimization, setting
\begin{equation}
\label{vr_sqz_zeta}
  \phi_{\mathrm{LO}} = \frac{\pi}{2} \,.
\end{equation}
In this case, the sum noise power (double-sided) spectral density and the optimal squeezing angle are equal to
\begin{equation}
\label{eq:vr_sqz_3}
  S^h(\Omega) = \frac{2\hbar}{ML^2\Omega^2}\left[
      \frac{e^{-2r} + \epsilon_d^2}{\mathcal{K}(\Omega)} + \mathcal{K}(\Omega)e^{-2r}
    \right]
\end{equation}
and
\begin{equation}
\label{eq:vr_sqz_theta}
  \tan\theta =  \mathcal{K}(\Omega) \,.
\end{equation}

The sum quantum noise power (double-sided) spectral density~\eqref{eq:vr_sqz_3} is plotted
in Figure~\ref{fig:LIGO_var} for the ideal lossless case and for
$\eta_d=0.95$ (dotted line). In both cases, the optical power and the squeezing
factor are equal to $J = J_{\mathrm{aLIGO}}$ and $e^{2r}=10$, respectively.

\epubtkImage{fig35.png}{
\begin{figure}[htb]
  \includegraphics[width=.48\textwidth]{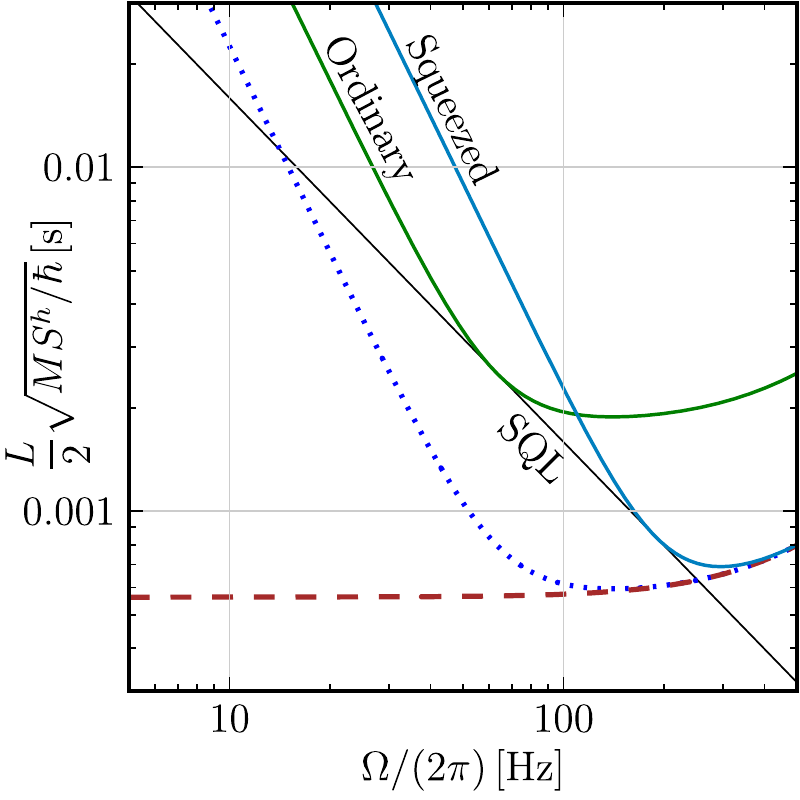}\hfill\includegraphics[width=.48\textwidth]{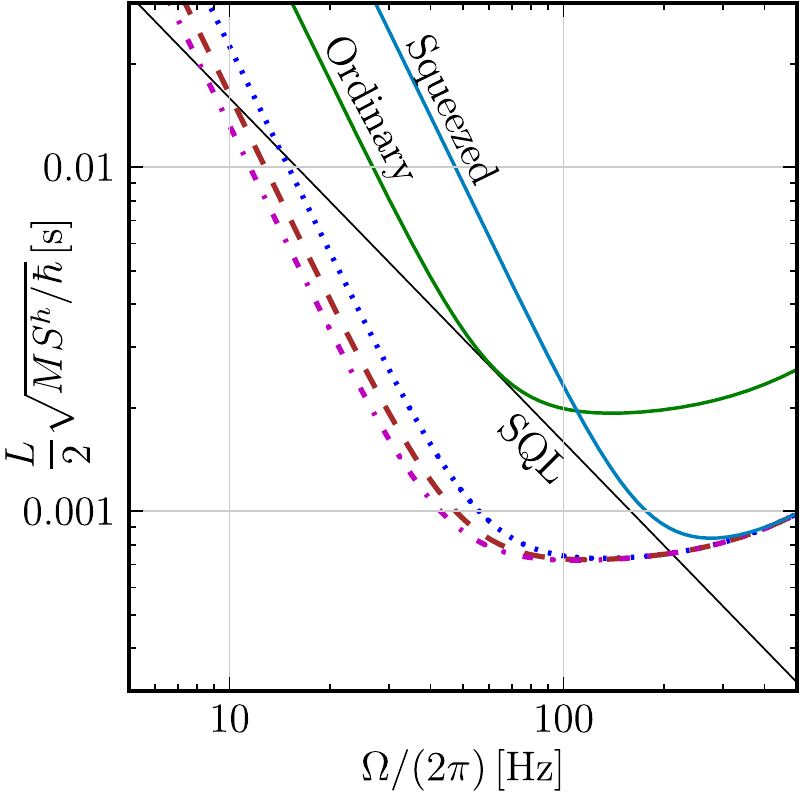}
  \caption{Examples of the sum quantum noise power (double-sided) spectral densities of the resonance-tuned interferometers with frequency-dependent squeezing and/or homodyne angles. Left: no optical losses, right: with optical losses, $\eta_d=0.95$. `Ordinary': no squeezing, $\phi_{\mathrm{LO}}=\pi/2$. `Squeezed': 10~dB squeezing, $\theta=0$, $\phi_{\mathrm{LO}}=\pi/2$ (these two plots are provided for comparison). Dots [pre-filtering, Eq.~\eqref{eq:vr_sqz_3}]: 10~dB squeezing, $\phi_{\mathrm{LO}}=\pi/2$, frequency-dependent squeezing angle. Dashes [post-filtering, Eq.~\eqref{eq:vr_post_opt}]: 10~dB squeezing, $\theta=0$, frequency-dependent homodyne angle. Dash-dots [pre- and post-filtering, Eq.~\eqref{eq:vr_opt_3}]: 10~dB squeezing, frequency-dependent squeeze and homodyne angles. For all plots, $J=J_{\mathrm{aLIGO}}$ and $\gamma=2\pi\times500\mathrm{\ s}^{-1}$.}
\label{fig:LIGO_var}
\end{figure}}

\paragraph*{Frequency dependent homodyne angle.}
\label{sec:var_post}

Suppose now that the squeezing angle corresponds to the classical optimization:
\begin{equation}
\label{eq:vr_post_theta}
  \theta = 0 \,
\end{equation}
and minimize the resulting sum noise spectral density:
\begin{equation}
  S^h(\Omega) = \frac{2\hbar}{ML^2\Omega^2}\left[
    \frac{\cosh2r + \sinh2r\cos2\phi_{\mathrm{LO}} + \epsilon_d^2}{\mathcal{K}(\Omega)\sin^2\phi_{\mathrm{LO}}}
   - 2e^{2r}\cot\phi_{\mathrm{LO}} + \mathcal{K}(\Omega)e^{2r}
  \right]
\end{equation}
with respect to $\phi_{\mathrm{LO}}$. The minimum is provided by the following dependence
\begin{equation}
\label{eq:vr_post_zeta}
  \cot\phi_{\mathrm{LO}} = \frac{\mathcal{K}(\Omega)}{1+\epsilon_d^2e^{-2r}}\,,
\end{equation}
and is equal to
\begin{equation}
\label{eq:vr_post_opt}
  S^h(\Omega) = \frac{2\hbar}{ML^2\Omega^2}\left[
    \frac{e^{-2r} + \epsilon_d^2}{\mathcal{K}(\Omega)}
    + \frac{\epsilon_d^2}{1 + \epsilon_d^2e^{-2r}}\,\mathcal{K}(\Omega)
  \right] .
\end{equation}
The sum quantum noise spectral density~\eqref{eq:vr_post_opt} is plotted in Figure~\ref{fig:LIGO_var} for the ideal lossless case and for $\eta_d=0.95$ (dashed lines).

Compare this spectral density with the one for the frequency-dependent squeezing angle (pre-filtering) case, see Eq.~\eqref{eq:vr_sqz_3}. The shot noise components in both cases are exactly equal to each other. Concerning the residual back-action noise, in the pre-filtering case it is limited by the available squeezing, while in the post-filtering case -- by the optical losses. In the latter case, were there no optical losses, the back-action noise could be removed completely, as shown in Figure~\ref{fig:LIGO_var}\,(left). For the parameters of the noise curves presented in Figure~\ref{fig:LIGO_var}\,(right), the post-filtering still has some advantage of about 40\% in the back-action noise amplitude $\sqrt{S}$.

Note that the required frequency dependences~\eqref{eq:vr_sqz_theta} and \eqref{eq:vr_post_zeta} in both cases are similar to each other (and become exactly equal to each other in the lossless case $\epsilon_d=0$). Therefore, similar setups can be used in both cases in order to create the necessary frequency dependences with about the same implementation cost. From this simple consideration, it is possible to conclude that pre-filtering is preferable if good squeezing is available, and the optical losses are relatively large, and vice versa. In particular, post-filtering can be used even without squeezing, $r=0$.

\paragraph*{Frequency dependent homodyne and squeezing angles.}
\label{sec:var_both}

And, finally, consider the most sophisticated configuration: double-filtering with both the homodyne angle $\phi_{\mathrm{LO}}$ and the squeezing angle $\theta$ being frequency dependent.

Concerning the squeezing angle, we can reuse Eqs.~(\ref{eq:vr_preopt_theta}) and (\ref{eq:vr_opt_2}). The minimum of the spectral density~\eqref{eq:vr_opt_2} in $\phi_{\mathrm{LO}}$ corresponds to
\begin{equation}
\label{vr_opt_zeta}
  \cot\phi_{\mathrm{LO}} = \frac{\mathcal{K}(\Omega)}{1 + \epsilon_d^2e^{2r}}
    \,,
\end{equation}
and is equal to
\begin{equation}
\label{eq:vr_opt_3}
  S^h(\Omega) = \frac{2\hbar}{ML^2\Omega^2}\left[
    \frac{e^{-2r} + \epsilon_d^2}{\mathcal{K}(\Omega)}
    + \frac{\epsilon_d^2}{1 + \epsilon_d^2e^{2r}}\,\mathcal{K}(\Omega)
    \right] .
\end{equation}
It also follows from Eqs.(\ref{eq:vr_preopt_theta}) and (\ref{vr_opt_zeta}) that the optimal squeezing angle in this case is given by
\begin{equation}
\label{eq:vr_opt_theta}
  \tan\theta =  \frac{\epsilon_d^2}{e^{-2r} + \epsilon_d^2}\,\mathcal{K}(\Omega) \,.
\end{equation}

It is easy to see that in the ideal lossless case the double-filtering configuration reduces to a post-filtering one. Really, if $\epsilon_d=0$, the spectral density~\eqref{eq:vr_opt_3} becomes exactly equal to that for the post-filtering case~\eqref{eq:vr_post_opt}, and the frequency dependent squeezing angle~\eqref{eq:vr_opt_theta} degenerates into a constant value~\eqref{eq:vr_post_theta}. However, if $\epsilon_d>0$, then the additional pre-filtering allows one to decrease more the residual back-action term. For example, if $e^{2r}=10$ and $\eta_d=0.95$ then the gain in the back-action noise amplitude $\sqrt{S}$ is equal to about 25\%.

We have plotted the sum quantum noise spectral density~\eqref{eq:vr_opt_3} in Figure~\ref{fig:LIGO_res}, right (dash-dots). This plot demonstrates the best sensitivity gain of about 3 in signal amplitude, which can be provided employing squeezing and filter cavities at the contemporary technological level.

Due to the presence of the residual back-action term in the spectral density~\eqref{eq:vr_opt_3}, there exists an optimal value of the coupling factor $\mathcal{K}(\Omega)$ (that is, the optical power) which provides the minimum to the sum quantum noise spectral density at any given frequency $\Omega$:
\begin{equation}
\label{vr_K_opt}
  \mathcal{K}(\Omega) = \frac{1}{\epsilon_de^r} + \epsilon_de^r \,,
\end{equation}
The minimum is equal to
\begin{equation}
\label{vr_loss_lim}
  S^h_{\mathrm{min}} = \frac{4\hbar}{ML^2\Omega^2}\,\epsilon_de^{-r} \,.
\end{equation}
This limitation is severe. The reasonably optimistic value of quantum efficiency $\eta_d=0.95$ that we use for our estimates corresponds to $\epsilon_d\approx0.23$. It means that without squeezing ($r=0$) one is only able to beat the SQL in amplitude by
\begin{equation}
  \xi_{\mathrm{min}} \equiv \sqrt{\frac{S^h_{\mathrm{min}}}{S^h_{\mathrm{SQL}}}}
    = \sqrt{\epsilon_d} \approx 0.5 \,.
\end{equation}
The gain can be improved using squeezing and, if $r\to\infty$ then, in principle, arbitrarily high sensitivity can be reached. But $\xi$ depends on $r$ only as $e^{-r/2}$, and for the 10~dB squeezing, only a modest value of
\begin{equation}
  \xi_{\mathrm{min}} \approx 0.27
\end{equation}
can be obtained.

In our particular case, the fact that the additional noise associated with the photodetector quantum inefficiency $\epsilon_d>0$ does not correlate with the quantum fluctuations of the light in the interferometer gives rise to this limit. This effect is universal for any kind of optical loss in the system, impairing the cross-correlation of the measurement and back-action noises and thus limiting the performance of the quantum measurement schemes, which rely on this cross-correlation.

Noteworthy is that Eq.~\eqref{eq:vr_opt_3} does not take into account optical losses in the filter cavities.  As we shall see below, the sensitivity degradation thereby depends on the ratio of the light absorption per bounce to the filter cavities length, $A_f/L_f$. Therefore, this method calls for long filter cavities. In particular, in the original paper~\cite{02a1KiLeMaThVy}, filter cavities with the same length as the main interferometer arm cavities (4\,km), placed side by side with them in the same vacuum tubes, were proposed. For such  long and expensive filter cavities, the influence of their losses indeed can be small. However, as we show below, in Section~\ref{sec:corr_real}, for the more practical short (up to tens of meters) filter cavities, optical losses thereof could be the main limiting factor in terms of sensitivity.

\paragraph*{Virtual rigidity for prototype interferometers.}
\label{sec:vr_short}

The optimization performed above can be viewed also in a different way, namely, as the minimization of the sum quantum noise spectral density of an ordinary interferometer with frequency-independent homodyne and squeezing angles, yet \emph{at some given frequency} $\Omega_0$. In Section~\ref{sec:toy_FI_correlation}, this kind of optimization was considered for a simple lossless system. It was shown capable of the narrow-band gain in sensitivity, similar to the one provided by the harmonic oscillator (thus the term `virtual rigidity').

This narrow-band gain could be more interesting not for the full-scale
GW detectors (where broadband optimization of the
sensitivity is required in most cases) but for smaller devices like the
10-m Hannover prototype interferometer~\cite{10m_site}, designed for
the development of the measurement methods with sub-SQL
sensitivity. Due to shorter arm length, the bandwidth $\gamma$ in
those devices is typically much larger than the mechanical frequencies
$\Omega$. If one takes, e.g., the power transmissivity value of
$T\gtrsim10^{-2}$ for the ITMs and length of arms equal to
$L\sim10\mathrm{\ m}$, then $\gamma\gtrsim10^5\mathrm{\ s}^{-1}$, which is
above the typical working frequencies band of such devices. In the
literature, this particular case is usually referred to as a \emph{bad
  cavity approximation}.

In this case, the coupling factor $\mathcal{K}(\Omega)$ can be approximated as:
\begin{equation}
\label{eq:short_calK}
  \mathcal{K}(\Omega) \approx \frac{\Omega_q^2}{\Omega^2} \,,
\end{equation}
where
\begin{equation}
\label{eq:short_Omega_q}
  \Omega_q^2 = \frac{2J}{\gamma} \,.
\end{equation}
Note that in this approximation, the noise spectral densities~\eqref{eq:vr_noises_x}, \eqref{eq:vr_noises_F} and \eqref{eq:vr_noises_xF} turn out to be frequency independent.

\epubtkImage{fig36.png}{
\begin{figure}[htbp]
  \centerline{
    \includegraphics[width=.48\textwidth]{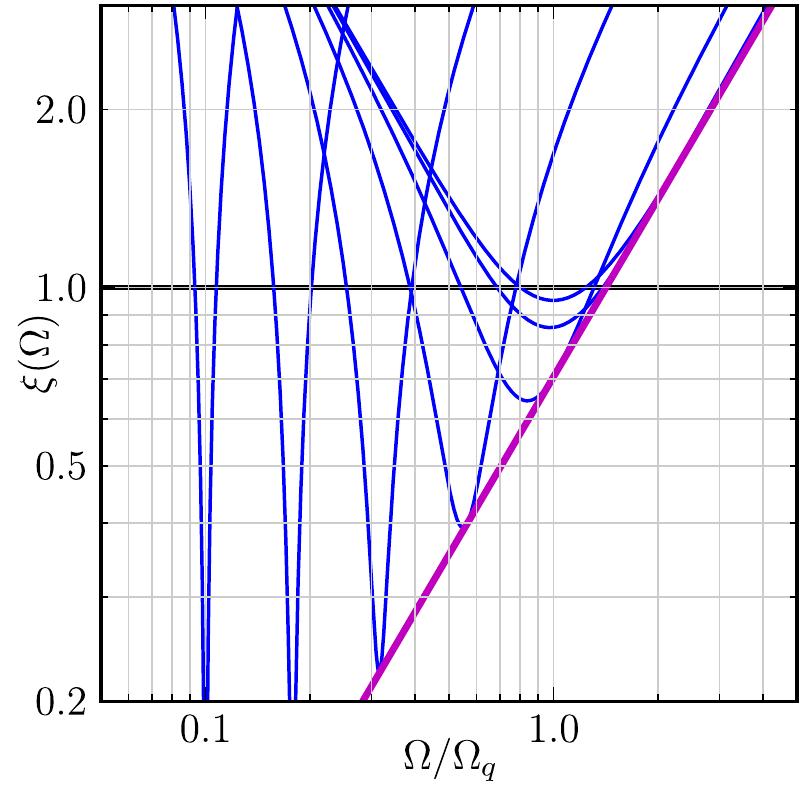}\hfill\includegraphics[width=.48\textwidth]{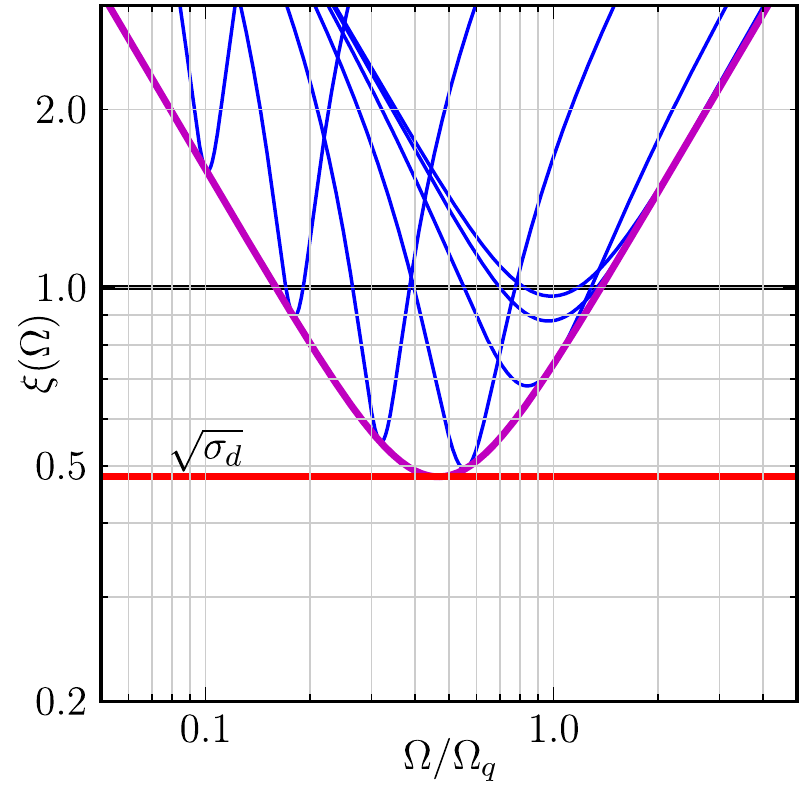}
  }
  \centerline{
    \includegraphics[width=.48\textwidth]{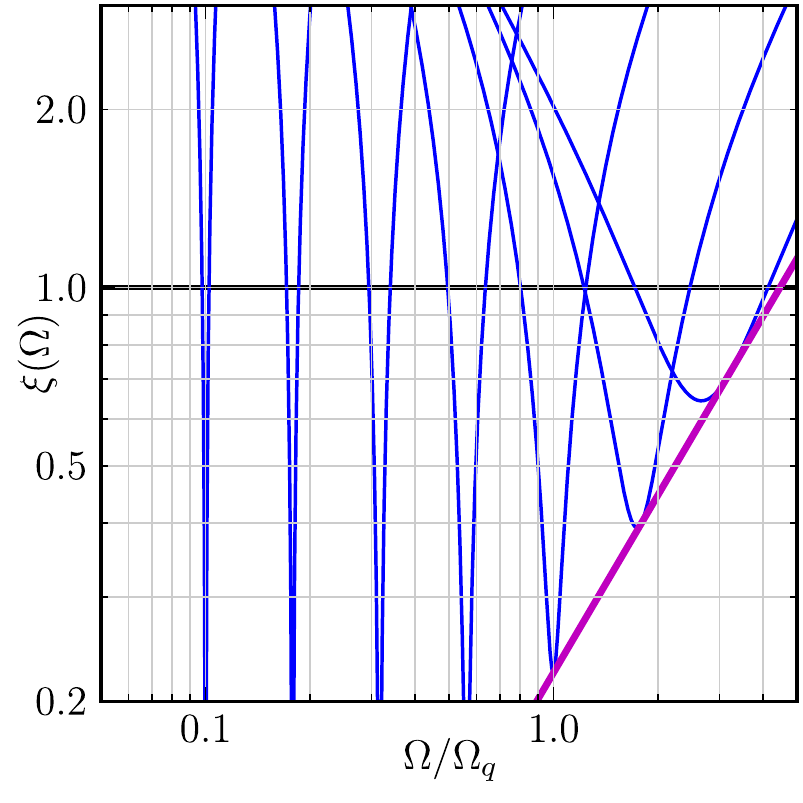}\hfill\includegraphics[width=.48\textwidth]{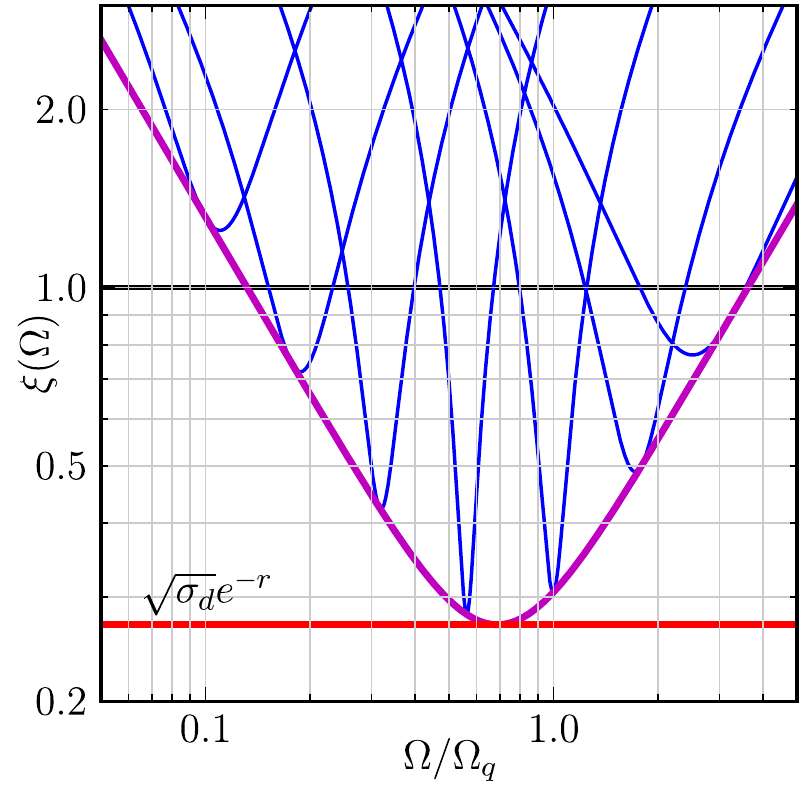}
  }
  \caption{Plots of the locally-optimized SQL beating factor $\xi(\Omega)$~\eqref{eq:xi2} of the interferometer with cross-correlated noises for the  ``bad cavity'' case $\Omega_0\ll\gamma$, for several different values of the optimization frequency $\Omega_0$ within the range $0.1\times\Omega_q\le\Omega_0\le\sqrt{10}\times\Omega_q$. Thick solid lines: the common envelopes of these plots; see Eq.~\eqref{vr_xi2_opt}. Left column: $\eta_d=1$; right column: $\eta_d=0.95$. Top row: no squeezing, $r=0$; bottom row: 10~dB squeezing, $e^{2r}=10$.}
\label{fig:short_vr_noises}
\end{figure}
}

In Figure~\ref{fig:short_vr_noises}, the SQL beating factor
\begin{equation}
\label{eq:xi2}
  \xi(\Omega) = \sqrt{\frac{S^h(\Omega)}{S^h_{\mathrm{SQL\,f.m.}}(\Omega)}}
\end{equation}
is plotted for the sum quantum noise spectral density $S^h(\Omega)$ with the following values of homodyne and squeezing angles
\begin{equation}
   \cot\phi_{\mathrm{LO}} = \frac{\mathcal{K}(\Omega_0)}{1 + \epsilon_d^2e^{2r}} \,, \qquad
   \tan\theta = \frac{\epsilon_d^2}{e^{-2r} + \epsilon_d^2}\,\mathcal{K}(\Omega_0) \,,
\end{equation}
and factoring in the  ``bad cavity''
condition~\eqref{eq:short_calK}. The four panes correspond to the
following four combinations: (upper left) no losses ($\eta_d=1$) and
no squeezing ($r=0$); (lower left) no losses ($\eta_d=1$) and 10~dB
squeezing ($e^{2r}=10$); (upper right) with losses ($\eta_d=0.95$) and
no squeezing ($r=0$); (lower right) with losses ($\eta_d=0.95$) and
10~dB squeezing($e^{2r}=10$). In each pane, the family of plots is
shown that corresponds to different values of the ratio
$\Omega_0/\Omega_q$, ranging from 0.1 to $\sqrt{10}$.

The minima of these plots form the common envelope, given by Eqs.~\eqref{eq:vr_opt_3} and \eqref{eq:short_calK}:
\begin{equation}
\label{vr_xi2_opt}
  \xi^2_{\mathrm{min}}(\Omega_0) = \frac{1}{2}\left[
    (e^{-2r} + \epsilon_d^2)\frac{\Omega_0^2}{\Omega_q^2}
    + \frac{\epsilon_d^2}{1 + \epsilon_d^2e^{2r}}\,\frac{\Omega_q^2}{\Omega_0^2}
    \right] ,
\end{equation}
which is also plotted in Figure~\ref{fig:short_vr_noises}. It is easy to see that in the ideal case of $\epsilon_d=0$, there is no limitation on the SQL beating factor, provided a sufficiently small ratio of $\Omega_0/\Omega_q$:
\begin{equation}
\label{vr_xi2_min0}
  \xi^2_{\mathrm{min}}(\Omega) = \frac{e^{-2r}}{2}\,\frac{\Omega_0^2}{\Omega_q^2} \,.
\end{equation}
However, if $\epsilon_d>0$, then function~\eqref{vr_xi2_opt} has a minimum in $\Omega_q$ at
\begin{equation}
  \Omega_q = \Omega_0\sqrt{\frac{1}{\epsilon_de^{r}} + \epsilon_de^{r}} \,,
\end{equation}
[compare with Eq.~\eqref{vr_K_opt}], equal to~\eqref{vr_loss_lim}.

\subsubsection{Filter cavities in GW interferometers}
\label{sec:corr_real}

\paragraph*{Input/output relations for the filter cavity.}

In essence, a filter cavity is an ordinary Fabry--P{\'e}rot cavity with one partly transparent input/output mirror. The technical problem of how to spatially separate the input and output beam can be solved in different ways. In the original paper~\cite{02a1KiLeMaThVy} the triangular cavities were considered. However, in this case, an additional mirror in each cavity is required, which adds to the optical loss per bounce. Another option is an ordinary linear cavity with additional optical circulator, which can be implemented, for example, by means of the polarization beamsplitter and Faraday rotator (note that while the typical polarization optics elements have much higher losses than the modern high-quality mirrors, the mirrors losses appear in the final expressions inflated by the filter cavity finesse).

In both cases, the filter cavity can be described by the input/output relation, which can be easily obtained from Eqs.~\eqref{FP_1_quad(a)} and \eqref{FP_1_quad(b)} by setting $\mathrm{E}_{c,s}=0$ (there is no classical pumping in the filter cavity and, therefore, there is no displacement sensitivity) and by some changes in the notations:
\begin{equation}
\label{eq:filter_io}
  \hat{\mathbf{o}}(\Omega)
  = \mathbb{R}_f(\Omega)\hat{\mathbf{i}}(\Omega) + \mathbb{T}_f(\Omega)\hat{\mathbf{q}}(\Omega)\,,
\end{equation}
where $\hat{\mathbf{i}}$ and $\hat{\mathbf{o}}$ are the two-photon quadrature amplitude vectors of the input and output beams, $\hat{\mathbf{q}}$ stands for noise fields entering the cavity due to optical losses (which are assumed to be in a vacuum state),
\begin{eqnarray}
  \mathbb{R}_f(\Omega) &=& \frac{1}{\mathcal{D}_f(\Omega)}
    \smatrix{\gamma_{f1}^2 - \gamma_{f2}^2 - \delta_f^2 + \Omega^2 + 2i\Omega\gamma_{f2}}
      {-2\gamma_{f1}\delta_f}{2\gamma_{f1}\delta_f}
      {\gamma_{f1}^2 - \gamma_{f2}^2 - \delta_f^2 + \Omega^2 + 2i\Omega\gamma_{f2}} ,
      \label{filter_bbR} \\\noalign{\smallskip}
  \mathbb{T}_f(\Omega) &=& \frac{2\sqrt{\gamma_{f1}\gamma_{f2}}}{\mathcal{D}_f(\Omega)}
    \smatrix{\gamma_f-i\Omega}{-\delta_f}{\delta_f}{\gamma_f-i\Omega} ,
      \label{filter_bbT} \\\noalign{\smallskip}
  \mathcal{D}_f(\Omega) &=& (\gamma_f-i\Omega)^2 + \delta_f^2 \,, \\
  \gamma_{f1} &=& \frac{cT_f}{4L_f} \\
  \gamma_{f2} &=& \frac{cA_f}{4L_f} \,,
\end{eqnarray}
where $T_f$ is the power transmittance of the input/output mirror, $A_f$ is the factor of power loss per bounce, $L_f$ is the filter cavity length,
\begin{equation}
  \gamma_f = \gamma_{f1} + \gamma_{f2}
\end{equation}
is its half-bandwidth, and $\delta_f$ is its detuning.

In order to demonstrate how the filter cavity works, consider the particular case of the lossless cavity. In this case,
\begin{equation}
  \label{filter_io_0}
  \hat{\mathbf{o}}(\Omega) = \mathbb{R}_f(\Omega)\hat{\mathbf{i}}(\Omega) \,,
\end{equation}
and the reflection matrix describes field amplitude rotations with the frequency-dependent rotation angle:
\begin{equation}
\label{eq:filter_bbR0}
  \mathbb{R}_f(\Omega)
    = \mathbb{P}\bigl[\theta_f(\Omega)\bigr]e^{i\beta_f(\Omega)} \,,
\end{equation}
where
\begin{eqnarray}
  \theta_f(\Omega) &=& \arctan\frac{2\gamma_f\delta_f}{\gamma_f^2 - \delta_f^2 + \Omega^2}
    \,, \label{theta_f} \\
  e^{i\beta_f(\Omega)} &=& \frac{|\mathcal{D}_f(\Omega)|}{\mathcal{D}_f(\Omega)} \,.
\end{eqnarray}
The phase factor $e^{i\beta_f(\Omega)}$ is irrelevant, for it does not appear in the final equations for the spectral densities.

Let us now analyze the influence of the filter cavities on the interferometer sensitivity in post- and pre-filtering variational schemes.
Start with the latter one. Suppose that the light, entering the interferometer from the signal port is in the squeezed state with fixed squeezing angle $\theta$ and squeezing factor $r$ and thus can be described by the following two-photon quadrature vector
\begin{equation}
  \hat{\boldsymbol{a}} = \mathbb{S}_{\sqz}[r,\theta]\hat{\boldsymbol{a}}^{\vac} \,,
\end{equation}
where the quadratures vector $\hat{\boldsymbol{a}}^{\vac}$ describes the vacuum state. After reflecting off the filter cavity, this light will be described with the following expression
\begin{equation}
  \hat{\mathbf{o}}(\Omega)
    = \mathbb{P}\bigl[\theta_f(\Omega)\bigr]\mathbb{S}_{\sqz}[r,\theta]\hat{\boldsymbol{a}}^{\vac}
        e^{i\beta_f(\Omega)}
  = \mathbb{S}_{\sqz}\bigl[r,\theta + \theta_f(\Omega)\bigr]\hat{\boldsymbol{a}}^{\vac'} \,,
\end{equation}
[see Eq.~\eqref{eq:SQZ_matrix_transform_def}], where $\hat{\boldsymbol{a}}^{vac'} = \mathbb{P}\bigl[\theta_f(\Omega)\bigr]\hat{\boldsymbol{a}}^{\vac}e^{i\beta_f(\Omega)}$ also describes the light field in a vacuum state. Thus, the pre-filtering indeed rotates the squeezing angle by a frequency-dependent angle $\theta_f(\Omega)$.

In a similar manner, we can consider the post-filtering schemes. Consider a homodyne detection scheme with losses, described by Eq.~\eqref{eq:homodyne_lossy}. Suppose that prior to detection, the light described by the quadrature vector $\hat{\boldsymbol{b}}$, reflects from the filter cavity. In this case, the photocurrent (in Fourier representation) is proportional to
\begin{eqnarray}
\label{HD_i_var}
  i_-(\Omega) &\propto& \boldsymbol{H}^{\mathsf{T}}[\phi_{\mathrm{LO}}]\bigl[
      \mathbb{P}\bigl[\theta_f(\Omega)\bigr]\hat{\boldsymbol{b}}(\Omega)
        e^{i\beta_f(\Omega)}
      + \epsilon_d\hat{\boldsymbol{n}}(\Omega)
    \bigr] \nonumber\\
  &=& \boldsymbol{H}^{\mathsf{T}}\bigl[\phi_{\mathrm{LO}} - \theta_f(\Omega)\bigr]
      \bigl[\hat{\boldsymbol{b}}(\Omega) + \epsilon_d\hat{\boldsymbol{n}}'(\Omega)\bigr]e^{i\beta_f(\Omega)}
    \,,
\end{eqnarray}
where $\hat{{\boldsymbol{n}}}' = \mathbb{P}\bigl[-\theta_f(\Omega)\bigr]\hat{{\boldsymbol{n}}}e^{-i\beta_f(\Omega)}$ again describes some new vacuum field. This formula  demonstrates that post-filtering is equivalent to the introduction of a frequency-dependent shift of the homodyne angle by $-\theta_f(\Omega)$.

It is easy to see that the necessary frequency dependencies of the homodyne and squeezing angles~\eqref{eq:vr_sqz_theta} or \eqref{eq:vr_post_zeta} (with the second-order polynomials in $\Omega^2$ in the r.h.s.\ denominators) cannot be implemented by the rotation angle~\eqref{theta_f} (with its first order in $\Omega^2$ polynomial in the r.h.s.\ denominator). As was shown in the paper~\cite{02a1KiLeMaThVy}, two filter cavities are required in both these cases. In the double pre- and post-filtering case, the total number of the filter cavities increases to four. Later it was also shown that, in principle, arbitrary frequency dependence of the homodyne and/or squeezing angle can be implemented, providing a sufficient number of filter cavities~\cite{Buonanno2004}.

However, in most cases, a more simple setup consisting of a single filter cavity might suffice. Really, the goal of the filter cavities is to compensate the back-action noise, which contributes significantly in the sum quantum noise only at low frequencies $\Omega\lesssim\Omega_q=\sqrt{2J/\gamma}$. However, when
\begin{equation}
\label{large_gamma}
  \gamma > J^{1/3},
\end{equation}
which is actually the case for the planned second and third generation GW detectors, the  factor $\mathcal{K}(\Omega)$ can be well approximated by Eq.~\eqref{eq:short_calK} in the low-frequency region. In such a case the single filter cavity can provide the necessary frequency dependence. Moreover, the second filter cavity could actually degrade sensitivity due to the additional optical losses it superinduces to the system.

Following this reasoning, we consider below two schemes, each based on a single filter cavity that realize pre-filtering and post-filtering, respectively.

\paragraph*{Single-filter cavity-based schemes.}
\label{sec:single_filter}

The schemes under consideration are shown in
Figure~\ref{fig:filter}. In the pre-filtering scheme drawn in the left
panel of Figure~\ref{fig:filter}, a squeezed light source emits
frequency-independent squeezed vacuum towards the filter cavity, where
it gets reflected, gaining a frequency-dependent phase shift
$\theta_f(\Omega)$, and then enters the dark port of the main
interferometer. The light going out of the dark port is detected by
the homodyne detector with fixed homodyne angle $\phi_{\mathrm{LO}}$ in the
usual way.

\epubtkImage{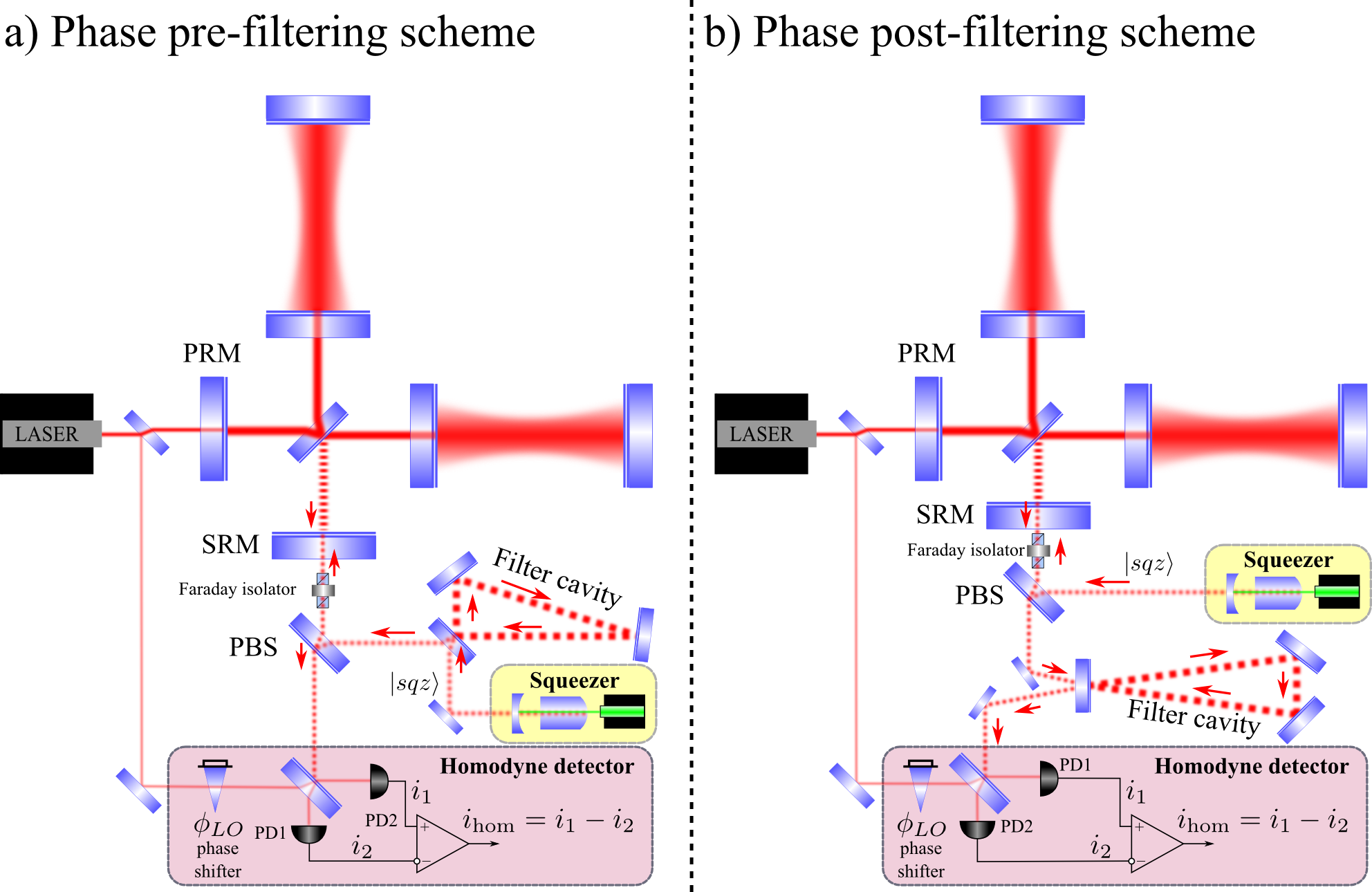}{
\begin{figure}[htb]
   \centerline{\includegraphics[width=\textwidth]{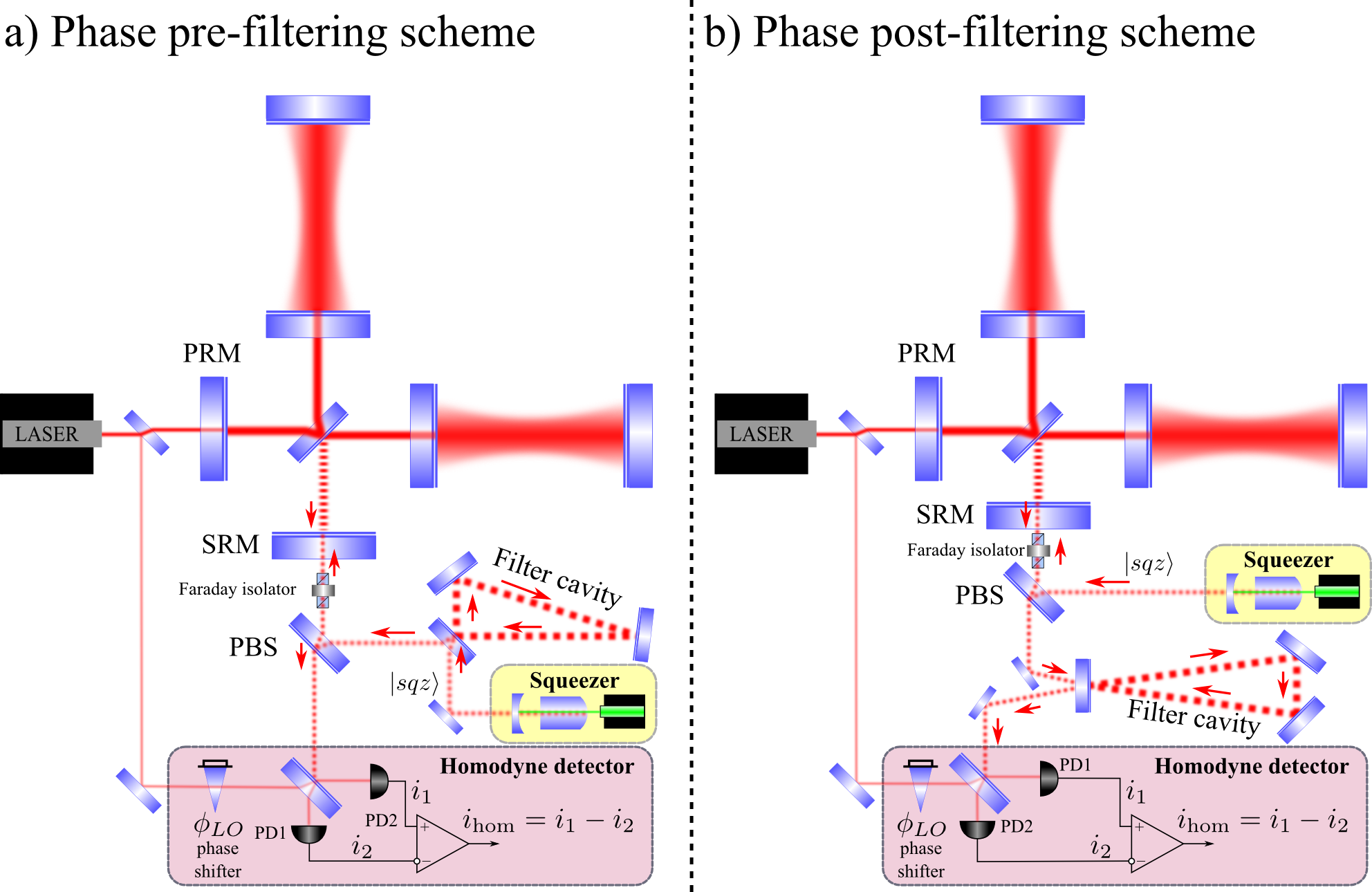}}
  \caption{Schemes of interferometer with the single filter cavity. \emph{Left:} In the pre-filtering scheme, squeezed vacuum from the squeezor is injected into the signal port of the interferometer after the reflection from the filter cavity; \emph{right:} in the post-filtering scheme, a squeezed vacuum first passes through the interferometer and, coming out, gets reflected from the filter cavity. In both cases the readout is performed by an ordinary homodyne detector with frequency independent homodyne angle $\phi_{\mathrm{LO}}$.}
\label{fig:filter}
\end{figure}}

Following the prescriptions of Section~\ref{sec:var_sqz}, we suppose the homodyne angle defined by Eq.~\eqref{vr_sqz_zeta}. The optimal squeezing angle should then be equal to zero at higher frequencies, see~\eqref{eq:vr_sqz_theta}. Taking into account that the phase shift introduced by the filter cavity goes to zero at high frequencies, we obtain that the squeezing angle $\theta$ of the input squeezed vacuum must be zero. Combining Eqs.~\eqref{eq:vr_X_sum} and \eqref{eq:filter_io} taking these assumptions into account, we obtain the following equation for the sum quantum noise of the pre-filtering scheme:
\begin{equation}
\label{eq:pre_X_sum}
  \hat{h}_{\mathrm{sum}}(\Omega) = -\frac{2}{L}\sqrt{\frac{\hbar}{2MJ\gamma}}\,
  \boldsymbol{H}^{\mathsf{T}}[\pi/2]\bigl\{
    (\gamma+i\Omega)\mathbb{K}(\Omega)\bigl[
      \mathbb{R}_f(\Omega)\mathbb{S}_{\sqz}[r,0]\hat{\boldsymbol{a}}^{\vac}(\Omega)
      + \mathbb{T}_f(\Omega)\hat{\mathbf{q}}(\Omega)
    \bigr]
    + (\gamma-i\Omega)\epsilon_d\hat{\boldsymbol{n}}(\Omega)
  \bigr\}\,,
\end{equation}
which yields the following expression for a power (double-sided) spectral density
\begin{equation}
\label{eq:pre_sum_noise}
  S^h(\Omega) = \frac{2\hbar}{ML^2\Omega^2\mathcal{K}(\Omega)}\bigl\{
    \boldsymbol{H}^{\mathsf{T}}[\pi/2]\mathbb{K}(\Omega)\bigl[
      \mathbb{R}_f(\Omega)\mathbb{S}_{\sqz}[2r,0]\mathbb{R}_f^\dagger(\Omega)
      + \mathbb{T}_f(\Omega)\mathbb{T}_f^\dagger(\Omega)
    \bigr]\mathbb{K}^\dagger(\Omega)\boldsymbol{H}[\pi/2]
    + \epsilon_d^2
  \bigr\} .
\end{equation}

In the ideal lossless filter cavity case, taking into account Eq.~\eqref{eq:filter_bbR0}, this spectral density can be simplified as follows:
\begin{equation}
\label{eq:pre_sum_noise0}
  S^h(\Omega) = \frac{2\hbar}{ML^2\Omega^2\mathcal{K}(\Omega)}\bigl[
    \boldsymbol{H}^{\mathsf{T}}[\pi/2]\mathbb{K}(\Omega)\mathbb{S}\bigl(2r,\theta_f(\Omega)\bigr)
      \mathbb{K}^\dagger(\Omega)\boldsymbol{H}[\pi/2]
    + \epsilon_d^2
  \bigr]
\end{equation}
[compare with Eq.~\eqref{eq:vr_sum_noise}]. In this case, the necessary frequency dependence of the squeezing angle~\eqref{eq:vr_sqz_theta} can be implemented by the following filter cavity parameters:
\begin{equation}
\label{filter0_pre}
  \gamma_f = \delta_f = \gamma_{f0} \,,
\end{equation}
where
\begin{equation}
\label{filter_g_f0}
  \gamma_{f0} = \sqrt{J/\gamma} \,.
\end{equation}

Along similar lines, the post-filtering scheme drawn in the right panel of Figure~\ref{fig:filter} can be considered. Here, the squeezed-vacuum produced by the squeezor first passes through the interferometer and then, coming out, gets reflected from the filter cavity, gaining a frequency-dependent phase shift, which is equivalent to introducing a frequency dependence into the homodyne angle, and then goes to the fixed angle homodyne detector. Taking into account that this equivalent homodyne angle at high frequencies has to be $\pi/2$, and that the phase shift introduced by the filter cavity goes to zero at high frequencies, we obtain that the real homodyne angle must also be $\pi/2$. Assuming that the squeezing angle is defined by Eq.~\eqref{eq:vr_post_theta} and again using Eqs.~\eqref{eq:vr_X_sum} and \eqref{eq:filter_io}, we obtain that the sum quantum noise and its power (double-sided) spectral density are equal to
\begin{eqnarray}
\label{eq:post_X_sum}
  \hat{h}_{\mathrm{sum}}(\Omega) &=& -\frac{2}{L}\sqrt{\frac{\hbar}{2MJ\gamma}}\,
    \frac{1}{\boldsymbol{H}^{\mathsf{T}}[\pi/2]\mathbb{R}_f(\Omega)\binom{0}{1}} \nonumber\\ \times
    \boldsymbol{H}^{\mathsf{T}}[\pi/2]\bigl\{
      (\gamma&+&i\Omega)
        \mathbb{R}_f(\Omega)\mathbb{K}(\Omega)\mathbb{S}_{\sqz}[r,0]\hat{\boldsymbol{a}}^{\vac}(\Omega)
      + (\gamma-i\Omega)\bigl[
          \mathbb{T}_f(\Omega)\hat{\mathbf{q}}(\Omega) + \epsilon_d\hat{\boldsymbol{n}}(\Omega)
        \bigr]
    \bigr\}
\end{eqnarray}
and
\begin{eqnarray}
\label{eq:post_sum_noise}
  S^h(\Omega) &=& \frac{2\hbar}{ML^2\Omega^2\mathcal{K}(\Omega)}
    \frac{1}{\left|\boldsymbol{H}^{\mathsf{T}}[\pi/2]\mathbb{R}_f(\Omega)\binom{0}{1}\right|^2}
    \nonumber\\ && \times
    \bigl\{
      \boldsymbol{H}^{\mathsf{T}}[\pi/2]\bigl[
        \mathbb{R}_f(\Omega)\mathbb{K}(\Omega)\mathbb{S}_{\sqz}[2r,0]\mathbb{K}^\dagger(\Omega)
          \mathbb{R}_f^\dagger(\Omega)
        + \mathbb{T}_f(\Omega)\mathbb{T}_f^\dagger(\Omega)
      \bigr]\boldsymbol{H}[\pi/2]
      + \epsilon_d^2
    \bigr\} .
\end{eqnarray}
In the ideal lossless filter cavity case, factoring in Eq.~\eqref{eq:filter_bbR0}, this spectral density takes a form similar to~\eqref{eq:vr_sum_noise}, but with the frequency-dependent homodyne angle:
\begin{equation}
\label{post_sum_noise0}
  S^h(\Omega) = \frac{2\hbar}{ML^2\Omega^2\mathcal{K}(\Omega)\sin^2\phi_{\mathrm{LO}}(\Omega)}
    \bigl[
      \boldsymbol{H}^{\mathsf{T}}\bigl[\phi_{\mathrm{LO}}(\Omega)\bigr]\mathbb{K}(\Omega)\mathbb{S}_{\sqz}[2r,0]
        \mathbb{K}^\dagger(\Omega)\boldsymbol{H}^{\mathsf{T}}\bigl[\phi_{\mathrm{LO}}(\Omega)\bigr]
      + \epsilon_d^2
    \bigr] ,
\end{equation}
where
\begin{equation}
  \phi_{\mathrm{LO}}(\Omega) = \pi/2-\theta_f(\Omega).
\end{equation}
The necessary frequency dependence~\eqref{eq:vr_post_zeta} of this effective homodyne angle can be implemented by the following parameters of the filter cavity:
\begin{equation}
\label{filter0_post}
  \gamma_f = \delta_f = \frac{\gamma_{f0}}{\sqrt{1+\epsilon_d^2e^{-2r}}} \,.
\end{equation}
Note that for reasonable values of loss and squeezing factors, these parameters differ only by a few percents from the ones for the pre-filtering.

It is easy to show that substitution of the conditions~\eqref{filter0_pre} and \eqref{filter0_post} into Eqs.~\eqref{eq:pre_sum_noise0} and \eqref{post_sum_noise0}, respectively, taking Condition~\eqref{large_gamma} into account, results in spectral densities for the ideal frequency dependent squeezing and homodyne angle, see Eqs.~\eqref{eq:vr_sqz_3} and \eqref{eq:vr_post_opt}.

In the general case of lossy filter cavities, the conditions (\ref{eq:vr_sqz_theta}) and~(\ref{eq:vr_post_zeta}) cannot be satisfied exactly by a single filter cavity at all frequencies. Therefore, the optimal filter cavity parameters should be determined using some integral sensitivity criterion, which will be considered at the end of this section.

However, it would be a reasonable assumption that the above consideration holds with good precision, if losses in the filter cavity are low compared to other optical losses in the system:
\begin{equation}
\label{filter_lim_1}
  \frac{\gamma_{f2}}{\gamma_f} \approx \frac{\gamma_{f2}}{\gamma_{f0}} \ll \epsilon_d^2 \,.
\end{equation}
This inequality can be rewritten as the following condition for the filter cavity specific losses:
\begin{equation}
\label{filter_lim_1_num}
  \frac{A_f}{L_f} \lesssim \frac{4\gamma_{f0}}{c}\,\epsilon_d^2 \,.
\end{equation}
In particular, for our standard parameters used for numerical estimates ($J=J_{\mathrm{aLIGO}}$, $\gamma=2\pi\times500\mathrm{\ s}^{-1}$, $\eta_d=0.95$), we obtain $\gamma_{f0}\approx280\mathrm{\ s}^{-1}$, and there should be
\begin{equation}
  \frac{A_f}{L_f} \lesssim 2\times10^{-7}\mathrm{\ m}^{-1},
\end{equation}
(the r.h.s.\ corresponds, in particular, to a 50~m filter cavity with the losses per bounce $A_f=10^{-5}$).

Another more crude limitation can be obtained from the condition that $\gamma_{f2}$ should be small compared to the filter cavity bandwidth $\gamma_f$:
\begin{equation}
\label{filter_lim_2}
  \frac{A_f}{L_f} \lesssim \frac{4\gamma_{f0}}{c} \,.
\end{equation}
Apparently, were it not the case, the filter cavity would just cease to work properly. For the same numerical values of $J$ and $\gamma$ as above, we obtain:
\begin{equation}
\label{filter_lim_2_num}
  \frac{A_f}{L_f} \lesssim 4\times10^{-6}\mathrm{\ m}^{-1},
\end{equation}
(for example, a very short 2.5~m filter cavity with $A_f=10^{-5}$ or 25~m cavity with $A_f=10^{-4}$).

\paragraph*{Numerical optimization of filter cavities.}
\label{sec:filter_snr_opt}

In the experiments devoted to detection of small forces, and, in particular, in the GW detection experiments, the main integral sensitivity measure is the probability to detect some calibrated signal. This probability, in turn, depends on the matched filtered SNR defined as
\begin{equation}
\label{6:snr}
  \rho^2= \intinfty\frac{|h_{s}(\Omega)|^2}{S^h(\Omega)}\,\frac{d\Omega}{2\pi}
\end{equation}
with $h_{s}(\Omega)$ the spectrum of this calibrated signal.

In the low and medium frequency range, where back-action noise dominates, and wherein our interest is focused, the most probable source of signal is the gravitational radiation of the inspiraling binary systems of compact objects such as neutron stars and/or black holes~\cite{lrr-2009-2,lrr-2006-6}. In this case, the SNR is equal to (see~\cite{Flanagan1998})
\begin{equation}
\label{snr_nsns}
  \rho^2 = k_0\int_0^{2\pi f_{\mathrm{max}}}\frac{\Omega^{-7/3}}{S^h(\Omega)}\,
    \frac{d\Omega}{2\pi} \,,
\end{equation}
where $k_0$ is a factor that does not depend on the interferometer parameters, and $f_{\mathrm{max}}$ is the cut-off frequency that depends on the binary system components' masses. In particular, for neutron stars with masses equal to 1.4 solar mass, $f_{\mathrm{max}}\approx1.5\mathrm{\ kHz}$.

Since our goal here is not the maximal value of the SNR itself, but rather the relative sensitivity gain offered by the filter cavity, and the corresponding optimal parameters $\gamma_{f1}$ and $\delta_f$, providing this gain, we choose to normalize the SNR by the value corresponding to the ordinary interferometer (without the filter cavities):
\begin{equation}
\label{filter_rho_0}
  \rho_0^2 = k_0\int_0^{2\pi f_{\mathrm{max}}}\frac{\Omega^{-7/3}}{S_0^h(\Omega)}\,
    \frac{d\Omega}{2\pi} \,,
\end{equation}
with power (double-sided) spectral density
\begin{equation}
  S_0^h(\Omega) = \frac{2\hbar}{ML^2\Omega^2}
    \left[\frac{1+\epsilon_d^2}{\mathcal{K}(\Omega)} + \mathcal{K}(\Omega)\right]
\end{equation}
[see Eq.~\eqref{eq:S_X_Caves}].

We optimized numerically the ratio $\rho^2/\rho_0^2$, with filter cavity half-bandwidth $\gamma_{f1}$ and detuning $\delta_f$ as the optimization parameters, for the values of the specific loss factor $A_f/L_f$ ranging from $10^{-9}$ (e.g., very long 10~km filter cavity with $A_f=10^{-5}$) to $10^{-5}$ (e.g., 10~m filter cavity with $A_f=10^{-4}$). Concerning the main interferometer parameters, we used the same values as in all our previous examples, namely, $J=J_{\mathrm{aLIGO}}$, $\gamma=2\pi\times500\mathrm{\ s}^{-1}$, and $\eta_d=0.95$.

The results of the optimization are shown in
Figure~\ref{fig:filter_opt}. In the left pane, the optimal values of
the filter cavity parameters $\gamma_{f1}$ and $\delta_f$ are plotted,
and in the right one the corresponding optimized values of the
SNR are. It follows from these plots that the optimal values of
$\gamma_{f1}$ and $\delta_f$ are virtually the same as $\gamma_{f0}$,
while the specific loss factor $A_f/L_f$ satisfies the condition
\eqref{filter_lim_1}, and starts to deviate sensibly from
$\gamma_{f0}$ only when $A_f/L_f$ approaches the limit
\eqref{filter_lim_2}. Actually, for such high values of specific
losses, the filter cavities only degrade the sensitivity, and the
optimization algorithm effectively turns them off, switching to the
ordinary frequency-independent squeezing regime (see the right-most
part of the right pane).

\epubtkImage{fig38.png}{
\begin{figure}[htbp]
  \includegraphics[width=.48\textwidth]{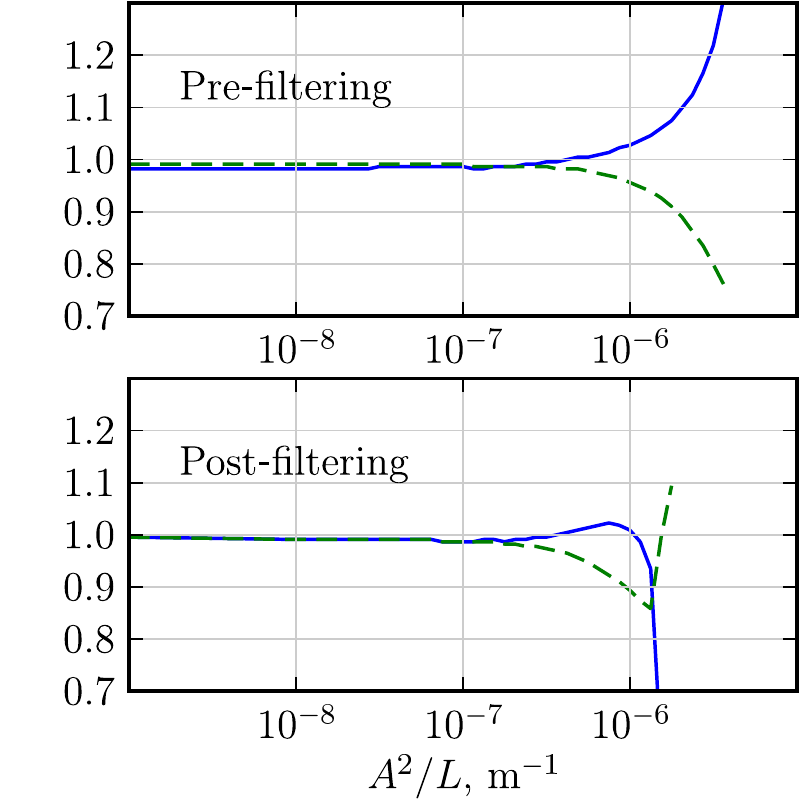}\hfill\includegraphics[width=.48\textwidth]{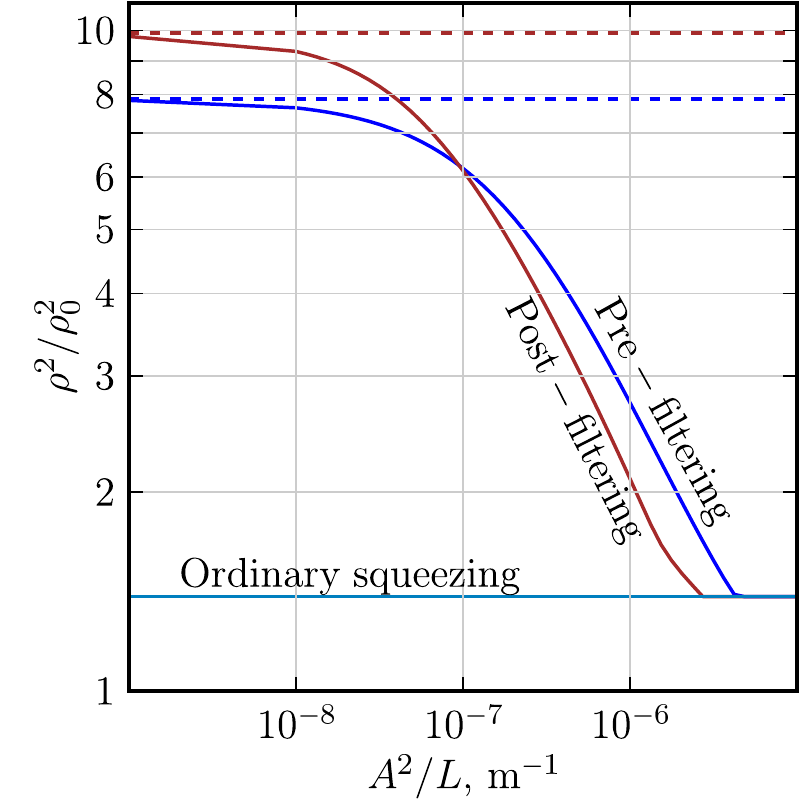}
  \caption{Left: Numerically-optimized filter-cavity parameters for a single cavity based pre- and post-filtering schemes: half-bandwidth $\gamma_{f1}$ (solid lines) and detuning $\delta_f$ (dashed lines), normalized by $\gamma_{f0}$ [see Eq.~\eqref{filter_g_f0}], as functions of the filter cavity specific losses $A_f/L_f$. Right: the corresponding optimal SNRs, normalized by the SNR for the ordinary interferometer [see Eq.~\eqref{filter_rho_0}]. Dashed lines: the normalized SNRs for the ideal frequency-dependent squeeze and homodyne angle cases, see Eqs.~\eqref{eq:vr_sqz_3} and \eqref{eq:vr_post_opt}. `Ordinary squeezing': frequency-independent 10~dB squeezing with $\theta=0$. In all cases, $J=J_{\mathrm{aLIGO}}$, $\gamma=2\pi\times500\mathrm{\ s}^{-1}$, and $\eta_d=0.95$.}
\label{fig:filter_opt}
\end{figure}}

It also follows from these plots that post-filtering provides slightly better sensitivity, if the optical losses in the filter cavity are low, while the pre-filtering has some advantage in the high-losses scenario. This difference can be explained in the following way~\cite{10a1Kh}. The post-filtration effectively rotates the homodyne angle from $\phi_{\mathrm{LO}}=\pi/2$ (phase quadrature) at high frequencies to $\phi_{\mathrm{LO}}\to0$ (amplitude quadrature) at low frequencies, in order to measure the back-action noise, which dominates the low frequencies. As a result, the optomechanical transfer function reduces at low frequencies, emphasizing all noises introduced after the interferometer [see the factor $\sin^2\phi_{\mathrm{LO}}(\Omega)$ in the denominator of Eq.~\eqref{post_sum_noise0}]. In the pre-filtering case there is no such effect, for the value of $\phi_{\mathrm{LO}}=\pi/2$, corresponding to the maximum of the optomechanical transfer function, holds for all frequencies (the squeezing angle got rotated instead).

The optimized sum quantum noise power (double-sided) spectral densities are plotted in
Figure~\ref{fig:LIGO_filter} for several typical values of the
specific loss factor, and for the same values of the rest of the
parameters, as in Figure~\ref{fig:filter_opt}. For comparison, the
spectral densities for the ideal frequency-dependent squeezing angle
Eqs.~\eqref{eq:vr_sqz_3} and homodyne angle~\eqref{eq:vr_post_opt} are
also shown. These plots clearly demonstrate that providing
sufficiently-low optical losses (say, $A_f/L_f\lesssim10^{-8}$), the
single filter cavity based schemes can provide virtually the same
result as the abstract ones with the ideal frequency dependence for
squeezing or homodyne angles.

\epubtkImage{fig39.png}{
\begin{figure}[htb]
  \includegraphics[width=.48\textwidth]{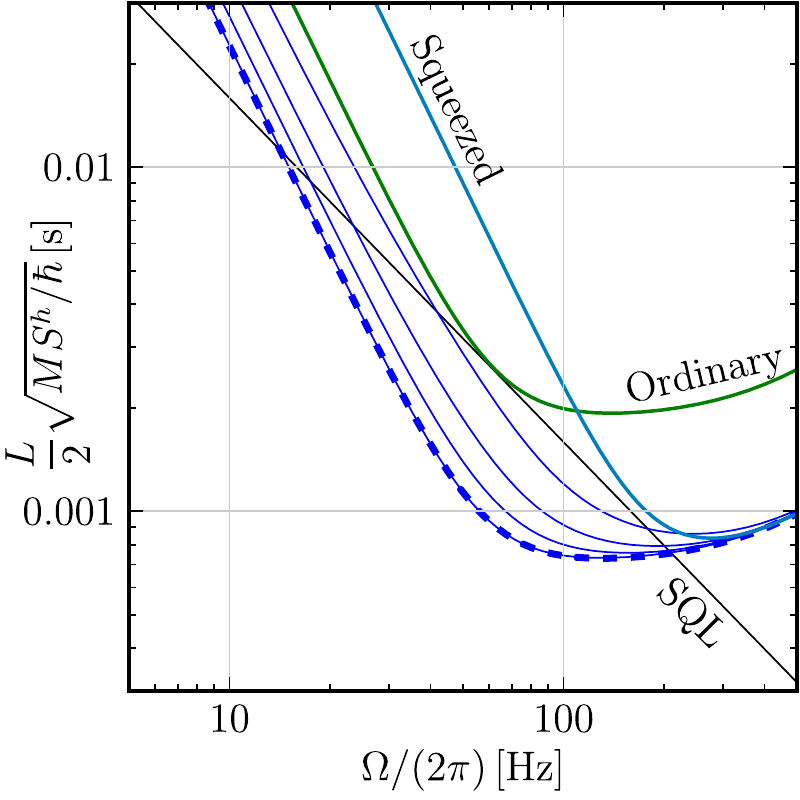}\hfill\includegraphics[width=.48\textwidth]{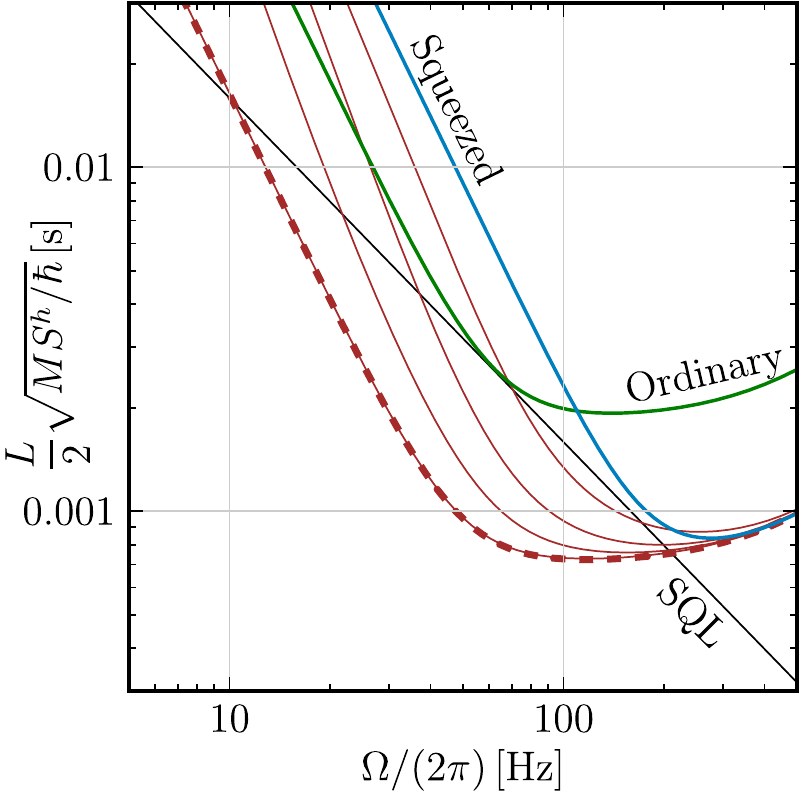}
  \caption{Examples of the sum quantum noise power (double-sided)
    spectral densities of the resonance-tuned interferometers with the
    single filter cavity based pre- and post-filtering. \emph{Left:}
    pre-filtering, see Figure~\ref{fig:filter}\,(left); dashes --
    10~dB squeezing, $\phi_{\mathrm{LO}}=\pi/2$, ideal
    frequency-dependent squeezing angle~\eqref{eq:vr_sqz_theta}; thin
    solid -- 10~dB squeezing, $\phi_{\mathrm{LO}}=\pi/2$, numerically-optimized lossy pre-filtering cavity with $A_f/L_f =
    10^{-9},\ 10^{-7}\,\ 10^{-6.5}\ \mbox{and}\ 10^{-6}$. \emph{Right:} post-filtering, see Figure~\ref{fig:filter}\,(right); dashes: 10~dB squeezing, $\theta=0$, ideal frequency-dependent homodyne angle~\eqref{eq:vr_post_zeta}; thin solid -- 10~dB squeezing, $\theta=0$, numerically optimized lossy post-filtering cavity with $A_f/L_f = 10^{-9},\ 10^{-8}\ \mbox{and}\ 10^{-7}$. In both panes (for the comparison): `Ordinary' -- no squeezing, $\phi_{\mathrm{LO}}=\pi/2$; `Squeezed': 10~dB squeezing, $\theta=0$, $\phi_{\mathrm{LO}}=\pi/2$}
\label{fig:LIGO_filter}
\end{figure}}

\subsection{Quantum speedmeter}
\label{sec:speedmeter}

\subsubsection{Quantum speedmeter topologies}

A quantum speedmeter epitomizes the approach to the broadband SQL beating, in some sense, opposite to the one based on the quantum noises cross-correlation tailoring with filter cavities, considered above. Here, instead of fitting the quantum noise spectral dependence to the Fabry--P{\'e}rot--Michelson interferometer optomechanical coupling factor~\eqref{eq:varK}, the interferometer topology is modified in such a way as to mold the new optomechanical coupling factor $\mathcal{K}_{\mathrm{SM}}(\Omega)$ so that it turns out frequency-independent in the low- and medium-frequency range, thus making the frequency-dependent cross-correlation not necessary.

The general approach to speed measurement is to use pairs of position measurements separated by a time delay $\tau\lesssim 1/\Omega$, where $\Omega$ is the characteristic signal frequency (cf.\ the simplified consideration in Section~\ref{sec:toy_speedmeter}). Ideally, the successive measurements should be coherent, i.e., they should be performed by the same photons. In effect, the velocity $v$ of the test mass is measured in this way, which gives the necessary frequency dependence of the $\mathcal{K}_{\mathrm{SM}}(\Omega)$.

In Section~\ref{sec:toy_speedmeter}, we have considered the simplest
toy scheme that implements this principle and which was first
proposed by Braginsky and Khalili in~\cite{90a1BrKh}. Also in this
paper, a modified version of this scheme, called the \emph{sloshing-cavity
  speedmeter}, was proposed. This version uses two coupled resonators
(e.g., microwave ones), as shown in Figure~\ref{figSM2}\,(left), one
of which (2), the \emph{sloshing cavity}, is pumped on resonance
through the input waveguide, so that another one (1) becomes excited
at its eigenfrequency $\omega_e$. The eigenfrequency of resonator~1
is modulated by the position $x$ of the test mass and puts a voltage
signal proportional to position $x$ into resonator~2, and a voltage
signal proportional to velocity $dx/dt$ into resonator~1. The velocity
signal flows from resonator~1 into an output waveguide, from which it
is monitored. One can understand the production of this velocity
signal as follows. The coupling between the resonators causes voltage
signals to slosh periodically from one resonator to the other at
frequency $\Omega$. After each cycle of sloshing, the sign of the
signal is reversed, so the net signal in resonator~1 is proportional
to the difference of the position at times $t$ and $t+2\pi/\Omega$,
thus implementing the same principle of the double position
measurement.

\epubtkImage{fig40.png}{
\begin{figure}[htbp]
\centerline{
  \includegraphics[width=0.38\textwidth]{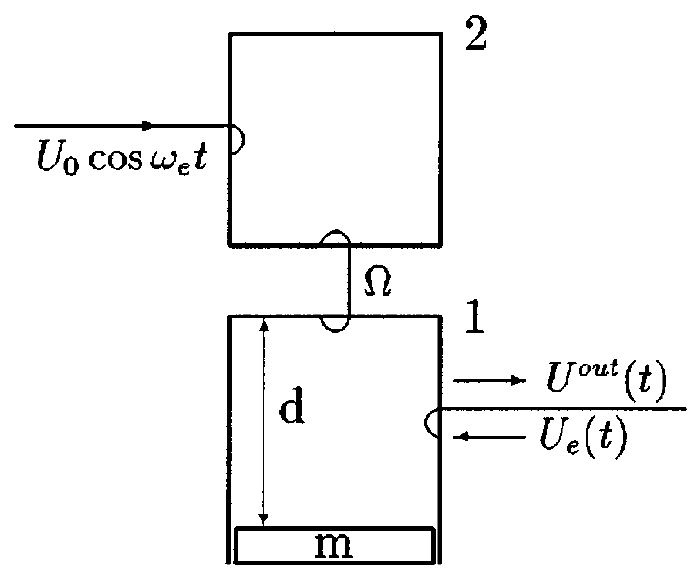}\hfill
  \includegraphics[width=0.58\textwidth]{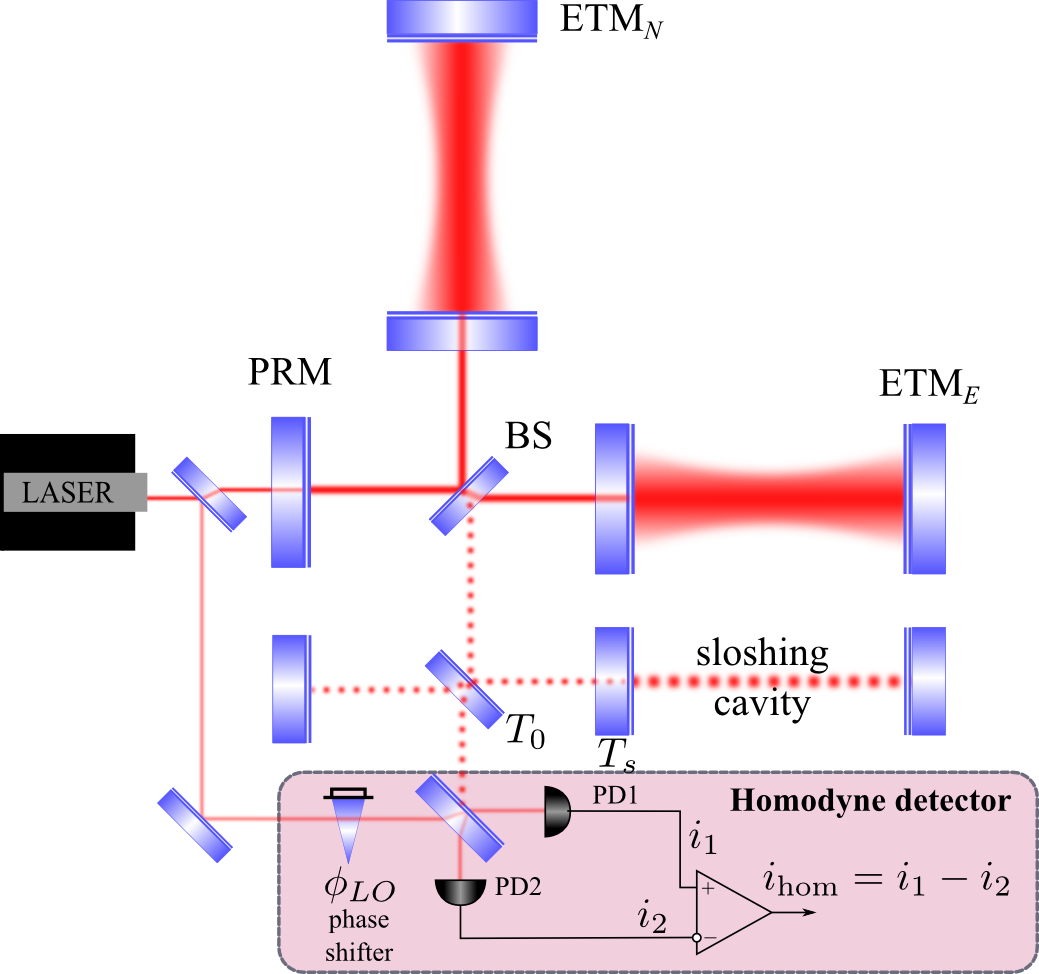}
}
  \caption{\emph{Left:} schematic diagram of the microwave speedmeter on coupled cavities as given in~\cite{90a1BrKh}. \emph{Right:} optical version of coupled-cavities speedmeter proposed in~\cite{Purdue2002}.}
\label{figSM2}
\end{figure}}

Later, the optical version of the sloshing-cavity speedmeter scheme suitable for large-scale laser GW detectors was developed~\cite{00a1BrGoKhTh, Purdue2001, Purdue2002}. The most elaborated variant proposed in~\cite{Purdue2002} is shown in Figure~\ref{figSM2}\,(right). Here, the differential mode of a Michelson interferometer serves as the resonator 1 of the initial scheme of~\cite{90a1BrKh}, and an additional kilometer-scale Fabry--P{\'e}rot cavity -- as the resonator 2, thus making a practical interferometer configuration.

\epubtkImage{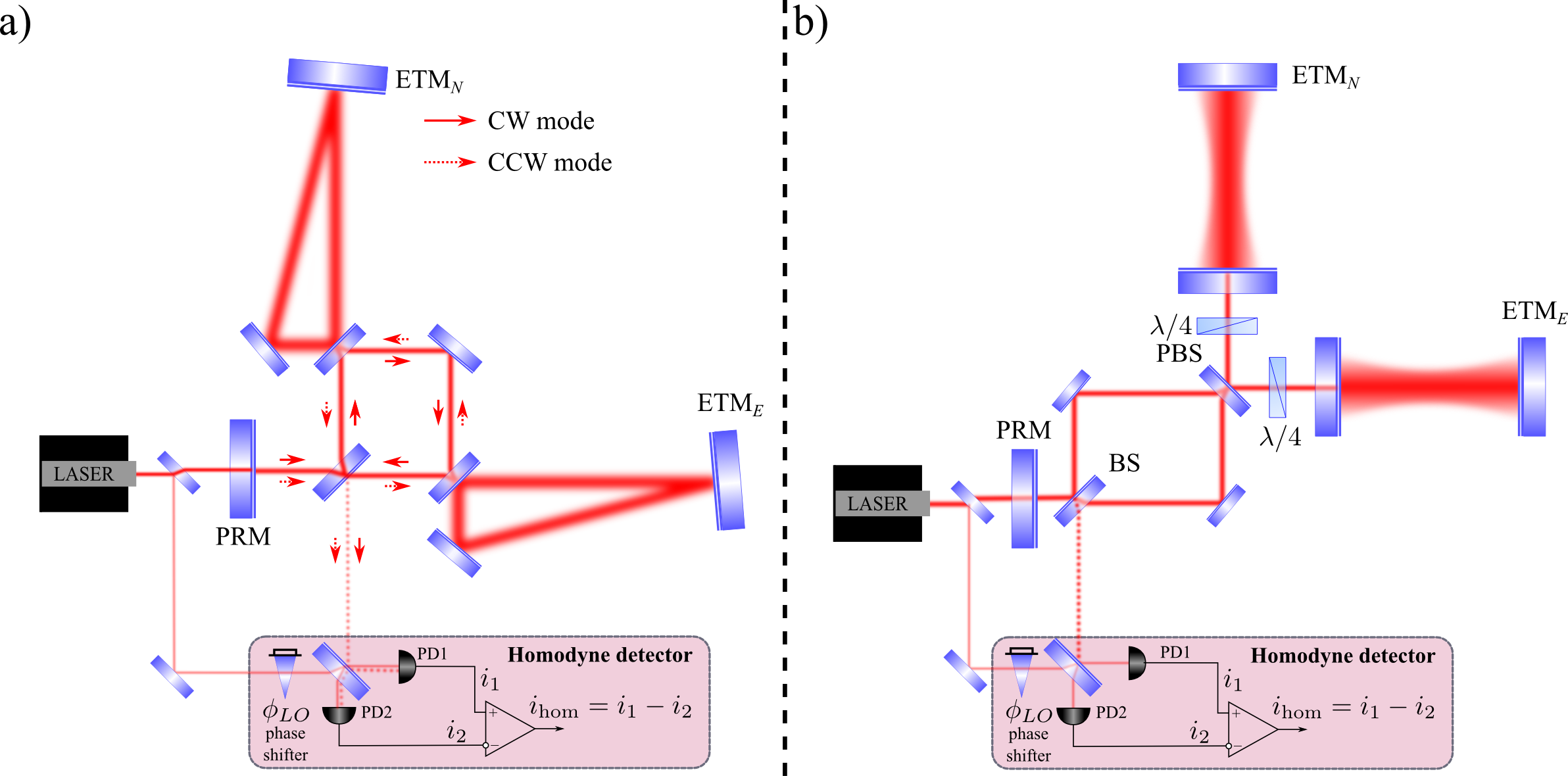}{
\begin{figure}[htbp]
\centerline{\includegraphics[width=\textwidth]{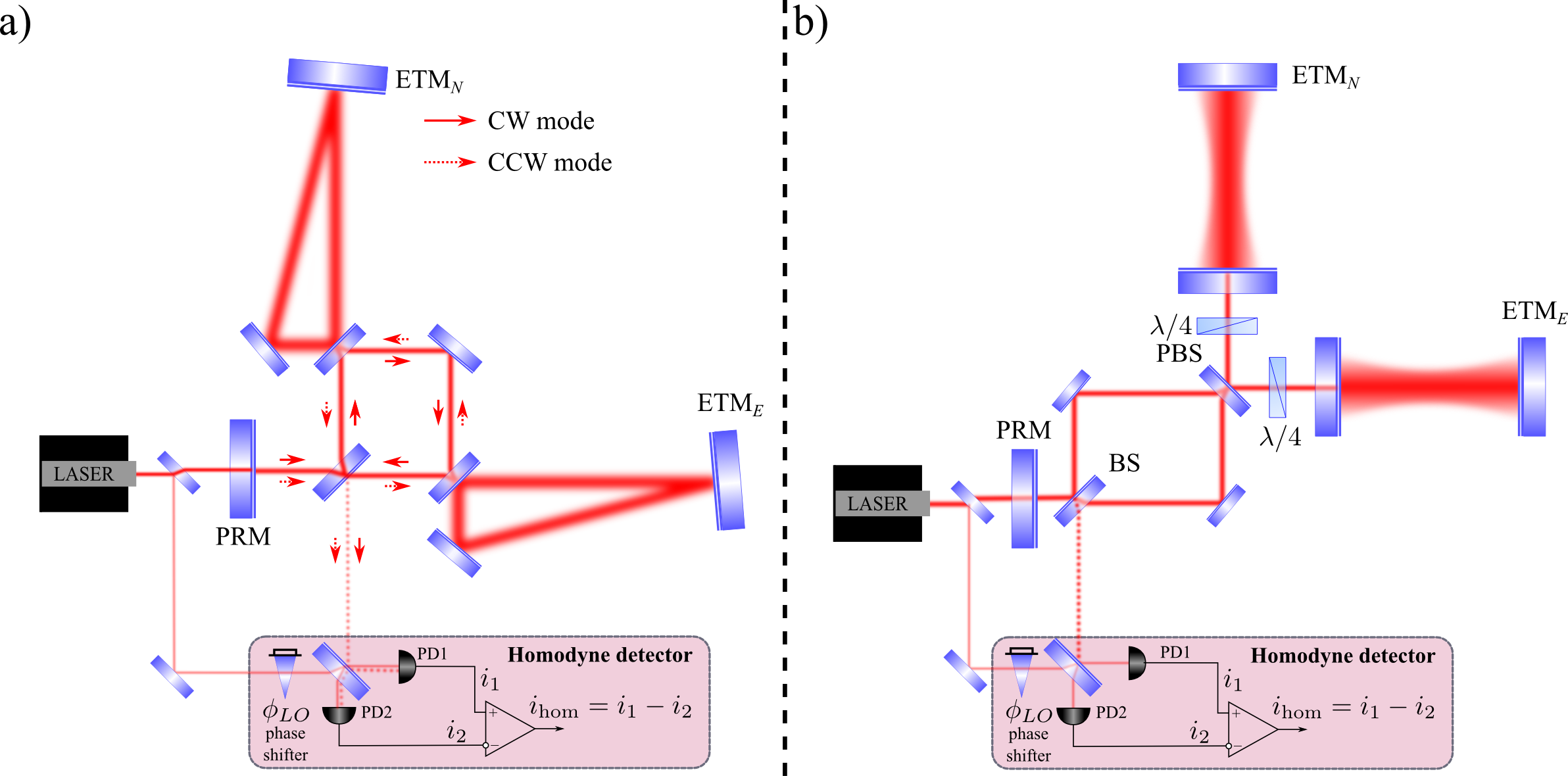}}
  \caption{Two possible optical realizations of zero area Sagnac speedmeter.
    \emph{Left panel:} The ring cavities can be used to spatially separate the ingoing fields from the  outgoing ones, in order to redirect output light from one arm to another~\cite{Chen2002}.
    \emph{Right panel:} The same goal can be achieved using an optical circulator consisting of the polarization beamsplitter (PBS) and two $\lambda/4$-plates~\cite{02a2Kh, 04a1Da}.
  }
\label{fig:Sagnacfig}
\end{figure}}

In parallel, it was realized by Chen and Khalili~\cite{Chen2002, 02a2Kh} that the zero area Sagnac interferometer~\cite{Byer1996, Byer1999, Byer2000} actually implements the initial double-measurement variant of the quantum speedmeter, shown in Figure~\ref{fig:speedmeter}. Further analysis with account for optical losses was performed in~\cite{04a1Da} and with detuned signal-recycling in~\cite{MuellerEbhardt2009}. Suggested configurations are pictured in Figure~\ref{fig:Sagnacfig}. The core idea is that light from the laser gets split by the beamsplitter (BS) and directed to Fabry--P\'erot cavities in the arms, exactly as in conventional Fabry--P{\'e}rot--Michelson interferometers. However, after it leaves the cavity, it does not go back to the beamsplitter, but rather enters the cavity in the other arm, and only afterwards returns to the beamsplitter, and finally to the photo detector at the dark port. The scheme of~\cite{Chen2002} uses ring Fabry--P{\'e}rot cavities in the arms to spatially separate ingoing and outgoing light beams to redirect the light leaving the first arm to the second one evading the output beamsplitter. The variant analyzed in~\cite{02a2Kh, 04a1Da} uses polarized optics for the same purposes: light beams after ordinary beamsplitter, having linear (e.g., vertical) polarization, pass through the polarized beamsplitter (PBS), then meet the $\lambda/4$ plates that transform their linear polarization into a circular one, and then enter the Fabry--P{\'e}rot  cavity. After reflection from the Fabry--P{\'e}rot  cavity, light passes through a $\lambda/4$-plate again, changing its polarization again to linear, but orthogonal to the initial one. As a result, the PBS reflects it and redirects to another arm of the interferometer where it passes through the same stages, restoring finally the initial polarization and comes out of the interferometer. With the exception of the implementation method for this round-robin pass of the light through the interferometer, both schemes have the same performance, and the same appellation \emph{Sagnac speedmeter} will be used for them below.

Visiting both arms, counter propagating light beams acquire phase shifts proportional to a sum of end mirrors displacements of both cavities taken with time delay equal to average single cavity storage time $\tau_{\mathrm{arm}}$:
\begin{equation}
    \delta\phi_R \propto x_N(t)+x_E(t+\tau_{\mathrm{arm}})\,,\quad
    \delta\phi_L \propto x_E(t)+x_N(t+\tau_{\mathrm{arm}})\,.
\end{equation}
After recombining at the beamsplitter and photo detection the output signal will be proportional to the phase difference of clockwise (R) and counter clockwise (L) propagating light beams:
\begin{equation}
\label{phi_speedmeter}
 \delta\phi_R - \delta\phi_L \propto [x_N(t)-x_N(t+\tau_{\mathrm{arm}})]
   -[x_E(t)-x_E(t+\tau_{\mathrm{arm}})]
  \propto \dot x_N(t) - \dot x_E(t) + O(\tau_{\mathrm{arm}})
\end{equation}
that, for frequencies $\ll\tau_{\mathrm{arm}}^{-1}$, are proportional to the relative velocity of the interferometer end test masses.

Both versions of the optical speedmeter, the sloshing cavity and the Sagnac ones, promise about the same sensitivity, and the choice between them depends mostly on the relative implementation cost of these schemes. Below we consider in more detail the Sagnac speedmeter, which does not require the additional long sloshing cavity.

We will not present here the full analysis of the Sagnac topology similar to the one we have provided for the Fabry--P{\'e}rot--Michelson one. The reader can find it in~\cite{Chen2002, 04a1Da}. We limit ourselves by the particular case of the resonance tuned interferometer (that is, no signal recycling and resonance tuned arm cavities). It seems that the detuned Sagnac interferometer can provide a quite interesting regime, in particular, the \emph{negative inertia} one~\cite{MuellerEbhardt2009}. However, for now (2011) the exhaustive analysis of these regimes is yet to be done. We assume that the squeezed light can be injected into the interferometer dark port, but consider only the particular case  of the classical optimization, $\theta=0$, which gives the best broadband sensitivity for a given optical power.

\subsubsection{Speedmeter sensitivity, no optical losses}
\label{sec:sm_noloss}

In order to reveal the main properties of the quantum speedmeter, start with the simplified case of the lossless interferometer and the ideal photodetector. In this case, the sum quantum noise power (double-sided) spectral density of the speedmeter can be written in a form similar to the one for the Fabry--P{\'e}rot--Michelson interferometer [see, e.g., Eqs.~\eqref{eq:vr_noises_x}, \eqref{eq:vr_noises_F} and \eqref{eq:vr_noises_xF}]:
\begin{equation}
\label{S_SM}
  S^h(\Omega) = \frac{S^h_{\mathrm{SQL}}(\Omega)}{2}\left[
    \frac{e^{-2r} + e^{2r}\cot^2\phi_{\mathrm{LO}}}{\mathcal{K}_{\mathrm{SM}}(\Omega)}
    - 2e^{2r}\cot\phi_{\mathrm{LO}} + \mathcal{K}_{\mathrm{SM}}(\Omega)e^{2r}
  \right] ,
\end{equation}
but with a different form of the optomechanical coupling factor, see~\cite{Chen2002}:
\begin{equation}
\label{KSM}
  \mathcal{K}_{\mathrm{SM}}(\Omega) = \frac{4J\gamma}{(\gamma^2+\Omega^2)^2} \,.
\end{equation}
The factor $J$ here is still defined by Eq.~\eqref{eq:FPMI_J_def}, but the circulating power is now twice as high as that of the position meter, for the given input power, because after leaving the beamsplitter, here each of the ``north'' and ``east'' beams visit both arms sequentially.

The key advantage of speedmeters over position meters is that at low  frequencies, $\Omega<\gamma$,
$\mathcal{K}_{\mathrm{SM}}$ is approximately constant and reaches the maximum there:
\begin{equation}
  \mathcal{K}_{\mathrm{SM}}(\Omega\ll\gamma) \approx \arccot\mathcal{K}_{\mathrm{SM}}(0)
  = \arccot\frac{4J}{\gamma^3} \,.
\end{equation}
As a consequence, a \emph{frequency-independent} readout quadrature optimized for low frequencies can be used:
\begin{equation}
  \phi_{\mathrm{LO}} = \arccot\mathcal{K}_{\mathrm{SM}}(0) = \arccot\frac{4J}{\gamma^3} \,,
\end{equation}
which gives the following power (double-sided) spectral density
\begin{equation}
\label{S_SM_LF}
  S^h(\Omega) = \frac{S^h_{\mathrm{SQL}}(\Omega)}{2}\left[
    \frac{e^{-2r}}{\mathcal{K}_{\mathrm{SM}}(\Omega)}
    + \frac{\Omega^4(2\gamma^2+\Omega^2)^2}{\gamma^8}\mathcal{K}_{\mathrm{SM}}(\Omega)e^{2r}
  \right] .
\end{equation}
Here, the radiation-pressure noise (the second term in brackets) is
significantly suppressed in low frequencies ($\Omega
\stackrel{<}{_\sim}\gamma$), and $  S^h_{\mathrm{SM\,LF}}$  can beat the
SQL in a broad frequency band.

This spectral density is plotted in Figure~\ref{fig:LIGO_sm}\,(left). For comparison, spectral densities for the lossless ordinary Fabry--P{\'e}rot--Michelson interferometer without and with squeezing, as well as for the ideal post-filtering configuration [see Eq.~\eqref{eq:vr_post_opt}] are also given. One might conclude from these plots that the Fabry--P{\'e}rot--Michelson interferometer with the additional filter cavities is clearly better than the speedmeter. However, below we demonstrate that optical losses change this picture significantly.

\epubtkImage{fig42.png}{
\begin{figure}[htbp]
\centerline{\includegraphics[width=.48\textwidth]{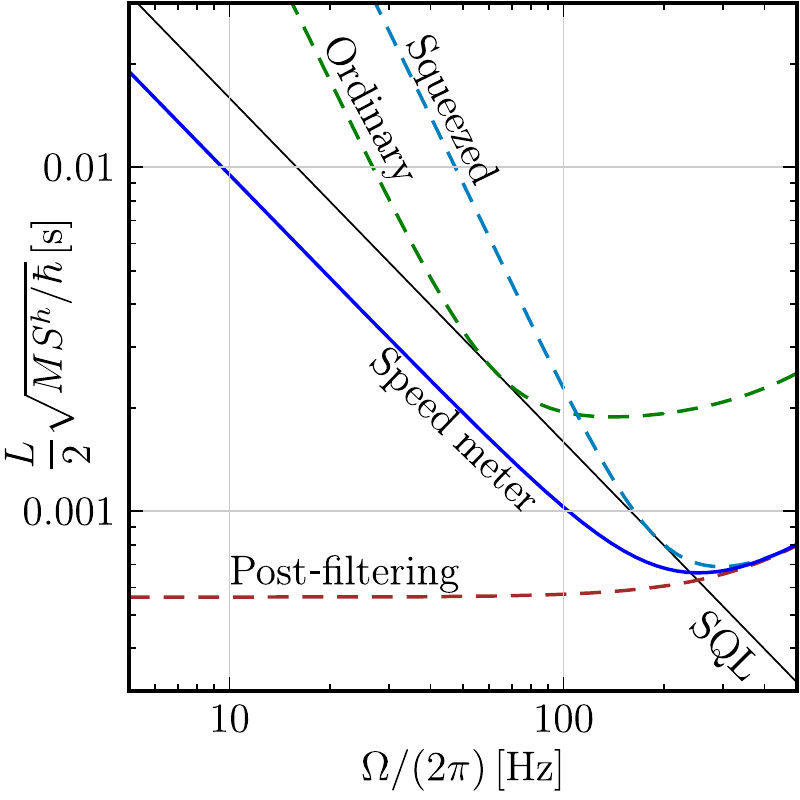}\hfill\includegraphics[width=.48\textwidth]{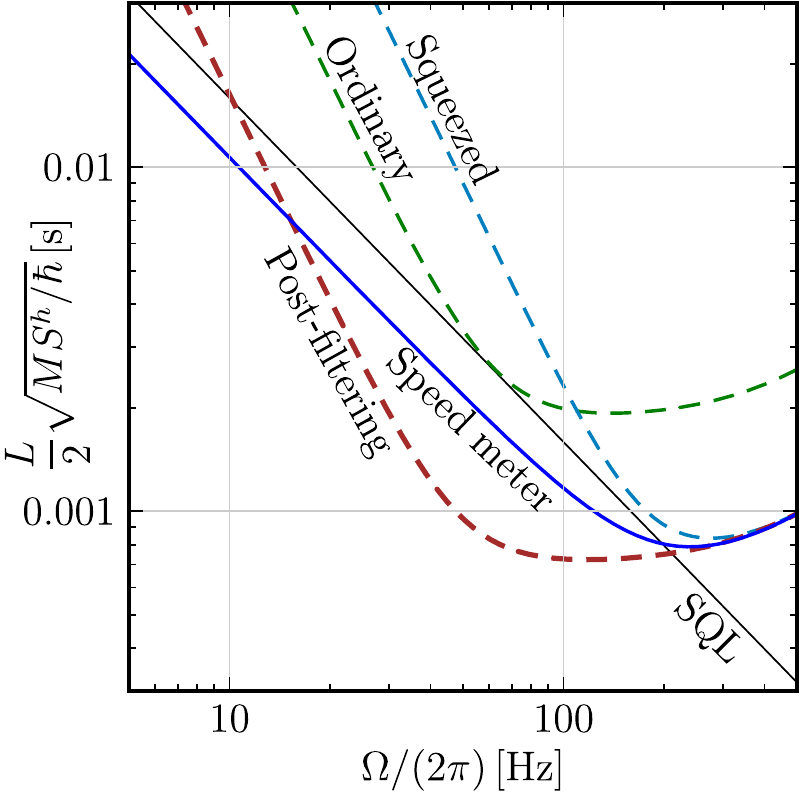}}
  \caption{Examples of the sum quantum noise power (double-sided) spectral densities of the Sagnac speedmeter interferometer (thick solid line) in comparison with the Fabry--P{\'e}rot--Michelson based topologies  considerd above (dashed lines). Left: no optical losses, right: with optical losses, $\eta_d=0.95$, the losses part of the bandwidth $\gamma_2=1.875\mathrm{\ s}^{-1}$ (which corresponds to the losses $A_{\mathrm{arm}}=10^{-4}$ per bounce in the 4~km length arms).  ``Ordinary'': no squeezing, $\phi_{\mathrm{LO}}=\pi/2$. ``Squeezed'': 10~dB squeezing, $\theta=0$, $\phi_{\mathrm{LO}}=\pi/2$. ``Post-filtering'': 10~dB squeezing, $\theta=0$, ideal frequency-dependent homodyne angle [see Eq.~\eqref{eq:vr_post_opt}]. For the Fabry--P{\'e}rot--Michelson-based topologies, $J=J_{\mathrm{aLIGO}}$ and $\gamma=2\pi\times500\mathrm{\ s}^{-1}$. In the speedmeter case, $J=2J_{\mathrm{aLIGO}}$ and the bandwidth is set to provide the same high-frequency noise as in the other plots ($\gamma=2\pi\times385\mathrm{\ s}^{-1}$ in the lossless case and $\gamma=2\pi\times360\mathrm{\ s}^{-1}$ in the lossy one).}
\label{fig:LIGO_sm}
\end{figure}}

\subsubsection{Optical losses in speedmeters}
\label{sec:sm_loss}

In speedmeters, optical losses in the arm cavities could noticeably affect the sum noise at low frequencies, even if
\begin{equation}
  \epsilon_{\mathrm{arm}}^2 \equiv \frac{\gamma_2}{\gamma_1} \ll \epsilon_d^2 \,,
\end{equation}
because the radiation pressure noise component created by the arm cavity losses has a frequency dependence similar to the one for position meters (remember that $\mathcal{K}/\mathcal{K}_{\mathrm{SM}}\to\infty$ if $\Omega\to0$; see Eqs.~\eqref{eq:varK}, \eqref{KSM}). In this paper, we will use the following expression for the lossy speedmeter sum noise, which takes these losses into account (more detailed treatment of the lossy speedmeter can be found in papers~\cite{Purdue2002, 04a1Da}):
\begin{eqnarray}
\label{S_SM_LF_loss}
  S^h(\Omega) = \frac{S^h_{\mathrm{SQL}}(\Omega)}{2}\biggl\{
    \frac{1}{\mathcal{K}_{\mathrm{SM}}(\Omega)}
      \left[e^{-2r} + e^{2r}\cot^2\phi_{\mathrm{LO}} +
        \frac{\epsilon_d^2}{\sin^2\phi_{\mathrm{LO}}}\right] \nonumber\\
    - 2e^{2r}\cot\phi_{\mathrm{LO}} + \mathcal{K}_{\mathrm{SM}}(\Omega)e^{2r} 
    + \epsilon_{\mathrm{arm}}^2\mathcal{K}(\Omega)
  \biggr\} .
\end{eqnarray}
The low-frequency optimized detection angle, in presence of loss, is
\begin{equation}
  \phi_{\mathrm{LO}} = \arccot\frac{\mathcal{K}_{\mathrm{SM}}(0)}{1 + \epsilon_d^2e^{-2r}}
  = \arccot\frac{4J/\gamma^3}{1 + \epsilon_d^2e^{-2r}} \,,
\end{equation}
which gives
\begin{equation}
\label{S_SM_LF_loss_opt}
  S^h(\Omega) = \frac{S^h_{\mathrm{SQL}}(\Omega)}{2}\biggl\{
    \frac{e^{-2r} + \epsilon_d^2}{\mathcal{K}_{\mathrm{SM}}(\Omega)}
    + \frac{\mathcal{K}_{\mathrm{SM}}(\Omega)}{1 + \epsilon_d^2e^{-2r}}\left[
          \frac{\Omega^4(2\gamma^2 + \Omega^2)^2}{\gamma^8}\,e^{2r} + \epsilon_d^2
        \right]
    + \epsilon_{\mathrm{arm}}^2\mathcal{K}(\Omega)
  \biggr\}
\end{equation}
[compare with Eq.~\eqref{S_SM_LF} and note the additional residual back-action term similar to one in Eq.~\eqref{eq:vr_post_opt}].

This spectral density is plotted in Figure~\ref{fig:LIGO_sm}\,(right), together with the lossy variants of the same configurations as in Figure~\ref{fig:LIGO_sm}\,(left), for the same moderately optimistic value of $\eta_d=0.95$, the losses part of the bandwidth and for $\gamma_2=1.875\mathrm{\ s}^{-1}$ [which corresponds to the losses $A_{\mathrm{arm}}=10^{-4}$ per bounce in the 4~km length arms, see Eq.~\eqref{sl_gammas}].
These plots demonstrate that the speedmeter in more robust with respect to optical losses than the filter cavities based configuration and is able to provide better sensitivity at very low frequencies.

It should also be noted that we have not taken into account here optical losses in the filter cavity. Comparison of Figure~\ref{fig:LIGO_sm} with Figure~\ref{fig:LIGO_filter}, where the noise spectral density for the more realistic lossy--filter-cavity cases are plotted, shows that the speedmeter has advantage over, at least, the short and medium length (tens or hundred of meters) filter cavities. In the choice between very long (and hence expensive) kilometer scale filter cavities and the speedmeter, the decision depends, probably, on the implementation costs of both configurations.

\subsection{Optical rigidity}
\label{sec:optical_rigidity}

\subsubsection{Introduction}
\label{sec:opt_rig_intro}

We have seen in Section~\ref{sec:SQL} that the harmonic oscillator, due to its strong response on near-resonance force, is characterized by the reduced values of the effective quantum noise and, therefore, by the SQL around the resonance frequency, see Eqs.~(\ref{eq:S_F_osc_nb}, \ref{eq:S_F_SQL_osc}) and Figure~\ref{fig:S_F_SQL}. However, practical implementation of this gain is limited by the following two shortcomings: (i) the stronger the sensitivity gain, the more narrow the frequency band in which it is achieved; see Eq.~\eqref{eq:osc_fm_SQL}; (ii) in many cases, and, in particular, in a GW detection scenario with its low signal frequencies and heavy test masses separated by the kilometers-scale distances, ordinary solid-state springs cannot be used due to unacceptably high levels of mechanical loss and the associated thermal noise.

At the same time, in detuned Fabry--P{\'e}rot cavities, as well as in the detuned configurations of the Fabry--P{\'e}rot--Michelson interferometer, the radiation pressure force depends on the mirror displacement (see Eqs.~\eqref{eq:FP_F_fl_K}), which is equivalent to the additional rigidity, called the \emph{optical spring}, inserted between the cavity mirrors. It does not introduce any additional thermal noise, except for the radiation pressure noise $\hat{F}_{\mathrm{b.a.}}$, and, therefore, is free from the latter of the above mentioned shortcomings. Moreover, as we shall show below, spectral dependence of the optical rigidity $K(\Omega)$ alleviates, to some extent, the former shortcoming of the `ordinary' rigidity and provides some limited sensitivity gain in a relatively broad band.

The electromagnetic rigidity was first discovered experimentally in radio-frequency systems~\cite{64a1eBrMi}. Then its existence was predicted for the optical Fabry--P{\'e}rot cavities~\cite{67a1eBrMa}. Much later it was shown that the excellent noise properties of the optical rigidity allows its use in quantum experiments with macroscopic mechanical objects~\cite{97a1BrGoKh, 99a1BrKh, 01a1BrKhVo}. The frequency dependence of the optical rigidity was explored in papers~\cite{Buonanno2001, 01a2Kh, PhysRevD.65.042001}. It was shown that depending on the interferometer tuning, either two resonances can exist in the system, \emph{mechanical} and \emph{optical} ones, or a single broader second-order resonance will exist.

In the last decade, the optical rigidity has been observed experimentally both in the table-top setup~\cite{Corbitt2007} and in the larger prototype interferometer~\cite{Miyakawa_PRD_74_022001_2006}.

\subsubsection{The optical noise redefinition}
\label{sec:opt_rig_noise_redef}

In detuned interferometer configurations, where the optical rigidity arises, the phase shifts between the input and output fields, as well as between the input fields and the field, circulating inside the interferometer, depend in sophisticated way on the frequency $\Omega$. Therefore, in order to draw full advantage from the squeezing, the squeezing angle of the input field should follow this frequency dependence, which is problematic from the implementation point of view. Due to this reason, considering the optical-rigidity--based regimes, we limit ourselves to the vacuum-input case only, setting $\mathbb{S}_{\sqz}[r,\theta]=\mathbb{I}$ in Eq.~\eqref{a_sqz}.

In this case, it is convenient to redefine the input noise operators as follows:
  \begin{eqnarray}
\label{eq:new_noises}
    \sqrt{\gamma}\hat{\boldsymbol{a}}_{\mathrm{new}}
      &=& \sqrt{\gamma_1}\hat{\boldsymbol{a}} + \sqrt{\gamma_2}\hat{\boldsymbol{g}} \,,\nonumber \\
    \sqrt{\gamma}\hat{\boldsymbol{g}}_{\mathrm{new}}
      &=& \sqrt{\gamma_1}\hat{\boldsymbol{g}} - \sqrt{\gamma_2}\hat{\boldsymbol{a}} \,, \nonumber\\
    \epsilon\hat{\boldsymbol{n}}_{\mathrm{new}}
      &=& \sqrt{\frac{\gamma_2}{\gamma_1}}\,\hat{\boldsymbol{g}}
        + \sqrt{\frac{\gamma}{\gamma_1}}\epsilon_d\hat{\boldsymbol{n}} \,,
  \end{eqnarray}
where
\begin{equation}
\label{vac_eta}
  \epsilon = \sqrt{\frac{1}{\eta}-1} \qquad\mbox{and}\qquad 
  \eta = \frac{\gamma_1}{\gamma}\eta_d
\end{equation}
is the \emph{unified quantum efficiency}, which accounts for optical losses both in the interferometer and in the homodyne detector.

Note that if the operators $\hat{\boldsymbol{a}}$, $\hat{\boldsymbol{g}}$, and $\hat{\boldsymbol{n}}$ describe mutually-uncorrelated vacuum noises, then the same is valid for the new $\hat{\boldsymbol{a}}_{\mathrm{new}}$, $\hat{\boldsymbol{g}}_{\mathrm{new}}$, and $\hat{\boldsymbol{n}}_{\mathrm{new}}$. Expressing Eqs.~(\ref{eq:ord_x_meas} and~\ref{eq:ord_F_pert}) in terms of new noises~\eqref{eq:new_noises} and renaming them, for brevity,
\begin{equation}
  \hat{\boldsymbol{a}}_{\mathrm{new}} \to \hat{\boldsymbol{a}} \,, \qquad
  \hat{\boldsymbol{n}}_{\mathrm{new}} \to \hat{\boldsymbol{n}} \,,
\end{equation}
we obtain:
\begin{eqnarray}
  \hat{\mathcal{X}}_{\mathrm{meas}}(\Omega) &=& \sqrt{\frac{\hbar}{2MJ\gamma}}\,
    \frac{\mathcal{D}(\Omega)}
      {\mathbf{H}^{\mathsf{T}}(\phi_{\mathrm{LO}})\mathbf{D}(\Omega)}
      \mathbf{H}^{\mathsf{T}}(\phi_{\mathrm{LO}})
        [\mathbb{R}(\Omega)\hat{\boldsymbol{a}}(\Omega) + \epsilon\hat{\boldsymbol{n}}(\Omega)] \,,
    \label{eq:ord0_x_meas} \\
  \hat{\mathcal{F}}_{\mathrm{b.a.}}(\Omega)
    &=& \sqrt{2\hbar MJ\gamma}\begin{bmatrix}1\\0\end{bmatrix}^{\mathsf{T}}\mathbb{L}(\Omega)\hat{\boldsymbol{a}}(\Omega) \,,
      \label{eq:ord0_F_pert}
\end{eqnarray}
where $\mathbb{R}(\Omega)$ is the lossless cavity reflection factor; see Eq.~\eqref{R_lossles}.

Thus, we have effectively reduced our lossy interferometer to the equivalent lossless one, but with less effective homodyne detector, described by the unified quantum efficiency $\eta<\eta_d$. Now we can write down explicit expressions for the interferometer quantum noises~\eqref{eq:FPMI_Sx}, \eqref{eq:FPMI_SF} and \eqref{eq:FPMI_SxF}, which can be calculated using Eqs.~\eqref{RL_props}:
  \begin{eqnarray}
\label{eq:sl_noises0}
    S_{\mathcal{X}\mathcal{X}}(\Omega) &=& \frac{\hbar}{4MJ\gamma\eta}\,
      \frac{|\mathcal{D}(\Omega)|^2}{\varGamma^2\sin^2\varphi + \Omega^2\sin^2\phi_{\mathrm{LO}}}\,,\nonumber\\
    S_{\mathcal{F}\mathcal{F}}(\Omega) &=& \frac{\hbar MJ\gamma(\varGamma^2 + \Omega^2)}{|\mathcal{D}(\Omega)|^2} \,, \nonumber\\
    S_{\mathcal{X}\mathcal{F}}(\Omega) &=& \frac{\hbar}{2}\,\frac{\varGamma\cos\varphi - i\Omega\cos\phi_{\mathrm{LO}}}
      {\varGamma\sin\varphi - i\Omega\sin\phi_{\mathrm{LO}}} \,,
  \end{eqnarray}
where
\begin{equation}
    \varGamma = \sqrt{\gamma^2+\delta^2} \,,  \qquad
    \varphi = \phi_{\mathrm{LO}} - \beta \,,  \qquad
   \beta = \arctan\frac{\delta}{\gamma}\,.
\end{equation}

\subsubsection{Bad cavities approximation}
\label{sec:opt_rig_short}

We start our treatment of the  optical rigidity with the  ``bad cavity'' approximation, discussed in Section~\ref{sec:vr_short} for the resonance-tuned interferometer case. This approximation, in addition to its importance for the smaller-scale prototype interferometers, provides a bridge between our idealized harmonic oscillator consideration of Section~\ref{sec:SQL_osc} and the frequency-dependent rigidity case specific to the large-scale GW detectors, which will be considered below, in Section~\ref{sec:fd_rigidity}.

In the ``bad cavity'' approximation $\varGamma\gg\Omega$, the Eqs.~\eqref{eq:sl_noises0} for the interferometer quantum noises, as well as the expression~\eqref{eq:sl_K} for the optical rigidity can be significantly simplified:
  \begin{equation}
\label{eq:SxSFK_short}
    S_{\mathcal{X}\mathcal{X}} = \frac{\hbar\varGamma^2}{4MJ\gamma\eta\sin^2\varphi} \,, \qquad
    S_{\mathcal{F}\mathcal{F}} =\frac{\hbar MJ\gamma}{\varGamma^2} \,, \qquad
    S_{\mathcal{X}\mathcal{F}} = \frac{\hbar}{2}\,\cot\varphi \,,
  \end{equation}
  and
  \begin{equation}
\label{eq:short_K}
    K = \frac{MJ\delta}{\varGamma^2} \,.
  \end{equation}
Substituting these equations into the equation for the sum quantum noise (cf.\ Eq.~\eqref{eq:gen_spdens}): 
\begin{eqnarray}
    S^F(\Omega) &=& |K-M\Omega^2|^2S_{\mathcal{X}\mathcal{X}}
      + 2\Re\bigl\{[K-M\Omega^2]S_{\mathcal{X}\mathcal{F}}\bigr\}
      + S_{\mathcal{F}\mathcal{F}} \nonumber\\
      &=& |K_{\mathrm{eff}}-M\Omega^2|^2S_{\mathcal{X}\mathcal{X}}+S_{\mathcal{F}\mathcal{F}}^{\mathrm{eff}}\,,
\end{eqnarray}
where
\begin{equation}
  S_{\mathcal{F}\mathcal{F}}^{\mathrm{eff}} = \frac{
      S_{\mathcal{X}\mathcal{X}}S_{\mathcal{F}\mathcal{F}}
      - |S_{\mathcal{X}\mathcal{F}}|^2
    }{S_{\mathcal{X}\mathcal{X}}}
  \qquad\mbox{and}\qquad
  K_{\mathrm{eff}} = K - \frac{S_{\mathcal{X}\mathcal{F}}}{S_{\mathcal{X}\mathcal{X}}}
\end{equation}
stand for the effective back-action noise and effective optical rigidity, respectively,
and dividing by $S^F_{\mathrm{SQL,\,f.m.}}$ defined by Eq.~\eqref{eq:S_F_SQL_fm}, we obtain the SQL beating factor~\eqref{eq:xi2}:
\begin{equation}
\label{short_gen}
  \xi^2(\Omega) = \frac{1}{\Omega^2}\left[
    (\Omega_m^2-\Omega^2)^2\frac{\varGamma^2}{4J\gamma\eta\sin^2\varphi}
    + \frac{J\gamma}{\varGamma^2}(1-\eta\cos^2\varphi)  \right] ,
\end{equation}
where
\begin{equation}
  \Omega_m^2 = \frac{K_{\mathrm{eff}}}{M}
  = \frac{J}{\varGamma^2}(\delta-\gamma\eta\sin2\varphi)
\end{equation}
is the effective resonance frequency (which takes into account both real and virtual parts of the effective rigidity $K_{\mathrm{eff}}$). Following the reasoning of Section~\ref{sec:SQL_osc}, it is easy to see that this spectral density allows for narrow-band sensitivity gain equal to
\begin{equation}
  \xi^2(\Omega_m+\nu) \le \xi^2(\Omega_m\pm\Delta\Omega/2)
  \approx \frac{2J\gamma}{\varGamma^2\Omega_m^2}(1-\eta\cos^2\varphi)
\end{equation}
within the bandwidth
\begin{equation}
  \frac{\Delta\Omega}{\Omega_m}
  = \frac{2J\gamma}{\varGamma^2\Omega_m^2}\sqrt{\eta\sin^2\varphi(1-\eta\cos^2\phi)}
  = \xi^2(\Omega_m\pm\Delta\Omega/2)
      \sqrt{\frac{\eta\sin^2\varphi}{1-\eta\cos^2\varphi}}  \,.
\end{equation}
In the ideal lossless case ($\eta=1$),
\begin{equation}
  \frac{\Delta\Omega}{\Omega_m} = \xi^2(\Omega_m\pm\Delta\Omega/2) \,,
\end{equation}
in accord with Eq.\eqref{eq:osc_fm_SQL}. However, if $\eta<1$, then the bandwidth, for a given $\xi$ lessens gradually as the homodyne angle $\varphi$ goes down. Therefore, the optimal case of the broadest bandwidth, for a given $\xi$, corresponds to $\varphi=\pi/2$, and, therefore, to $S_{xF}=0$ [see Eqs.~\eqref{eq:SxSFK_short}], that is, to the pure `real' rigidity case with non-correlated radiation-pressure and shot noises. This result naturally follows from the above conclusion concerning the amenability of the quantum noise sources cross-correlation to the influence of optical loss.

Therefore, setting $\varphi=\pi/2$ in Eq.~\eqref{short_gen} and taking into account that
\begin{equation}
  \frac{J\gamma}{\varGamma^2} = \frac{\Omega_q^2}{2}\cos^2\beta \,, \qquad
  \frac{K}{M} = \frac{J\delta}{\varGamma^2} = \frac{\Omega_q^2}{4}\sin2\beta \,,
\end{equation}
where $\Omega_q^2$ is the normalized optical power defined in Eq.~\eqref{eq:short_Omega_q},
we obtain that
\begin{equation}
\label{eq:short_or_sigma}
  \xi^2(\Omega) = \frac{1}{2\Omega^2}\biggl[
      \left(\frac{\Omega_q^2}{4}\sin2\beta - \Omega^2\right)^2
        \frac{1}{\Omega_q^2\eta\cos^2\beta}
      + \Omega_q^2\cos^2\beta
    \biggr] .
\end{equation}

Consider now the local minimization of this function at some given frequency $\Omega_0$, similar to one discussed in Section~\ref{sec:vr_short}. Now, the optimization parameter is $\beta$, that is, the detuning $\delta$ of the interferometer. It is easy to show that the optimal $\beta$ is given by the following equation:
\begin{equation}
\label{eq:or_short_env}
  \frac{4\Omega_0^2}{\Omega_q^2} - \frac{2}{\tan\beta}
  - \frac{\Omega_q^2}{\Omega_0^2}(4\eta-1)\cos^4\beta = 0 \,.
\end{equation}
This fifth-order equation for $\tan\beta$ cannot be solved in radicals. However, in the most interesting case of $\Omega_0\ll\Omega_q$, the following asymptotic solution can easily be obtained:
\begin{equation}
  \beta \approx \frac{\pi}{2} - \frac{2\Omega_0^2}{\Omega_q^2} \,,
\end{equation}
thus yielding
\begin{equation}
\label{eq:short_or_sigma_app}
  \xi^2(\Omega) \approx \frac{1}{2}\biggl[
      \left(1 - \frac{\Omega^2}{\Omega_0^2}\right)^2\frac{\Omega_q^2}{\Omega_0^2\eta}
      + \frac{4\Omega_0^2}{\Omega_q^2}
    \biggr] .
\end{equation}

\epubtkImage{fig43.png}{
\begin{figure}
  \centerline{\includegraphics[width=.5\textwidth]{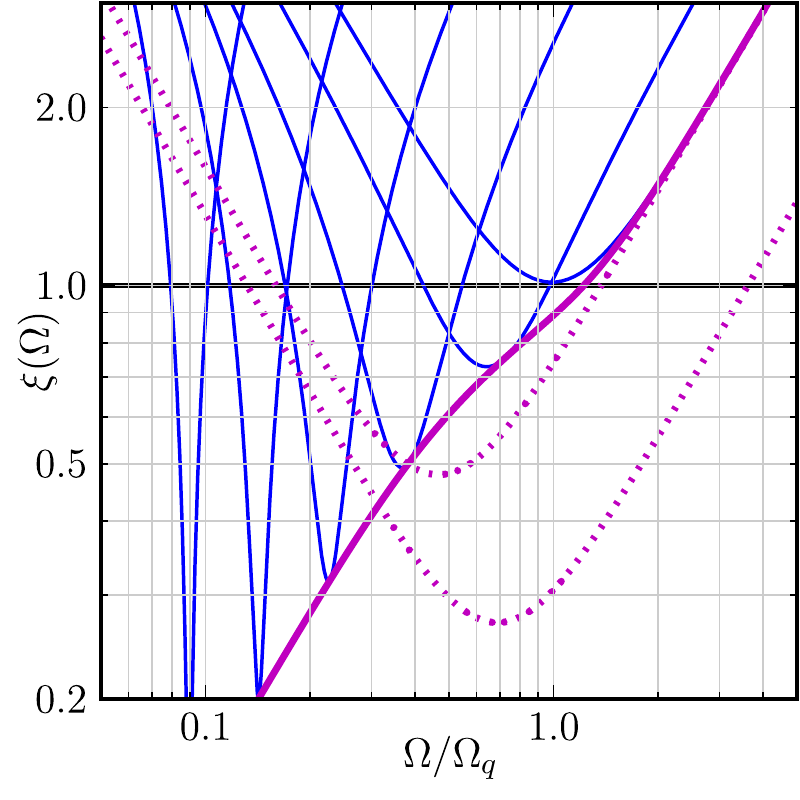}}
  \caption{Plots of the SQL beating factor~\eqref{eq:short_or_sigma} of the detuned interferometer, for different values of the normalized detuning: $0\le\beta\equiv\arctan(\delta/\gamma)<\pi/2$, and for unified quantum efficiency  $\eta=0.95$. \emph{Thick solid line:} the common envelope of these plots. \emph{Dashed lines:} the common envelopes~\eqref{vr_xi2_opt} of the SQL beating factors for the virtual rigidity case, without squeezing, $r=0$, and with 10~dB squeezing, $e^{2r}=10$ (for comparison).}
\label{fig:short_or_noises}
\end{figure}}

The function~\eqref{eq:short_or_sigma}, with optimal values of $\beta$ defined by the condition~\eqref{eq:or_short_env}, is plotted in Figure~\ref{fig:short_or_noises} for several values of the normalized detuning. We assumed in these plots that the unified quantum efficiency is equal to $\eta=0.95$. In the ideal lossless case $\eta=1$, the corresponding curves do not differ noticeably from the plotted ones. It means that in the real rigidity case, contrary to the virtual one, the sensitivity is not affected significantly by optical loss. This conclusion can also be derived directly from Eqs.~(\ref{eq:short_or_sigma}) and (\ref{eq:short_or_sigma_app}). It stems from the fact that quantum noise sources cross-correlation, amenable to the optical loss, has not been used here. Instead, the sensitivity gain is obtained by means of signal amplification using the resonance character of the effective harmonic oscillator response, provided by the optical rigidity.

The common envelope of these plots (that is, the optimal SQL-beating factor), defined implicitly by Eqs.~\eqref{eq:short_or_sigma} and \eqref{eq:or_short_env}, is also shown in Figure~\ref{fig:short_or_noises}. Note that at low frequencies, $\Omega\ll\Omega_q$, it can be approximated as follows:
\begin{equation}
\label{or_short_env_app}
  \xi^2_{\mathrm{env}}(\Omega) = \frac{2\Omega^2}{\Omega_q^2}
  \approx \frac{\gamma}{\delta}\,,
\end{equation}
(actually, this approximation works well starting from $\Omega\lesssim0.3\Omega_q$). It follows from this equation that in order to obtain a sensitivity significantly better than the SQL level, the interferometer should be detuned far from the resonance, $\delta\gg\gamma$.

For comparison, we reproduce here the common envelopes of the plots of $\xi^2(\Omega)$ for the virtual rigidity case with $\eta=0.95$; see Figure~\ref{fig:short_vr_noises} (the dashed lines). It follows from Eqs.~\eqref{or_short_env_app} and \eqref{vr_xi2_opt} that in absence of the optical loss, the sensitivity of the real rigidity case is inferior to that of the virtual rigidity one. However, even a very modest optical loss value changes the situation drastically.  The noise cancellation (virtual rigidity) method proves to be advantageous only for rather moderate values of the SQL beating factor  of $\xi\gtrsim0.5$ in the absence of squeezing and $\xi\gtrsim0.3$ with 10~dB squeezing. The conclusion is forced upon you that in order to dive really deep under the SQL, the use of real rather than virtual rigidity is inevitable.

Noteworthy, however, is the fact that optical rigidity has an inherent feature that can complicate its experimental implementation. It is dynamically unstable. Really, the expansion of the optical rigidity~\eqref{eq:sl_K} into a Taylor series in $\Omega$ gives
\begin{equation}
\label{eq:short_K_dump}
  K(\Omega) \approx \frac{MJ\delta}{\varGamma^2}
    + \frac{2MJ\gamma\delta}{\varGamma^4}i\Omega + \dots
\end{equation}
The second term, proportional to $i\Omega$, describes an optical friction, and the positive sign of this term (if $\delta>0$) means that this friction is negative.

The corresponding characteristic instability time is equal to
\begin{equation}
  \tau_{\mathrm{inst}} = \frac{\varGamma^4}{2J\gamma\delta} \,.
\end{equation}
In principle, this instability can be damped by some feedback control system as analyzed in~\cite{Buonanno2001, 03pth1Ch}. However, it can be done without significantly affecting the system dynamics, only if the instability is slow in the timescale of the mechanical oscillations frequency:
\begin{equation}
  \Omega_m\tau_{\mathrm{inst}} = \frac{\varGamma^2}{2\gamma\Omega_m}
  \approx \frac{J}{2\Omega_m^3\xi^2} \gg 1\,.
\end{equation}
Taking into account that in real-life experiments the normalized optical power $J$ is limited for technological reasons, the only way to get a more stable configuration is to decrease $\xi$, that is, to \emph{improve} the sensitivity by means of increasing the detuning. Another way to vanquish the instability is to create a stable optical spring by employing the second pumping carrier light with opposite detuning as proposed in~\cite{Corbitt2007, PhysRevD.78.062003_2008_Rehbein}. The parameters of the second carrier should be chosen so that the total optical rigidity must have both positive real and imaginary parts in Eq.~\eqref{eq:short_K_dump}:
\begin{equation}
  K_{\mathrm{sum}}(\Omega) = K_1(\Omega) + K_2(\Omega) \approx \left(\frac{MJ_1\delta_1}{\varGamma_1^2}+\frac{MJ_2\delta_2}{\varGamma_2^2}\right)
    + i\Omega\left(\frac{2MJ_1\gamma_1\delta_1}{\varGamma_1^4}+\frac{2MJ_2\gamma_2\delta_2}{\varGamma_2^4}\right) + \dots
\end{equation}
that can always be achieved by a proper choice of the parameters $J_{1,2}$, $\gamma_{1,2}$ and $\delta_{1,2}$ ($\delta_1\delta_2<0$).

\subsubsection{General case}
\label{sec:fd_rigidity}

\paragraph*{Frequency-dependent rigidity.}

In the large-scale laser GW detectors with kilometer-scale arm cavities, the interferometer bandwidth can easily be made comparable or smaller than the GW signal frequency $\Omega$. In this case, frequency dependences of the quantum noise spectral densities~\eqref{eq:FPMI_Sx}, \eqref{eq:FPMI_SF} and \eqref{eq:FPMI_SxF} and of the optical rigidity~\eqref{eq:sl_K} influence the shape of the sum quantum noise and, therefore, the detector sensitivity.

Most quantum noise spectral density is affected by the effective mechanical dynamics of the probe bodies, established by the frequency-dependent optical rigidity~\eqref{eq:sl_K}. Consider the characteristic equation for this system:
\begin{equation}
\label{eq:long_cheq}
  -\Omega^2[(\gamma-i\Omega)^2+\delta^2] + J\delta = 0 \,.
\end{equation}
In the asymptotic case of $\gamma=0$, the roots of this equation are equal to
\begin{equation}
\label{long_roots0}
  \Omega_m^{(0)} = \sqrt{\frac{\delta^2}{2}-\sqrt{\frac{\delta^4}{4}-J\delta}} \,, \qquad
  \Omega_o^{(0)} = \sqrt{\frac{\delta^2}{2}+\sqrt{\frac{\delta^4}{4}-J\delta}},
\end{equation}
(hereafter we omit the roots with negative-valued real parts). The corresponding maxima of the effective mechanical susceptibility:
\begin{equation}
\label{mech_Green}
  \chi_{xx}^{\mathrm{eff}}(\Omega) = \frac{1}{K(\Omega)-M\Omega^2}
\end{equation}
are, respectively, called the \emph{mechanical resonance} ($\Omega_m$) and the \emph{optical resonance} ($\Omega_o$) of the interferometer~\cite{Buonanno2001}.  In order to clarify their origin, consider an asymptotic case of the weak optomechanical coupling, $J\ll\delta^3$. In this case,
\begin{equation}
  \Omega_m \approx \frac{J}{\delta^3} = \sqrt{\frac{K(0)}{M}}\,, \qquad
  \Omega_o \approx \delta \,.
\end{equation}
It is easy to see that $\Omega_m$ originates from the ordinary resonance of the mechanical oscillator consisting of the test mass $M$ and the optical spring $K$ [compare with Eq.~\eqref{eq:short_K}]. At the same time, $\Omega_o$, in this approximation, does not depend on the optomechanical coupling and, therefore, has a pure optical origin -- namely, sloshing of the optical energy between the carrier power and the differential optical mode of the interferometer, detuned from the carrier frequency by $\delta$.

In the realistic general case of $\gamma\ne0$, the characteristic equation roots are complex. For small values of $\gamma$, keeping only linear in $\gamma$ terms, they can be approximated as follows:
\begin{equation}
\label{eq:long_roots1}
   \Omega_m = \Omega_m^{(0)}
      \biggl[1 + \frac{i\gamma\Omega_m^{(0)}}{\sqrt{\delta^4-4J\gamma}}\biggr] , \qquad
   \Omega_o = \Omega_o^{(0)}
      \biggl[1 - \frac{i\gamma\Omega_o^{(0)}}{\sqrt{\delta^4-4J\gamma}}\biggr] .
\end{equation}
Note that the signs of the imaginary parts correspond to a positive dumping for the optical resonance, and to a negative one (that is, to instability) for the mechanical resonance (compare with Eq.~\eqref{eq:short_K_dump}).

\epubtkImage{fig44.png}{
\begin{figure}
  \centerline{\includegraphics[width=.8\textwidth]{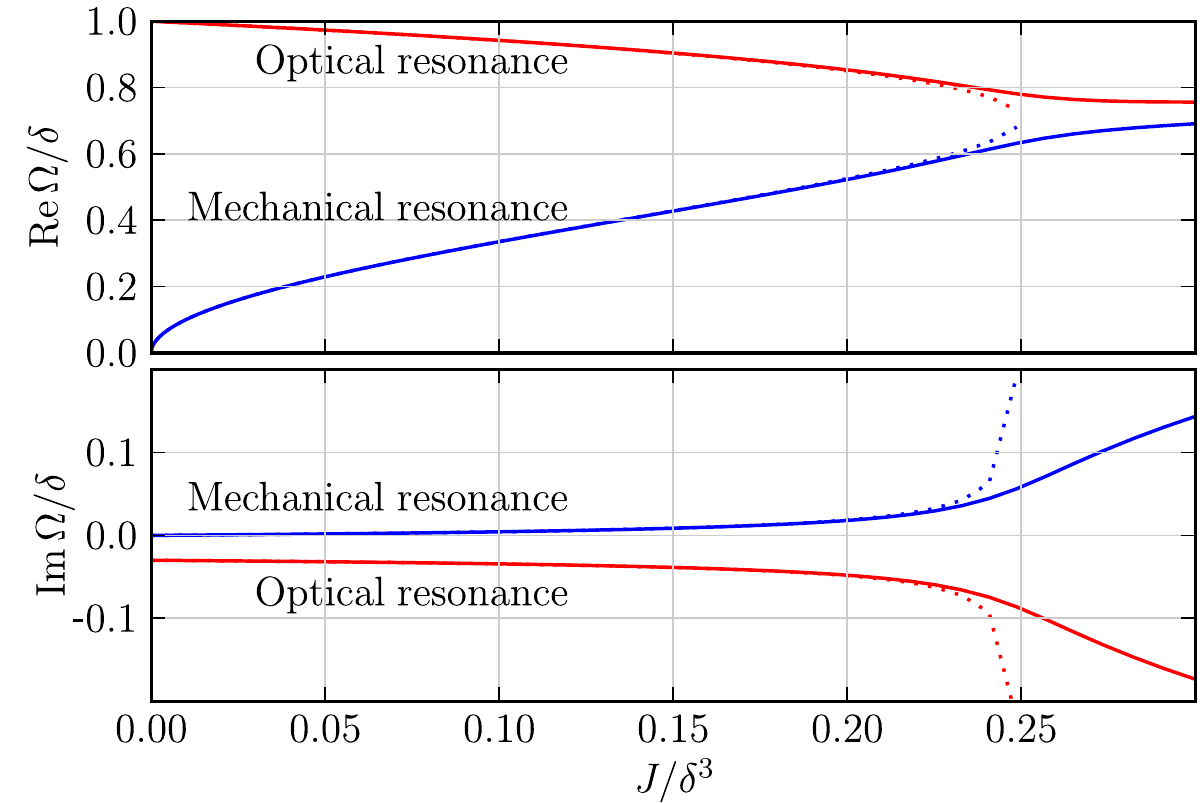}}
  \caption{Roots of the characteristic equation~\eqref{eq:long_cheq} as functions of the optical power, for $\gamma/\delta=0.03$. Solid lines: numerical solution. Dashed lines: approximate solution, see Eqs.~\eqref{eq:long_roots1}}
\label{fig:long_roots}
\end{figure}}

In Figure~\ref{fig:long_roots}, the numerically-calculated roots of Eq.~\eqref{eq:long_cheq} are plotted as a function of the normalized optical power $J/\delta^3$, together with the analytical approximate solution~\eqref{eq:long_roots1}, for the particular case of $\gamma/\delta=0.03$. These plots demonstrate the peculiar feature of the parametric optomechanical interaction, namely, the \emph{decrease} of the separation between the eigenfrequencies of the system  as the optomechanical coupling strength goes up. This behavior is opposite to that of the ordinary coupled linear oscillators, where the separation between the eigenfrequencies increases as the coupling strength grows (the well-known avoided crossing feature).

As a result, if the optomechanical coupling reaches the critical value:
\begin{equation}
\label{long_J_crit}
  J = \frac{\delta^3}{4} \,,
\end{equation}
then, in the asymptotic case of $\gamma=0$, the eigenfrequencies become equal to each other:
\begin{equation}
\label{long_Omega_0}
  \Omega_m^{(0)} = \Omega_o^{(0)} = \Omega_0 \equiv \frac{\delta}{\sqrt{2}} \,.
\end{equation}
If $\gamma>0$, then some separation retains, but it gets smaller than $2\gamma$, which means that the corresponding resonance curves effectively merge, forming a single, broader resonance. This \emph{second-order pole regime}, described for the first time in~\cite{01a2Kh}, promises some significant advantages for high-precision mechanical measurements, and we shall consider it in more detail below.

If $J<J_{\mathrm{crit}}$, then two resonances yield two more or less
separated minima of the sum quantum noise spectral density, whose
location on the frequency axis mostly depends on the detuning
$\delta$, and their depth (inversely proportional to their width)
hinges on the bandwidth $\gamma$.The choice of the preferable
configuration depends on the criterion of the optimization, and also
on the level of the technical (non-quantum) noise in the
interferometer.

\epubtkImage{fig45.png}{
\begin{figure}[htb]
  \centerline{\includegraphics[width=.8\textwidth]{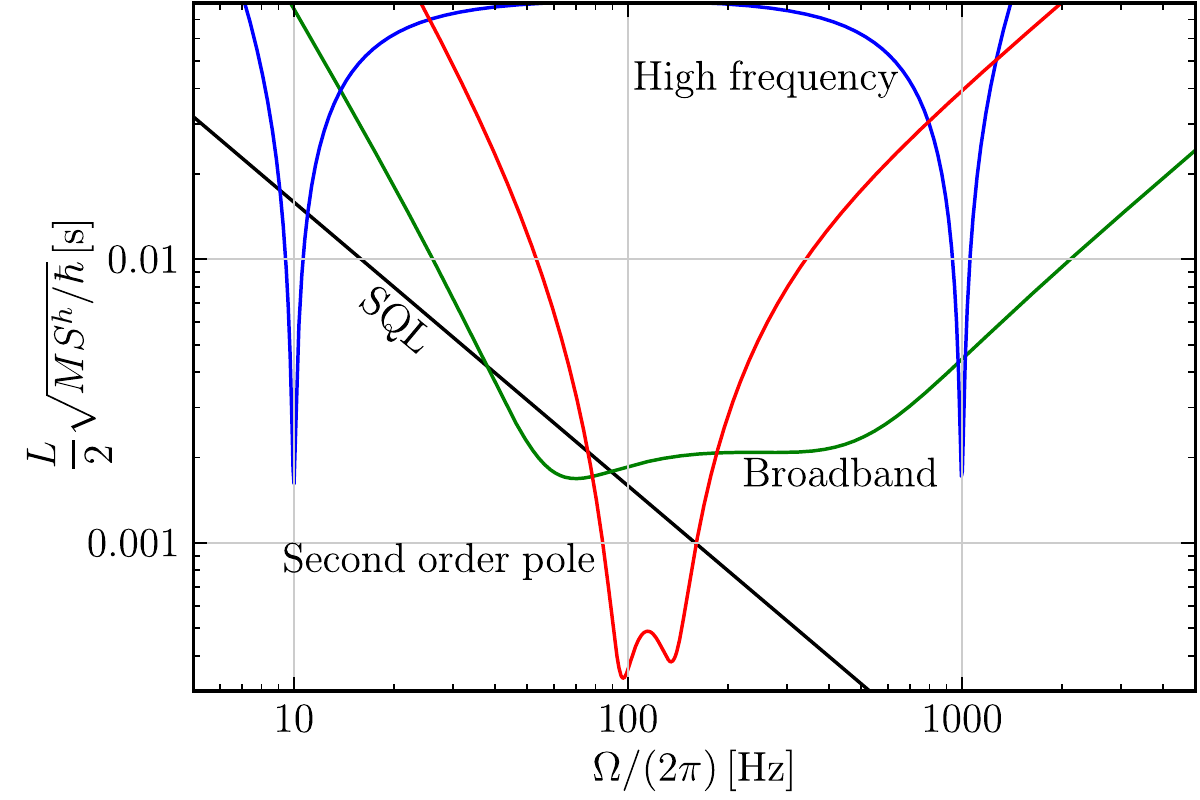}}
  \caption{Examples of the sum noise power (double-sided) spectral densities of the detuned interferometer. `Broadband': double optimization of the Advanced LIGO interferometer for NS-NS inspiraling and burst sources in presence of the classical noises~\cite{08a1KoSiKhDa} ($J=J_{\mathrm{aLIGO}}\equiv(2\pi\times100)^3\mathrm{\ s}^{-3}$, $\varGamma=3100\mathrm{\ s}^{-1}$, $\beta=0.80$, $\phi_{\mathrm{LO}}=\pi/2-0.44$). `High-frequency': low-power configuration suitable for detection of the GW signals from the millisecond pulsars, similar to one planned for GEO HF~\cite{Willke2006} [$J=0.1J_{\mathrm{aLIGO}}$, $\varGamma=2\pi\times1000\mathrm{\ s}^{-1}$, $\beta=\pi/2-0.01$, $\phi_{\mathrm{LO}}=0$]. `Second-order pole': the regime close to the second-order pole one, which provides a maximum of the SNR for the GW burst sources given that technical noise is smaller than the SQL [$S_{\mathrm{tech}}=0.1S_{\mathrm{SQL}}$, $J=J_{\mathrm{aLIGO}}$, $\varGamma=1050\mathrm{\ s}^{-1}$, $\beta=\pi/2-0.040$, $\phi_{\mathrm{LO}}=0.91$].
  In all cases, $\eta_d=0.95$ and the losses part of the bandwidth $\gamma_2=1.875\mathrm{\ s}^{-1}$ (which corresponds to the losses $A_{\mathrm{arm}}=10^{-4}$ per bounce in the 4~km long arms).}
\label{fig:LIGO_det}
\end{figure}}

Two opposite examples are drawn in Figure~\ref{fig:LIGO_det}. The first one features the sensitivity of a broadband configuration, which provides the best SNR for the GW radiation from the inspiraling neutron-star--neutron-star binary and, at the same time, for broadband radiation from the GW burst sources, for the parameters planned for the Advanced LIGO interferometer (in particular, the circulating optical power $\mathcal{I}_{\mathrm{arm}}=840\mathrm{\ kW}$, $L=4\mathrm{\ km}$, and $M=40\mathrm{\ kg}$, which translates to $J=J_{\mathrm{aLIGO}}\equiv(2\pi\times100)^3\mathrm{\ s}^{-3}$, and the planned technical noise). The optimization performed in~\cite{08a1KoSiKhDa} gave the quantum noise spectral density, labeled as `Broadband' in Figure~\ref{fig:LIGO_det}. It is easy to notice two (yet not discernible) minima on this plot, which correspond to the mechanical and the optical resonances.

Another example is the configuration suitable for detection of the narrow-band GW radiation from millisecond pulsars. Apparently, one of two resonances should coincide with the signal frequency in this case. It is well to bear in mind that in order to create an optical spring with mechanical resonance in a kHz region in contemporary and planned GW detectors, an enormous amount of optical power might be required. This is why the optical resonance, whose frequency depends mostly on the detuning $\delta$, should be used for this purpose. This is, actually, the idea behind the GEO HF project~\cite{Willke2006}. The example of this regime is represented by the curve labeled as `High-frequency' in Figure~\ref{fig:LIGO_det}. Here, despite one order of magnitude less optical power used ($J=0.1J_{\mathrm{aLIGO}}$), several times better sensitivity at frequency 1~kHz, than in the `Broadband' regime, can be obtained. Note that the mechanical resonance in this case corresponds to 10~Hz only and therefore is virtually useless.

\paragraph*{The second-order pole regime.}
\label{sec:double_pole}

In order to clarify the main properties of the second-order pole regime, start with the asymptotic case of $\gamma\to0$. In this case, the optical rigidity and the mechanical susceptibility~\eqref{mech_Green} read
\begin{eqnarray}
  K(\Omega) &=& \frac{MJ\delta}{\delta^2-\Omega^2} \,, \\
  \chi^{\mathrm{eff,\,dbl}}_{xx}(\Omega) &=& \frac{1}{K(\Omega)-M\Omega^2}
    = \frac{1}{M}\frac{\delta^2-\Omega^2}{J\delta - \delta^2\Omega^2 + \Omega^4} \,.
\end{eqnarray}
If condition~\eqref{long_J_crit} is satisfied, then in the close vicinity of the frequency $\Omega_0$ (see Eq.~\eqref{long_Omega_0}):
\begin{equation}
  |\Omega-\Omega_0| \ll \Omega_0 \,,
\end{equation}
the susceptibility can be approximated as follows:
\begin{equation}
   \chi^{\mathrm{eff,\,dbl}}_{xx}(\Omega) \approx \frac{\Omega_0^2}{M(\Omega_0^2-\Omega^2)^2} \,.
\end{equation}
Note that this susceptibility is proportional to the square of the susceptibility of the ordinary oscillator,
\begin{equation}
  \chi_{xx}^{\mathrm{osc}}(\Omega) = \frac{1}{M(\Omega_0^2-\Omega^2)} \,,
\end{equation}
and has the second-order pole at the frequency $\Omega_0$ (thus the name of this regime).

This `double-resonance' feature creates a stronger response to the external forces with spectra concentrated near the frequency $\Omega_0$, than in the ordinary harmonic oscillator case. Consider, for example, the resonance force $F_0\sin\Omega_0t$. The response of the ordinary harmonic oscillator with eigenfrequency $\Omega_0$ on this force increases linearly with time:
\begin{equation}
  x_{\mathrm{osc}}(t) = \frac{F_0t}{2M\Omega_0}\sin\Omega_0t \,,
\end{equation}
while that of the second-order pole object grows quadratically:
\begin{equation}
  x_{\mathrm{dbl}}(t)
    = -\frac{F_0}{8M}\left(t^2\cos\Omega_0t - \frac{\sin\Omega_0t}{\Omega_0}\right) .
\end{equation}
It follows from Eq.~\eqref{eq:SQL_force} that due to this strong response, the second-order pole test object has a reduced value of the SQL around $\Omega_0$ by contrast to the harmonic oscillator.

Consider the quantum noise of the system, consisting of this test object and the SQM (that is, the Heisenberg's-uncertainty-relation--limited quantum meter with frequency-independent and non-correlated measurement and back-action noises; see Section~\ref{sec:linear_toy}), which monitors its position. Below we show that the real-life long-arm interferometer, under some assumptions, can be approximated by this model.

The sum quantum noise power (double-sided) spectral density of this system is equal to
\begin{equation}
  S^h(\Omega) = \frac{4}{M^2L^2\Omega^4}(|\chi^{\mathrm{eff,\,dbl}}_{xx}(\Omega)|^2S_{\mathcal{X}\mathcal{X}} + S_{\mathcal{F}\mathcal{F}}) \,.
\end{equation}
If the frequency $\Omega$ is close to $\Omega_0$:
\begin{equation}
  \Omega = \Omega_0+\nu \,, \qquad |\nu|\ll\Omega_0 \,,
\end{equation}
then
\begin{equation}
\label{eq:S_X_dbl}
  S^h(\Omega_0+\nu)
  \approx \frac{2\hbar}{ML^2\Omega_0^4}\left(\frac{16\nu^4}{\Omega_q^2} + \Omega_q^2\right),
\end{equation}
and
\begin{equation}
\label{eq:xi2_dbl}
  \xi^2(\Omega_0+\nu)
  \equiv \frac{S^h(\Omega_0+\nu)}{S^h_{\mathrm{SQL\,f.m.}}(\Omega_0+\nu)}
  \approx
    \frac{1}{2\Omega_0^2}\left(\frac{16\nu^4}{\Omega_q^2} + \Omega_q^2\right),
\end{equation}
(compare with Eq.~\eqref{eq:S_F_osc_nb}), where the frequency $\Omega_q$ is defined by Eq.~\eqref{eq:Omega_q_def}.

The same minimax optimization as performed in Section~\ref{sec:SQL_osc} for the harmonic oscillator case gives that the optimal value of $\Omega_q$ is equal to
\begin{equation}
  \Omega_q = \Delta\Omega \,,
\end{equation}
and in this case,
\begin{equation}
\label{eq:dbl_fm_SQL}
  \xi^2(\Omega_0+\nu)\Bigr|_{|\nu|\le\Delta\Omega/2}
  \le \xi^2(\Omega_0\pm\Delta\Omega/2) = \left(\frac{\Delta\Omega}{\Omega_0}\right)^2.
\end{equation}
Comparison of Eqs.~\eqref{eq:dbl_fm_SQL} and \eqref{eq:osc_fm_SQL} shows that for a given SQL-beating bandwidth $\Delta\Omega$, the second-order pole system can provide a much stronger sensitivity gain (i.e., much smaller value of $\xi^2$), than the harmonic oscillator, or, alternatively, much broader bandwidth $\Delta\Omega$ for a given value of $\xi^2$. It is noteworthy that the factor~\eqref{eq:dbl_fm_SQL} can be made smaller than the normalized oscillator SQL $2|\nu|/\Omega_0$ (see Eq.~\eqref{eq:osc_fm_SQL}), which means beating not only the free mass SQL, but also the harmonic oscillator one.

\epubtkImage{fig44.png}{%
\begin{figure}
  \centerline{\includegraphics[width=.48\textwidth]{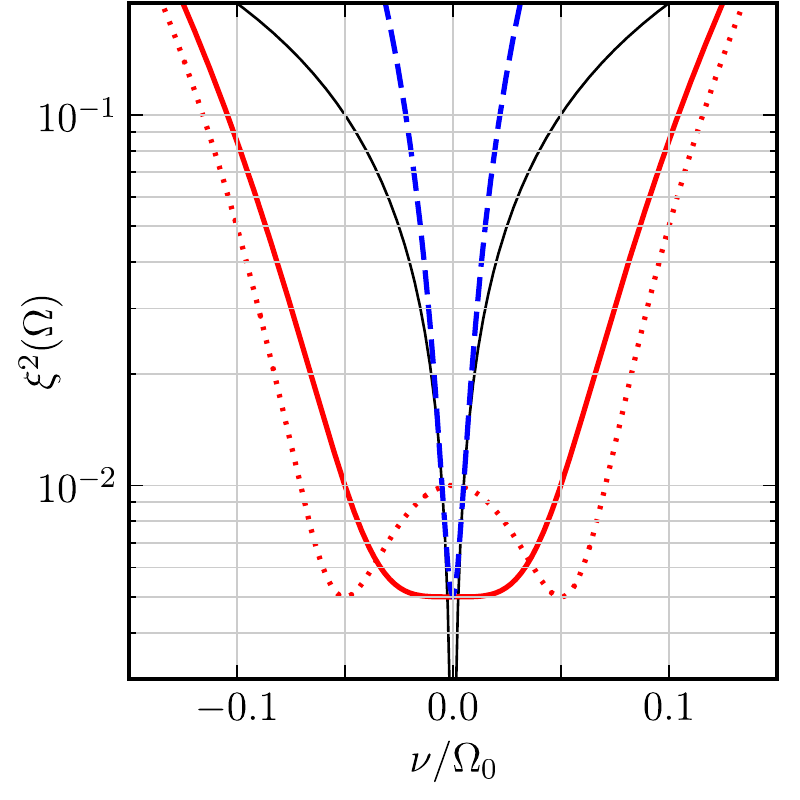}\hfill\includegraphics[width=.48\textwidth]{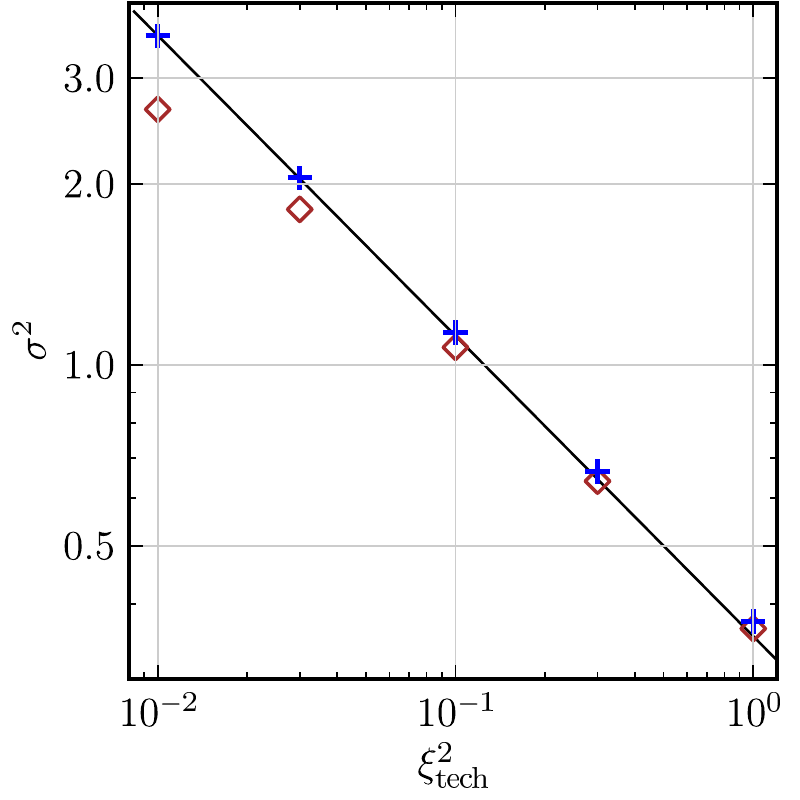}}
  \caption{\emph{Left panel:} the SQL beating factors $\xi^2$ for $\Omega_q/\Omega_0=0.1$. Thick solid: the second-order pole system~\eqref{eq:xi2_dbl}; dots: the two-pole system with optimal separation between the poles~\eqref{eq:dbl_sigma2}, \eqref{dbl_optimal}; dashes: the  harmonic oscillator~\eqref{eq:xi2_osc}; thin solid -- SQL of the harmonic oscillator~\eqref{eq:osc_fm_SQL}. \emph{Right:} the normalized SNR~\eqref{eq:dbl_rho2_tech}. Solid line: analytical optimization, Eq.~\eqref{eq:dbl_rho2_opt}; pluses: numerical optimization of the spectral density~\eqref{eq:sl_sum0} in the lossless case ($\eta=1$); diamonds: the same for the interferometer with $J=(2\pi\times100)^3\mathrm{\ s}^{-3}$, $\eta_d=0.95$ and the losses part of the bandwidth $\gamma_2=1.875\mathrm{\ s}^{-1}$ (which corresponds to the losses $A_{\mathrm{arm}}=10^{-4}$ per bounce in the 4~km long arms).}
\label{fig:sigma2_dbl}
\end{figure}}

This consideration is illustrated by the left panel of Figure~\ref{fig:sigma2_dbl}, where the factors $\xi^2$ for the harmonic oscillator~\eqref{eq:xi2_osc} and of the second-order pole system~\eqref{eq:xi2_dbl} are plotted for the same value of the normalized back-action noise spectral density $(\Omega_q/\Omega_0)^2=0.01$, as well as the normalized oscillator SQL~\eqref{eq:osc_fm_SQL}.

Now return to the quantum noise of a real interferometer. With account of the noises redefinition~\eqref{eq:new_noises}, Eq.~\eqref{eq:S_sum_BC} for the sum quantum noise power (double-sided) spectral density takes the following form:
\begin{eqnarray}
\label{eq:sl_sum0}
  S^h(\Omega) = \frac{\hbar}{ML^2J\gamma}
    \frac{1}{\Omega^2\sin^2\phi_{\mathrm{LO}} + \varGamma^2\sin^2\varphi}
    \biggl\{
      \left[
        \gamma^2 - \delta^2 + \Omega^2 +\frac{J}{\Omega^2}(\delta-\gamma\sin2\phi_{\mathrm{LO}})
      \right]^2 \nonumber\\
      + 4\gamma^2\left(\delta - \frac{J}{\Omega^2}\sin^2\phi_{\mathrm{LO}}\right)^2
      + \epsilon^2\left|\mathcal{D}(\Omega) - \frac{J\delta}{\Omega^2}\right|^2
    \biggr\} .
\end{eqnarray}
Suppose that the interferometer parameters satisfy approximately the second-order pole conditions. Namely, introduce a new parameter $\Lambda$ defined by the following equation:
\begin{equation}
  J(\delta-\gamma\sin2\phi_{\mathrm{LO}}) = \Omega_0^4 - \Omega_0^2\Lambda^2 \,,
\end{equation}
where the frequency $\Omega_0$ is defined by Eq.~\eqref{long_Omega_0}, and assume that
\begin{equation}
\label{dbl_conds}
  \nu^2 \sim \Lambda^2 \sim \Omega_0\gamma \ll \Omega_0^2 \,.
\end{equation}
Keeping only the first non-vanishing terms in $\nu^2$, $\Lambda^2$, and $\gamma$ in Eq.~\eqref{eq:sl_sum0}, we obtain that
\begin{equation}
\label{eq:S_X_nb}
  S^h(\Omega_0+\nu) \approx \frac{2\hbar}{ML^2\Omega_0^4}\Biggl(
    \frac{1}{\Omega_q^2}\biggl\{
        (4\nu^2 - \Lambda^2)^2
        + \epsilon^2\biggl[
            \left(4\nu^2-\Lambda^2+\frac{\Omega_0\gamma}{\sqrt{2}}\sin2\phi_{\mathrm{LO}}\right)^2
            + 4\gamma^2\Omega_0^2
          \biggr]
      \biggr\}
    + \Omega_q^2
  \Biggr) ,
\end{equation}
where
\begin{equation}
\label{eq:Omega_q_nb}
  \Omega_q^2 = \sqrt{2}\gamma\Omega_0(1+\cos^2\phi_{\mathrm{LO}}) \,.
\end{equation}
It follows from Eq.~\eqref{eq:S_X_nb} that the parameter $\Lambda$ is equal to the separation between the two poles of the susceptibility $\chi_{xx}^{\mathrm{eff,\,dbl}}$.

It is evident that the spectral density~\eqref{eq:S_X_nb} represents a direct generalization of Eq.~\eqref{eq:S_X_dbl} in two aspects. First, it factors in optical losses in the interferometer. Second, it includes the case of $\Lambda\ne0$. We show below that a small yet non-zero value of $\Lambda$ allows one to further increase the sensitivity.

\paragraph*{Optimization of the signal-to-noise ratio.}

The peculiar feature of the second-order pole regime is that, while being, in essence, narrow-band, it can provide an arbitrarily-high SNR for the broadband signals, limited only by the level of the additional noise of non-quantum (technical) origin. At the same time, in the ordinary harmonic oscillator case, the SNR is fundamentally limited.

In both the harmonic oscillator and the second-order pole test object cases, the quantum noise spectral density has a deep and narrow minimum, which makes the major part of the SNR integral. If the bandwidth of the signal force exceeds the width of this minimum (which is typically the case in GW experiments, save to the narrow-band signals from pulsars), then the SNR integral~\eqref{6:snr} can be approximated as follows:
\begin{equation}
  \rho^2 = \frac{|h_{\mathrm{s}}(\Omega_0)|^2}{\pi}
    \intinfty\frac{d\nu}{S^h(\Omega_0+\nu)} \,.
\end{equation}
It is convenient to normalize both the signal force and the noise spectral density by the corresponding SQL values, which gives:
\begin{equation}
  \rho^2 = \frac{ML^2\Omega_0^3}{4\hbar}\,|h_{\mathrm{s}}(\Omega_0)|^2\sigma^2 \,,
\end{equation}
where
\begin{equation}
\label{eq:snr_rho2}
  \sigma^2 = \frac{1}{\pi\Omega_0}\intinfty\frac{d\nu}{\xi^2(\Omega_0+\nu)}
\end{equation}
is the dimensionless integral sensitivity measure, which we shall use here, and
\begin{equation}
  \xi^2(\Omega_0+\nu) = \frac{ML^2\Omega_0^2}{4\hbar}\,S^h(\Omega_0+\nu) \,.
\end{equation}

For a harmonic oscillator, using Eq.~\eqref{eq:xi2_osc}, we obtain
\begin{equation}
  \sigma^2 = 1 \,.
\end{equation}
This result is natural, since the depth of the sum quantum-noise spectral-density minimum (which makes the dominating part of the integral~\eqref{eq:snr_rho2}) in the harmonic oscillator case is inversely proportional to its width $\Delta\Omega$, see Eq.~\eqref{eq:osc_fm_SQL}. As a result, the integral does not depend on how small the minimal value of $\xi^2$ is.

The situation is different for the second-order pole-test object. Here, the minimal value of $\xi^2$ is proportional to $(\Delta\Omega)^{-2}$ (see Eq.~\eqref{eq:dbl_fm_SQL}) and, therefore, it is possible to expect that the SNR will be proportional to
\begin{equation}
  \sigma^2 \propto \frac{1}{(\Delta\Omega)^2}\times\Delta\Omega
  \propto \frac{1}{\Delta\Omega} \propto \frac{1}{\xi} \,.
\end{equation}
Indeed, after substitution of Eq.~\eqref{eq:xi2_dbl} into~\eqref{eq:snr_rho2}, we obtain:
\begin{equation}
  \sigma^2 = \frac{1}{\sqrt{2}}\frac{\Omega_0}{\Omega_q} = \frac{1}{2\xi(\Omega_0)} \,.
\end{equation}
Therefore, decreasing the width of the dip in the sum quantum noise spectral density and increasing its depth, it is possible, in principle, to obtain an arbitrarily high value of the SNR.

Of course, it is possible only if there are no other noise sources in the interferometer except for the quantum noise. Consider, though, a more realistic situation. Let there be an additional (technical) noise in the system with the spectral density  $S_{\mathrm{tech}}(\Omega)$. Suppose also that this spectral density does not vary much within our frequency band of interest $\Delta\Omega$. Then the factor $\sigma^2$ can be approximated as follows:
\begin{equation}
\label{eq:dbl_rho2_tech}
  \sigma^2 = \frac{1}{\pi\Omega_0}
    \intinfty\frac{d\nu}{\xi^2(\Omega_0+\nu) + \xi^2_{\mathrm{tech}}} \,,
\end{equation}
where
\begin{equation}
\label{dbl_sigma2_tech}
  \xi^2_{\mathrm{tech}} = \frac{S_{\mathrm{tech}}(\Omega_0)}{S_{\mathrm{SQL\,f.m.}}(\Omega_0)} \,.
\end{equation}
Concerning quantum noise, we consider the regime close (but not necessarily
exactly equal) to that yielding the second-order pole, that is, we suppose $0\le\Lambda\ll\Omega_0$. In order to simplify our calculations, we neglect the contribution from optical loss into the sum spectral density (we show below that it does not affect the final sensitivity much). Thus, as follows from Eq.~\eqref{eq:S_X_nb}, one gets
\begin{equation}
\label{eq:dbl_sigma2}
  \xi^2(\Omega_0+\nu) = \frac{1}{2\Omega_0^2}
    \biggl[\frac{(4\nu^2 - \Lambda^2)^2}{\Omega_q^2} + \Omega_q^2\biggr] .
\end{equation}
In the Appendix~\ref{app:SOP}, we calculate integral Eq.~\eqref{eq:dbl_rho2_tech} and optimize it over  $\Lambda$ and $\Omega_q$. The optimization gives the best sensitivity, for a given value of $\xi^2_{\mathrm{tech}}$, is provided by
\begin{equation}
\label{dbl_optimal}
  \Lambda = \Omega_q = \Omega_0\xi_{\mathrm{tech}} \,.
\end{equation}
In this case,
\begin{equation}
\label{eq:dbl_rho2_opt}
  \sigma^2 = \frac{1}{2\sqrt{2}\xi_{\mathrm{tech}}} \,.
\end{equation}
The pure second-order pole regime ($\Lambda=0$), with the same optimal value of $\Omega_q$, provides slightly worse sensitivity:
\begin{equation}
\label{eq:dbl_rho2_sub}
  \sigma^2 = \frac{1}{\sqrt{6\sqrt{3}}\xi_{\mathrm{tech}}} \,.
\end{equation}
The optimized function~\eqref{eq:dbl_sigma2} is shown in Figure~\ref{fig:sigma2_dbl}\,(left) for the particular case of $\Lambda=\Omega_q=0.1\Omega_0$. In Figure~\ref{fig:sigma2_dbl}\,(right), the optimal SNR~\eqref{eq:dbl_rho2_opt} is plotted as a function of the normalized technical noise $\xi_{\mathrm{tech}}^2$.

In order to verify our narrow-band model, we optimized numerically the general normalized SNR for the broadband burst-type signals:
\begin{equation}
\label{eq:dbl_snr_burst}
  \sigma^2_{\mathrm{burst}} = \frac{2\hbar}{\pi ML^2\Omega_0^2}
    \intinfty\frac{d\Omega/\Omega}{S^h(\Omega) + S^h_{\mathrm{tech}}} \,,
\end{equation}
where $S^h$ is the sum quantum noise of the interferometer defined by Eq.~\eqref{eq:sl_sum0}. The only assumption we have made here is that the technical noise power (double-sided) spectral density
\begin{equation}
  S^h_{\mathrm{tech}} = \frac{2\hbar}{ML^2\Omega_0^2}\,\xi^2_{\mathrm{tech}}
\end{equation}
does not depend on frequency, which is reasonable, since only the narrow frequency region around $\Omega_0$ contributes noticeably to the integral~\eqref{eq:dbl_snr_burst}. The result is shown in Figure~\ref{fig:sigma2_dbl}\,(right) for two particular cases: the ideal (no loss) case with $\eta=1$, and the realistic case of the interferometer with $J=(2\pi\times100)^3\mathrm{\ s}^{-3}$, $\eta_d=0.95$ and $\gamma_2=1.875\mathrm{\ s}^{-1}$ (which corresponds to the loss factor of $A_{\mathrm{arm}}=10^{-4}$ per bounce in the $4\mathrm{\ km}$ long arms; see Eq.~\eqref{sl_gammas}). The typical optimized quantum noise spectral density (for the particular case of $\xi_{\mathrm{tech}}^2=0.1$) is plotted in Figure~\ref{fig:LIGO_det}.

It is easy to see that the approximations~\eqref{eq:dbl_sigma2} work very well, even if $\xi_{\mathrm{tech}}\sim1$ and, therefore, the assumptions~\eqref{dbl_conds} cease to be valid. One can conclude, looking at these plots, that optical losses do not significantly affect the sensitivity of the interferometer, working in the second-order pole regime. The reason behind it is apparent. In the optical rigidity based systems, the origin of the sensitivity gain is simply the resonance increase of the probe object dynamical response to the signal force, which is, evidently, immune to the optical loss.

The only noticeable discrepancy between the analytical model and the numerical calculations for the lossless case, on the one hand, and the numerical calculations for the lossy case, on the other hand, appears only for very small values of $\xi_{\mathrm{tech}}^2\sim0.01$. It follows from Eqs.~\eqref{eq:Omega_q_nb} and \eqref{dbl_optimal} that this case corresponds to the proportionally reduced bandwidth of the interferometer,
\begin{equation}
  \gamma \sim \Omega_0\xi_{\mathrm{tech}}^2 \sim 10\mathrm{\ s}^{-1}
\end{equation}
(for a typical value of $\Omega_0\sim10^3\mathrm{\ s}^{-1}$). Therefore, the loss-induced part of the total bandwidth $\gamma_2$, which has no noticeable effect on the unified quantum efficiency $\eta$ [see Eq.~\eqref{vac_eta}] for the `normal' broadband values of $\gamma\sim10^3\mathrm{\ s}^{-1}$, degrades it in this narrow-band case. However, it has to be emphasized that the degradation of $\sigma^2$, for the reasonable values of $\xi_{\mathrm{tech}}^2$, is only about a few percent, and even for the quite unrealistic case of $\xi_{\mathrm{tech}}^2 = 0.01$, does not exceed $\sim$~25\%.

\newpage
\section{Conclusion and Future Directions}
\label{sec:Conclusion}

In this review, our primary goal was to tell in a clear and understandable way what is meant by quantum measurement in GW detectors. It was conceived as a comprehensive introduction to the quantum noise calculation techniques that are employed currently for the development of advanced interferometric detectors. The target audience are the young researchers, students and postdocs, who have just started their way in this field and need a guide that provides a step-by-step tutorial into the techniques and covers all the current achievements in the field. At the same time, we tried to make this manuscript interesting to all our colleagues from the GW community and, perhaps, from other branches of physics, who might be interested in getting themselves familiar with this area, not necessarily close to their own research field.

However, the reality is crude and such a lofty ambition is always a pot of gold at the end of the rainbow. Thus, we could not claim this review to be a complete and comprehensive description of the field of quantum measurement. We present here a pretty detailed analysis of the quantum noise features in the first and second generation of GW interferometers, contemplating the techniques considered robust and established. However, many hot topics, related to the planned third generation of GW interferometers~\cite{GRG.43.2.671_2011_Chen, CQG.27.19.194002_2010_Punturo, CQG.28.9.094013_2011_Hild} remained uncovered. Here are only some of them: (i) xylophone configurations~\cite{CQG.27.1.015003_2010_Hild}, (ii) multiple-carrier detectors~\cite{PhysRevD.76.062002_2007_Rehbein, PhysRevD.78.062003_2008_Rehbein}, (iii) negative optical inertia~\cite{PhysRevD.83.062003_2011_Khalili}, (iv) intracavity detection schemes~\cite{98a1BrGoKh, 97a1BrGoKh, Phys.Lett.A.298.5.308_2002_Khalili, Phys.Lett.A.317.3.169_2003_Khalili, PhysRevD.73.022002_2006_Danilishin}, etc. It is our determined intention to enjoy the great advantages of the format of \emph{living} reviews and include those topics in future revisions of this review.

We would like to conclude our review by pointing out how the new swiftly-developing areas of modern science and technology, not directly related to GW astronomy and detector science, turn out to be deeply rooted in the quantum measurement theory developed by the GW community.
It is amazing how sinuous the ways of scientific progress are. The history of how GW detection and quantum-measurement theory developed and interwove might serve as an example thereof. Indeed, from the very first steps towards the experimental observation of GWs made by Weber in the early 1960s~\cite{weber1961general, PRL.18.498_1967_Weber}, it was realized that the extreme weakness of interaction between the ripples of space-time and matter appeals for unprecedentedly precise measurement. And almost at the same time, Braginsky realized that the expected amplitude of the GW-induced oscillations of the bar detector signal mode would be on the order of the zero point oscillations of this mode, as predicted by quantum mechanics; that is, in order to observe GWs, one has to treat a detector quantum-mechanically and as a consequence there will be a quantum back action, setting a limitation on the achievable sensitivity, the SQL~\cite{Sov.Phys.JETP_26.831_1968}.

This serendipity had a powerful impact on the quantum measurement theory development, for it set an objective to contrive some ways to overcome this limitation. For decades up to this point, it was a purely theoretical discipline having little in common with experimental science, and, fancy that, become a vital necessity for GW astronomy. And again, for several decades, GW detection has been perhaps the only field where the results of quantum measurement theory were applied, mainly in the struggle with quantum noise, considered as a hindrance towards the noble goal of the detection of GWs. And only recently, the same optomechanical interaction, begetting quantum noise and the SQL in the interferometric GW detectors, has aroused a keen interest among wide circles of researchers studying the quantum behavior of macroscopic objects and testing the very foundations of quantum mechanics in the macroscopic world~\cite{Sci.321.5893.1172_2008_Kippenberg, JOSAB.27.6.A189_2010_Aspelmeyer}.

All the techniques and concepts developed in the GW community turn out to
be highly sought by this new
field~\cite{RevModPhys.82.1155_2010_Clerk}. Such methods, initially
developed for future GW detectors, as back-action evasion via properly
constructed cross correlation between the measurement and back-action
noise sources~\cite{94a1VyZuMa, 96a2eVyMa, 96a1eVyMa, 94a1VyZu,
  95a1VyZu, Phys.Lett.A.300.547_2002_Danilishin,
  Phys.Lett.A.278.3.123_2000_Danilishin}, find a use in the
optomechanical experiments with micro- and nanoscale mechanical
oscillators~\cite{NJP.10.9.095010_2008_Clerk,
  PhysRevA.80.043802_2009_Mueller-Ebhardt,
  PhysRevA.81.012114_2010_Miao, PhysRevA.81.052307_2010_Miao,
  PhysRevLett.103.100402_2009_Miao,
  PhysRevA.81.033849_2010_Yamamoto,arXiv:1104.3251_2011_Friedrich}. It
turns out that GW detectors themselves fit extremely well for testing the
fundamental principles of quantum mechanics just for the record low
values of the noise, having non-quantum origin, that owes to the
ingenuity, patience and dedication by an entire generation of
experimental physicists~\cite{lrr-2009-2}. The very fact that the
mechanical differential mode of the km-scale LIGO detector has been
cooled down to $T_{\mathrm{eff}} = 1.4\mathrm{\ \mu\,K}$ without
  any special arrangement, just by modifying the standard feedback
  kernel of the actuators to provide a virtual rigidity, shifting the
  10-kg suspended mirrors oscillation frequency from $\Omega_m/2\pi =
  0.74\mathrm{\ Hz}$ to $150\mathrm{\ Hz}$, where the GW detector is most
  sensitive~\cite{NJP.11.7.073032_2009_Abbott}, tells its own
  tale. Noteworthy also is the experiment on cooling a several-ton
  AURIGA bar detector mechanical oscillation mode to $T_{\mathrm{eff}}
  = 0.17\mathrm{\ mK}$~\cite{PhysRevLett.101.033601_2008_Vinante}. In
  principle, some dedicated efforts might yield even cooling to ground
  state of these definitely macroscopic
  oscillators~\cite{arXiv:0809.2024_2008_Danilishin,
    NJP.12.8.083032_2010_Miao}.

One might foresee even more striking, really quantum phenomena, to be demonstrated experimentally by future GW detectors, whose sensitivity will be governed by quantum noise and not limited by the SQL. It is possible, e.g., to prepare the mechanical degree of freedom of the interferometer in a close-to-pure squeezed quantum state~\cite{PhysRevA.80.043802_2009_Mueller-Ebhardt}, entangle the differential and common motion of the kg-scale mirrors in the EPR-like fashion~\cite{PhysRev.47.777_1935_Einstein, PhysRevLett.100.013601_2008_Mueller-Ebhardt}, or even prepare it in a  highly non-classical Schr\"odinger-cat state~\cite{NatWiss.23.48.807_1935_Schroedinger, PhysRevLett.105.070403_2010_Miao}.

\newpage
\section{Acknowledgements}
\label{sec:Acknowledgements}

This review owes its existence to the wholehearted support and sound
advice of our colleagues and friends. We would like to express our
special thanks to our friends, Yanbei Chen and Haixing Miao for
enlightening discussions, helpful suggestions and encouragement we
enjoyed in the course of writing this review. Also we are greatly thankful to the referees and to our younger colleagues, Mikhail Korobko and Nikita Voronchev, who went to the trouble of thoroughly reading the manuscript and pointing out many imperfections, typos, and misprints. We greatly acknowledge as well
our fellow researchers from LIGO-Virgo Scientific Collaboration for
all the invaluable experience and knowledge they shared with us over
the years. Especially, we want to say thank you to Gregg Harry,
Innocenzo Pinto and Roman Schnabel for consulting with us on literature in
the areas of their expertise. And finally, we would like to thank
\textit{Living Reviews in Relativity} and especially Bala Iyer for the
rewarding opportunity to prepare this manuscript.

\newpage
\appendix
\section{Appendices}

\subsection{Input/Output relations derivation for a Fabry--P{\'e}rot cavity}
\label{app:FP_I/O-relations}

Here we consider the derivation of the I/O-relations for a Fabry--P{\'e}rot cavity given by Eqs.~\eqref{eq:FP_I/O_time} in Section~\ref{sec:Fabry-Perot}. We start by writing down the input fields in terms of reduced complex amplitudes $A_{1,2}$ for the classical part of the field and annihilation operators $\hat{a}_{1,2}$ for quantum corrections, respectively, as prescribed by Eqs.~\eqref{eq:EMW_quantum} and \eqref{eq:2photon_Eplus}:
\begin{eqnarray}
    \hat{\mathrm{A}}_1(t-x/c) = \mathrm{A}_1e^{-i\omega_p(t-x/c)} + \mbox{\ c.c.}
    + \intOinfty\sqrt{\frac{\omega}{\omega_p}}\,\hat{\mathrm{a}}_1(\omega)e^{-i\omega(t-x/c)}\,
        \frac{d\omega}{2\pi} + \hc \,, \\
    \hat{\mathrm{A}}_2(t+x/c) = \mathrm{A}_2e^{-i\omega_p(t+x/c)} + \mbox{\ c.c.}
    + \intOinfty\sqrt{\frac{\omega}{\omega_p}}\,\hat{\mathrm{a}}_2(\omega)e^{-i\omega(t+x/c)}\,
        \frac{d\omega}{2\pi} + \hc \,,
  \end{eqnarray}
with c.c. standing for ``complex conjugate'' and h.c. --- for ``Hermitian conjugate''. It would be convenient for us to make the preliminary calculations in terms of complex amplitudes before going for two-photon quadrature amplitudes.

Start with the equations that connect the classical field amplitudes of the fields shown in Figure~\ref{fig:Fabry-Perot}. At each mirror, the corresponding fields are related according to Eqs.~\eqref{eq:mirror_IOrel} with mirror transfer matrix $\mathbb{M}_{\mathrm{real}}$:
  \begin{eqnarray}
    \mathrm{B}_{1,2} &=& -\sqrt{R_{1,2}}\mathrm{A}_{1,2} + \sqrt{T_{1,2}}\mathrm{E}_{1,2} \,, \label{FP_eq_0_B} \\
    \mathrm{F}_{1,2} &=& \sqrt{T_{1,2}}\mathrm{A}_{1,2} + \sqrt{R_{1,2}}\mathrm{E}_{1,2} \,, 
  \end{eqnarray}
and two equations for the running waves inside the cavity:
  \begin{eqnarray}
    \mathrm{E}_{1,2} &=& \mathrm{F}_{2,1}e^{i\omega_p\tau} \,
  \end{eqnarray}
describes the free propagation of light between the mirrors.
The solution to these equations is the following:
  \begin{eqnarray}
\label{eq:FP_0}
    \mathrm{B}_{1,2} &=& \frac{
        [\sqrt{R_{2,1}}e^{2i\omega_p\tau} - \sqrt{R_{1,2}}]\mathrm{A}_{1,2}
        + \sqrt{T_{1}T_{2}}\mathrm{A}_{2,1}e^{i\omega_p\tau}
      }{1-\sqrt{R_{1}R_{2}}e^{2i\omega_p\tau}} \,,\nonumber \\
    \mathrm{E}_{1,2} &=& \frac{
        \sqrt{R_{2,1}}\sqrt{T_{1,2}}\mathrm{A}_{1,2}e^{2i\omega_p\tau}
        + \sqrt{T_{2,1}}\mathrm{A}_{2,1}e^{i\omega_p\tau}
      }{1-\sqrt{R_{1}R_{2}}e^{2i\omega_p\tau}} \,,\nonumber \\
    \mathrm{F}_{1,2}
      &=& \frac{\sqrt{T_{1,2}}\mathrm{A}_{1,2} + \sqrt{R_{1,2}}\sqrt{T_{2,1}}\mathrm{A}_{2,1}e^{i\omega_p\tau}}{1-\sqrt{R_{1}R_{2}}e^{2i\omega_p\tau}}\,.
  \end{eqnarray}

The equations set for the quantum fields has the same structure, but with more sophisticated boundary conditions, which include mirrors' motion as described in Section~\ref{sec:Mirror_motion}:
  \begin{eqnarray}
\label{FP_eq_1}
    \hat{\mathrm{b}}_{1,2}(\omega)
      &=& -\sqrt{R_{1,2}}\bigl[
          \hat{\mathrm{a}}_{1,2}(\omega) - 2i\sqrt{kk_p}\mathrm{A}_{1,2}\hat{x}_{1,2}(\Omega)
        \bigr]
      + \sqrt{T_{1,2}}\hat{\mathrm{e}}_{1,2}(\omega), \\
    \hat{\mathrm{f}}_{1,2}(\omega)
      &=& \hat{\mathrm{s}}_{1,2}(\omega) + \sqrt{R_{1,2}}\hat{\mathrm{e}}_{1,2}(\omega) \,,\nonumber \\
    \hat{\mathrm{e}}_{1,2}(\omega) &=& \hat{\mathrm{f}}_{2,1}(\omega)e^{i\omega\tau} \,,
  \end{eqnarray}
where
\begin{equation}
  \hat{\mathrm{s}}_{1,2}(\omega) = \sqrt{T_{1,2}}\hat{\mathrm{a}}_{1,2}(\omega)
    + 2i\sqrt{kk_p}\sqrt{R_{1,2}}\mathrm{E}_{1,2}\hat{x}_{1,2}(\Omega) \,,
\end{equation}
and
\begin{equation}
  \Omega = \omega-\omega_p\,, \qquad k=\omega/c\,, \qquad k_p=\omega_p/c\,.
\end{equation}
The solution to these equations reads:
  \begin{eqnarray}
\label{eq:FP_1}
 \,,\hat{\mathrm{b}}_{1,2}(\omega) &=& \frac{
        [\sqrt{R_{2,1}}e^{2i\omega\tau} - \sqrt{R_{1,2}}]\hat{\mathrm{a}}_{1,2}(\omega)
        + \sqrt{T_{1,2}}e^{i\omega\tau}[
            2i\sqrt{kk_p}\sqrt{R_{1}R_{2}}\mathrm{E}_{1,2}\hat{x}_{1,2}(\Omega)
            + \hat{\mathrm{s}}_{2,1}(\omega)
          ]
      }{1-\sqrt{R_{1}R_{2}}e^{2i\omega\tau}} \nonumber \\
      && + 2i\kappa \sqrt{R_{1,2}}A_{1,2}x_{1,2}(\Omega) \\
    \hat{\mathrm{e}}_{1,2}(\omega) &=& \frac{
        \sqrt{R_{2,1}}\hat{\mathrm{s}}_{1,2}(\omega)e^{2i\omega\tau}
        + \hat{\mathrm{s}}_{2,1}(\omega)e^{i\omega\tau}
      }{1-\sqrt{R_{1}R_{2}}e^{2i\omega\tau}} \,, \nonumber\\
    \hat{\mathrm{f}}_{1,2}(\omega) &=& \frac{
          \hat{\mathrm{s}}_{1,2}(\omega) + \sqrt{R_{1,2}}\hat{\mathrm{s}}_{2,1}(\omega)e^{i\omega\tau}
        }{1-\sqrt{R_{1}R_{2}}e^{2i\omega\tau}} \,.
  \end{eqnarray}

\subsection{Proof of Eq.~\eqref{eq:FPMI_Sx}, \eqref{eq:FPMI_SF} and \eqref{eq:FPMI_SxF}}
\label{app:SxSFSxf}

In the lossless case [see Eq.~\eqref{no_loss}], the noises (double-sided) power spectral densities~\eqref{eq:FPMI_Sx}, \eqref{eq:FPMI_SF} and \eqref{eq:FPMI_SxF} are equal to
  \begin{eqnarray}
\label{sl_noises_noloss}
    S_x(\Omega)
      &=& \frac{\hbar}{4MJ\gamma}\bm{\psi}^\dagger\mathbb{S}_{\sqz}[2r,0]\bm{\psi} \,,\nonumber \\
    S_F(\Omega) &=& \hbar MJ\gamma\,\bm{\phi}^\dagger\mathbb{S}_{\sqz}[2r,0]\bm{\phi} \nonumber\,, \\
    S_{xF}(\Omega) &=& \frac{\hbar}{2}\bm{\psi}^\dagger\mathbb{S}_{\sqz}[2r,0]\bm{\phi} \,,
  \end{eqnarray}
were
  \begin{eqnarray}
    \bm{\psi} &=& \svector{\psi_c}{\psi_s}
      = \frac{\mathbb{P}(\theta)\mathbb{R}^\dagger\mathbf{H}}
          {(0\ 1)\mathbb{L}^\dagger\mathbf{H}} \,, \\
    \bm{\phi} &=& \svector{\phi_c}{\phi_s} = \mathbb{P}(\theta)\mathbb{L}^\dagger\svup \,,
  \end{eqnarray}
and
\begin{equation}
  \mathbb{R}(\Omega) = 2\gamma\mathbb{L}(\Omega) - 1 \,. \label{R_lossles}
\end{equation}
Therefore,
\begin{eqnarray}
  S_xS_F - |S_{xF}|^2
  &=& \frac{\hbar^2}{4}\left[
      \bm{\psi}^\dagger\mathbb{S}_{\sqz}[2r,0]\bm{\psi}\cdot
        \bm{\phi}^\dagger\mathbb{S}_{\sqz}[2r,0]\bm{\phi}
      - |\bm{\psi}^\dagger\mathbb{S}_{\sqz}[2r,0]\bm{\phi}|^2
    \right] \nonumber\\
  &=& \frac{\hbar^2}{4}\left(|\psi_c\phi_s - \psi_s\phi_c|^2\right)
  = \frac{\hbar^2}{4}\left(
      \bm{\psi}^\dagger\bm{\psi}\cdot\bm{\phi}^\dagger\bm{\phi}
      - |\bm{\psi}^\dagger\bm{\phi}|^2
    \right) .
\end{eqnarray}
Note that the squeezing factor $r$ has gone.

Taking into account that
\begin{equation}
  \mathbb{R}\mathbb{R}^\dagger = \mathbb{I}\,, \qquad
  \mathbb{R}\mathbb{L}^\dagger = \mathbb{L} \label{RL_props}
\end{equation}
we obtain that
\begin{equation}
  \bm{\psi}^\dagger\bm{\psi} = \frac{1}{\left|\mathbf{H}\mathbb{L}\svup\right|^2} \,, \qquad
  \bm{\phi}^\dagger\bm{\phi} = \begin{bmatrix}1\\0\end{bmatrix}^{\mathsf{T}}\mathbb{L}\mathbb{L}^\dagger\svup \,, \qquad
  \bm{\psi}^\dagger\bm{\phi} = \frac{\boldsymbol{H}^{\mathsf{T}}\mathbb{L}\svup}
      {\boldsymbol{H}^{\mathsf{T}}\mathbb{L}\svdown}
\end{equation}
and
\begin{equation}
  S_xS_F - |S_{xF}|^2 = \frac{\hbar^2}{4}\frac{
      \begin{bmatrix}1\\0\end{bmatrix}^{\mathsf{T}}\mathbb{L}\mathbb{L}^\dagger\svup
      - \left|\boldsymbol{H}^{\mathsf{T}}\mathbb{L}\svup\right|^2
    }{\left|\boldsymbol{H}^{\mathsf{T}}\mathbb{L}\svdown\right|^2} \,.
\end{equation}
It is easy to show by direct calculation that
\begin{equation}
  \left|\boldsymbol{H}^{\mathsf{T}}\mathbb{L}\textstyle\svup\right|^2
    + \left|\boldsymbol{H}^{\mathsf{T}}\mathbb{L}\textstyle\svdown\right|^2
  = \begin{bmatrix}1\\0\end{bmatrix}^{\mathsf{T}}\mathbb{L}\mathbb{L}^\dagger\textstyle\svup \,,
\end{equation}
which proves Eq.~\eqref{SxSFSxf}.

\subsection{SNR in second-order pole regime}
\label{app:SOP}

Substituting Eqs.~(\ref{eq:dbl_sigma2}) and (\ref{dbl_sigma2_tech}) into Eq.~\eqref{eq:dbl_rho2_tech}, we obtain that
\begin{equation}
\label{dbl_rho}
  \sigma^2 = \frac{2}{\pi}\,\frac{\Omega_0}{\Omega_q}
    \intinfty\frac{dy}{(4y^2-\lambda^2)^2 + 1 + 2s^2} 
  = \frac{\Omega_0}{\sqrt{2}\Omega_q}
      \frac{1}{
        \sqrt{(1 + 2s^2 + \lambda^4)(\sqrt{1 + 2s^2 + \lambda^4} - \lambda^2)}
      } \,,
\end{equation}
where
\begin{equation}
  y = \frac{\nu}{\Omega_q} \,, \qquad \lambda = \frac{\Lambda}{\Omega_q} \,, \qquad
  s = \frac{\Omega_0}{\Omega_q}\,\xi_{\mathrm{tech}} \,.
\end{equation}
If $\lambda=0$ (the sub-optimal case), then
\begin{equation}
\label{dbl_rho_sub_1}
  \sigma^2 = \frac{\Omega_0}{\sqrt{2}\Omega_q}(1+2s^2)^{-3/4}.
\end{equation}
The maximum of~\eqref{dbl_rho} in $\lambda$ corresponds to
\begin{equation}
  \lambda^2 = \sqrt{\frac{1+2s^2}{3}} \,,
\end{equation}
and is equal to
\begin{equation}
\label{dbl_rho_opt_1}
  \sigma^2 = \frac{\Omega_0}{2\sqrt{2}\Omega_q}\left(\frac{1+2s^2}{3}\right)^{-3/4} .
\end{equation}
In both cases~\eqref{dbl_rho_sub_1} and \eqref{dbl_rho_opt_1}, we should minimize the expression
\begin{equation}
  \frac{(1+2s^2)^{-3/4}}{\Omega_q} = \left[
    \Omega_q\left(1+2\frac{\Omega_0^2}{\Omega_q^2}\,\xi^2_{\mathrm{tech}}\right)^{3/4}
  \right]^{-1}
\end{equation}
in $\Omega_q$. It is easy to see that the optimal $\Omega_q$ is equal to
\begin{equation}
  \Omega_q = \Omega_0\xi_{\mathrm{tech}} \,,
\end{equation}
which gives the SNR~\eqref{eq:dbl_rho2_opt} for the optimal case~\eqref{dbl_optimal} and \eqref{eq:dbl_rho2_sub} -- for the sub-optimal one $\Lambda=0$.

\newpage
\bibliography{refs}

\end{document}